\documentclass{article}%
\usepackage{hyperref}
\usepackage{amsmath}
\usepackage{amsfonts}
\usepackage{amssymb}
\usepackage{graphicx}%
\providecommand{\U}[1]{\protect\rule{.1in}{.1in}}
\newtheorem{theorem}{Theorem}

\newtheorem{lemma}[theorem]{Lemma}

\newtheorem{proposition}[theorem]{Proposition}

\begin{document}

\title{Finite Field-Dependent BRST-antiBRST Transformations: Jacobians and
Application to the Standard Model}
\author{\textsc{Pavel Yu. Moshin${}^{a}$\thanks{moshin@rambler.ru \hspace{0.5cm}
${}^{\dagger}$reshet@ispms.tsc.ru}\ \ and Alexander A. Reshetnyak${}%
^{b\dagger}$}\\\textit{${}^{a}$Tomsk State University, Department of Physics, 634050, Tomsk,
Russia,}\\\textit{${}^{b}$Institute of Strength Physics and Materials Science}\\\textit{Siberian Branch of Russian Academy of Sciences, 634021, Tomsk,
Russia,}}
\maketitle

\begin{abstract}
We continue our research Nucl.Phys B888, 92 (2014); Int. J. Mod. Phys. A29,
1450159 (2014); Phys. Lett. B739, 110 (2014); Int. J. Mod. Phys. A30, 1550021
(2015) and extend the class of finite BRST-antiBRST transformations with
odd-valued parameters $\lambda_{a}$, $a=1,2$, introduced in these works. In
doing so, we evaluate the Jacobians induced by finite BRST-antiBRST
transformations linear in functionally-dependent parameters, as well as those
induced by finite BRST-antiBRST transformations with arbitrary functional
parameters. The calculations cover the cases of gauge theories with a closed
algebra, dynamical systems with first-class constraints, and general gauge
theories. The resulting Jacobians in the case of linearized transformations
are different from those in the case of polynomial dependence on the
parameters. Finite BRST-antiBRST transformations with arbitrary parameters
induce an extra contribution to the quantum action, which cannot be absorbed
into a change of the gauge. These transformations include an extended case of
functionally-dependent parameters that implies a modified compensation
equation, which admits non-trivial solutions leading to a Jacobian equal to
unity. Finite BRST-antiBRST transformations with functionally-dependent
parameters are applied to the Standard Model, and an explicit form of
functionally-dependent parameters $\lambda_{a}$ is obtained, providing the
equivalence of path integrals in any $3$-parameter $R_{\boldsymbol{\xi}}$-like
gauges. The Gribov--Zwanziger theory is extended to the case of the Standard
Model, and a form of the Gribov horizon functional is suggested in the Landau
gauge, as well as in $R_{\boldsymbol{\xi}}$-like gauges, in a
gauge-independent way using field-dependent BRST-antiBRST transformations, and
in $R_{\boldsymbol{\xi}}$-like gauges using transverse-like non-Abelian gauge fields.

\end{abstract}


\section{Introduction}

Recently, in the articles \cite{MRnew,MRnew1,MRnew2, MRnew3}, we have proposed
an extension of BRST-antiBRST transformations
\cite{aBRST1,aBRST2,aBRST3,aBRST4} to the case of finite (both global and
field-dependent) parameters for Yang--Mills and general gauge theories in the
framework of the generalized Hamiltonian \cite{BLT1h,BLT2h} -- see also
\cite{GH1} -- and Lagrangian \cite{BLT1,BLT2,Hull} BRST-antiBRST quantization
schemes. The idea of \textquotedblleft finiteness\textquotedblright%
\ incorporates into finite transformations a new term being quadratic in the
transformation parameters $\lambda_{a}$, thereby lifting BRST-antiBRST
transformations from the algebraic level to the group level, which has been
discussed also in \cite{BLThf, BLTlf}. BRST transformations \cite{BRST1,
BRST2, BRST3} in both the Lagrangian \cite{deWittH, BV} and generalized
Hamiltonian \cite{BRST3, BFV, Henneaux1} quantization schemes -- described by
a single odd-valued parameter $\mu$ and trivially lifted from the algebraic
form $\delta\phi^{A}=\phi^{A}\overleftarrow{s}\mu$ to the finite (group) form
$\Delta\phi^{A}=\phi^{A}[1-\exp(\overleftarrow{s}\mu)]$, with $\overleftarrow
{s}{}^{2}=0$, in view of the nilpotency property $\mu^{2}=0$ -- have first
been suggested in Yang--Mills theories for field-dependent parameters in
\cite{JM, RM}; see also \cite{Upadhyay1, Upadhyay3}. The introduction of such
transformations is based on a functional equation for the infinitesimal
parameter, providing the invariance of the integrand of the vacuum functional
(in the path integral representation based on the Faddeev--Popov rules
\cite{FP}) under a change of variables $\phi^{A}\rightarrow\phi^{\prime
A}\left(  \phi;\mu(\phi)\right)  $ which preserves the quantum action in
gauges described by different gauge Fermions $\psi(\phi)$ and $(\psi
+\Delta\psi)(\phi)$, related by the given change. The problem of finding a
relation between the different forms of the Faddeev--Popov quantum action in
different gauges, expressed by an exact solution $\mu=\mu(\phi;\Delta\psi)$ to
a functional equation for a finite field-dependent odd-valued parameter (which
ensures the preservation of the integrand) has been solved for the Yang--Mills
theory in the article \cite{LL1}. The respective problem for constrained
dynamical systems has been solved in \cite{BLThfbrst}, and for general gauge
theories, in \cite{Reshetnyak,BLTfin}, on the basis of finding the Jacobian of
a change of variables induced by the respective field-dependent BRST transformations.

As we return to the approach of \cite{MRnew,MRnew1,MRnew2, MRnew3} -- reviewed
and extended in \cite{MRnew4, MRnew5} -- we notice, in the first place, that
it allows one to realize the complete BRST-antiBRST invariance of the
integrand in the vacuum functional. The functionally-dependent parameters
$\lambda_{a}=s_{a}\Lambda$, induced by an even-valued functional $\Lambda$ and
by an $\mathrm{Sp}(2)$-doublet of BRST-antiBRST generators $s_{a}$, provide an
explicit correspondence (due to the compensation equation for the
corresponding Jacobian) between a choice of $\Lambda$ and a transition from
the vacuum functional of a given theory in a certain gauge induced by a gauge
Boson $F_{0}$ to the same theory in a different gauge induced by another gauge
Boson $F$. This becomes a key instrument of a BRST-antiBRST approach that
allows one to consistently examine the notion of \textquotedblleft soft
BRST-antiBRST symmetry breaking\textquotedblright\ \cite{MRnew3}, extending
the concept of \textquotedblleft soft BRST symmetry breaking\textquotedblright%
\ \cite{llr1, lrr, rl} in the framework of Lagrangian BRST quantization
\cite{BV}, which implies an extension of the quantum action given by the
Lagrangian BRST-antiBRST recipe \cite{BLT1,BLT2,Hull} by a BRST-antiBRST
non-invariant term, which is then employed in the concept of effective average
action in the \emph{functional renormalization group} approach
\cite{Wett-Reu-1, Wett-Reu-2, Wett-1, Polch} in \cite{Reshetnyak,LS}, as well
as the interacting Fermi systems \cite{Salmhofer} and in the elimination of
residual gauge invariance in the deep IR region, known as Gribov copies
\cite{Gribov}. Finite field-dependent BRST and BRST-antiBRST transformations,
respectively, in soft BRST and BRST-antiBRST symmetry breaking allow one to
solve the consistency problem for the Lagrangian quantization methods from the
viewpoint of gauge-independence for the conventional $S$-matrix in non-Abelian
gauge theories, namely, in determining a BRST(-antiBRST) non-invariant
addition to the corresponding quantum action -- known as the Gribov horizon
functional \cite{Gribov} which is initially given by the Landau gauge in the
Gribov--Zwanziger theory \cite{Zwanziger1, Zwanziger2} -- by using any other
gauge, including the one-parameter $R_{\xi}$-gauges in the BRST
\cite{Reshetnyak,LL2,Reshetnyak2} and BRST-antiBRST \cite{MRnew} settings.

In the case of finite BRST-antiBRST transformations, the
functionally-dependent parameters $\lambda_{a}$ chosen as solutions to the
compensation equation, relating the Jacobian to a finite change of the gauge
condition, turn out to establish a coincidence of vacuum functionals also in
first-class constraint dynamical systems in different gauges. This has been
shown explicitly in the case of Yang--Mills theories, thereby providing the
unitarity of the conventional $S$-matrix in Lagrangian formalism within
different gauges \cite{MRnew1}. At the same time, we have examined
\cite{MRnew2} the Freedman--Townsend model \cite{FRTnsend}, being the case of
a first-stage reducible gauge theory (of a non-Abelian antisymmetric tensor
field), in the path integral representation, starting from a reference frame
with a certain gauge Boson $F_{0}$, and reaching the same integrand, by using
finite field-dependent BRST-antiBRST transformations, in a different reference
frame with another gauge Boson $F$, depending on $3$ gauge parameters.

It should be noted that we have so far examined the finite field-dependent
BRST-antiBRST transformations with functionally-dependent parameters of the
form $\lambda_{a}=s_{a}\Lambda$. However, in the conclusions of \cite{MRnew,
MRnew3} we have announced that an interesting problem, left outside the scope
of \cite{MRnew, MRnew3}, is the evaluation of Jacobians for finite
field-dependent BRST-antiBRST transformations with a functionally-independent
$\mathrm{Sp}(2)$-doublet of arbitrary odd-valued parameters $\lambda_{a}$,
i.e., not being induced by any even-valued functional $\Lambda$, $\lambda
_{a}\not \equiv s_{a}\Lambda$. Such Jacobians have not been found explicitly
in \cite{BLThf, BLTlf} by using solutions of the equations involved. Another
interesting task is the evaluation of Jacobians of linearized transformations,
i.e., those without the term being quadratic in the parameters $\lambda_{a}$.
It then appears to be important to apply the results involving the study of
finite BRST(-antiBRST) field-dependent transformations to the realistic
physical model being an example of the Yang--Mills theory interacting with
scalar and spinor matter fields and known as the Standard Model
\cite{SM1,SM2,SM3} -- see also \cite{book1,book2,book3,book4} -- which
describes the known spectrum of the elementary particles corresponding to the
three fundamental interactions: electromagnetic, weak and strong, whose
cornerstone, the Higgs Boson \cite{BE,Higgs1,Higgs2,Gural'nik}, has been
discovered \cite{CMS, Atlas} at the LHC in July 2012, with the present
estimation \cite{CMSATlas} of its mass being $m_{\mathrm{H}}=(125,09\pm
0,24)\mathrm{GeV}$.

Based on the above reasons, we examine the following problems related to gauge
theories in the Lagrangian and generalized Hamiltonian descriptions:

\begin{enumerate}
\item evaluation of the Jacobian for a change of variables in the vacuum
functional corresponding to \emph{linearized finite field-dependent
BRST-antiBRST transformations} in Yang--Mills theories and first-class
constraint dynamical systems;

\item evaluation of the Jacobian for a change of variables in the vacuum
functional corresponding to \emph{finite field-dependent BRST-antiBRST
transformations} with arbitrary functional parameters, $\lambda_{a}(\phi)$,
$s^{2}\lambda_{a}(\phi)\not \equiv 0$, in Yang--Mills theories, first-class
constraint dynamical systems, and general gauge theories, and investigation of
its influence on the structure of the quantum action;

\item construction of the parameters $\lambda_{a}$ of finite field-dependent
BRST-antiBRST transformations in the Lagrangian action of the Standard Model,
which generates a change of the gauge in the path integral within a class of
linear $3$-parameter $R_{\boldsymbol{\xi}}$-like gauges, realized in terms of
an even-valued gauge functionals $F_{\boldsymbol{\xi}}$, with $(\xi_{1}%
,\xi_{2},\xi_{3})=\mathbf{0},\mathbf{1}$, corresponding to the Landau and
Feynman (covariant) gauges, respectively;

\item construction of the Gribov--Zwanziger theory for the BRST-antiBRST
Lagrangian formulation of the Standard Model, including the horizon functional
$h_{\boldsymbol{\xi}}$ in arbitrary $R_{\boldsymbol{\xi}}$-like gauges by
means of finite field-dependent BRST-antiBRST transformations, starting from
the BRST-antiBRST non-invariant functional $h_{\boldsymbol{0}}$ given in the
Landau gauge and realized in terms of the even-valued functional
$F_{\boldsymbol{0}}$.

\item construction of an horizon functional $h_{\boldsymbol{\xi}}^{T}$ for the
Standard Model in the Gribov--Zwanziger theory with arbitrary
$R_{\boldsymbol{\xi}}$-like gauges by means of a Hermitian extension of the
corresponding Faddeev--Popov operator (or, equivalently, in terms of
transverse-like components of non-Abelian gauge fields), following the recipe
of \cite{LRquarks2012} for a Yang--Mills theory with an $SU(N)$ gauge group.
\end{enumerate}

The work is organized as follows. In Section~\ref{finitesec}, we give an
overview of the ingredients of finite field-dependent BRST-antiBRST
transformations \cite{MRnew,MRnew1,MRnew2, MRnew3} in theories with a closed
gauge algebra, as well as in first-class constraint dynamical systems and
general gauge theories. In Section~\ref{linear}, we consider an evaluation of
the Jacobian for a change of variables in the vacuum functional given by
{finite field-dependent BRST-antiBRST transformations being \emph{linear} in
functionally-dependent parameters }$\lambda_{a}={{{s}}}_{a}\Lambda$ for
Yang--Mills theories and first-class constraint dynamical systems in
generalized Hamiltonian formalism. {In Section~\ref{arbitrary}, we consider an
evaluation of the Jacobian for a change of variables in the vacuum functional
given by {finite field-dependent BRST-antiBRST transformations with arbitrary
parameters }}$\lambda_{a}${{. In Section~\ref{AppA}, we consider an
application of finite BRST-antiBRST transformations to the Standard Model. In
Appendices~\ref{AppB} and \ref{AppC}, we examine the respective details of
calculations for linearized finite {BRST-antiBRST transformations with
functionally-dependent parameters}, as well as for finite BRST-antiBRST
transformations with {arbitrary parameters}. In Discussion, we suggest another
form of the Gribov horizon functional in the covariant gauge and make
concluding remarks. }}As a rule, we use the conventions of our previous works
\cite{MRnew,MRnew1,MRnew2, MRnew3} and the generally accepted definition
\cite{SlavnovFI} of functional integrals for quasi-Gaussian functionals, which
is well justified in perturbation theory, see, e.g., \cite{GitmanTyutin}.
Notice that Sections \ref{linear}, \ref{arbitrary} do not need to use the
operation of functional integration in itself, but only the definition of a
functional Jacobian. Unless otherwise specified, derivatives with respect to
the fields are taken from the right, and those with respect to the
corresponding sources and antifields are taken from the left. Left-handed
derivatives with respect to the fields are labelled by the subscript
\textquotedblleft$l$\textquotedblright, whereas right-handed derivatives with
respect to the antifields are labelled by the subscript \textquotedblleft%
$r$\textquotedblright. Derivatives with respect to the phase-space variables
and the variables of the triplectic manifold are understood as taken from the
right. Depending on the convenience, we use two forms of notation for the
BRST-antiBRST generators: $s_{a}$ and $\overleftarrow{s}_{a}$, which are
related by $s_{a}A\equiv A\overleftarrow{s}_{a}$, where $A$ is an arbitrary
functional. The raising and lowering of $\mathrm{Sp}\left(  2\right)  $
indices, $s^{a}=\varepsilon^{ab}s_{b}$, $s_{a}=\varepsilon_{ab}s^{b}$, is
carried out with the help of a constant antisymmetric tensor $\varepsilon
^{ab}$, $\varepsilon^{ac}\varepsilon_{cb}=\delta_{b}^{a}$, subject to the
normalization condition $\varepsilon^{12}=1$. The Grassmann parity of any
homogeneous quantity $B$ is denoted by $\varepsilon(B)$.

\section{Finite BRST-antiBRST Transformations}

\label{finitesec}
\renewcommand{\theequation}{\arabic{section}.\arabic{equation}} \setcounter{equation}{0}

In this section, we examine the case of finite BRST-antiBRST transformations
realized in different spaces of quantum field theory: the configuration space
of Yang--Mills theories, the phase space of arbitrary dynamical systems with
first-class constraints, and the triplectic space of general gauge theories in
Lagrangian formalism.

\subsection{Yang--Mills Theories in Lagrangian Formalism}

\label{YMgen}

The generating functional of Green's functions corresponding to irreducible
gauge theories with a closed algebra in BRST-antiBRST Lagrangian quantization
\cite{BLT1,BLT2} is given by
\begin{equation}
Z(J)=\int d\phi\ \exp\left\{  \frac{i}{\hbar}\left[  S_{F}\left(  \phi\right)
+J_{A}\phi^{A}\right]  \right\}  \equiv\int\mathcal{I}_{\phi}^{F}\exp\left(
\frac{i}{\hbar}J_{A}\phi^{A}\right)  \,. \label{z(j)}%
\end{equation}
Here, $\hbar$ is the Planck constant, whereas the quantum action $S_{F}\left(
\phi\right)  $,
\begin{equation}
S_{F}\left(  \phi\right)  =S_{0}\left(  A\right)  +F_{,A}Y^{A}-\left(
1/2\right)  \varepsilon_{ab}X^{Aa}F_{,AB}X^{Bb}\ , \label{action}%
\end{equation}
the classical action $S_{0}\left(  A\right)  $, the (admissible) even-valued
gauge-fixing functional $F\left(  \phi\right)  $, and the
functions\footnote{By functions we understand those of the space-time
coordinates.} $X^{Aa}\left(  \phi\right)  $, $Y^{A}\left(  \phi\right)  $ are
defined in the configuration space $\mathcal{M}_{\phi}=\{\phi^{A}%
\}=\{A^{i},C^{\alpha a},B^{\alpha}\}$, parameterized by the initial classical
fields $A^{i}$, $i=1,\ldots,n$, the Nakanishi--Lautrup fields $B^{\alpha}$,
$\alpha=1,\ldots,m<n$, and the ghost-antighost fields $C^{\alpha a}$,
organized in $\mathrm{Sp}\left(  2\right)  $-doublets with the identification
$C^{\alpha a}=(C^{\alpha1},C^{\alpha2})\equiv(C^{\alpha},\bar{C}^{\alpha})$.
The Grassmann parity is given by%
\begin{equation}
\varepsilon(\phi^{A})=\varepsilon(A^{i},B^{\alpha},C^{\alpha a})=(\varepsilon
_{i},\varepsilon_{\alpha},\varepsilon_{\alpha}+1)=\varepsilon_{A}\ .
\label{grasspar}%
\end{equation}
The classical action $S_{0}\left(  A\right)  $ is invariant with respect to
the infinitesimal gauge transformations
\begin{equation}
\delta A^{i}=R_{\alpha}^{i}(A)\xi^{\alpha}\Longrightarrow S_{0,i}(A)R_{\alpha
}^{i}(A)=0\ , \label{Nid}%
\end{equation}
with $R_{\alpha}^{i}(A)$ being the generators of the gauge transformations,
$\varepsilon(R_{\alpha}^{i})=\varepsilon_{i}+\varepsilon_{\alpha}$, and
$\xi^{\alpha}$ being arbitrary functions. The generators $R_{\alpha}^{i}(A)$
form a closed gauge algebra, with structure constants $F_{\alpha\beta}%
^{\gamma}\left(  A\right)  =\mathrm{const}$ and vanishing quantities
$M_{\alpha\beta}^{ij}\left(  A\right)  $ in the general relations \cite{BV}%
\begin{align}
&  R_{\alpha,j}^{i}(A)R_{\beta}^{j}(A)-\left(  -1\right)  ^{\varepsilon
_{\alpha}\varepsilon_{\beta}}R_{\beta,j}^{i}(A)R_{\alpha}^{j}(A)=-R_{\gamma
}^{i}(A)F_{\alpha\beta}^{\gamma}\left(  A\right)  -S_{0,j}(A)M_{\alpha\beta
}^{ij}\left(  A\right)  \ ,\nonumber\\
&  F_{\alpha\beta}^{\gamma}=-\left(  -1\right)  ^{\varepsilon_{\alpha
}\varepsilon_{\beta}}F_{\beta\alpha}^{\gamma}\ ,\ \ \ M_{\alpha\beta}%
^{ij}=-\left(  -1\right)  ^{\varepsilon_{i}\varepsilon_{j}}M_{\alpha\beta
}^{ji}=-\left(  -1\right)  ^{\varepsilon_{\alpha}\varepsilon_{\beta}}%
M_{\beta\alpha}^{ij}\,. \label{gauge_alg}%
\end{align}
In a first-rank gauge theory with a closed algebra, the functions
$X^{Aa}\left(  \phi\right)  $, $Y^{A}\left(  \phi\right)  $ in (\ref{action})
are given by \cite{BLT1}
\begin{align}
&  X^{Aa}=\left(  X_{1}^{ia},X_{2}^{\alpha a},X_{3}^{\alpha ab}\right)  \ , &
&  Y^{A}=\left(  Y_{1}^{i},Y_{2}^{\alpha},Y_{3}^{\alpha a}\right)
\ ,\label{solxy}\\
&  X_{1}^{ia}=R_{\alpha}^{i}C^{\alpha a}\ , &  &  X_{2}^{\alpha a}=-\frac
{1}{2}F_{\gamma\beta}^{\alpha}B^{\beta}C^{\gamma a}-\frac{1}{12}\left(
-1\right)  ^{\varepsilon_{\beta}}\left(  2F_{\gamma\beta,j}^{\alpha}R_{\rho
}^{j}+F_{\gamma\sigma}^{\alpha}F_{\beta\rho}^{\sigma}\right)  C^{\rho
b}C^{\beta a}C^{\gamma c}\varepsilon_{cb}\ ,\nonumber\\
&  X_{3}^{\alpha ab}=-\varepsilon^{ab}B^{\alpha}-\frac{1}{2}\left(  -1\right)
^{\varepsilon_{\beta}}F_{\beta\gamma}^{\alpha}C^{\gamma b}C^{\beta a}\ , &  &
Y_{1}^{i}=R_{\alpha}^{i}B^{\alpha}+\frac{1}{2}\left(  -1\right)
^{\varepsilon_{\alpha}}R_{\alpha,j}^{i}R_{\beta}^{j}C^{\beta b}C^{\alpha
a}\varepsilon_{ab}\ ,\nonumber\\
&  Y_{2}^{\alpha}=0\ , &  &  Y_{3}^{\alpha a}=-2X_{3}^{\alpha a} \label{xy}%
\end{align}
and in Yang--Mills theories they assume the following representation:%
\begin{align}
&  X_{1}^{\mu ma}=D^{\mu mn}C^{na}\ , &  &  Y_{1}^{\mu m}=D^{\mu mn}%
B^{n}+\frac{1}{2}f^{mnl}C^{la}D^{\mu nk}C^{kb}\varepsilon_{ba}\ ,\nonumber\\
&  X_{2}^{ma}=-\frac{1}{2}f^{mnl}B^{l}C^{na}-\frac{1}{12}f^{mnl}f^{lrs}%
C^{sb}C^{ra}C^{nc}\varepsilon_{cb}\ , &  &  Y_{2}^{m}=0\ ,\label{xyYM}\\
&  X_{3}^{mab}=-\varepsilon^{ab}B^{m}-\frac{1}{2}f^{mnl}C^{lb}C^{na}\ , &  &
Y_{3}^{ma}=f^{mnl}B^{l}C^{na}+\frac{1}{6}f^{mnl}f^{lrs}C^{sb}C^{ra}%
C^{nc}\varepsilon_{cb}\ ,\nonumber
\end{align}
corresponding to the generators $R_{\alpha}^{i}$ and structure functions
$F_{\alpha\beta}^{\gamma}$,
\begin{equation}
R_{\mu}^{mn}(x;y)=D_{\mu}^{mn}(x)\delta(x-y)\,,\ \ \ D_{\mu}^{mn}=\delta
^{mn}\partial_{\mu}+f^{mln}A_{\mu}^{l}\,,\ \ \ F_{\alpha\beta}^{\gamma
}=f^{lmn}\delta(x-z)\delta(y-z)\,, \label{R(A)}%
\end{equation}
written down in terms of a covariant derivative $D_{\mu}^{mn}$ and completely
antisymmetric structure constants $f^{lmn}$ related to a compact subalgebra of
an $su(N)$ Lie algebra.

The quantum action $S_{F}$, the integration measure $d\phi$, and thereby also
the integrand $\mathcal{I}_{\phi}^{F}$, are invariant under BRST-antiBRST
transformations, which are infinitesimal transformations with an
$\mathrm{Sp}\left(  2\right)  $-doublet of constant odd-valued parameters
$\mu_{a}$,
\begin{equation}
\delta\phi^{A}=\left(  s^{a}\phi^{A}\right)  \mu_{a}\ ,\ \ \ s^{a}\phi
^{A}=X^{Aa}\ , \label{brst-antibrst}%
\end{equation}
where $s^{a}$ are the generators of BRST-antiBRST transformations. Starting
from this point, the invariance of the integrand $\mathcal{I}_{\phi}^{F}$ in
the case of finite constant values of the corresponding anticommuting
parameters $\lambda_{a}$ is achieved by solving the equation $G(\phi
+\Delta\phi)=G(\phi)$ for an arbitrary regular functional $G(\phi)$ subject to
BRST-antiBRST invariance, $s^{a}G=0$. This solution has the form of a finite
(polynomial in $\lambda_{a}$) BRST-antiBRST transformation \cite{MRnew}%
\begin{equation}
\Delta\phi^{A}=\phi^{A}+X^{Aa}\lambda_{a}-\frac{1}{2}Y^{A}\lambda^{2}=\left(
s^{a}\phi^{A}\right)  \lambda_{a}+\frac{1}{4}\left(  s^{2}\phi^{A}\right)
\lambda^{2} \label{finite}%
\end{equation}
and implies that a finite variation $\Delta\phi^{A}$ includes the generators
of BRST-antiBRST transformations $\left(  s^{1},s^{2}\right)  $, as well as
their commutator $s^{2}=\varepsilon_{ab}s^{b}s^{a}=s^{1}s^{2}-s^{2}s^{1}$,
being the generator of mixed BRST-antiBRST transformations. Equivalently,
(\ref{finite}) can be represented as a group transformation in the
configuration space $\mathcal{M}_{\phi}$,%
\begin{equation}
\phi^{A}\rightarrow\phi^{\prime A}=\phi^{A}\left(  1+\overleftarrow{s}%
^{a}\lambda_{a}+\frac{1}{4}\overleftarrow{s}^{2}\lambda^{2}\right)  =\phi
^{A}\exp\left(  \overleftarrow{s}^{a}\lambda_{a}\right)  \ ,\mathrm{\ \ \ }%
s^{a}\phi^{A}=\phi^{A}\overleftarrow{s}{}^{a},\ \ \ \overleftarrow{s}%
^{2}\equiv\overleftarrow{s}^{a}\overleftarrow{s}_{a}={s}^{2}\equiv{s}_{a}%
{s}^{a}\,, \label{finite1}%
\end{equation}
where the set of elements $\{g(\lambda)\}=\{\exp\left(  \overleftarrow{s}%
^{a}\lambda_{a}\right)  \}$ forms an Abelian two-parameter supergroup with
odd-valued generating elements $\lambda_{a}$. This circumstance can also be
justified by the Frobenius theorem, which deals with an implementation of
anticommuting generators $\overleftarrow{s}^{a}$ in terms of vector fields
$\overleftarrow{s}^{a}\left(  \Gamma\right)  =\frac{\overleftarrow{\delta}%
}{\delta\Gamma^{p}}(\Gamma^{p}\overleftarrow{s}^{a})$ in a certain
configuration space $\mathcal{M}_{\Gamma}$. The BRST-antiBRST invariance of
$\mathcal{I}_{\phi}^{F}$ implies the relation%
\begin{equation}
\mathcal{I}_{\phi g(\lambda)}^{F}=\mathcal{I}_{\phi}^{F}. \label{finbab1}%
\end{equation}
which can be established by the fact \cite{MRnew} that the global finite
transformations (corresponding to $\lambda_{a}=\mathrm{const}$) respect the
integration measure:%
\begin{equation}
\mathrm{Sdet}\left(  \frac{\delta\phi^{\prime}}{\delta\phi}\right)
=1\Longrightarrow d\phi^{\prime}=d\phi\ . \label{constJ}%
\end{equation}
For finite field-dependent transformations, it has been established
\cite{MRnew} that in the particular case of functionally-dependent parameters
$\lambda_{a}=\Lambda\overleftarrow{s}_{a}$, $s^{1}\lambda_{1}+s^{2}\lambda
_{2}=-s^{2}\Lambda$, with a certain even-valued potential, $\Lambda
=\Lambda\left(  \phi\right)  $, whose introduction has been inspired by
infinitesimal field-dependent BRST-antiBRST transformations induced by the
parameters \cite{BLT1}%
\begin{equation}
\mu_{a}=\frac{i}{2\hbar}\varepsilon_{ab}\left(  \Delta F\right)  _{,A}%
X^{Ab}=\frac{i}{2\hbar}\left(  s_{a}\Delta F\right)  \,, \label{partic_case}%
\end{equation}
the vacuum functional $Z_{F}(0)$ is gauge-independent: $Z_{F+\Delta
F}(0)=Z_{F}(0)$. Namely, in the case of finite {field-dependent}
transformations with group-like elements $g(\Lambda\overleftarrow{s}_{a})$
whose set forms a nonlinear non-Abelian group-like structure\footnote{For
BRST-antiBRST-closed (in particular, BRST-antiBRST-exact) functional
parameters $\lambda_{a}(\phi)=\Lambda_{a}(\phi)\overleftarrow{s}^{2}$ with
odd-valued functionals $\Lambda_{a}(\phi)$, the subset $g(\Lambda
_{a}\overleftarrow{s}^{2})$ forms an Abelian subgroup in $g(\Lambda
\overleftarrow{s}_{a})$ and thereby in $g(\lambda_{a}(\phi))$. Indeed, the
choice $\Lambda=2s^{a}\Lambda_{a}$, in view of $\lambda_{b}(\phi
)\overleftarrow{s}^{a}=0$ provided by $\overleftarrow{s}^{a}\overleftarrow
{s}^{b}\overleftarrow{s}^{c}\equiv0$, implies that $g(\Lambda_{a}%
^{1}\overleftarrow{s}^{2})g(\Lambda_{a}^{2}\overleftarrow{s}^{2}%
)=g(\Lambda_{a}^{2}\overleftarrow{s}^{2})g(\Lambda_{a}^{1}\overleftarrow
{s}^{2})$ and $g(\Lambda_{a}^{1}\overleftarrow{s}^{2})g(-\Lambda_{a}%
^{1}\overleftarrow{s}^{2})=1$, for any odd-valued functionals $\Lambda_{a}%
^{i}\ ,$ $i=1,2$, with the unit element \textquotedblleft$1$\textquotedblright%
.\label{footnote-label}} the superdeterminant of a change of variables is
given by%
\begin{align}
&  \mathrm{Sdet}\left[  \frac{\delta(\phi g(\Lambda\overleftarrow{s}_{a}%
)}{\delta\phi}\right]  =\exp\left[  \Im\left(  \phi\right)  \right]
\ ,\ \ \ \mathrm{where}\ \ \ \Im\left(  \phi\right)  =-2\mathrm{\ln}\left[
1-\frac{1}{2}s^{2}\Lambda\left(  \phi\right)  \right]  \ ,\label{superJaux}\\
&  d\phi^{\prime}=d\phi\ \exp\left[  \frac{i}{\hbar}\left(  -i\hbar\Im\right)
\right]  =d\phi\ \exp\left\{  \frac{i}{\hbar}\left[  i\hbar\,\mathrm{\ln
}\left(  1-\frac{1}{2}s^{2}\Lambda\right)  ^{2}\right]  \right\}  \ .
\label{superJ1}%
\end{align}
The invariance of the quantum action $S_{F}\left(  \phi\right)  $ with respect
to (\ref{finite}) implies that the change $\phi^{A}\rightarrow\phi^{\prime
A}=\phi^{A}g(\lambda(\phi))$ induces in (\ref{z(j)}) the following
transformation of the integrand $\mathcal{I}_{\phi}^{F}$ :%
\begin{equation}
\mathcal{I}_{\phi g(\lambda(\phi))}^{F}=d\phi\ \exp\left[  \Im\left(
\phi\right)  \right]  \exp\left[  \left(  i/\hbar\right)  S_{F}\left(
\phi\left(  g\lambda\left(  \phi\right)  \right)  \right)  \right]
=d\phi\ \exp\left\{  \left(  i/\hbar\right)  \left[  S_{F}\left(  \phi\right)
-i\hbar\Im\left(  \phi\right)  \right]  \right\}  \ , \label{superJ2}%
\end{equation}
whence
\begin{equation}
\mathcal{I}_{\phi g(\lambda(\phi))}^{F}=d\phi\ \exp\left\{  \left(
i/\hbar\right)  \left[  S_{F}\left(  \phi\right)  +i\hbar\ \mathrm{\ln}\left(
1-\Lambda\overleftarrow{s}{}^{2}/2\right)  ^{2}\right]  \right\}  \ .
\label{superJ3}%
\end{equation}
Next, due to the explicit form of the initial quantum action $S_{F}%
=S_{0}-\left(  1/2\right)  F\overleftarrow{s}^{2}$, the BRST-antiBRST-exact
contribution $i\hbar\,\mathrm{\ln}\left(  1+s^{a}s_{a}\Lambda/2\right)  ^{2}$
to the quantum action $S_{F}$ can be interpreted as a change of the
gauge-fixing functional made in the original integrand $\mathcal{I}_{\phi}%
^{F}$,%
\begin{align}
&  i\hbar\ \mathrm{\ln}\left(  1+s^{a}s_{a}\Lambda/2\right)  ^{2}=s^{a}%
s_{a}\left(  \Delta F/2\right) \label{superJ3m}\\
&  \Longrightarrow\mathcal{I}_{\phi g(\lambda(\phi))}^{F}=d\phi\ \exp\left\{
\left(  i/\hbar\right)  \left[  S_{0}+\left(  1/2\right)  s^{a}s_{a}\left(
F+\Delta F\right)  \right]  \right\}  =\mathcal{I}_{\phi}^{F+\Delta F},
\label{superJ41}%
\end{align}
with a certain $\Delta F\left(  \phi|\Lambda\right)  $, whose correspondence
to $\Lambda\left(  \phi\right)  $ is established by the relation
(\ref{superJ3m}), which is also known as the compensation equation for an
unknown parameter $\Lambda(\phi)$ and which thereby provides the
gauge-independence of the vacuum functional, $Z_{F}(0)=Z_{F+\Delta F}(0)$. An
explicit solution of (\ref{superJ3m}), satisfying the solvability condition
due to the BRST-antiBRST-exactness of both sides (up to BRST-antiBRST-exact
terms), is given by%
\begin{equation}
\Lambda\left(  \phi|\Delta F\right)  =2\Delta F\left(  s^{a}s_{a}\Delta
F\right)  ^{-1}\left[  \exp\left(  \frac{1}{4i\hbar}s^{b}s_{b}\Delta F\right)
-1\right]  =\frac{1}{2i\hbar}\Delta F\sum_{n=0}^{\infty}\frac{1}{\left(
n+1\right)  !}\left(  \frac{1}{4i\hbar}s^{a}s_{a}\Delta F\right)  ^{n}\ .
\label{Lambda-Fsol1}%
\end{equation}
Conversely, having considered the equation (\ref{superJ3m}) for an unknown
$\Delta F$ with a given $\Lambda$, we obtain%
\begin{equation}
\Delta F\left(  \phi\right)  =4i\hbar\ \Lambda\left(  \phi\right)  \left(
s^{a}s_{a}\Lambda\left(  \phi\right)  \right)  ^{-1}\ln\left(  1+s^{a}%
s_{a}\Lambda\left(  \phi\right)  /2\right)  \ , \label{Lambda-Fsol}%
\end{equation}
and therefore a field-dependent transformation with the parameters
$\lambda_{a}=s_{a}\Lambda$,%
\begin{equation}
\lambda_{a}=\frac{1}{2i\hbar}\left(  \Delta F\overleftarrow{s}_{a}\right)
\sum_{n=0}^{\infty}\frac{1}{\left(  n+1\right)  !}\left(  \frac{1}{4i\hbar
}\Delta F\overleftarrow{s}^{2}\right)  ^{n}\,, \label{lambda-Fsol}%
\end{equation}
amounts to a precise change of the gauge-fixing functional.

In view of (\ref{Lambda-Fsol1}), the property (\ref{superJ41}) implies a
so-called modified Ward identity \cite{MRnew3}, depending on field-dependent
parameters $\lambda_{a}=\Lambda\overleftarrow{s}_{a}$ and thereby also on a
finite change of the gauge:%
\begin{equation}
\left\langle \left\{  1+\frac{i}{\hbar}J_{A}\left[  X^{Aa}\lambda_{a}%
(\Lambda)-\frac{1}{2}Y^{A}\lambda^{2}(\Lambda)\right]  -\frac{1}{4}\left(
\frac{i}{\hbar}\right)  {}^{2}\varepsilon_{ab}J_{A}X^{Aa}J_{B}X^{Bb}%
\lambda^{2}(\Lambda)\right\}  \left(  1-\frac{1}{2}\Lambda\overleftarrow
{s}^{2}\right)  {}^{-2}\right\rangle _{F,J}=1\ . \label{mWIclalg}%
\end{equation}
The property (\ref{superJ41}) also provides a relation which describes the
gauge-dependence of $Z_{F}(J)$ for a finite change $F\rightarrow F+\Delta F$:%
\begin{align}
Z_{F+\Delta F}(J)-Z_{F}(J)  &  =Z_{F}(J)\left\langle \frac{i}{\hbar}%
J_{A}\left[  X^{Aa}\lambda_{a}\left(  \phi|-\Delta{F}\right)  -\frac{1}%
{2}Y^{A}\lambda^{2}\left(  \phi|-\Delta{F}\right)  \right]  \right.
\nonumber\\
&  -\left.  (-1)^{\varepsilon_{B}}\left(  \frac{i}{2\hbar}\right)  ^{2}%
J_{B}J_{A}\left(  X^{Aa}X^{Bb}\right)  \varepsilon_{ab}\lambda^{2}\left(
\phi|-\Delta{F}\right)  \right\rangle _{F,J}. \label{GDInew1}%
\end{align}
In (\ref{mWIclalg}), (\ref{GDInew1}), the symbol \textquotedblleft%
$\langle\mathcal{A}\rangle_{F,J}$\textquotedblright\ for a certain functional
$\mathcal{A}(\phi)$ denotes a source-dependent average expectation value
corresponding to a gauge-fixing functional $F(\phi)$:%
\begin{equation}
\left\langle \mathcal{A}\right\rangle _{F,J}=Z_{F}^{-1}(J)\int d\phi
\ \mathcal{A}\left(  \phi\right)  \exp\left\{  \frac{i}{\hbar}\left[  {S}%
_{F}\left(  \phi\right)  +J_{A}\phi^{A}\right]  \right\}  \ ,\ \ \left\langle
1\right\rangle _{F,J}=1\ . \label{aexv}%
\end{equation}
In the case of constant $\lambda_{a}$, the relation (\ref{mWIclalg}) implies
an $\mathrm{Sp}(2)$-doublet of the usual Ward identities (at the first order
in $\lambda_{a}$) and a derivative identity (at the second order in
$\lambda_{a}$), namely,%
\begin{equation}
J_{A}\left\langle X^{Aa}\right\rangle _{F,J}=0\ ,\ \ \ \left\langle
J_{A}\left[  2Y^{A}+\left(  i/\hbar\right)  \varepsilon_{ab}X^{Aa}J_{B}%
X^{Bb}\right]  \right\rangle _{F,J}=0\ . \label{WIlag3}%
\end{equation}
Below, we intend to study the case of finite field-dependent BRST-antiBRST
transformations for Yang--Mills theories in Lagrangian formalism with
arbitrary functional parameters, generally assumed to be
functionally-independent, $\lambda_{a}\not \equiv s_{a}\Lambda$. It is also
intended to study the case of finite field-dependent BRST-antiBRST
transformations being linear in functionally-dependent parameters of the form
$\lambda_{a}=s_{a}\Lambda$.

\subsection{Dynamical Systems in Generalized Hamiltonian Formalism}

\label{hamgen}

The generating functional of Green's functions for dynamical systems with
first-class constraints has the form \cite{BLT1h, BLT2h}%
\begin{equation}
Z_{\Phi}\left(  I\right)  =\int d\Gamma\exp\left\{  \frac{i}{\hbar}\int
dt\left[  \frac{1}{2}\Gamma^{p}(t)\omega_{pq}\dot{\Gamma}^{q}(t)-H_{\Phi
}(t)+I(t)\Gamma(t)\right]  \right\}  \equiv\int\mathcal{I}_{\Gamma}^{\Phi}%
\exp\left\{  \frac{i}{\hbar}\int dt\,I(t)\Gamma(t)\right\}  \label{ZPI}%
\end{equation}
and determines the vacuum functional $Z_{\Phi}=Z_{\Phi}\left(  0\right)  $ at
the vanishing external sources $I_{p}(t)$ to the phase-space variables
${\Gamma}^{p}(t)$. In (\ref{ZPI}), integration over time is taken over the
range $t_{\mathrm{in}}\leq t\leq t_{\mathrm{out}}$; the functions of time
$\Gamma^{p}(t)\equiv\Gamma_{t}^{p}$ for $t_{\mathrm{in}}\leq t\leq
t_{\mathrm{out}}$\ are trajectories, $\dot{\Gamma}^{p}(t)\equiv d{\Gamma}%
^{p}(t)/dt$; the quantities $\omega_{pq}=(-1)^{(\varepsilon_{p}+1)(\varepsilon
_{q}+1)}\omega_{qp}$ compose an even supermatrix inverse to that with the
elements $\omega^{pq}$; the unitarizing Hamiltonian $H_{\Phi}(t)=H_{\Phi
}(\Gamma(t))$ is determined by four $t$-local functions: an even-valued
function $\mathcal{H}(t)$, with $\mathrm{gh}(\mathcal{H})=0$, an
$\mathrm{Sp}(2)$-doublet of odd-valued functions $\Omega^{a}(t)$, with
$\mathrm{gh}(\Omega^{a})=-(-1)^{a}$, and an even-valued function $\Phi(t)$,
with $\mathrm{gh}(\Phi)=0$, known as the gauge-fixing Boson,
\[
H_{\Phi}(t)=\mathcal{H}(t)+\frac{1}{2}\varepsilon_{ab}\left\{  \left\{
\Phi(t),\Omega^{a}(t)\right\}  _{t},\Omega^{b}(t)\right\}  _{t}\ ,
\]
where the functions $\mathcal{H}$, $\Omega^{a}$ are defined in the phase space
$\mathcal{M}_{\Gamma}$ parameterized by the canonical coordinates $\Gamma
^{p}=\left(  \eta,\Gamma_{\mathrm{gh}}\right)  $, $\varepsilon(\Gamma
^{p})=\varepsilon(I_{p})=\varepsilon_{p}$, and obey the following generating
equations in terms of the Poisson superbracket, $\{\Gamma^{p},\Gamma
^{q}\}=\omega^{pq}=\mathrm{const}$, related to the even supermatrix
$\omega^{pq}$ , with $\omega^{pq}=-(-1)^{\varepsilon_{p}\varepsilon_{q}}%
\omega^{qp}$:
\begin{equation}
\left\{  \Omega^{a}\left(  t\right)  ,\Omega^{b}\left(  t\right)  \right\}
_{t}=0\ ,\ \ \left\{  \mathcal{H}\left(  t\right)  ,\Omega^{b}\left(
t\right)  \right\}  _{t}=0\ , \label{HOmega}%
\end{equation}
with account taken of the rule $\left\{  A(t),B(t)\right\}  _{t}$ = $\left\{
A(\Gamma),B(\Gamma)\right\}  |_{\Gamma=\Gamma(t)}$ for any $A,B$. The
functions $\mathcal{H}$, $\Omega^{a}$ are subject to the boundary conditions%
\begin{equation}
\left.  \mathcal{H}\right\vert _{\Gamma_{\mathrm{gh}}=0}=H_{0}\left(
\eta\right)  \ ,\ \ \ \left.  \frac{\delta\Omega^{a}}{\delta C^{\alpha b}%
}\right\vert _{\Gamma_{\mathrm{gh}}=0}=\delta_{b}^{a}T_{\alpha}\left(
\eta\right)  \ , \label{bcond}%
\end{equation}
where the classical Hamiltonian $H_{0}=H_{0}(\eta)$ and the set of first-class
constraints $T_{\alpha}=T_{\alpha}(\eta)$, $\varepsilon(T_{\alpha
})=\varepsilon_{\alpha}$, of a given dynamical system depend on the classical
phase-space variables $\eta$, with the involution relations%
\begin{equation}
\left\{  H_{0},T_{\alpha}\right\}  =T_{\gamma}V_{\alpha}^{\gamma
}\ ,\ \ \ \left\{  T_{\alpha},T_{\beta}\right\}  =T_{\gamma}U_{\alpha\beta
}^{\gamma}\ ,\ \ \ \mathrm{where}\ \ \ \ U_{\alpha\beta}^{\gamma
}=-(-1)^{\varepsilon_{\alpha}\varepsilon_{\beta}}U_{\beta\alpha}^{\gamma}\ .
\label{invrel}%
\end{equation}
In (\ref{bcond}), the variables $\Gamma_{\mathrm{gh}}$ contain the entire set
of auxiliary variables that correspond to the towers \cite{BFV} of
ghost-antighost coordinates $C$ and Lagrangian multipliers $\pi$, as well as
their respective conjugate momenta $\mathcal{P}$ and $\lambda$, whose
structure depends on the reducibility or irreducibility of a given dynamical
system and is arranged into $\mathrm{Sp}(2)$-symmetric tensors
\cite{BLT1h,BLT2h}.

In virtue of the generating equations (\ref{HOmega}), the integrand with
vanishing sources $\mathcal{I}_{\Gamma}^{\Phi}$ in (\ref{ZPI}) is invariant
under the infinitesimal BRST-antiBRST transformations \cite{BLT1h}
\begin{equation}
\Gamma^{p}\rightarrow\check{\Gamma}^{p}=\Gamma^{p}+\left(  s^{a}\Gamma
^{p}\right)  \mu_{a}\ , \label{BABinf}%
\end{equation}
which are realized on phase-space trajectories $\Gamma^{p}(t)$,%
\begin{equation}
\Gamma^{p}(t)\rightarrow\check{\Gamma}^{p}(t)=\Gamma^{p}(t)+\left\{
\Gamma^{p}(t),\Omega^{a}(t)\right\}  _{t}\mu_{a}=\Gamma^{p}(t)+\left(
s^{a}\Gamma^{p}\right)  \left(  t\right)  \mu_{a}\ , \label{BABinftr}%
\end{equation}
where $\mu_{a}$ form an $\mathrm{Sp}(2)$-doublet of infinitesimal
anticommuting constant parameters, and the generators $s^{a}$ of BRST-antiBRST
transformations, $s^{a}=\left\{  \bullet,\Omega^{a}\right\}  $, are
anticommuting, nilpotent, and obey the Leibnitz rule when acting on the
product and the Poisson superbracket:%
\begin{equation}
s^{a}s^{b}+s^{b}s^{a}=0\ ,\ s^{a}s^{b}s^{c}=0\ ,\ s^{a}\left(  AB\right)
=\left(  s^{a}A\right)  B\left(  -1\right)  ^{\varepsilon_{B}}+A\left(
s^{a}B\right)  \ ,\ s^{a}\left\{  A,B\right\}  =\left\{  s^{a}A,B\right\}
\left(  -1\right)  ^{\varepsilon_{B}}+\left\{  A,s^{a}B\right\}  .
\label{algPbr}%
\end{equation}
The first three relations for $s^{a}$ are also valid in the case of Lagrangian
BRST-antiBRST transformations in Yang--Mills theories. Once again, the
achievement of BRST-antiBRST invariance of $\mathcal{I}_{\Gamma}^{\Phi}$ in
(\ref{ZPI}) with finite constant values of the parameters (now denoted by
$\lambda_{a}$) leads to finite transformations \cite{MRnew1} of the canonical
variables $\Gamma^{p}$,%
\begin{equation}
\Gamma^{p}\rightarrow\check{\Gamma}^{p}=\Gamma^{p}+\Delta\Gamma^{p}%
\ ,\ \ \ \mathrm{where}\ \ \ \Delta\Gamma^{p}=\left(  s^{a}\Gamma^{p}\right)
\lambda_{a}+\frac{1}{4}\left(  s^{2}\Gamma^{p}\right)  \lambda^{2}\ ,
\label{defin}%
\end{equation}
with the same interpretation of both the terms in $\Delta\Gamma^{p}$ as in the
comments that follow the relation (\ref{finite}) of Subsection~\ref{YMgen}. In
particular, the transformations (\ref{defin}) may be represented as group
transformations, defined this time in the phase space $\mathcal{M}_{\Gamma}$
and realized on the canonical coordinates:%
\begin{equation}
\Gamma^{p}\rightarrow\check{\Gamma}^{p}=\Gamma^{p}\left(  1+\overleftarrow
{s}^{a}\lambda_{a}+\frac{1}{4}\overleftarrow{s}^{2}\lambda^{2}\right)
\equiv\Gamma^{p}\exp\left(  \overleftarrow{s}^{a}\lambda_{a}\right)  \,,
\label{finite2}%
\end{equation}
where the operators $\overleftarrow{s}^{a}$ obey the same notation
(\ref{finite1}) that takes place for their Lagrangian counterparts. The set of
elements $\{g(\lambda)\}=\{\exp\left(  \overleftarrow{s}^{a}\lambda
_{a}\right)  \}$ forms an Abelian two-parameter supergroup with odd-valued
generating elements $\lambda_{a}$, acting this time in $\mathcal{M}_{\Gamma}$,
instead of the configuration space $\mathcal{M}$. The transformations
(\ref{defin}) are realized on phase-space trajectories $\Gamma^{p}(t)$ as
follows:%
\begin{equation}
\check{\Gamma}^{p}\left(  t\right)  =\Gamma^{p}\left(  t\right)  \exp\left(
\overleftarrow{s}^{a}\lambda_{a}\right)  \Longleftrightarrow\Delta\Gamma
^{p}\left(  t\right)  =\Gamma^{p}\left(  t\right)  \left[  \exp\left(
\overleftarrow{s}^{a}\lambda_{a}\right)  -1\right]  =\left(  s^{a}\Gamma
^{p}\right)  \left(  t\right)  \lambda_{a}+\frac{1}{4}\left(  s^{2}\Gamma
^{p}\right)  \left(  t\right)  \lambda^{2}\,. \label{deffin_traj}%
\end{equation}
The BRST-antiBRST invariance of $\mathcal{I}_{\Gamma}^{\Phi}$ implies the
relation%
\begin{equation}
\mathcal{I}_{\Gamma g(\lambda)}^{\Phi}=\mathcal{I}_{\Gamma}^{\Phi}\ ,
\label{finbab}%
\end{equation}
in view of the fact that, due to Liuville's theorem, the measure $d\Gamma$ in
(\ref{ZPI}) is right-invariant with respect to the action of the Abelian
supergroup, which plays the role of finite canonical transformations,
$d\check{\Gamma}=d\Gamma$, and the fact that the Hamiltonian action
$S_{H}(\Gamma)=\int dt\left[  \frac{1}{2}\Gamma^{p}(t)\omega_{pq}\dot{\Gamma
}^{q}(t)-H_{\Phi}(t)\right]  $ is also invariant, $S_{H}(\Gamma)=S_{H}%
(\check{\Gamma})$.

The finite field-dependent transformations (\ref{deffin_traj}) with parameters
$\lambda_{a}=\lambda_{a}(\Gamma)$ having no dependence on $t$ and $\Gamma^{p}$
as functions, $(d\lambda_{a})/(dt)$ = $(\partial\lambda_{a})/(\partial
\Gamma^{p})=0$, make it possible \cite{MRnew1} to establish the
gauge-independence of the the vacuum functional, $Z_{\Phi+\Delta\Phi
}(0)=Z_{\Phi}(0)$, in the particular case of functionally-dependent
parameters, $\lambda_{a}\left(  \Gamma\right)  =\int dt\ \left(  s^{a}%
\Lambda\left(  t\right)  \right)  =\varepsilon_{ab}\int dt\ \left\{
\Lambda\left(  t\right)  ,\Omega^{b}\left(  t\right)  \right\}  _{t}$ with a
certain even-valued potential function $\Lambda(t)=\Lambda\left(
\Gamma(t)\right)  $, which is inspired by infinitesimal field-dependent
BRST-antiBRST transformations with the parameters \cite{BLT1h}%
\begin{equation}
\mu_{a}=\frac{i}{2\hbar}\varepsilon_{ab}\int dt\left\{  \Delta\Phi
,\ \Omega^{b}\right\}  _{t}=\frac{i}{2\hbar}\int dt\ \left(  s_{a}\Delta
\Phi\right)  \left(  t\right)  \ . \label{inffdpar}%
\end{equation}
The gauge-independence of the vacuum functional $Z_{\Phi}(0)$ implies the
gauge-independence of the $S$-matrix, due to the equivalence theorem
\cite{KalloshTyutin}.

In the case of finite field-dependent transformations with group-like elements
$g({\widehat{\Lambda}}\overleftarrow{s}_{a})$, ${\widehat{\Lambda}}%
(\Gamma)=\int dt\,\Lambda(t)$, whose set forms a nonlinear non-Abelian
group-like structure,\footnote{For BRST-antiBRST-closed (in particular,
BRST-antiBRST-exact) parameters $\lambda_{a}(\Gamma)={\widehat{\Lambda}}%
_{a}\overleftarrow{s}^{2}$ with ${\widehat{\Lambda}}_{a}=\int dt\,{\Lambda
}_{a}(\Gamma(t),t)$, the subset $g({\widehat{\Lambda}}_{a}\overleftarrow
{s}^{2})$ forms an Abelian subgroup in $g({\widehat{\Lambda}}\overleftarrow
{s}_{a})$, and thereby in $g({\lambda_{a}}(\Gamma))$; for details, see
Footnote~\ref{footnote-label}.} the superdeterminant of a change of variables
reads
\begin{align}
&  \mathrm{Sdet}\left\{  \frac{\delta\left[  (\Gamma(t^{\prime})g({\widehat
{\Lambda}}\overleftarrow{s}_{a})\right]  }{\delta\Gamma(t^{\prime\prime}%
)}\right\}  =\exp\left[  \Im\left(  \Gamma\right)  \right]
,\ \ \ \mathrm{where}\ \ \ \Im\left(  \Gamma\right)  =-2\mathrm{\ln}\left[
1-\frac{1}{2}\int dt\ \left(  s^{2}\Lambda\right)  _{t}\right]
\ ,\label{superJaux2}\\
&  d\check{\Gamma}=d\Gamma\exp\left[  \frac{i}{\hbar}\left(  -i\hbar
\Im\right)  \right]  =d\Gamma\exp\left\{  \frac{i}{\hbar}\left[
i\hbar\,\mathrm{\ln}\left(  1-\frac{1}{2}\varepsilon_{ab}\int dt\left\{
\left\{  \Lambda,\Omega^{a}\right\}  _{t},\Omega^{b}\right\}  _{t}\right)
^{2}\right]  \right\}  \ , \label{superJ32}%
\end{align}
with account taken of $\left(  s^{2}\Lambda\right)  _{t}=\varepsilon
_{ab}\left\{  \left\{  \Lambda,\Omega^{a}\right\}  _{t},\Omega^{b}\right\}
_{t}$. In view of the invariance of the quantum action $S_{H}\left(
\Gamma\right)  $ with respect to (\ref{deffin_traj}), the change $\Gamma
^{p}(t)\rightarrow\check{\Gamma}^{p}(t)=[\Gamma^{p}g(\lambda(\Gamma))](t)$
leads to the following transformation of the integrand $\mathcal{I}_{\Gamma
}^{\Phi}$ in (\ref{ZPI}):%
\begin{equation}
\mathcal{I}_{\Gamma g(\lambda(\Gamma))}^{\Phi}\ =\ d\Gamma\ \exp\left[
\Im\left(  \Gamma\right)  \right]  \exp\left[  \left(  i/\hbar\right)
S_{H}\left(  \Gamma\left(  g\lambda\left(  \Gamma\right)  \right)  \right)
\right]  \ =\ d\Gamma\ \exp\left\{  \left(  i/\hbar\right)  \left[
S_{H}\left(  \Gamma\right)  -i\hbar\Im\left(  \Gamma\right)  \right]
\right\}  \ , \label{superJ4}%
\end{equation}
and thereby implies%
\begin{equation}
\mathcal{I}_{\Gamma g(\lambda(\Gamma))}^{\Phi}\ =\ d\Gamma\ \exp\left\{
\left(  i/\hbar\right)  \left[  S_{H}\left(  \Gamma\right)  +i\hbar
\ \mathrm{\ln}\left(  1-{\widehat{\Lambda}}\overleftarrow{s}{}^{2}/2\right)
^{2}\right]  \right\}  \ . \label{superJ5}%
\end{equation}
Because of the fact that the Jacobian-induced contribution $i\hbar
\,\mathrm{\ln}\left(  1-{\widehat{\Lambda}}\overleftarrow{s}^{2}/2\right)
^{2}$ to the action $S_{H}\left(  \Gamma\right)  $ is a BRST-antiBRST-exact
term, it can be compensated by another BRST-antiBRST-exact addition to
$S_{H}\left(  \Gamma\right)  $ related to a change of the gauge-fixing
function, $\Phi(t)\rightarrow(\Phi+\Delta\Phi)(t)$, made in the original
integrand $\mathcal{I}_{\Gamma}^{\Phi}$,%
\begin{align}
&  i\hbar\ \mathrm{\ln}\left(  1-\frac{1}{2}{\widehat{\Lambda}}\overleftarrow
{s}{}^{2}\right)  ^{2}=-\frac{1}{2}\left(  \Delta\widehat{\Phi}\right)
\overleftarrow{s}^{2}\ ,\ \ \ \mathrm{where}\ \ \ \Delta\widehat{\Phi}=\int
dt\,\Delta\Phi(t)\ ,\label{superJ3mm}\\
&  \Longrightarrow\mathcal{I}_{\Gamma g(\lambda(\Gamma))}^{\Phi}=d\Gamma
\exp\left\{  \frac{i}{\hbar}\left[  S_{H,\Phi}(\Gamma)-\frac{1}{2}\left(
\Delta\widehat{\Phi}\right)  \overleftarrow{s}^{2}\right]  \right\}
=\mathcal{I}_{\Gamma}^{\Phi+\Delta\Phi}\ . \label{superJ4m}%
\end{align}
The relation of $\Delta\Phi\left(  \Gamma(t)|\Lambda\right)  $ to the
field-dependent parameter $\Lambda\left(  \Gamma(t)\right)  $ is established
by (\ref{superJ3mm}), also known as the compensation relation for an unknown
parameter $\Lambda(\Gamma(t))$, which provides the gauge-independence of the
vacuum functional, $Z_{\Phi}(0)=Z_{\Phi+\Delta\Phi}(0)$. An explicit solution
of (\ref{superJ3mm}), satisfying the solvability condition, due to the
BRST-antiBRST exactness (up to BRST-antiBRST-exact terms) of both of its
sides, is given by%
\begin{equation}
\Lambda(\Gamma(t)|\Delta{\Phi})=2\Delta{\Phi}(t)\left(  \Delta\widehat{\Phi
}\overleftarrow{s}^{2}\right)  ^{-1}\left[  1-\exp\left(  \frac{1}{4i\hbar
}\Delta\widehat{\Phi}\overleftarrow{s}^{2}\right)  \right]  =-\frac{1}%
{2i\hbar}\Delta\Phi(t)\sum_{n=0}^{\infty}\frac{1}{\left(  n+1\right)
!}\left(  \frac{1}{4i\hbar}\Delta{\widehat{\Phi}}\overleftarrow{s}^{2}\right)
^{n}\ . \label{solcompeq2}%
\end{equation}
Conversely, having considered the equation (\ref{superJ3mm}) for an unknown
$\Delta\Phi(t)$ with a given $\Lambda(t)$, we obtain%
\begin{equation}
\Delta\Phi\left(  \Gamma(t)\right)  =-2i\hbar\ \Lambda\left(  \Gamma
(t)\right)  \left(  \widehat{\Lambda}\left(  \Gamma\right)  \overleftarrow
{s}^{2}\right)  ^{-1}\ln\left(  1-\widehat{\Lambda}\left(  \Gamma\right)
\overleftarrow{s}^{2}/2\right)  ^{2}\ . \label{Lambda-Fsol2}%
\end{equation}
Therefore, the field-dependent transformations with the parameters
$\lambda_{a}(\Gamma)={\widehat{\Lambda}}\overleftarrow{s}_{a}$,
\[
\lambda_{a}=-\frac{1}{2i\hbar}\left(  \Delta\widehat{\Phi}\overleftarrow
{s}_{a}\right)  \sum_{n=0}^{\infty}\frac{1}{\left(  n+1\right)  !}\left(
\frac{1}{4i\hbar}\Delta\widehat{\Phi}\overleftarrow{s}^{2}\right)  ^{n}\,,
\]
amount to a precise change of the gauge-fixing function.

In virtue of (\ref{Lambda-Fsol2}), the property (\ref{superJ4m}) leads to a
so-called modified Ward identity \cite{MRnew1} in generalized Hamiltonian
formalism, depending on field-dependent parameters, $\lambda_{a}(\Gamma
|\Delta\Phi)=\int dt\Lambda\overleftarrow{s}_{a}$, and thereby also on a
finite change of the gauge:
\begin{align}
&  \left\langle \left\{  1+\frac{i}{\hbar}\int dtI_{p}(t)\Gamma^{p}(t)\left(
\overleftarrow{s}^{a}\lambda_{a}(\Lambda)+\frac{1}{4}\overleftarrow{s}%
^{2}\lambda^{2}(\Lambda)\right)  -\frac{1}{4}\left(  \frac{i}{\hbar}\right)
{}^{2}\int dt\ dt^{\prime}\ I_{p}(t)\Gamma^{p}(t)\overleftarrow{s}^{a}%
I_{q}(t^{\prime})\Gamma^{q}(t^{\prime})\overleftarrow{s}_{a}\lambda
^{2}(\Lambda)\right\}  \right. \nonumber\\
&  \quad\left.  \times\left\{  1-\frac{1}{2}\left[  \int dt\Lambda(t)\right]
\overleftarrow{s}^{2}\right\}  {}^{-2}\right\rangle _{\Phi,I}=1\,, \label{mWI}%
\end{align}
where the symbol \textquotedblleft$\langle\mathcal{A}\rangle_{\Phi,I}%
$\textquotedblright\ for any quantity $\mathcal{A}=\mathcal{A}(\Gamma)$
denotes a source-dependent average expectation value corresponding to a gauge
$\Phi(\Gamma)$, namely,%
\begin{equation}
\left\langle \mathcal{A}\right\rangle _{\Phi,I}=Z_{\Phi}^{-1}(I)\int
d\Gamma\ \mathcal{A}(\Gamma)\exp\left\{  \frac{i}{\hbar}\left[  S_{H,\Phi
}(\Gamma)+\int dt\,I(t)\Gamma(t)\right]  \right\}  \ ,\ \ \left\langle
1\right\rangle _{\Phi,I}=1\ . \label{aexv2}%
\end{equation}
The property (\ref{superJ4m}) implies a relation which describes the
gauge-dependence of the generating functional $Z_{\Phi}(I)$,
\begin{align}
Z_{\Phi+\Delta\Phi}(I)  &  =Z_{\Phi}(I)\left\{  1+\left\langle \frac{i}{\hbar
}\int dt\ I_{p}(t)\left[  (s^{a}\Gamma^{p}(t))\lambda_{a}\left(
\Gamma|-\Delta{\Phi}\right)  +\frac{1}{4}(s^{2}\Gamma^{p}(t))\lambda
^{2}\left(  \Gamma|-\Delta{\Phi}\right)  \right]  \right.  \right. \nonumber\\
&  -\left.  \left.  (-1)^{\varepsilon_{q}}\left(  \frac{i}{2\hbar}\right)
^{2}\int dt\ dt^{\prime}I_{q}(t^{\prime})I_{p}(t)(s^{a}\Gamma^{p}%
(t))(s_{a}\Gamma^{q}(t^{\prime}))\lambda^{2}\left(  \Gamma|-\Delta{\Phi
}\right)  \right\rangle \right\}  \,, \label{GDI}%
\end{align}
and extends the result (\ref{superJ5}) to non-vanishing external sources
$I_{p}(t)$.

For constant parameters, $\lambda_{a}=\mathrm{const}$, the identity
(\ref{mWI}) implies two independent usual Ward identities at the first degree
in powers of $\lambda_{a}$, as well as a new (derivative) Ward identity at the
second degree in powers of $\lambda_{a}$,%
\begin{equation}
\left\langle \int dt\ I_{p}(t)\Gamma^{p}(t)\overleftarrow{s}^{a}\right\rangle
_{\Phi,I}=0\ ,\ \ \ \left\langle \int dt\,I_{p}(t)\Gamma^{p}(t)\left[
\overleftarrow{s}^{2}-\overleftarrow{s}^{a}\left(  \frac{i}{\hbar}\right)
\int dt^{\prime}\ I_{q}(t^{\prime})\left(  \Gamma^{q}(t^{\prime}%
)\overleftarrow{s}_{a}\right)  \right]  \right\rangle _{\Phi,I}=0\,.
\label{ushWI}%
\end{equation}
Below, we intend to study the more general case of finite field-dependent
BRST-antiBRST transformations in Hamiltonian formalism with arbitrary
functional parameters, generally assumed to be functionally-independent,
$\lambda_{a}\not \equiv \int dt\,\Lambda\overleftarrow{s}_{a}$. It is also
intended to study the case of finite field-dependent BRST-antiBRST
transformations being linear in functionally-dependent parameters of the form
$\lambda_{a}\left(  \Gamma\right)  =\int dt\ s_{a}\Lambda\left(
\Gamma(t)\right)  $.

\subsection{General Gauge Theories in Lagrangian Formalism}

\label{laggen}

The generating functional of Green's functions $Z_{F}(J)$, depending on
external sources $J_{A}$, $\varepsilon(J_{A})=\varepsilon_{A}$,
\begin{equation}
Z_{F}(J)=\int d\mathsf{\Gamma}\;\exp\left\{  \left(  i/\hbar\right)  \left[
\mathcal{S}_{F}\left(  \mathsf{\Gamma}\right)  +J_{A}\phi^{A}\right]
\right\}  \ ,\ \ \ \mathcal{S}_{F}=S+\phi_{Aa}^{\ast}\pi^{Aa}+\left(
\bar{\phi}_{A}-F_{,A}\right)  \lambda^{A}-\left(  1/2\right)  \varepsilon
_{ab}\pi^{Aa}F_{,AB}\pi^{Bb}\ , \label{z(0)}%
\end{equation}
and the corresponding vacuum functional $Z_{F}\equiv Z_{F}(0)$ are defined on
the triplectic \cite{BSemikhatov} manifold\footnote{Amongst the ingredients of
\cite{BSemikhatov}, we only use differential operations in local coordinates,
and therefore our description of triplectic geometry reduces to a description
of BRST-antiBRST quantization \cite{BLT1,BLT2,Hull}.} $\mathcal{M}%
_{\mathsf{\Gamma}}$ locally parameterized by the coordinates
\begin{equation}
\mathsf{\Gamma}^{\mathsf{p}}=\left(  \phi^{A},\phi_{Aa}^{\ast},\bar{\phi}%
_{A},\pi^{Aa},\lambda^{A}\right)  \ , \label{gencoor}%
\end{equation}
where $\phi^{A}$ are the fields of the total configuration space of the BV
formalism \cite{BV}, which is larger in reducible gauge theories, being more
general than the theories examined in Section~\ref{YMgen}, and is organized
into $\mathrm{Sp}(2)$-symmetric tensors, according to the rules of
$\mathrm{Sp}(2)$-covariant Lagrangian quantization \cite{BLT1, BLT2}. The
manifold $\mathcal{M}_{\mathsf{\Gamma}}$ also contains the triplets of
antifields $\phi_{Aa}^{\ast}$, $\bar{\phi}_{A}$ and auxiliary fields $\pi
^{Aa}$, $\lambda^{A}$, with the following distribution of Grassmann parity:%
\[
\varepsilon\left(  \phi^{A},\ \phi_{Aa}^{\ast},\ \bar{\phi}_{A},\ \pi
^{Aa},\ {\lambda}^{A}\right)  =\left(  \varepsilon_{A},\ \varepsilon
_{A}+1,\ \varepsilon_{A},\ \varepsilon_{A}+1,\ \varepsilon_{A}\right)  .
\]
The functional $Z_{F}(J)$ is determined by an even-valued functional
$S=S(\phi,\phi^{\ast},\bar{\phi})$ and by an even-valued gauge-fixing
functional $F=F(\phi)$, where $S$ is subject to the generating equations%
\begin{equation}
\frac{1}{2}(S,S)^{a}+V^{a}S=i\hbar\Delta^{a}S\Longleftrightarrow\left(
\Delta^{a}+\frac{i}{\hbar}V^{a}\right)  \exp\left(  \frac{i}{\hbar}S\right)
=0\ , \label{3.3}%
\end{equation}
with the classical action $S_{0}(A)$ being the boundary condition for $S$ in
the case of vanishing antifields, $\phi_{a}^{\ast}=\bar{\phi}=0$. The extended
antibracket $(\bullet,\bullet)^{a}$ and the operators $\Delta^{a}$, $V^{a}$
are given by%
\begin{equation}
(\bullet,\bullet)^{a}=\frac{\delta\bullet}{\delta\phi^{A}}\frac{\delta\bullet
}{\delta\phi_{Aa}^{\ast}}-\frac{\delta_{r}\bullet}{\delta\phi_{Aa}^{\ast}%
}\frac{\delta_{l}\bullet}{\delta\phi^{A}}\ ,\ \ \ \Delta^{a}=(-1)^{\varepsilon
_{A}}\frac{\delta_{l}}{\delta\phi^{A}}\frac{\delta}{\delta\phi_{Aa}^{\ast}%
}\ ,\ \ \ V^{a}=\varepsilon^{ab}\phi_{Ab}^{\ast}\frac{\delta}{\delta\bar{\phi
}_{A}}\ . \label{abrack}%
\end{equation}
The classical action is invariant under the infinitesimal gauge
transformations (\ref{Nid}) with the generators $R_{\alpha}^{i}(A)$ satisfying
the general relations (\ref{gauge_alg}) of a gauge algebra.

The integrand $\mathcal{I}_{\mathsf{\Gamma}}^{F}=d\mathsf{\Gamma}\exp\left[
\left(  i/\hbar\right)  \mathcal{S}_{F}\left(  \mathsf{\Gamma}\right)
\right]  $ is invariant under the global infinitesimal BRST-antiBRST
transformations (\ref{Gamma}), with the corresponding generators
$\mathsf{s}^{a}$ being different from $s^{a}$ of Subsections~\ref{YMgen}%
,~\ref{hamgen},%
\begin{equation}
\delta\mathsf{\Gamma}^{\mathsf{p}}=\left(  \mathsf{s}^{a}\mathsf{\Gamma
}^{\mathsf{p}}\right)  \mu_{a}=\mathsf{\Gamma}^{\mathsf{p}}\overleftarrow
{\mathsf{s}}{}^{a}\mu_{a}=\left(  \pi^{Aa},\ \delta_{b}^{a}S_{,A}\left(
-1\right)  ^{\varepsilon_{A}},\ \varepsilon^{ab}\phi_{Ab}^{\ast}\left(
-1\right)  ^{\varepsilon_{A}+1},\ \varepsilon^{ab}\lambda^{A},\ 0\right)
\mu_{a}\ , \label{Gamma}%
\end{equation}
where the invariance at the first order in $\mu_{a}$ is established by using
the generating equations (\ref{3.3}).

Despite the fact that the generators $\mathsf{s}^{a}$ do not obey,
$\mathsf{s}^{a}\mathsf{s}^{b}+\mathsf{s}^{b}\mathsf{s}^{a}\neq0$, the
BRST-antiBRST algebra in the sector of the antifields $\phi_{Aa}^{\ast}$,
$\bar{\phi}_{A}$, the mentioned infinitesimal invariance is sufficient to
determine \emph{finite BRST-antiBRST transformations}, $\mathsf{\Gamma
}^{\mathsf{p}}\rightarrow\mathsf{\Gamma}^{\mathsf{p}}+\Delta\mathsf{\Gamma
}^{\mathsf{p}}$, with anticommuting parameters $\lambda_{a}$, $a=1,2$,
introduced in \cite{MRnew} according to%
\begin{equation}
\mathcal{I}_{\mathsf{\Gamma}+\Delta\mathsf{\Gamma}}^{{F}}=\mathcal{I}%
_{\mathsf{\Gamma}}^{{F}}\ ,\ \ \ \left.  \Delta\mathsf{\Gamma}^{\mathsf{p}%
}\frac{\overleftarrow{\partial}}{\partial\lambda_{a}}\right\vert _{\lambda
=0}=\mathsf{\Gamma}^{\mathsf{p}}\overleftarrow{\mathsf{s}}{}^{a}%
\ \ \ \mathrm{and}\mathtt{\ \ \ }\Delta\mathsf{\Gamma}^{\mathsf{p}}%
\frac{\overleftarrow{\partial}}{\partial\lambda_{b}}\frac{\overleftarrow
{\partial}}{\partial\lambda_{a}}=\frac{1}{2}\varepsilon^{ab}\mathsf{\Gamma
}^{\mathsf{p}}\overleftarrow{\mathsf{s}}{}^{2}\,,\ \ \ \mathrm{where}%
\ \ \ \mathsf{s}^{2}=\mathsf{s}_{a}\mathsf{s}^{a}=\overleftarrow{\mathsf{s}}%
{}^{2}=\overleftarrow{\mathsf{s}}{}^{a}\overleftarrow{\mathsf{s}}_{a}\ .
\label{Gamma_fin_def}%
\end{equation}
The finite BRST-antiBRST transformations for the integrand $\mathcal{I}%
_{\Gamma}^{{F}}$ in a general gauge theory are established, once again, by
solving the functional equation $G(\mathsf{\Gamma}+\Delta\mathsf{\Gamma
})=G(\mathsf{\Gamma})$ for any regular functional $G(\mathsf{\Gamma})$ defined
in $\mathcal{M}_{\mathsf{\Gamma}}$ and subject to infinitesimal BRST-antiBRST
invariance, $G\overleftarrow{\mathsf{s}}{}^{a}=0$, which may be considered as
the integrability condition for the above functional equation. The resulting
finite BRST-antiBRST transformations are given by
\begin{equation}
\Delta\mathsf{\Gamma}^{\mathsf{p}}=\mathsf{\Gamma}^{\mathsf{p}}\left(
\overleftarrow{\mathsf{s}}{}^{a}\lambda_{a}+\frac{1}{4}\overleftarrow
{\mathsf{s}}{}^{2}\lambda^{2}\right)  \ , \label{Gamma_fin}%
\end{equation}
or, equivalently, in a group-like form%
\begin{equation}
\mathsf{\Gamma}^{\prime\mathsf{p}}=\mathsf{\Gamma}^{\mathsf{p}}\left(
1+\overleftarrow{\mathsf{s}}^{a}\lambda_{a}+\frac{1}{4}\overleftarrow
{\mathsf{s}}^{2}\lambda^{2}\right)  =\mathsf{\Gamma}^{\mathsf{p}}\exp\left(
\overleftarrow{\mathsf{s}}^{a}\lambda_{a}\right)  \equiv\mathsf{\Gamma
}^{\mathsf{p}}\mathsf{g}(\lambda)\ , \label{Gamma_fineq}%
\end{equation}
so that there holds the exact relation%
\begin{equation}
\mathcal{I}_{\mathsf{\Gamma}\mathsf{g}(\lambda)}^{{F}}=\mathcal{I}%
_{\mathsf{\Gamma}}^{{F}}\ , \label{babinv2}%
\end{equation}
considering that the functional $S$ meets the generating equations
(\ref{3.3}). To establish the relation (\ref{babinv2}), we need to take into
account the change of the integration measure under the global finite
transformations (corresponding to $\lambda_{a}=\mathrm{const}$) and the
respective change of the functional $\mathcal{S}_{F}(\mathsf{\Gamma})$ in
(\ref{z(0)}), according to the rules \cite{MRnew2}
\begin{align}
&  d\mathsf{\Gamma}^{\prime}=d\mathsf{\Gamma}\ \mathrm{Sdet}\left[
\frac{\delta\left(  \mathsf{\Gamma}\mathsf{g}(\lambda)\right)  }%
{\delta\mathsf{\Gamma}}\right]  =d\mathsf{\Gamma}\exp\left[  -\left(
\Delta^{a}S\right)  \lambda_{a}-\frac{1}{4}\left(  \Delta^{a}S\right)
\overleftarrow{\mathsf{s}}_{a}\lambda^{2}\right]  \,,\label{ansjacob}\\
&  \exp\left[  \frac{i}{\hbar}\mathcal{S}_{F}\left(  \mathsf{\Gamma}^{\prime
}\right)  \right]  =\exp\left\{  \frac{i}{\hbar}\left[  \mathcal{S}_{F}\left(
\mathsf{\Gamma}\right)  +{\mathcal{S}_{F}}\left(  \mathsf{\Gamma}\right)
\overleftarrow{\mathsf{s}}^{a}\lambda_{a}+\frac{1}{4}\mathcal{S}_{F}\left(
\mathsf{\Gamma}\right)  \overleftarrow{\mathsf{s}}^{2}\lambda^{2}\right]
\right\}  \,, \label{qacttrans}%
\end{align}
so that, due to the relations
\begin{equation}
\mathcal{S}_{F}\overleftarrow{\mathsf{s}}^{a}=-\frac{1}{2}(S,S)^{a}-V^{a}S\,,
\label{identsg}%
\end{equation}
implied by (\ref{3.3}), the finite BRST-antiBRST invariance (\ref{babinv2}) of
$\mathcal{I}_{\mathsf{\Gamma}}^{{F}}$ does indeed take place.

The set of elements $\left\{  \mathsf{g}(\lambda)\right\}  =\{\exp
(\overleftarrow{\mathsf{s}}^{a}\lambda_{a})\}$, in contrast to the respective
sets of finite BRST-antiBRST transformations (\ref{finite1}), (\ref{finite2})
in Yang--Mills theories and first-class constraint dynamical systems, does not
form a supergroup with respect to multiplication, denoted by the symbol
\textquotedblleft$\cdot$\textquotedblright, being an associative composition
law. Indeed, for any elements $\mathsf{g}(\lambda_{\left(  1\right)  })$,
$\mathsf{g}(\lambda_{\left(  2\right)  })$\ their composition is given
by\footnote{In case the parameters $\lambda_{\left(  i\right)  }^{a}$,
$i=1,2$, belong to a vector space with some anticommuting basis elements
$\nu^{a}$, $\mathrm{gh}\left(  \nu^{a}\right)  =\left(  -1\right)  ^{a+1}$,
namely, $\lambda_{\left(  i\right)  }^{a}=c_{\left(  i\right)  }\left(
a\right)  \cdot\nu^{a}$, with certain $c$\textbf{-}numbers $c_{\left(
i\right)  }\left(  a\right)  $ and no summation over $a$, it follows that
$\lambda_{\left(  i\right)  }^{2}\lambda_{\left(  j\right)  a}=\lambda
_{\left(  i\right)  }^{2}\lambda_{\left(  j\right)  }^{2}\equiv0$; however,
the deviation $\mathrm{dev}(\lambda_{\left(  1\right)  },\lambda_{\left(
2\right)  })$ remains non-vanishing, $\mathrm{dev}(\lambda_{\left(  1\right)
},\lambda_{\left(  2\right)  })=-\frac{1}{2}\overleftarrow{s}^{b}%
\overleftarrow{s}^{a}\left[  c_{\left(  2\right)  }\left(  b\right)
c_{\left(  1\right)  }\left(  a\right)  -c_{\left(  1\right)  }\left(
b\right)  c_{\left(  2\right)  }\left(  a\right)  \right]  \nu_{b}\nu_{a}$,
which is readily seen in components: $\mathrm{dev}(\lambda_{\left(  1\right)
},\lambda_{\left(  2\right)  })=\left(  1/2\right)  \left(  \overleftarrow
{s}^{1}\overleftarrow{s}^{2}+\overleftarrow{s}^{2}\overleftarrow{s}%
^{1}\right)  \left[  c_{\left(  1\right)  }\left(  1\right)  c_{\left(
2\right)  }\left(  2\right)  -c_{\left(  1\right)  }\left(  2\right)
c_{\left(  2\right)  }\left(  1\right)  \right]  \nu_{1}\nu_{2}\not =0$.}%
\begin{align}
\mathsf{g}(\lambda_{\left(  1\right)  })\cdot\mathsf{g}(\lambda_{\left(
2\right)  })  &  =\mathsf{g}(\lambda_{\left(  1\right)  }+\lambda_{\left(
2\right)  })+\mathrm{dev}(\lambda_{\left(  1\right)  },\lambda_{\left(
2\right)  })\ ,\label{complaw}\\
\mathrm{dev}(\lambda_{\left(  1\right)  },\lambda_{\left(  2\right)  })=  &
-\frac{1}{2}\overleftarrow{s}^{b}\overleftarrow{s}^{a}\left[  \lambda_{\left(
2\right)  b}\lambda_{\left(  1\right)  a}-\lambda_{\left(  1\right)  b}%
\lambda_{\left(  2\right)  a}\right] \nonumber\\
&  +\frac{1}{4}\left[  \overleftarrow{s}^{2}\overleftarrow{s}^{a}%
\lambda_{\left(  2\right)  }^{2}\lambda_{\left(  1\right)  a}+\overleftarrow
{s}^{a}\overleftarrow{s}^{2}\lambda_{\left(  1\right)  }^{2}\lambda_{\left(
2\right)  a}\right]  +\frac{1}{16}\overleftarrow{s}^{2}\overleftarrow{s}%
^{2}\lambda_{\left(  2\right)  }^{2}\lambda_{\left(  1\right)  }^{2}\not =0\ ,
\label{deviator}%
\end{align}
and therefore contains non-vanishing operator structures, $\overleftarrow
{\mathsf{s}}^{2}\overleftarrow{\mathsf{s}}^{a}$, $\overleftarrow{\mathsf{s}%
}^{a}\overleftarrow{\mathsf{s}}^{2}$, $\overleftarrow{\mathsf{s}}%
^{2}\overleftarrow{\mathsf{s}}^{2}$, which are absent from a group element
$\mathsf{g}\left(  \lambda\right)  $. Notice that the non-vanishing deviation
$\mathrm{dev}(\lambda_{\left(  1\right)  },\lambda_{\left(  2\right)  })$\ of
the action of $\{\mathsf{g}(\lambda)\}$\ from that of an Abelian two-parameter
supergroup is not symmetric with respect to the permutation of the arguments,
$\lambda_{1}\leftrightarrow\lambda_{2}$: $\mathrm{dev}(\lambda_{\left(
1\right)  },\lambda_{\left(  2\right)  })\not =\mathrm{dev}(\lambda_{\left(
2\right)  },\lambda_{\left(  1\right)  })$. This implies that the commutator
of any $\mathsf{g}(\lambda_{\left(  1\right)  })$, $\mathsf{g}(\lambda
_{\left(  2\right)  })$\ in the set $\{\mathsf{g}(\lambda)\}$\ is
non-vanishing:%
\[
\lbrack\mathsf{g}(\lambda_{\left(  1\right)  }),\mathsf{g}(\lambda_{\left(
2\right)  })]\equiv\mathsf{g}(\lambda_{\left(  1\right)  })\cdot
\mathsf{g}(\lambda_{\left(  2\right)  })-\mathsf{g}(\lambda_{\left(  2\right)
})\cdot\mathsf{g}(\lambda_{\left(  1\right)  })\not =0\,.
\]
At the same time, the set $\{\mathsf{g}(\lambda)\}$, being considered as
right-hand transformations realized on regular functionals in
$M_{\mathsf{\Gamma}}$\ restricted to $\tilde{G}=\tilde{G}(\phi,\pi,\lambda)$,
$\tilde{G}=\left.  G(\mathsf{\Gamma})\right\vert {}_{\phi^{\ast}=\bar{\phi}%
=0}$, turns into an Abelian supergroup $\{\mathsf{\tilde{g}}(\lambda)\}$\ with
the elements%
\begin{equation}
\{\mathsf{\tilde{g}}(\lambda)\}=\left\{  \mathsf{\tilde{g}}(\lambda
)\in\{\mathsf{g}(\lambda)\}|\mathsf{\tilde{g}}(\lambda)=\exp(\overleftarrow
{U}^{a}\lambda_{a})\,,\ \ \ \overleftarrow{U}^{a}=\left.  \overleftarrow
{\mathsf{s}}^{a}\right\vert _{\phi,\pi,\lambda}\right\}  \ ,
\label{usbabgroup}%
\end{equation}
where the operators $\overleftarrow{U}{}^{a}$\ are anticommuting and thereby
nilpotent \cite{MRnew3}, namely,%
\begin{equation}
\overleftarrow{U}^{a}=\frac{\overleftarrow{\delta}}{\delta\phi^{A}}\pi
^{Aa}+\varepsilon^{ab}\frac{\overleftarrow{\delta}}{\delta\pi^{Ab}}\lambda
^{A}\ ,\ \ \ \overleftarrow{U}{}^{a}\overleftarrow{U}{}^{b}+\overleftarrow
{U}{}^{b}\overleftarrow{U}{}^{a}=0\ ,\ \ \ \overleftarrow{U}{}^{a}%
\overleftarrow{U}{}^{b}\overleftarrow{U}{}^{c}=0\ . \label{nilp}%
\end{equation}
Indeed, due to the nilpotency of $\overleftarrow{U}{}^{a}$, it follows that%
\begin{align}
\mathsf{\tilde{g}}(\lambda_{\left(  1\right)  })\cdot\mathsf{\tilde{g}%
}(\lambda_{\left(  2\right)  })  &  =\mathsf{\tilde{g}}(\lambda_{\left(
1\right)  }+\lambda_{\left(  2\right)  })+\mathrm{dev}(\lambda_{\left(
1\right)  },\lambda_{\left(  2\right)  })\ ,\label{tilde}\\
\mathrm{dev}(\lambda_{\left(  1\right)  },\lambda_{\left(  2\right)  })  &
=-\frac{1}{2}\overleftarrow{U}^{b}\overleftarrow{U}^{a}\left[  \lambda
_{\left(  2\right)  b}\lambda_{\left(  1\right)  a}-\lambda_{\left(  1\right)
b}\lambda_{\left(  2\right)  a}\right]  =0\ , \label{dev_tilde}%
\end{align}
since%
\begin{equation}
\overleftarrow{U}^{b}\overleftarrow{U}^{a}\left[  \lambda_{\left(  2\right)
b}\lambda_{\left(  1\right)  a}-\lambda_{\left(  1\right)  b}\lambda_{\left(
2\right)  a}\right]  =-\frac{1}{2}\overleftarrow{U}^{2}\left[  \lambda
_{\left(  2\right)  a}\lambda_{\left(  1\right)  }^{a}-\lambda_{\left(
1\right)  a}\lambda_{\left(  2\right)  }^{a}\right]  =-\frac{1}{2}%
\overleftarrow{U}^{2}\left[  \lambda_{\left(  2\right)  a}\lambda_{\left(
1\right)  }^{a}-\lambda_{\left(  2\right)  a}\lambda_{\left(  1\right)  }%
^{a}\right]  \equiv0\ , \label{since}%
\end{equation}
which proves the Abelian nature of the supergroup $\{\mathsf{\tilde{g}%
}(\lambda)\}$, namely, $\mathsf{\tilde{g}}(\lambda_{\left(  1\right)  }%
)\cdot\mathsf{\tilde{g}}(\lambda_{\left(  2\right)  })=\mathsf{\tilde{g}%
}(\lambda_{\left(  1\right)  }+\lambda_{\left(  2\right)  })=\mathsf{\tilde
{g}}(\lambda_{\left(  2\right)  })\cdot\mathsf{\tilde{g}}(\lambda_{\left(
1\right)  })$.

For finite field-dependent transformations, it has been shown
\cite{MRnew2,MRnew3} that in the case of functionally-dependent parameters
$\lambda_{a}=\Lambda\overleftarrow{U}_{a}$ with an even-valued potential
$\Lambda=\Lambda\left(  \phi,\pi,\lambda\right)  $, inspired by infinitesimal
field-dependent BRST-antiBRST transformations with the parameters \cite{MRnew,
MRnew2}%
\begin{equation}
\mu_{a}(\phi,\pi,\lambda|\Delta F)=-\frac{i}{2\hbar}\varepsilon_{ab}\left(
\Delta F\right)  _{,A}\pi^{Ab}=-\frac{i}{2\hbar}\varepsilon_{ab}\Delta
F\overleftarrow{U}{}^{b}\,, \label{partic_case1}%
\end{equation}
there holds the gauge-independence of the vacuum functional: $Z_{F+\Delta
F}(0)=Z_{F}(0)$. Indeed, a finite transformation with a group-like element
$\mathsf{\tilde{g}}(\Lambda\overleftarrow{U}_{a})$ leads to the
superdeterminant of a change of variables $\mathsf{\Gamma}^{\mathsf{p}%
}\rightarrow\mathsf{\Gamma}{}^{\mathsf{p}}\mathsf{\tilde{g}}(\Lambda
\overleftarrow{U}_{a})$ and implies the corresponding change of the
integration measure given by \cite{MRnew2, MRnew3}:%
\begin{align}
&  \mathrm{Sdet}\left\{  \frac{\delta\left[  \mathsf{\Gamma}\check{g}%
(\Lambda\overleftarrow{U}_{a})\right]  }{\delta\mathsf{\Gamma}}\right\}
=\exp\left[  -\left(  \Delta^{a}S\right)  \lambda_{a}-\frac{1}{4}\left(
\Delta^{a}S\right)  \overleftarrow{U}_{a}\lambda^{2}\right]  \exp\left[
\ln\left(  1-\frac{1}{2}\Lambda\overleftarrow{U}{}^{2}\right)  ^{-2}\right]
\ ,\label{superJaux6}\\
&  d\mathsf{\Gamma}{}^{\prime}=d\mathsf{\Gamma}\,\mathrm{Sdet}\left\{
\frac{\delta\left[  \mathsf{\Gamma}\check{g}(\Lambda\overleftarrow{U}%
_{a})\right]  }{\delta\mathsf{\Gamma}}\right\}  =d\mathsf{\Gamma}%
\ \exp\left\{  \frac{i}{\hbar}\left[  {i}{\hbar}\left(  \Delta^{a}S\right)
\lambda_{a}+\frac{i\hbar}{4}\left(  \Delta^{a}S\right)  \overleftarrow{U}%
_{a}\lambda^{2}+i\hbar\,\mathrm{\ln}\left(  1-\frac{1}{2}\Lambda
\overleftarrow{U}{}^{2}\right)  ^{2}\right]  \right\}  \,. \label{superJ7}%
\end{align}
Using the Jacobian (\ref{superJaux6}), the transformation of the action
$\mathcal{S}_{F}$ according to (\ref{identsg}), the equations (\ref{3.3}) with
their consequence resulting from applying $\overleftarrow{\mathsf{s}}^{a}$,
and the BRST-antiBRST exactness of the term $F\overleftarrow{U}{}^{2}$, we
arrive at \cite{MRnew2, MRnew3}
\begin{align}
Z_{F}  &  \overset{\mathsf{\Gamma}\rightarrow{\mathsf{\Gamma}{}^{\prime}}}%
{=}\int d\mathsf{\Gamma}\;\exp\left\{  \frac{i}{\hbar}\left[  \mathcal{S}%
_{F}+\left(  \mathcal{S}_{F}\overleftarrow{\mathsf{s}}^{a}+{i}{\hbar}%
\Delta^{a}S\right)  \lambda_{a}+\frac{1}{4}\left(  \mathcal{S}_{F}%
\overleftarrow{\mathsf{s}}{}^{2}+{i}{\hbar}\Delta^{a}S\overleftarrow
{\mathsf{s}}_{a}\right)  \lambda^{2}+i\hbar\,\mathrm{\ln}\left(  1-\frac{1}%
{2}\Lambda\overleftarrow{U}{}^{2}\right)  ^{2}\right]  \right\} \nonumber\\
&  =\int d\mathsf{\Gamma}\;\exp\left\{  \frac{i}{\hbar}\left[  \mathcal{S}%
_{F+\Delta F}+i\hbar\,\mathrm{\ln}\left(  1-\frac{1}{2}\Lambda\overleftarrow
{\mathsf{s}}{}^{2}\right)  ^{2}+\frac{1}{2}\Delta F\overleftarrow{U}{}%
^{2}\right]  \right\}  \,. \label{zftrans}%
\end{align}
The coincidence of the vacuum functionals $Z_{F}$ and $Z_{F+\Delta F}$,
evaluated for the respective even-valued functionals $F$ and $F+\Delta F$, is
valid, together with a compensation equation for an unknown even-valued
functional $\Lambda$:%
\begin{equation}
i\hbar\,\mathrm{\ln}\left(  1-\frac{1}{2}\Lambda\overleftarrow{U}{}%
^{2}\right)  ^{2}=-\frac{1}{2}\Delta F\overleftarrow{U}{}^{2}\,.
\label{eqexpl}%
\end{equation}
An explicit solution of (\ref{eqexpl}) satisfying the solvability condition
(that both sides should be BRST-antiBRST-exact) has the usual form -- up to
$\overleftarrow{U}{}^{a}$-exact terms -- identical with (\ref{Lambda-Fsol1})
for the similar equations (\ref{superJ3m}), (\ref{superJ3m}) in Yang--Mills
theories and first-class constraint dynamical systems:
\begin{equation}
\Lambda\left(  \phi,\pi,\lambda|\Delta F\right)  =2\Delta F\left(  \Delta
F\overleftarrow{U}{}^{2}\right)  ^{-1}\left[  \exp\left(  -\frac{1}{4i\hbar
}\Delta F\overleftarrow{U}{}^{2}\right)  -1\right]  =\frac{1}{2i\hbar}\Delta
F\sum_{n=0}^{\infty}\frac{1}{\left(  n+1\right)  !}\left(  -\frac{1}{4i\hbar
}\Delta F\overleftarrow{U}{}^{2}\right)  ^{n}\ . \label{Lambda-Fsol14}%
\end{equation}
Conversely, the equation (\ref{eqexpl}) examined for a certain unknown change
$\Delta F$ of the gauge-fixing functional for a given functional
$\Lambda\left(  \phi,\pi,\lambda\right)  $ has the following solution, with
accuracy up to $\overleftarrow{U}{}^{a}$-exact terms:%
\begin{equation}
\Delta F\left(  \phi,\pi,\lambda\right)  =-2i\hbar\ \Lambda\left(  \phi
,\pi,\lambda\right)  \left(  \Lambda\left(  \phi,\pi,\lambda\right)
\overleftarrow{U}{}^{2}\right)  ^{-1}\ln\left(  1-\Lambda\left(  \phi
,\pi,\lambda\right)  \overleftarrow{U}{}^{2}/2\right)  ^{2}\ .
\label{Lambda-Fsol5}%
\end{equation}
Field-dependent transformations with the functional-dependent parameters
$\lambda_{a}=\Lambda\overleftarrow{U}_{a}$ given by
\[
\lambda_{a}=\frac{1}{2i\hbar}\left(  \Delta F\overleftarrow{U}_{a}\right)
\sum_{n=0}^{\infty}\frac{1}{\left(  n+1\right)  !}\left(  \frac{1}{4i\hbar
}\Delta F\overleftarrow{U}^{2}\right)  ^{n}%
\]
amount to a precise change of the gauge-fixing functional in a general gauge theory.

It has been shown \cite{MRnew2} that the relation $Z_{F}=Z_{F+\Delta F}$ in
(\ref{zftrans}) leads to the presence of a modified Ward identity,
\begin{equation}
\left\langle \left\{  1+\frac{i}{\hbar}J_{A}\phi^{A}\left[  \overleftarrow
{U}^{a}\lambda_{a}(\Lambda)+\frac{1}{4}\overleftarrow{U}^{2}\lambda
^{2}(\Lambda)\right]  -\frac{1}{4}\left(  \frac{i}{\hbar}\right)  {}^{2}%
J_{A}\phi^{A}\overleftarrow{U}^{a}J_{B}(\phi^{B})\overleftarrow{U}_{a}%
\lambda^{2}(\Lambda)\right\}  \left(  1-\frac{1}{2}\Lambda\overleftarrow
{U}^{2}\right)  {}^{-2}\right\rangle _{F,J}=1\ , \label{mWInew}%
\end{equation}
and allows one to study the gauge dependence of $Z_{F}(J)$ in (\ref{z(0)}) for
a finite change of the gauge $F\rightarrow F+\Delta F$,%
\begin{align}
Z_{F+\Delta F}(J)  &  =Z_{F}(J)\left\{  1+\left\langle \frac{i}{\hbar}%
J_{A}\phi^{A}\left[  \overleftarrow{U}^{a}\lambda_{a}\left(  \mathsf{\Gamma
}|-\Delta{F}\right)  +\frac{1}{4}\overleftarrow{U}^{2}\lambda^{2}\left(
\mathsf{\Gamma}|-\Delta{F}\right)  \right]  \right.  \right. \nonumber\\
&  -\left.  \left.  (-1)^{\varepsilon_{B}}\left(  \frac{i}{2\hbar}\right)
^{2}J_{B}J_{A}\left(  \phi^{A}\overleftarrow{U}{}^{a}\right)  \left(  \phi
^{B}\overleftarrow{U}_{a}\right)  \lambda^{2}\left(  \mathsf{\Gamma}%
|-\Delta{F}\right)  \right\rangle _{F,J}\right\}  \ , \label{GDInew}%
\end{align}
where the symbol \textquotedblleft$\langle\mathcal{A}\rangle_{F,J}%
$\textquotedblright\ for a quantity $\mathcal{A}=\mathcal{A}(\mathsf{\Gamma})$
stands for a source-dependent average expectation value corresponding to a
gauge-fixing $F(\phi,\pi,\lambda)$:
\begin{equation}
\left\langle \mathcal{A}\right\rangle _{F,J}=Z_{F}^{-1}(J)\int d\mathsf{\Gamma
}\ \mathcal{A}\left(  \mathsf{\Gamma}\right)  \exp\left\{  \frac{i}{\hbar
}\left[  \mathcal{S}_{F}\left(  \mathsf{\Gamma}\right)  +J_{A}\phi^{A}\right]
\right\}  \ ,\ \ \ \mathrm{where\ \ \ }\left\langle 1\right\rangle _{F,J}=1\ .
\label{aexvnew}%
\end{equation}
In the case of constant $\lambda_{a}$, the modified Ward identity
(\ref{mWInew}) contains an $\mathrm{Sp}(2)$-doublet of the usual Ward
identities at the first order in $\lambda_{a}$ and a derivative identity at
the second order in $\lambda_{a}$:
\begin{equation}
J_{A}\left\langle \phi^{A}\overleftarrow{U}{}^{a}\right\rangle _{F,J}%
=0\ ,\ \ \ \left\langle J_{A}\left[  \phi^{A}\overleftarrow{U}{}^{2}+\left(
i/\hbar\right)  \varepsilon_{ab}\phi^{A}\overleftarrow{U}{}^{a}J_{B}(\phi
^{B}\overleftarrow{U}{}^{b})\right]  \right\rangle _{F,J}=0\ .
\label{WIlag3fin}%
\end{equation}
In the case of first-rank gauge theories with a closed gauge algebra,
$M_{\alpha\beta}^{ij}=0$, in (\ref{gauge_alg}), provided that the solution to
the generating equations (\ref{3.3}) is linear in the antifields $\phi
_{Aa}^{\ast}$, $\bar{\phi}_{A}$, the representation (\ref{z(0)}) for
$Z_{F}(J)$ reduces to (\ref{z(j)}), with the action $S_{F}(\phi)$ being
identical to (\ref{action}) in irreducible gauge theories (\ref{solxy}),
(\ref{xy}), in particular, Yang--Mills theories (\ref{xyYM}). The study of
finite (field-dependent) BRST-antiBRST transformations and their consequences
to the quantum properties of a theory is then reduced to the results of
Section~\ref{YMgen}. Below, we intend to study the case of finite
field-dependent BRST-antiBRST transformations for general gauge theories with
arbitrary functional parameters, generally assumed to be
functionally-independent, $\lambda_{a}\not \equiv s_{a}\Lambda$.

\section{Linearized Finite BRST-antiBRST Transformations}

\label{linear} \renewcommand{\theequation}{\arabic{section}.\arabic{equation}} \setcounter{equation}{0}

In this section, we examine the calculation of the Jacobian for the linear
part of finite field-dependent BRST-antiBRST transformations, i.e., the part
being linear in functionally-dependent parameters of the form $\lambda
_{a}\left(  \phi\right)  =s_{a}\Lambda\left(  \phi\right)  $ and $\lambda
_{a}\left(  \Gamma\right)  =\int dt\ s_{a}\Lambda\left(  \Gamma\right)  $,
respectively, in Yang--Mills theories and arbitrary dynamical systems with
first-class constraints. We shall carry out the explicit calculations in the
Yang--Mills case and then translate the resulting Jacobian to the case of
dynamical systems in question, using the anticommutativity, $s^{a}s^{b}%
+s^{b}s^{a}=0$, of the corresponding generators $s^{a}$ and the invariance of
the functional integration measure, $d\phi$ and $d\Gamma$, under global
BRST-antiBRST transformations, $\lambda_{a}=\mathrm{const}$, in both these cases.

\subsection{Yang--Mills Theories}

\label{linearYM}

In the Yang--Mills case, the linear part of finite field-dependent
BRST-antiBRST transformations\ in question has the form
\begin{equation}
\phi^{A}\rightarrow\phi^{\prime A}=\phi^{A}+\Delta\phi^{A}%
\ ,\ \ \ \mathrm{where}\ \ \ \Delta\phi^{A}=\left(  s^{a}\phi^{A}\right)
\lambda_{a}=X^{Aa}\lambda_{a}~,\ \ \ \lambda_{a}=s_{a}\Lambda\ .
\label{liearized}%
\end{equation}
Let us examine the even matrix $M=\|M_{B}^{A}\|$%
\[
\frac{\delta\left(  \Delta\phi^{A}\right)  }{\delta\phi^{B}}=M_{B}%
^{A}\ ,\ \ \ \varepsilon\left(  M_{B}^{A}\right)  =\varepsilon_{A}%
+\varepsilon_{B}\ ,
\]
and the corresponding Jacobian $\exp\left(  \Im\right)  $%
\[
\Im=\mathrm{Str}\ln\left(  \mathbb{I}+M\right)  =-\sum_{n=1}^{\infty}%
\frac{\left(  -1\right)  ^{n}}{n}\mathrm{Str}\left(  M^{n}\right)
\ ,\ \ \ \mathrm{where}\ \ \ \mathrm{Str}\left(  M^{n}\right)  =\left(
M^{n}\right)  _{A}^{A}\left(  -1\right)  ^{\varepsilon_{A}}\ ,\ \ \ \mathbb{I}%
_{B}^{A}=\delta_{B}^{A}\ .
\]
Explicitly, the matrix $M_{B}^{A}$ is given by the sum of two even matrices:%
\begin{align}
M_{B}^{A}  &  =X^{Aa}\frac{\delta\lambda_{a}}{\delta\phi^{B}}+\frac{\delta
X^{Aa}}{\delta\phi^{B}}\lambda_{a}\left(  -1\right)  ^{\varepsilon_{B}}\equiv
P_{B}^{A}+Q_{B}^{A}\ ,\nonumber\\
P_{B}^{A}  &  =X^{Aa}\frac{\delta\lambda_{a}}{\delta\phi^{B}}\ ,\ \ \ Q_{B}%
^{A}=\frac{\delta X^{Aa}}{\delta\phi^{B}}\lambda_{a}\left(  -1\right)
^{\varepsilon_{B}}\ . \label{notation1}%
\end{align}
Further considerations are based on the following statements, established in
our previous work \cite{MRnew}, using the properties%
\[
X_{,A}^{Aa}=0\ ,\ \ \ s^{b}s^{a}\phi^{A}=s^{b}X^{Aa}=\varepsilon^{ab}Y^{A}\ ,
\]
which take place in Yang--Mills theories, and the supertrace property%
\begin{equation}
\mathrm{Str}\left(  AB\right)  =\mathrm{Str}\left(  BA\right)  \,,
\label{transp}%
\end{equation}
which takes place for arbitrary even matrices $A$ and $B$:

\begin{proposition}
\label{StrQ}The matrices (\ref{notation1}) with arbitrary odd-valued
$\lambda_{a}\not \equiv s_{a}\Lambda$ obey the properties%
\begin{align}
&  \mathrm{Str}\left(  P+Q\right)  ^{n}=\mathrm{Str}\left(  P^{n}%
+nP^{n-1}Q+C_{n}^{2}P^{n-2}Q^{2}\right)  \ ,\ \ \ \mathrm{where\ \ \ }%
C_{n}^{k}=\frac{n!}{k!\left(  n-k\right)  !}\ ,\ \ \ n=2,3\ ,\label{2,3}\\
&  \mathrm{Str}\left(  Q\right)  =0\ ,\ \ \ \mathrm{Str}\left(  Q^{2}\right)
=2\mathrm{Str}\left(  R\right)  \ ,\ \ \ \mathrm{where}\ \ \ R_{B}^{A}%
\equiv-\frac{1}{2}\lambda^{2}\frac{\delta Y^{A}}{\delta\phi^{B}}\ .
\label{lemmaStr(Q)}%
\end{align}

\end{proposition}

\begin{proposition}
\label{P^2}Let us suppose that the condition $\lambda_{a}=s_{a}\Lambda$ is
fulfilled. Then there hold the properties\footnote{Further on, we will use
different forms of the same matrices: $Q_{B}^{A}=X_{,B}^{Aa}\lambda_{a}\left(
-1\right)  ^{\varepsilon_{B}}=\lambda_{a}X_{,B}^{Aa}\left(  -1\right)
^{\varepsilon_{A}+1}$,$\ \left(  Q_{2}\right)  _{B}^{A}=\lambda_{a}%
Y^{A}\lambda_{,B}^{a}\left(  -1\right)  ^{\varepsilon_{A}+1}=-\left(
1/2\right)  Y^{A}\lambda_{,B}^{2}$.}%
\begin{align}
&  P^{2}=f\cdot P\Longrightarrow P^{n}=f\cdot P^{n-1}\ ,\ \mathrm{where}%
\ s^{a}\lambda_{b}=\delta_{b}^{a}f\ \Longrightarrow f=-\frac{1}{2}s^{2}%
\Lambda=-\frac{1}{2}\mathrm{Str}\left(  P\right)  \ ,\label{contracprop}\\
&  QP=\left(  1+f\right)  \cdot Q_{2}\ ,\ \mathrm{where}\ \left(
Q_{2}\right)  _{B}^{A}\equiv\lambda_{a}Y^{A}\frac{\delta\lambda^{a}}%
{\delta\phi^{B}}\left(  -1\right)  ^{\varepsilon_{A}+1}\ ,\label{QP}\\
&  \mathrm{Str}\left(  P+Q\right)  ^{n}=\mathrm{Str}\left(  P^{n}%
+nP^{n-1}Q^{1}+nP^{n-2}Q^{2}+K_{n}P^{n-3}QPQ\right)  ,\ \mathrm{where}%
\ K_{n}\equiv C_{n}^{2}-C_{n}^{1}=\frac{n\left(  n-3\right)  }{2}%
\ ,\ n\geq4\ . \label{Str(P+Q)^n}%
\end{align}

\end{proposition}

\noindent Note: the equality (\ref{2,3}) is entirely due to the Grassmann
parity of $P$, $Q$ and the character of dependence of these matrices on
$\lambda_{a}$; the property $\mathrm{Str}\left(  Q\right)  =0$ in
(\ref{lemmaStr(Q)}) translates to the invariance, $d\phi^{\prime}=d\phi$, of
functional integration measure under global BRST-antiBRST transformations
$\delta\phi^{A}=\left(  s^{a}\phi^{A}\right)  \lambda_{a}$, $\lambda
_{a}=\mathrm{const}$, while the property $\mathrm{Str}\left(  Q^{2}\right)
=2\mathrm{Str}\left(  R\right)  $ is implied by the anticommutativity,
$s^{a}s^{b}+s^{b}s^{a}=0$, of the generators $s^{a}$, as well as by the
above-mentioned invariance of functional integration measure, encoded in
$\mathrm{Str}\left(  Q\right)  =0$;\ the properties (\ref{contracprop}),
(\ref{QP}), substantially related to $\lambda_{a}=s_{a}\Lambda$, are implied
by the anticommutativity of the generators $s^{a}$; the combinatorial
coefficient $K_{n}$\ in (\ref{Str(P+Q)^n}) corresponds to the decomposition of
the binomial coefficient $C_{n}^{2}$ into two parts:\ $C_{n}^{1}$ and
$K_{n}=C_{n}^{2}-C_{n}^{1}$; in fact, the coefficient $K_{n}$ is the number of
monomials in $\left(  P+Q\right)  ^{n}$\ for $n\geq4$ that contain two
matrices $Q$\ and cannot be transformed by cyclic permutations under the
symbol $\mathrm{Str}$ of supertrace to the form $\mathrm{Str}(P^{n-2}Q^{2})$
by using (\ref{transp}); in virtue of the contraction property $P^{n}=f\cdot
P^{n-1}$ in (\ref{contracprop}) the supertrace of all such monomials is equal
to $\mathrm{Str}\left(  P^{n-3}QPQ\right)  $; Propositions \ref{StrQ},
\ref{P^2} will be proved independently in Subsection \ref{Yang-Mills}, which
deals with the case of arbitrary parameters $\lambda_{a}\not \equiv
s_{a}\Lambda$.

From the above properties (\ref{transp})--(\ref{Str(P+Q)^n}), it follows that
the quantity $\Im$ takes the form (see Appendix \ref{AppB})%
\begin{equation}
\Im=-2\ln\left(  1+f\right)  +\Re\ ,\ \ \ \Re=-\mathrm{Str}\left[
R+Q_{2}+\left(  1/2\right)  Q_{2}^{2}-\left(  1+f\right)  QQ_{2}\right]  \ ,
\label{J_lin_final}%
\end{equation}
where%
\[
f=-\frac{1}{2}s^{2}\Lambda=\frac{1}{2}s^{a}s_{a}\Lambda\ .
\]
Substituting the explicit form of the matrices (\ref{notation1}),
(\ref{lemmaStr(Q)}), (\ref{QP}), we have%
\begin{equation}
\Re=\frac{1}{2}\left(  -1\right)  ^{\varepsilon_{A}}\left[  \frac
{\delta\left(  Y^{A}\lambda^{2}\right)  }{\delta\phi^{A}}-\frac{1}{4}%
Y^{A}\frac{\delta\lambda^{2}}{\delta\phi^{B}}Y^{B}\frac{\delta\lambda^{2}%
}{\delta\phi^{A}}-\left(  1+f\right)  \frac{\delta X^{Aa}}{\delta\phi^{B}%
}Y^{B}\lambda_{a}\frac{\delta\lambda^{2}}{\delta\phi^{A}}\right]  \ .
\label{R_final1}%
\end{equation}
Using the relations $X^{Aa}=s^{a}\phi^{A}$,$\ Y^{A}=-\left(  1/2\right)
s^{2}\phi^{A}$ and bearing in mind that $\lambda_{a}=s_{a}\Lambda$, one can
represent the contribution (\ref{R_final1}) in terms of BRST-antiBRST
variations:%
\begin{equation}
\Re=-\frac{1}{4}\left(  -1\right)  ^{\varepsilon_{A}}\left\{  \left[  \left(
s^{2}\phi^{A}\right)  \lambda^{2}\right]  _{,A}+\frac{1}{8}\left(  s^{2}%
\phi^{A}\right)  \lambda_{,B}^{2}\left(  s^{2}\phi^{B}\right)  \lambda
_{,A}^{2}-\left(  1-\frac{1}{2}s^{2}\Lambda\right)  \left(  s^{a}\phi
^{A}\right)  _{,B}\left(  s^{2}\phi^{B}\right)  \lambda_{a}\lambda_{,A}%
^{2}\right\}  \ . \label{R_final2}%
\end{equation}
Let us now examine the transformation of the integrand%
\[
d\phi\exp\left[  \left(  i/\hbar\right)  S_{F}\left(  \phi\right)  \right]
\]
in the Yang--Mills path integral under the linearized finite BRST-antiBRST
transformations (\ref{liearized}):%
\begin{align}
\left.  d\phi\exp\left[  \left(  i/\hbar\right)  S_{F}\left(  \phi\right)
\right]  \right\vert _{\phi\rightarrow\phi^{\prime}}  &  =d\phi\ J\left(
\phi\right)  \exp\left[  \left(  i/\hbar\right)  S_{F}\left(  \phi+\Delta
\phi\right)  \right]  =d\phi\ \exp\left\{  \left(  i/\hbar\right)  \left[
S_{F}\left(  \phi+\Delta\phi\right)  -i\hbar\Im\left(  \phi\right)  \right]
\right\} \label{integrand_linear}\\
&  =d\phi\ \exp\left\{  \left(  i/\hbar\right)  \left[  S_{F}\left(
\phi+\Delta\phi\right)  +i\hbar\ln\left[  1-\left(  1/2\right)  s^{2}%
\Lambda\left(  \phi\right)  \right]  ^{2}-i\hbar\Re\left(  \phi\right)
\right]  \right\}  \ ,\nonumber
\end{align}
where%
\begin{align}
S_{F}\left(  \phi+\Delta\phi\right)   &  =S_{F}\left(  \phi\right)
+\frac{\delta S_{F}}{\delta\phi^{A}}\left(  \phi\right)  \Delta\phi^{A}%
+\frac{1}{2}\frac{\delta^{2}S_{F}}{\delta\phi^{A}\delta\phi^{B}}\left(
\phi\right)  \Delta\phi^{B}\Delta\phi^{A}\nonumber\\
&  =S_{F}\left(  \phi\right)  +\frac{1}{2}\left(  -1\right)  ^{\varepsilon
_{A}}\frac{\delta^{2}S_{F}}{\delta\phi^{A}\delta\phi^{B}}\left(  \phi\right)
X^{Bb}\left(  \phi\right)  X^{Aa}\left(  \phi\right)  \lambda_{a}\left(
\phi\right)  \lambda_{b}\left(  \phi\right)  \ . \label{Taylor}%
\end{align}
Here, the first order of expansion in $\Delta\phi^{A}=X^{Aa}\lambda_{a}$ drops
out due to the invariance property $s^{a}S_{F}=0$,%
\begin{equation}
s^{a}S_{F}=\frac{\delta S_{F}}{\delta\phi^{A}}X^{Aa}=0\ . \label{invar_prop}%
\end{equation}
Then, differentiating the above relation,%
\begin{equation}
\frac{\delta S_{F}}{\delta\phi^{A}}\frac{\delta X^{Aa}}{\delta\phi^{B}}%
+\frac{\delta^{2}S_{F}}{\delta\phi^{B}\delta\phi^{A}}X^{Aa}\left(  -1\right)
^{\varepsilon_{B}}=0\ , \label{differentiate}%
\end{equation}
multiplying the result from the right by the quantity $X^{Bb}\lambda
_{b}\lambda_{a}$ and using the property $X_{,B}^{Aa}X^{Bb}=\varepsilon
^{ab}Y^{A}$, we obtain%
\begin{equation}
S_{F}\left(  \phi+\Delta\phi\right)  =S_{F}\left(  \phi\right)  +\frac{1}%
{2}\frac{\delta S_{F}}{\delta\phi^{A}}\left(  \phi\right)  Y^{A}\left(
\phi\right)  \lambda^{2}\left(  \phi\right)  =S_{F}\left(  \phi\right)
-\frac{1}{4}\frac{\delta S_{F}}{\delta\phi^{A}}\left(  \phi\right)  \left(
s^{2}\phi^{A}\right)  \lambda^{2}\left(  \phi\right)  \ , \label{Taylor2}%
\end{equation}
which implies the following transformation of the integrand:%
\begin{equation}
\left.  d\phi\exp\left[  \frac{i}{\hbar}S_{F}\left(  \phi\right)  \right]
\right\vert _{\phi\rightarrow\phi^{\prime}}=d\phi\exp\left\{  \frac{i}{\hbar
}\left[  S_{F}\left(  \phi\right)  +i\hbar\ln\left(  1-\frac{1}{2}s^{2}%
\Lambda\left(  \phi\right)  \right)  ^{2}+S_{F}^{\mathrm{add}}\left(
\phi\right)  \right]  \right\}  \equiv d\phi\exp\left[  \frac{i}{\hbar
}S^{\prime}\left(  \phi\right)  \right]  \ , \label{integrand_trans}%
\end{equation}
where%
\begin{equation}
S_{F}^{\mathrm{add}}=\frac{1}{2}\frac{\delta S_{F}}{\delta\phi^{A}}%
Y^{A}\lambda^{2}-i\hbar\Re\ . \label{S_add}%
\end{equation}
The above expression is obviously not BRST-antiBRST-invariant: $s^{a}%
S_{F}^{\mathrm{add}}\not \equiv 0$. As a consequence, the corresponding
quantum action $S^{\prime}\left(  \phi\right)  $ fails to be
BRST-antiBRST-invariant, $s^{a}S^{\prime}\not \equiv 0$, and therefore it does
not amount to an exact change of the gauge-fixing functional:%
\begin{equation}
S^{\prime}\left(  \phi\right)  \not =S_{0}\left(  A\right)  -\frac{1}{2}%
s^{2}F^{\prime}\left(  \phi\right)  \ . \label{Sprime}%
\end{equation}
Finally, it should be noted that the integrand fails to be invariant under
global linearized finite BRST-antiBRST transformations, $\lambda
_{a}=\mathrm{const}$:%
\begin{equation}
\left.  d\phi\exp\left[  \frac{i}{\hbar}S_{F}\left(  \phi\right)  \right]
\right\vert _{\phi\rightarrow\phi^{\prime}}-d\phi\exp\left[  \frac{i}{\hbar
}S_{F}\left(  \phi\right)  \right]  =d\phi\exp\left[  \frac{i}{\hbar}%
S_{F}\left(  \phi\right)  \right]  \left\{  \exp\left[  \frac{i}{\hbar}%
S_{F}^{\mathrm{add}}\left(  \phi\right)  \right]  -1\right\}  \not =0\ ,
\label{difference_measure}%
\end{equation}
where $S_{F}^{\mathrm{add}}$ reduces to\footnote{Even though in Yang--Mills
theories there hold the properties $Y_{,i}^{i}=Y_{,\alpha}^{\alpha}=Y_{,\alpha
a}^{\alpha a}=0\Longrightarrow\left(  -1\right)  ^{\varepsilon_{A}}Y_{,A}%
^{A}=0$, the quantity $S_{F}^{\mathrm{add}}$ does not vanish identically,
$S_{F}^{\mathrm{add}}\not \equiv 0$, so that the invariance of the integrand
in the vacuum functional under global linearized finite BRST-antiBRST
transformations can only take place on solutions of the equation $S_{F,A}%
Y^{A}=0$.}%
\begin{equation}
S_{F}^{\mathrm{add}}=\frac{1}{2}\frac{\delta S_{F}}{\delta\phi^{A}}%
Y^{A}\lambda^{2}-\frac{i\hbar}{2}\left(  -1\right)  ^{\varepsilon_{A}}%
\frac{\delta Y^{A}}{\delta\phi^{A}}\lambda^{2}\not \equiv 0\ , \label{Sadd2}%
\end{equation}
which implies that linearized finite BRST-antiBRST transformations can be
interpreted neither as global symmetry transformations of the integrand nor as
field-dependent transformations inducing an exact change of the gauge-fixing
functional. Therefore, they do not possess the properties of finite
BRST-antiBRST transformations.

\subsection{Constrained Dynamical Systems\label{constrained}}

The case of arbitrary dynamical systems with first-class constraints can be
examined in complete analogy with the Yang--Mills case and is based on the
propositions and considerations of Subsection \ref{linearYM}. Namely, in the
case of dynamical systems in question, the linear part of finite
field-dependent BRST-antiBRST transformations for phase-space trajectories
$\Gamma_{t}^{p}$ has the form
\begin{equation}
\Gamma_{t}^{p}\rightarrow\check{\Gamma}_{t}^{p}=\Gamma_{t}^{p}+\Delta
\Gamma_{t}^{p}\ ,\ \ \ \mathrm{where}\ \ \ \Delta\Gamma_{t}^{p}=\left(
s^{a}\Gamma_{t}^{p}\right)  \lambda_{a}=X_{t}^{pa}\lambda_{a}\ ,\ \ \ \lambda
_{a}\left(  \Gamma\right)  =\int dt\ s_{a}\Lambda\left(  \Gamma\right)  \ .
\label{linearized_Ham}%
\end{equation}
Let us examine the even matrix $M=\|M_{q|t^{\prime},t^{\prime\prime}}^{p}\|$%
\[
\frac{\delta\left(  \Delta\Gamma_{t^{\prime}}^{p}\right)  }{\delta
\Gamma_{t^{\prime\prime}}^{q}}=M_{q|t^{\prime},t^{\prime\prime}}%
^{p}\ ,\ \ \ \varepsilon( M_{q|t^{\prime},t^{\prime\prime}}^{p})
=\varepsilon_{p}+\varepsilon_{q}%
\]
and the corresponding Jacobian $\exp\left(  \Im\right)  $%
\begin{align}
\Im &  =\mathrm{Str}\ln\left(  \mathbb{I}+M\right)  =-\sum_{n=1}^{\infty}%
\frac{\left(  -1\right)  ^{n}}{n}\,\,\mathrm{Str}\left(  M^{n}\right)
,\,\,\,\mathrm{\ Str}\left(  M^{n}\right)  =\left(  -1\right)  ^{\varepsilon
_{p}}\int dt\ \left(  M^{n}\right)  _{p}^{p}\left(  t,t\right)  \,,\nonumber\\
\mathbb{I}  &  \mathbb{=\delta}_{q}^{p}\delta\left(  t^{\prime}-t^{\prime
\prime}\right)  \ ,\ \ \ \left(  M\right)  _{q}^{p}\left(  t^{\prime
},t^{\prime\prime}\right)  =\frac{\delta\Delta\Gamma^{p}\left(  t^{\prime
}\right)  }{\delta\Gamma^{q}\left(  t^{\prime\prime}\right)  }\ ,\ \ \ \left(
AB\right)  _{q}^{p}\left(  t^{\prime},t^{\prime\prime}\right)  =\int
dt\ \left(  A\right)  _{r}^{p}\left(  t^{\prime},t\right)  B_{q}^{r}\left(
t,t^{\prime\prime}\right)  \ . \label{superJ}%
\end{align}
Explicitly, the matrix $M$ is given by the sum of two even matrices:%
\begin{align}
&  M_{q|t^{\prime},t^{\prime\prime}}^{p}=X_{t^{\prime}}^{pa}\frac
{\delta\lambda_{a}}{\delta\Gamma_{t^{\prime\prime}}^{q}}+\frac{\delta
X_{t^{\prime}}^{pa}}{\delta\Gamma_{t^{\prime\prime}}^{q}}\lambda_{a}\left(
-1\right)  ^{\varepsilon_{p}}\equiv U_{q|t^{\prime},t^{\prime\prime}}%
^{p}+V_{q|t^{\prime},t^{\prime\prime}}^{p}\ ,\nonumber\\
&  U_{q|t^{\prime},t^{\prime\prime}}^{p}=X_{t^{\prime}}^{pa}\frac
{\delta\lambda_{a}}{\delta\Gamma_{t^{\prime\prime}}^{q}}\ ,\ \ \left(
V\right)  _{q|t^{\prime},t^{\prime\prime}}^{p}=\frac{\delta X_{t^{\prime}%
}^{pa}}{\delta\Gamma_{t^{\prime\prime}}^{q}}\lambda_{a}\left(  -1\right)
^{\varepsilon_{q}}\ . \label{defU,V}%
\end{align}
The matrices $U$, $V$ correspond to the matrices $P$, $Q$\ of Subsection
\ref{linearYM}. This correspondence is given explicitly by Table
\ref{table in}.{\ \begin{table}[ptbh]
{\ }
\par
\begin{center}%
\begin{tabular}
[c]{||l|l||}\hline\hline
First-class constraint{\ systems} & {Yang--Mills theories}\\\hline\hline
$\Gamma_{t}^{p}\ ,\ \Delta\Gamma_{t}^{p}=\left(  s^{a}\Gamma_{t}^{p}\right)
\lambda_{a}\ ,\ \lambda^{a}\left(  \Gamma\right)  =\int dt\ s^{a}%
\Lambda\left(  \Gamma\right)  $ & $\phi^{A}\ ,\ \Delta\phi^{A}=\left(
s^{a}\phi^{A}\right)  \lambda_{a},\ \lambda^{a}\left(  \phi\right)
=s^{a}\Lambda\left(  \phi\right)  $\\
$s^{a}\Gamma_{t}^{p}=X_{t}^{pa},\ s^{b}s^{a}\Gamma_{t}^{p}=\int dt^{\prime
}\ \frac{\delta X_{t}^{pa}}{\delta\Gamma_{t^{\prime}}^{q}}X_{t^{\prime}}%
^{qb}=\varepsilon^{ab}Y_{t}^{p}$ & $s^{a}\phi^{A}=X^{Aa},\ s^{b}s^{a}\phi
^{A}=\frac{\delta X^{Aa}}{\delta\phi^{B}}X^{Bb}=\varepsilon^{ab}Y^{A}$\\
$\int dt\ \frac{\delta X_{t}^{pa}}{\delta\Gamma_{t}^{p}}=0,\ Y_{t}^{p}%
=-\frac{1}{2}\varepsilon_{ab}\int dt^{\prime}\frac{\delta X_{t}^{pa}}%
{\delta\Gamma_{t^{\prime}}^{q}}X_{t^{\prime}}^{Bb}$ & $\frac{\delta X^{Aa}%
}{\delta\phi^{A}}=0,\ Y^{A}=-\frac{1}{2}\varepsilon_{ab}\frac{\delta X^{Aa}%
}{\delta\phi^{B}}X^{Bb}$\\
$\frac{\delta\left(  \Delta\Gamma_{t^{\prime}}^{p}\right)  }{\delta
\Gamma_{t^{\prime\prime}}^{q}}=M_{q|t^{\prime},t^{\prime\prime}}%
^{p}=U_{q|t^{\prime},t^{\prime\prime}}^{p}+V_{q|t^{\prime},t^{\prime\prime}%
}^{p}$ & $\frac{\delta\left(  \Delta\phi^{A}\right)  }{\delta\phi^{B}}%
=M_{B}^{A}=P_{B}^{A}+Q_{B}^{A}$\\
$U_{q|t^{\prime},t^{\prime\prime}}^{p}=X_{t^{\prime}}^{pa}\frac{\delta
\lambda_{a}}{\delta\Gamma_{t^{\prime\prime}}^{q}}\ ,\ \left(  V\right)
_{q|t^{\prime},t^{\prime\prime}}^{p}=\frac{\delta X_{t^{\prime}}^{pa}}%
{\delta\Gamma_{t^{\prime\prime}}^{q}}\lambda_{a}\left(  -1\right)
^{\varepsilon_{q}}$ & $P_{B}^{A}=X^{Aa}\frac{\delta\lambda_{a}}{\delta\phi
^{B}}\ ,\ \left(  Q\right)  _{B}^{A}=\frac{\delta X^{Aa}}{\delta\phi^{B}%
}\lambda_{a}\left(  -1\right)  ^{\varepsilon_{B}}$\\\hline\hline
\end{tabular}
\end{center}
\par
\vspace{-2ex}\caption{Correspondence of the matrix elements in arbitrary
first-class constraint systems and Yang--Mills theories. Linearized
field-dependent BRST-antiBRST transformations.}%
\label{table in}%
\end{table}}

In this connection, due to the property $\mathrm{Str}\left(  AB\right)
=\mathrm{Str}\left(  BA\right)  $ for even matrices, Propositions \ref{StrQ},
\ref{P^2} of Subsection\textbf{\ }\ref{linearYM} remain formally the
same\footnote{One should, of course, take into account that formal summation
over the time variable included in the index $A$ is replaced by explicit
integration over $t$ in terms of $\left(  p,t\right)  $. \label{footnote}} in
terms of $U_{q|t^{\prime},t^{\prime\prime}}^{p}$, $V_{q|t^{\prime}%
,t^{\prime\prime}}^{p}$, substituted instead of the respective matrices
$P_{B}^{A}$, $Q_{B}^{A}$, which establishes the following

\begin{proposition}
\label{StrV}The matrices (\ref{notation1}) with arbitrary odd-valued
$\lambda_{a}\not \equiv s_{a}\int dt\ \Lambda$ obey the properties%
\begin{align}
&  \mathrm{Str}\left(  U+V\right)  ^{n}=\mathrm{Str}\left(  U^{n}%
+nU^{n-1}V+C_{n}^{2}U^{n-2}V^{2}\right)  \ ,\ \ \ \mathrm{where\ \ \ }%
C_{n}^{k}=\frac{n!}{k!\left(  n-k\right)  !}\ ,\ \ \ n=2,3\ ,\label{(U+V)^n}\\
&  \mathrm{Str}\left(  V\right)  =0\ ,\ \ \ \mathrm{Str}\left(  V^{2}\right)
=2\mathrm{Str}\left(  W\right)  \ ,\ \ \ \mathrm{where}\ \ \ W_{q}^{p}%
\equiv-\frac{1}{2}\lambda^{2}\frac{\delta Y^{p}}{\delta\Gamma^{q}}\ .
\label{W}%
\end{align}

\end{proposition}

\begin{proposition}
\label{U^2}Let us suppose that the condition $\lambda_{a}=\int dt\ s_{a}%
\Lambda$ is fulfilled. Then there hold the properties\footnote{Further on, we
will use different forms of the same matrices: $V_{q}^{p}=X_{,q}^{pa}%
\lambda_{a}\left(  -1\right)  ^{\varepsilon_{q}}=\lambda_{a}X_{,q}^{pa}\left(
-1\right)  ^{\varepsilon_{p}+1}$,$\ \left(  V_{2}\right)  _{q}^{p}=\lambda
_{a}Y^{p}\lambda_{,q}^{a}\left(  -1\right)  ^{\varepsilon_{p}+1}=-\left(
1/2\right)  Y^{p}\lambda_{,p}^{2}$.}%
\begin{align}
&  U^{2}=f\cdot U\Longrightarrow U^{n}=f\cdot U^{n-1}\ ,\ \mathrm{where}%
\ s^{a}\lambda_{b}=\delta_{b}^{a}f\ \Longrightarrow f=-\frac{1}{2}\int
dt\ s^{2}\Lambda=-\frac{1}{2}\mathrm{Str}\left(  U\right)  \ ,\label{U^2=fU}\\
&  VU=\left(  1+f\right)  \cdot V_{2}\ ,\ \mathrm{where}\ \left(
V_{2}\right)  _{q}^{p}\equiv\lambda_{a}Y^{p}\frac{\delta\lambda^{a}}%
{\delta\Gamma^{q}}\left(  -1\right)  ^{\varepsilon_{p}+1}\ ,
\label{VU=(1+f)V2}\\
&  \mathrm{Str}\left(  U+V\right)  ^{n}=\mathrm{Str}\left(  U^{n}%
+nU^{n-1}V^{1}+nU^{n-2}V^{2}+K_{n}U^{n-3}VUV\right)  ,\ \mathrm{where}%
\ K_{n}\equiv C_{n}^{2}-C_{n}^{1}=\frac{n\left(  n-3\right)  }{2}%
\ ,\ n\geq4\ . \label{str(U+V)^n}%
\end{align}

\end{proposition}

\noindent Note: these statements may be supplied by the same remarks that
follow Propositions \ref{StrQ}, \ref{P^2}, with the replacement of $\phi^{A}$,
$P_{B}^{A}$, $Q_{B}^{A}$,$\ R_{B}^{A}$ by $\Gamma^{p}$, $U_{q|t^{\prime
},t^{\prime\prime}}^{p}$, $V_{q|t^{\prime},t^{\prime\prime}}^{p}$,
$W_{q|t^{\prime},t^{\prime\prime}}^{p}$, respectively, and with the
replacement of $\lambda^{a}\left(  \phi\right)  =s^{a}\Lambda\left(
\phi\right)  $ by $\lambda^{a}\left(  \Gamma\right)  =\int dt\ s^{a}%
\Lambda\left(  \Gamma\right)  $; in particular, it may be emphasized that the
properties (\ref{W}), (\ref{U^2=fU}), (\ref{VU=(1+f)V2}) are implied by the
invariance, $d\check{\Gamma}=d\Gamma$, of functional integration measure under
global BRST-antiBRST transformations $\delta\Gamma^{p}=\left(  s^{a}\Gamma
^{p}\right)  \lambda_{a}$, $\lambda_{a}=\mathrm{const}$, being canonical
transformations of phase-space variables, as well as by the anticommutativity,
$s^{a}s^{b}+s^{b}s^{a}=0$, of the corresponding generators $s^{a}$.

From (\ref{(U+V)^n})--(\ref{str(U+V)^n}), with allowance for $\mathrm{Str}%
\left(  AB\right)  =\mathrm{Str}\left(  BA\right)  $, it follows that $\Im$
acquires the form, cf. (\ref{J_lin_final_Ham}),%
\begin{equation}
\Im=-2\ln\left(  1+f\right)  +\Re\ ,\ \ \ \Re=-\mathrm{Str}\left[
W+V_{2}+\left(  1/2\right)  V_{2}^{2}-\left(  1+f\right)  VV_{2}\right]  \ ,
\label{J_lin_final_Ham}%
\end{equation}
where%
\[
f=-\frac{1}{2}\int dt\ s^{2}\Lambda=\frac{1}{2}\int dt\ s^{a}s_{a}\Lambda\ .
\]
Substituting the explicit form of the matrices (\ref{defU,V}), (\ref{W}),
(\ref{VU=(1+f)V2}), we have, cf. (\ref{R_final1}),%
\begin{equation}
\Re=\frac{1}{2}\left(  -1\right)  ^{\varepsilon_{p}}\int dt\left[
\frac{\delta\left(  Y_{t}^{p}\lambda^{2}\right)  }{\delta\Gamma_{t}^{p}}%
-\frac{1}{4}Y_{t}^{p}\frac{\delta\lambda^{2}}{\delta\Gamma_{t}^{q}}Y_{t}%
^{q}\frac{\delta\lambda^{2}}{\delta\Gamma_{t}^{p}}-\left(  1+f\right)
\frac{\delta X_{t}^{pa}}{\delta\Gamma_{t}^{q}}Y_{t}^{q}\lambda_{a}\frac
{\delta\lambda^{2}}{\delta\Gamma_{t}^{p}}\right]  \ . \label{R_final1_Ham}%
\end{equation}
Using the relations $X^{pa}=s^{a}\Gamma^{p}$,$\ Y^{p}=-\left(  1/2\right)
s^{2}\Gamma^{p}$ and bearing in mind that $\lambda_{a}=\int dt\ s_{a}\Lambda$,
one can represent the contribution (\ref{R_final1_Ham}) in terms of
BRST-antiBRST variations, cf. (\ref{R_final2}),%
\begin{align}
\Re=  &  -\frac{1}{4}\left(  -1\right)  ^{\varepsilon_{p}}\int dt\left[
\left(  s^{2}\Gamma_{t}^{p}\right)  \lambda^{2}\right]  _{,\left(  p,t\right)
}-\frac{1}{32}\left(  -1\right)  ^{\varepsilon_{p}}\int dt^{\prime}%
dt^{\prime\prime}\left(  s^{2}\Gamma_{t^{\prime}}^{p}\right)  \lambda
_{,\left(  q,t^{\prime\prime}\right)  }^{2}\left(  s^{2}\Gamma_{t^{\prime
\prime}}^{q}\right)  \lambda_{,\left(  p,t^{\prime}\right)  }^{2}\nonumber\\
&  +\frac{1}{4}\left(  -1\right)  ^{\varepsilon_{p}}\left(  1-\frac{1}{2}\int
dt\ s^{2}\Lambda\right)  \int dt^{\prime}dt^{\prime\prime}\left(  s^{a}%
\Gamma_{t^{\prime}}^{p}\right)  _{,\left(  q,t^{\prime\prime}\right)  }\left(
s^{2}\Gamma_{t^{\prime\prime}}^{q}\right)  \lambda_{a}\lambda_{,\left(
p,t^{\prime}\right)  }^{2}\ , \label{R_final1_Ham2}%
\end{align}
where%
\[
A_{,\left(  p,t\right)  }\equiv\frac{\delta A}{\delta\Gamma_{t}^{p}}%
=A\frac{\overleftarrow{\delta}}{\delta\Gamma_{t}^{p}}\ .
\]
By analogy with Subsection\textbf{\ }\ref{linearYM}, one can state that the
linearized finite BRST-antiBRST transformations (\ref{linearized_Ham}) for
dynamical systems with first-class constraints can be interpreted neither as
global symmetry transformations of the integrand, nor as field-dependent
transformations inducing an exact change of the gauge-fixing functional.
Therefore, linearized finite BRST-antiBRST transformations in Hamiltonian
formalism do not possess the properties of finite BRST-antiBRST transformations.

\section{Finite BRST-antiBRST Transformations with Arbitrary Parameters}

\label{arbitrary} \label{FT}%
\renewcommand{\theequation}{\arabic{section}.\arabic{equation}} \setcounter{equation}{0}\setcounter{theorem}{0}

In this section, we examine the calculation of the Jacobian for finite
field-dependent BRST-antiBRST transformations in the case of arbitrary, i.e.,
generally independent parameters, $\lambda_{a}\not \equiv s_{a}\Lambda$. Once
again, we shall carry out the explicit calculations in the Yang--Mills case
and then make a relation of the resulting Jacobian to the case of arbitrary
dynamical systems with first-class constraints. Furthermore, as long as the
case of general gauge theories in Lagrangian formalism proves similar to the
Yang--Mills case, the corresponding general considerations will be provided as
well. The calculations in Yang--Mills theories and first-class constraint
systems will partially repeat the case of linearized BRST-antiBRST
transformations and will therefore effectively use some of the corresponding
statements given by the above propositions. At the same time, we will slightly
change the notation (\ref{notation1}), (\ref{defU,V}) of the matrix objects
for the sake of convenience.

\subsection{Yang--Mills Theories\label{Yang-Mills}}

In the Yang--Mills case, the finite field-dependent BRST-antiBRST
transformations in question have the form%
\[
\phi^{A}\rightarrow\phi^{\prime A}=\phi^{A}+\Delta\phi^{A}%
\ ,\ \ \ \mathrm{where}\ \ \ \Delta\phi^{A}=\left(  s^{a}\phi\right)
\lambda_{a}+\frac{1}{4}\left(  s^{2}\phi\right)  \lambda^{2}=X^{Aa}\lambda
_{a}-\frac{1}{2}Y^{A}\lambda^{2}\ ,\ \ \ \lambda_{a}\not \equiv s_{a}%
\Lambda\ .
\]
Let us examine the corresponding even matrix $M=\|M_{B}^{A}\|$ and the related
quantity $\Im$%
\[
M_{B}^{A}=\frac{\delta\left(  \Delta\phi^{A}\right)  }{\delta\phi^{B}%
}\ ,\ \ \ \Im=\mathrm{Str}\ln\left(  \mathbb{I}+M\right)  =-\sum_{n=1}%
^{\infty}\frac{\left(  -1\right)  ^{n}}{n}\mathrm{Str}\left(  M^{n}\right)
\ .
\]
Explicitly, the matrix $M_{B}^{A}$ is given by the sum of three even matrices:%
\begin{align}
&  M_{B}^{A}=P_{B}^{A}+Q_{B}^{A}+R_{B}^{A}\,,\,\,\,\mathrm{where}%
\,\,\,Q_{B}^{A}=\left(  Q_{1}\right)  _{B}^{A}+\left(  Q_{2}\right)  _{B}%
^{A}\ ,\label{MABext}\\
&  \,P_{B}^{A}=X^{Aa}\frac{\delta\lambda_{a}}{\delta\phi^{B}}\ ,\ \left(
Q_{1}\right)  _{B}^{A}=\lambda_{a}\frac{\delta X^{Aa}}{\delta\phi^{B}}\left(
-1\right)  ^{\varepsilon_{A}+1}\ ,\ \left(  Q_{2}\right)  _{B}^{A}=\lambda
_{a}Y^{A}\frac{\delta\lambda^{a}}{\delta\phi^{B}}\left(  -1\right)
^{\varepsilon_{A}+1}\ ,\ R_{B}^{A}=-\frac{1}{2}\lambda^{2}\frac{\delta Y^{A}%
}{\delta\phi^{B}}\ . \label{PRQABext}%
\end{align}
Here, the matrix $Q_{B}^{A}$ of Subsection \ref{linearYM} has been naturally
extended by its summation with the matrix $\left(  Q_{2}\right)  _{B}^{A}$,
which has already emerged in the relation (\ref{QP}) of the mentioned
subsection. The additional matrix $R_{B}^{A}$ has also emerged
(\ref{lemmaStr(Q)}) in Subsection \ref{linearYM}.

Using the property $\mathrm{Str}\left(  AB\right)  =\mathrm{Str}\left(
BA\right)  $ for arbitrary even matrices and the fact that the occurrence of
$R\sim\lambda^{2}$ in $\mathrm{Str}\left(  M^{n}\right)  $ more than once
yields zero, $\lambda^{4}\equiv0$, we have%
\begin{equation}
\mathrm{Str}\left(  M^{n}\right)  =\mathrm{Str}\left(  P+Q+R\right)  ^{n}%
=\sum_{k=0}^{1}C_{n}^{k}\mathrm{Str}\left[  \left(  P+Q\right)  ^{n-k}%
R^{k}\right]  \ ,\ \ \ C_{n}^{k}=\frac{n!}{k!\left(  n-k\right)  !}\ .
\label{Strgen}%
\end{equation}
Moreover,%
\begin{equation}
\mathrm{Str}\left(  P+Q+R\right)  ^{n}=\mathrm{Str}\left(  P+Q\right)
^{n}+n\mathrm{Str}\left[  \left(  P+Q\right)  ^{n-1}R\right]  =\mathrm{Str}%
\left(  P+Q\right)  ^{n}+n\mathrm{Str}\left(  P^{n-1}R\right)  \ , \label{R}%
\end{equation}
since any occurrence of $R\sim\lambda^{2}$ and $Q\sim\lambda_{a}$
simultaneously entering $\mathrm{Str}\left(  M\right)  ^{n}$\ yields zero,
owing to $\lambda_{a}\lambda^{2}=0$, as a consequence of which $R$ can only be
coupled with $P^{n-1}$.

Further considerations are based on the following statements, proved in
Appendices \ref{proof_lemma M^n}-- \ref{proof_lemma QP^n}, respectively:

\begin{lemma}
\label{lemma M^n} The expressions $\mathrm{Str}\left(  M^{n}\right)  $ for
$n\geq1$ are given by%
\begin{equation}
\mathrm{Str}\left(  M^{n}\right)  =\mathrm{Str}\left(  P+Q\right)
^{n}+n\mathrm{Str}\left(  P^{n-1}R\right)  =\left\{
\begin{array}
[c]{ll}%
\mathrm{Str}\left(  P+Q\right)  +\mathrm{Str}\left(  R\right)  \ , & n=1\ ,\\
\mathrm{Str}\left(  P+Q\right)  ^{n}\ , & n>1\ .
\end{array}
\right.  \label{P+Q}%
\end{equation}

\end{lemma}

\noindent Note:\ the relation (\ref{P+Q}) uses the nilpotency $s^{a}s^{b}%
s^{c}\equiv0$ of the generators $s^{a}$, as a consequence of their
anticommutativity, and implies that the matrix $R$ drops out of $\mathrm{Str}%
\left(  M^{n}\right)  $, $n>1$, and enters the quantity $\Im$ only as
$\mathrm{Str}\left(  R\right)  $.

\begin{lemma}
\label{lemma (P+Q)^n}The expressions $\mathrm{Str}\left(  P+Q\right)  ^{n}$
for $n>1$ are given by%
\begin{align}
\mathrm{Str}\left(  P+Q\right)  ^{n}  &  =\sum_{k=0}^{n}C_{n}^{k}%
\mathrm{Str}\left(  P^{n-k}Q^{k}\right)  =\mathrm{Str}\left(  P^{n}+C_{n}%
^{1}P^{n-1}Q+C_{n}^{2}P^{n-2}Q^{2}\right)  \ ,\ \ \ n=2,3\ ,\label{(2,3)}\\
\mathrm{Str}\left(  P+Q\right)  ^{2k}  &  =\sum_{l=0}^{1}C_{2k}^{l}%
\mathrm{Str}\left(  P^{2k-l}Q^{l}\right)  +C_{2k}^{1}\sum_{l=0}^{k-2}%
\mathrm{Str}\left[  P^{2(k-l-1)}\left(  P^{l}Q\right)  ^{2}\right]  +C_{k}%
^{1}\mathrm{Str}\left[  \left(  P^{k-1}Q\right)  ^{2}\right]  ,\ \ \ k\geq
2\ ,\label{(2k)}\\
\mathrm{Str}\left(  P+Q\right)  ^{2k+1}  &  =\sum_{l=0}^{1}C_{2k+1}%
^{l}\mathrm{Str}\left(  P^{2k+1-l}Q^{l}\right)  +C_{2k+1}^{1}\sum_{l=0}%
^{k-1}\mathrm{Str}\left[  P^{2(k-l)-1}\left(  P^{l}Q\right)  ^{2}\right]
\ ,\ \ \ k\geq2\ . \label{(2k+1)}%
\end{align}

\end{lemma}

\noindent Note:\ the relation (\ref{(2,3)}) coincides with the formula
(\ref{2,3}) of Proposition (\ref{StrQ}), whereas the relations (\ref{(2k)}),
(\ref{(2k+1)}) generalize the formula (\ref{Str(P+Q)^n}) of Proposition
(\ref{P^2}) to the case $\lambda_{a}\not \equiv s_{a}\Lambda$. Indeed, let us
suppose that the case $\lambda_{a}=s_{a}\Lambda$, with the implied condition
$P^{n}=f\cdot P^{n-1}$, does indeed take place. Then it is straightforward to
verify the equalities%
\begin{equation}
\ \left.
\begin{array}
[c]{l}%
\mathrm{Str}\left(  P+Q\right)  ^{2k}\\
\mathrm{Str}\left(  P+Q\right)  ^{2k+1}%
\end{array}
\right\}  =\mathrm{Str}\left(  P^{n}\right)  +n\mathrm{Str}\left(
P^{n-1}Q\right)  +n\mathrm{Str}\left(  P^{n-2}Q^{2}\right)  +K_{n}%
\mathrm{Str}\left(  P^{n-3}QPQ\right)  \ ,\ \ \ k\geq2\ , \label{reduc}%
\end{equation}
where the coefficients $K_{n}\,$are given by (\ref{Str(P+Q)^n}), with
allowance for%
\begin{equation}
K_{2k}=\left(  k-2\right)  C_{2k}^{1}+C_{k}^{1}\ ,\ \ \ K_{2k+1}=\left(
k-1\right)  C_{2k+1}^{1}\ , \label{K_even_odd}%
\end{equation}
which shows that in the respective cases $n=\left(  2k,2k+1\right)  $ the
above relations for $\mathrm{Str}\left(  P+Q\right)  ^{2k}$ and $\mathrm{Str}%
\left(  P+Q\right)  ^{2k+1}$ are reduced to the formula (\ref{Str(P+Q)^n}) for
$\mathrm{Str}\left(  P+Q\right)  ^{n}$, when $\lambda_{a}=s_{a}\Lambda$.

\begin{lemma}
\label{lemma R-1/2Q^2} There hold the properties%
\begin{equation}
\mathrm{Str}\left(  Q_{1}\right)  =0\ ,\ \ \ \mathrm{Str}\left(  R\right)
-\frac{1}{2}\mathrm{Str}\left(  Q_{1}^{2}\right)  =0\ . \label{StrQ=0}%
\end{equation}

\end{lemma}

\noindent Note:\ the relations (\ref{StrQ=0}) repeat, in different notation,
the formulas (\ref{lemmaStr(Q)}) of Proposition \ref{StrQ}, established in our
paper \cite{MRnew}; for the sake of completeness of the present subsection, we
will provide the corresponding proof in Appendix \ref{proof R-1/2Q^2}.

\begin{lemma}
\label{lemma P^n} There hold the properties%
\begin{equation}
\mathrm{Str}\left(  P^{n}\right)  =-\mathrm{tr}\left[  \left(  m^{n}\right)
_{b}^{a}\right]  \equiv-\mathrm{tr}\left(  m^{n}\right)  =-\left(
m^{n}\right)  _{a}^{a}\ ,\ \ \mathrm{where}\ \ m_{b}^{a}\equiv s^{a}%
\lambda_{b}\ , \label{P^ngen}%
\end{equation}
where powers in $m=m_{b}^{a}$ are understood in the sense of matrix
multiplication with respect to $\mathrm{Sp}(2)$ indices.
\end{lemma}

\begin{lemma}
\label{lemma QP^n} There hold the properties%
\begin{equation}
QP^{n}=\mathrm{tr}\left[  m^{n-1}\left(  e+m\right)  Y\right]  \ ,\ \ \ n\geq
1\ , \label{QPgen}%
\end{equation}
where $e=\left(  e\right)  _{b}^{a}$ is the unit matrix $\left(  e\right)
_{b}^{a}\equiv\delta_{b}^{a}$, and the matrix $Y=(Y_{b}^{a})_{B}^{A}$ is given
by%
\begin{equation}
(Y_{b}^{a})_{B}^{A}\equiv\left(  -1\right)  ^{\varepsilon_{A}}\lambda^{a}%
Y^{A}\frac{\delta\lambda_{b}}{\delta\phi^{B}}\Longrightarrow(Y_{a}^{a}%
)_{B}^{A}=(Q_{2})_{B}^{A}\ . \label{matrY}%
\end{equation}

\end{lemma}

\noindent Note:\ the relations (\ref{P^ngen}), (\ref{QPgen}) generalize the
respective formulae (\ref{contracprop}), (\ref{QP}) of Proposition \ref{P^2}
to the case $\lambda_{a}\not \equiv s_{a}\Lambda$, which is readily
established by inserting the particular form of the matrix $m_{b}^{a}%
\sim\delta_{b}^{a}$,%
\begin{equation}
m_{b}^{a}=s^{a}\lambda_{b}=s^{a}s_{b}\Lambda=-\frac{1}{2}\delta_{b}^{a}\left(
s^{2}\Lambda\right)  \ , \label{matr_m_part}%
\end{equation}
corresponding to the case $\lambda_{a}=s_{a}\Lambda$, in the relations
(\ref{P^ngen}), (\ref{QPgen}), with the resulting formulae (\ref{contracprop}%
), (\ref{QP}); due to the natural appearance of the matrices $m_{b}^{a}$ and
$\left(  Y_{b}^{a}\right)  _{B}^{A}$ in (\ref{P^ngen}), (\ref{QPgen}), we
shall evaluate the quantity $\Im$ as a series in powers of these objects.

Proceeding to the calculation of $\Im$ on the basis of the above lemmas and
collecting the relations (\ref{Strgen})--(\ref{(2k+1)}), (\ref{StrQ=0}%
)--(\ref{matrY}), we arrive (see Appendix\textbf{\ }\ref{Jacobian_arb}) at the
following result:%
\begin{equation}
\Im=-\mathrm{tr\ln}\left(  e+m\right)  \ , \label{J-arb}%
\end{equation}
where the operation $\mathrm{\ln}$ is to be understood in the sense of an
expansion in powers with respect to the multiplication of matrices carrying
$\mathrm{Sp}(2)$ indices:%
\begin{equation}
\mathrm{\ln}\left(  e+m\right)  =\left[  \mathrm{\ln}\left(  e+m\right)
\right]  _{b}^{a}=-\sum_{n=1}^{\infty}\frac{\left(  -1\right)  ^{n}}{n}\left(
m^{n}\right)  _{b}^{a}\ . \label{J-exp}%
\end{equation}
It should be emphasized that the considerations of Appendix \ref{Jacobian_arb}
do not utilize the anticommutativity of the BRST-antiBRST generators $s_{a}$,
except for the treatment of $\mathrm{Str}\left(  P^{n-1}R\right)
$,$\ \mathrm{Str}\left(  R\right)  $,$\ \mathrm{Str}\left(  Q_{1}^{2}\right)
$ in (\ref{P+Q}) and (\ref{StrQ=0}). In the remaining part of this subsection,
we examine the consequences implied in (\ref{J-arb}), (\ref{J-exp}) by the
anticommutativity of $s_{a}$. Namely, in the particular case $\lambda
_{a}=s_{a}\Lambda$, the quantity $\Im$ reduces, in accordance with
(\ref{matr_m_part}),%
\begin{equation}
\left(  m^{n}\right)  _{b}^{a}=f^{n}\cdot\delta_{b}^{a}\ ,\ \ \ \mathrm{tr}%
\left(  m^{n}\right)  =2f^{n}\ ,\ \ \ \ f=-\frac{1}{2}s^{2}\Lambda\ ,
\label{mab}%
\end{equation}
to the BRST-antiBRST-exact expression \cite{MRnew}%
\begin{equation}
\Im=\sum_{n=1}^{\infty}\frac{\left(  -1\right)  ^{n}}{n}\mathrm{tr}\left(
m^{n}\right)  =2\sum_{n=1}^{\infty}\frac{\left(  -1\right)  ^{n}}{n}%
f^{n}=-2\ln\left(  1+f\right)  =\ln\left(  1-\frac{1}{2}s^{2}\Lambda\right)
^{-2}\ . \label{J-lambda}%
\end{equation}
In the general case, however, $\lambda_{a}\not \equiv s_{a}\Lambda$, the
quantity $\Im$ fails to be BRST-antiBRST-invariant,%
\begin{align}
s^{a}\Im &  =\sum_{n=1}^{\infty}\frac{\left(  -1\right)  ^{n}}{n}s^{a}\left(
m^{n}\right)  _{b}^{b}=\sum_{n=1}^{\infty}\left(  -1\right)  ^{n}\left(
s^{a}m_{c}^{b}\right)  \left(  m^{n-1}\right)  _{b}^{c}\nonumber\\
&  =-\left(  s^{a}m_{c}^{b}\right)  \sum_{k=0}^{\infty}\left(  -1\right)
^{k}\left(  m^{k}\right)  _{b}^{c}=-\left(  s^{a}m_{c}^{b}\right)  \left[
\left(  e+m\right)  ^{-1}\right]  _{b}^{c}\not \equiv 0\ , \label{non-inv}%
\end{align}
whence it is generally no longer BRST-antiBRST-exact and does not amount to an
exact change of the gauge-fixing Boson:%
\begin{equation}
\Im\not \equiv \frac{1}{2i\hbar}s^{2}\Delta F\ . \label{non-exact}%
\end{equation}
The condition of BRST-antiBRST-invariance of $\Im$ therefore reads%
\begin{equation}
\left(  s^{a}m_{c}^{b}\right)  \left[  \left(  e+m\right)  ^{-1}\right]
_{b}^{c}=\frac{1}{2}\varepsilon^{ab}\left(  s^{2}\lambda_{c}\right)  \left[
\left(  e+m\right)  ^{-1}\right]  _{b}^{c}=0\ , \label{condit}%
\end{equation}
which is a necessary condition of BRST-antiBRST-exactness of $\Im$.
Furthermore, if we impose on $\Im\left(  \lambda\right)  $, given by an
expansion in powers of $\lambda_{a}$,%
\begin{equation}
\Im\left(  \lambda\right)  =-s^{a}\lambda_{a}+\frac{1}{2}\left(  s^{a}%
\lambda_{b}\right)  \left(  s^{b}\lambda_{a}\right)  -\frac{1}{3}\left(
s^{a}\lambda_{b}\right)  \left(  s^{b}\lambda_{c}\right)  \left(  s^{c}%
\lambda_{a}\right)  +\cdots\ , \label{M(0,psi)=exp}%
\end{equation}
the requirement of BRST-antiBRST-exactness at the first order, $s^{a}%
\lambda_{a}=s^{a}s_{a}\Lambda$, for a certain even-valued functional $\Lambda
$, then this requirement meets the condition (\ref{condit}) and turns out to
provide the corresponding exactness at the succeeding orders, which implies
the following (see Appendix \ref{proof_lemma psi})

\begin{lemma}
\label{lemma psi} If there exists an even-valued functional $\Lambda$ such
that $s^{a}\lambda_{a}=s^{a}s_{a}\Lambda$, then there also exists a sequence
of $\Lambda_{n}\,$such that%
\begin{equation}
\mathrm{tr}\left(  m^{n}\right)  =s^{a}s_{a}\Lambda_{n}\ ,\ \ \ n\geq
2\ \Longrightarrow\ \Im=\sum_{n=1}^{\infty}\frac{\left(  -1\right)  ^{n}}%
{n}s^{a}s_{a}\Lambda_{n}\ . \label{lemma psi form}%
\end{equation}
\
\end{lemma}

\noindent This implies the following criterion: the quantity $\Im\left(
\lambda\right)  $ is BRST-antiBRST-exact to all orders of its expansion in
powers of $\lambda_{a}$ if and only if there exists such an even-valued
functional $\Lambda$ that $s^{a}\lambda_{a}=-s^{2}\Lambda$. Such a choice of
$\lambda_{a}$ obviously corresponds to the case of functionally-dependent
parameters,%
\begin{equation}
s^{1}\lambda_{1}+s^{2}\lambda_{2}=-s^{2}\Lambda\ , \label{slsmbda=exact}%
\end{equation}
which we have previously examined \cite{MRnew} in the particular case
$\lambda_{a}=s_{a}\Lambda$. Since an arbitrary set of functional parameters
$\lambda_{a}\left(  \phi\right)  $ is generally\ not functionally-dependent,
$s^{a}\lambda_{a}\not \equiv -s^{2}\Lambda$, it is obvious that the
corresponding quantum action induced by a finite BRST-antiBRST transformation
with such parameters $\lambda_{a}\left(  \phi\right)  $ \emph{cannot be
reproduced} by the conventional Lagrangian BRST-antiBRST quantization scheme.

It has been previously established \cite{MRnew} that the particular case
$\lambda_{a}=s_{a}\Lambda$ of functionally-dependent parameters $\lambda_{a}$
allows one to obtain a unique solution of the corresponding compensation
equation%
\[
\ln\left(  1-\frac{1}{2}s^{2}\Lambda\right)  ^{-2}=\frac{1}{2i\hbar}%
s^{2}\Delta F\ ,
\]
with accuracy up to BRST-antiBRST-exact terms. This is a consequence of the
fact that the resulting quantity $\Im$ is actually controlled by a single
functional parameter $\Lambda$, which is in one-to-one correspondence (up to
the above-mentioned accuracy) with a change $\Delta F$ of the gauge Boson. In
this respect, it is natural to examine the most general case of solutions to
$s^{a}\lambda_{a}=s^{a}s_{a}\Lambda$, parameterized by an additional
odd-valued doublet $\psi_{a}$,%
\begin{equation}
s^{a}\left(  \lambda_{a}-s_{a}\Lambda\right)  =0\Longrightarrow\lambda
_{a}=s_{a}\Lambda+\psi_{a}\ ,\ \ \ s^{a}\psi_{a}=0\ , \label{solut}%
\end{equation}
which may, in particular, be constant, $\psi_{a}=\mathrm{const}$. In virtue of
(\ref{solut}), the additional parameters $\psi_{a}$ are functionally-dependent
and obey (see Appendix \ref{proof_lemma psi2})

\begin{lemma}
\label{lemma psi 2} The condition $s^{a}\psi_{a}=0$ implies%
\begin{equation}
\mathrm{tr}\left(  m_{\psi}^{2}\right)  =-\frac{1}{2}s^{2}\left(  \psi
^{2}\right)  \ ,\ \ \ \mathrm{tr}\left(  m_{\psi}^{n}\right)
=0\ ,\ \ \ \mathrm{for\ \ \ }n\geq3\ ,\ \ \ \mathrm{where}\ \ \ \psi^{2}%
\equiv\psi_{a}\psi^{a}\ ,\ \ \ \left(  m_{\psi}\right)  _{b}^{a}\equiv
s^{a}\psi_{b}\ , \label{implies}%
\end{equation}
whence the corresponding quantity $\Im$, parameterized by the functional
parameters $\left(  \Lambda,\psi_{a}\right)  $, is BRST-antiBRST-exact and
reads as follows:%
\begin{equation}
\Im\left(  \Lambda,\psi\right)  =\ln\left(  1-\frac{1}{2}s^{2}\Lambda\right)
^{-2}-\frac{1}{4}s^{2}\left(  \psi^{2}\right)  \left[  \left(  1-\frac{1}%
{2}s^{2}\Lambda\right)  ^{-2}\right]  \ . \label{reads}%
\end{equation}

\end{lemma}

\noindent As a consequence, the resulting compensation equation takes the form%
\begin{equation}
\ln\left(  1-\frac{1}{2}s^{2}\Lambda\right)  ^{-2}-\frac{1}{4}s^{2}\left(
\psi^{2}\right)  \left[  \left(  1-\frac{1}{2}s^{2}\Lambda\right)
^{-2}\right]  =\frac{1}{2i\hbar}s^{2}\Delta F\ . \label{comp_mod}%
\end{equation}
For a given gauge variation $\Delta F$ and a certain given solution $\psi_{a}$
of the subsidiary condition (\ref{solut}), the modified compensation equation
(\ref{comp_mod}) may be considered as an equation for some unknown functional
$\Lambda$, whose solution may be sought as $\Lambda=\Lambda\left(  \Delta
F,\psi\right)  $. More explicitly, there holds (see Appendix
\ref{proof_psi 2 solution}) the following

\begin{lemma}
\label{lemma psi 2 solution} The solutions $\Lambda$ of the modified
compensation equation (\ref{comp_mod}) have the form%
\begin{align}
s^{2}\Delta F  &  \not =0:\ \ \ \Lambda\left(  \Delta F,\psi\right)
=\frac{2\Delta F}{s^{2}\Delta F}\left.  \left\{  1-\left[  1+\vartheta\left(
\gamma X_{0}\right)  \right]  ^{-\frac{1}{2}}X_{0}^{-\frac{1}{2}}\right\}
\right\vert _{X_{0}=\exp\left(  \frac{1}{2i\hbar}s^{2}\Delta F\right)
,\gamma=\frac{1}{4}s^{2}(\psi^{2})}\ ,\label{delta_Fnot0}\\
s^{2}\Delta F  &  =0\ ,\ s^{2}\left(  \psi^{2}\right)  =0:\ \ \ \Lambda
=s^{a}\tilde{\lambda}_{a}+s^{2}\tilde{\Lambda}\ ,\label{1delta_F=0}\\
s^{2}\Delta F  &  =0,\ s^{2}\left(  \psi^{2}\right)  \not =0:\ \ \ \Lambda
\left(  \psi\right)  =\frac{2\psi^{2}}{s^{2}\left(  \psi^{2}\right)  }\left.
\left\{  1-\left[  1+\vartheta\left(  \gamma\right)  \right]  ^{-\frac{1}{2}%
}\right\}  \right\vert _{\gamma=\frac{1}{4}s^{2}(\psi^{2})}\ ,
\label{2delta_F=0}%
\end{align}
where the function $\vartheta\left(  y\right)  $ is defined by%
\begin{align*}
\theta\left(  x\right)   &  =\frac{\ln\left(  1+x\right)  }{\left(
1+x\right)  }\ ,\ \ \ \theta\left(  0\right)  =0\ ,\\
\vartheta\left(  y\right)   &  :\vartheta\left(  \theta\left(  x\right)
\right)  =x\ ,\ \ \ \vartheta\left(  0\right)  =0\ .
\end{align*}

\end{lemma}

\noindent Therefore, in the case $s^{2}\Delta F\not =0$, $s^{2}\left(
\psi^{2}\right)  =0$, we find the solution (\ref{Lambda-Fsol1}),
\[
\Lambda\left(  \Delta F,0\right)  =\frac{2\Delta F}{s^{2}\Delta F}\left\{
1-\exp\left[  \left(  i/4\hbar\right)  s^{2}\Delta F\right]  \right\}  \ ,
\]
of the usual compensation equation (\ref{superJ3m}), whereas in the case
$s^{2}\Delta F=0$,$\ s^{2}\left(  \psi^{2}\right)  \not =0$ we arrive at a
finite BRST-antiBRST\ transformation, with the parameters $\lambda_{a}%
=s_{a}\Lambda\left(  \psi\right)  +\psi_{a}$ given by (\ref{solut}),
(\ref{2delta_F=0}), which induces a Jacobian equal to unity:%
\begin{equation}
\Im=\frac{1}{2i\hbar}s^{2}\Delta F=0\Longrightarrow\exp\left(  \Im\right)
=1\ . \label{Jacob_unity}%
\end{equation}
On the other hand, given the functionals $\Lambda$ and $\psi_{a}$, we obtain a
change of the gauge $\Delta F$, according to (\ref{comp_mod}), which, in the
case $\Lambda=0$, takes the form%
\begin{equation}
s^{2}\Delta F=-\frac{i\hbar}{2}s^{2}\left(  \psi^{2}\right)  \label{reduction}%
\end{equation}
and corresponds to the quantity $\Im\left(  \psi\right)  $ given by%
\begin{equation}
\Im\left(  \psi\right)  =-\frac{1}{4}s^{2}\left(  \psi^{2}\right)  \ ,
\label{Jac-psi}%
\end{equation}
which implies a non-trivial Jacobian, $\exp\left(  \Im\right)  \not =1$, in
the case $s^{2}\left(  \psi^{2}\right)  \not =0$. In order to investigate this
possibility in more detail, let us notice that the solutions of the subsidiary
condition (\ref{solut}) can be presented in the form%
\begin{align}
&  \psi_{a}=\mu_{a}+\frac{1}{2}s_{a}\Psi_{b}^{b}+s^{b}\Psi_{ba}+s_{b}%
s^{b}\varphi_{a}\ ,\ \ \ \Psi_{b}^{a}\equiv\varepsilon^{ac}\Psi_{cb}%
\ ,\nonumber\\
&  \varepsilon(\mu_{a})=\varepsilon(\varphi_{a})=1\ ,\ \ \ \varepsilon
(\Psi_{ab})=0\ , \label{psi}%
\end{align}
parameterized by a constant \textrm{Sp}$(2)$-doublet, $\mu_{a}=\mathrm{const}%
$, an \textrm{Sp}$(2)$-doublet of arbitrary functionals, $\varphi_{a}\left(
\phi\right)  $, and an \textrm{Sp}$(2)$-tensor of arbitrary functionals,
$\Psi_{ab}\left(  \phi\right)  $, with the corresponding Grassmann parities
(\ref{psi}). The above solutions can be found from the following Ansatz:%
\begin{equation}
\psi_{a}=\mu_{a}+s_{a}\Psi+s^{b}\Psi_{ba}+s_{b}s^{b}\varphi_{a}\ ,
\label{Ansatz}%
\end{equation}
expanding the functionals $\psi_{a}$ in powers of the operators $s_{a}$. Once
a certain solution (\ref{psi}) is given, one can decompose the corresponding
tensor $\Psi_{ab}$ into its symmetric and antisymmetric components,%
\[
\Psi_{ab}=\Psi_{\left\{  ab\right\}  }+\Psi_{\left[  ab\right]  }\ ,
\]
and notice that the antisymmetric component, $\Psi_{\left[  ab\right]  }%
\equiv\varepsilon_{ab}\Psi$, $\Psi=\left(  1/2\right)  \varepsilon^{ba}%
\Psi_{\left[  ab\right]  }$, actually vanishes from $\psi_{a}$:%
\[
\frac{1}{2}\varepsilon^{bc}s_{a}\Psi_{\left[  cb\right]  }+s^{b}\Psi_{\left[
ba\right]  }=\frac{1}{2}\varepsilon^{bc}\varepsilon_{cb}s_{a}\Psi
+\varepsilon_{ba}s^{b}\Psi=s_{a}\Psi-s_{a}\Psi\equiv0\ .
\]
Therefore, regular solutions of the equation $s^{a}\psi_{a}=0$ in
(\ref{solut}) vanishing in the case $\phi^{A}=0$ have the form
\begin{equation}
\psi_{a}=s^{b}\Psi_{\{ba\}}+s^{2}\varphi_{a}\ , \label{Ansatzfin}%
\end{equation}
which is a particular case ($n=1$) of a regular solution, vanishing in the
case $\phi^{A}=0$, of a more general equation for an unknown completely
symmetric $\mathrm{Sp}(2)$-tensor of rank $n$,%
\begin{equation}
s^{a_{1}}\psi_{\{a_{1}a_{2}...a_{n}\}}=0\Longleftrightarrow\psi_{\{a_{1}%
a_{2}...a_{n}\}}=s^{b}\Psi_{\{ba_{1}a_{2}...a_{n}\}}+s^{2}\varphi
_{\{a_{1}a_{2}...a_{n}\}}\ , \label{gAnsatzfin}%
\end{equation}
with certain rank-$n$ and rank-$(n+1)$ symmetric $\mathrm{Sp}(2)$-tensors
$\varphi_{\{a_{1}a_{2}...a_{n}\}}$,$\ \Psi_{\{ba_{1}a_{2}...a_{n}\}}$,
$\varepsilon(\varphi)=\varepsilon(\psi)=\varepsilon(\Psi)+1$. It can next be
noticed that the components $\mu_{a}$ and $\varphi_{a}$ in (\ref{Ansatz}) do
not contribute to $(m_{\psi})_{b}^{a}=s^{a}\psi_{b}$, whereas the symmetric
component $\Psi_{\left\{  ab\right\}  }$ (once non-vanishing) does,%
\[
(m_{\psi})_{b}^{a}=s^{a}\psi_{b}=s^{a}\left(  \frac{1}{2}\varepsilon^{cd}%
s_{b}\Psi_{\{dc\}}+s^{c}\Psi_{\{cb\}}\right)  =\frac{1}{2}\varepsilon
^{ac}s^{2}\Psi_{\{cb\}}\not \equiv 0\ ,
\]
and furthermore it provides a non-vanishing contribution to $\mathrm{tr}%
(m_{\psi})=(m_{\psi})_{b}^{a}(m_{\psi})_{a}^{b}$,%
\[
\left(  m_{\psi}\right)  _{b}^{a}\left(  m_{\psi}\right)  _{a}^{b}=\frac{1}%
{4}\varepsilon^{ac}\varepsilon^{bd}\left(  s^{2}\Psi_{\{cb\}}\right)  \left(
s^{2}\Psi_{\{da\}}\right)  =-\frac{1}{2}s^{2}\left(  \psi^{2}\right)  \ ,
\]
which makes it possible to express the quantity $\Im$ in (\ref{Jac-psi})
entirely in terms of the symmetric component:%
\begin{equation}
\Im=\frac{1}{8}\varepsilon^{ac}\varepsilon^{bd}\left(  s^{2}\Psi
_{\{ad\}}\right)  \left(  s^{2}\Psi_{\{bc\}}\right)  \ . \label{Jacob-sym}%
\end{equation}
Finally, in the most general case of arbitrary functionals $\lambda_{a}\left(
\phi\right)  $, the condition (\ref{solut}) is\ not fulfilled, making it
thereby impossible to present the quantity $\Im$ in a BRST-antiBRST-exact form
(\ref{reads}) and to relate it with some change of the gauge (\ref{comp_mod}).
This means that the corresponding quantity $\Im$ acquires some extra
contributions w.r.t. (\ref{reads}), which can be related to a decomposition of
the parameters $\lambda_{a}$ into the following components:%
\[
\lambda_{a}=s_{a}\Lambda+\psi_{a}+\sigma_{a}\ ,
\]
where%
\[
s^{a}\psi_{a}=0\ ,\ \ \ s^{a}\sigma_{a}\not =0\ ,\ \ \ s^{a}\sigma_{b}%
\not =\delta_{b}^{a}f^{\prime}\ .
\]
Using the notation%
\[
\left(  m_{\Lambda}\right)  _{b}^{a}=s^{a}s_{b}\Lambda=\delta_{b}%
^{a}f\ ,\ \ \ \left(  m_{\psi}\right)  _{b}^{a}=s^{a}\psi_{b}\ ,\ \ \ \left(
m_{\sigma}\right)  _{b}^{a}=s^{a}\sigma_{b}\ ,
\]
and considerations similar to the relations (\ref{trmm}), (\ref{J-lambdapsi}),
(\ref{M(Lambda,psi)}) of Appendix \ref{proof_lemma psi2}, we have%
\begin{equation}
\mathrm{tr}\left(  m_{\Lambda}+m_{\psi}+m_{\sigma}\right)  ^{n}=\mathrm{tr}%
\sum_{k=0}^{n}C_{n}^{k}f^{n-k}\left(  m_{\psi}+m_{\sigma}\right)  ^{k}\ ,
\label{trsigma}%
\end{equation}
whence the corresponding quantity $\Im=\Im\left(  \Lambda,\psi,\sigma\right)
$ reads%
\begin{equation}
\Im\left(  \Lambda,\psi,\sigma\right)  =\ln\left(  1-\frac{1}{2}s^{2}%
\Lambda\right)  ^{-2}+M\left(  \Lambda,\psi,\sigma\right)  \ ,\ \ \ M\left(
\Lambda,\psi,\sigma\right)  =\sum_{n=1}^{\infty}\frac{\left(  -1\right)  ^{n}%
}{n}\sum_{k=1}^{n}C_{n}^{k}f^{n-k}\mathrm{tr}\left[  \left(  m_{\psi
}+m_{\sigma}\right)  ^{k}\right]  _{\ f=-\frac{1}{2}s^{2}\Lambda}\ .
\label{J-Lambdasigma}%
\end{equation}
Using the fact that $\mathrm{tr}(m_{\psi})=0$, we find%
\begin{equation}
M\left(  \Lambda,\psi,\sigma\right)  =-\mathrm{tr}\left(  m_{\sigma}\right)
\left(  1+\frac{1}{2}s^{2}\Lambda\right)  +\frac{1}{2}\mathrm{tr}\left(
m_{\psi}+m_{\sigma}\right)  ^{2}+\sum_{n=3}^{\infty}\frac{\left(  -1\right)
^{n}}{n}\sum_{k=1}^{n}C_{n}^{k}f^{n-k}\mathrm{tr}\left[  \left(  m_{\psi
}+m_{\sigma}\right)  ^{k}\right]  _{\ f=-\frac{1}{2}s^{2}\Lambda}\ ,
\label{M(Lambda,sigma)}%
\end{equation}
where account is to be taken of%
\begin{equation}
\mathrm{tr}\left(  m_{\psi}^{2}\right)  =-\frac{1}{2}s^{2}\left(  \psi
^{2}\right)  \ ,\ \ \ \mathrm{tr}\left(  m_{\psi}^{k}\right)  \equiv
0\ ,\ \ \ k\geq3\ . \label{account}%
\end{equation}
Accordingly, the corresponding quantity $\Im\left(  \Lambda,\psi
,\sigma\right)  $ is given by%
\begin{equation}
\Im\left(  \Lambda,\psi,\sigma\right)  =\Im\left(  \Lambda,\psi\right)
+\Re\left(  \Lambda,\psi,\sigma\right)  \ , \label{Jacob-Lambda-sigma}%
\end{equation}
where the quantity $M\left(  \Lambda,\psi,\sigma\right)  $, given by
(\ref{M(Lambda,sigma)}), has been decomposed as%
\begin{equation}
M\left(  \Lambda,\psi,\sigma\right)  =M\left(  \Lambda,\psi\right)
+\Re\left(  \Lambda,\psi,\sigma\right)  \ ,\ \ \ M\left(  \Lambda,\psi\right)
\equiv\left.  M\left(  \Lambda,\psi,\sigma\right)  \right\vert _{\sigma=0}\ .
\label{MR}%
\end{equation}
In (\ref{Jacob-Lambda-sigma}), $\Im\left(  \Lambda,\psi\right)  $ has the form
(\ref{reads}) and thereby represents the BRST-antiBRST-exact contribution,
whereas $\Re\left(  \Lambda,\psi,\sigma\right)  $ represents the contribution%
\begin{equation}
\Re\left(  \Lambda,\psi,\sigma\right)  =M\left(  \Lambda,\psi,\sigma\right)
-M\left(  \Lambda,\psi\right)  \ , \label{addact1}%
\end{equation}
which is not BRST-antiBRST-exact and cannot be, therefore, reproduced by the
conventional BRST-antiBRST quantization scheme; instead, it should be regarded
as an addition to the transformed quantum action in the integrand of
(\ref{z(j)})
\begin{equation}
\mathcal{I}_{\phi g(\lambda(\phi))}^{F}=d\phi\ \exp\left\{  \left(
i/\hbar\right)  \left[  S_{0}+\left(  1/2\right)  s^{a}s_{a}\left(  F+\Delta
F\right)  -i\hbar\Re\left(  \Lambda,\psi,\sigma\right)  \right]  \right\}  \ ,
\label{addact}%
\end{equation}
calculated in the reference frame with the gauge Boson $F+\Delta
F(\Lambda,\psi)$.

\subsection{Constrained Dynamical Systems}

\setcounter{theorem}{4}

The case of arbitrary dynamical systems with first-class constraints can be
examined in complete analogy with the case of Yang--Mills theories. It is
based on the propositions and considerations of Subsection \ref{Yang-Mills}
and repeats, in part, the considerations of Subsection \ref{constrained}.
Namely, in the case of dynamical systems in question, the finite
field-dependent BRST-antiBRST transformations with arbitrary parameters have
the form
\[
\Gamma_{t}^{p}\rightarrow\check{\Gamma}_{t}^{p}=\Gamma_{t}^{p}+\Delta
\Gamma_{t}^{p}\ ,\ \ \ \mathrm{where\ \ \ }\Delta\Gamma_{t}^{p}=\left(
s^{a}\Gamma_{t}^{p}\right)  \lambda_{a}+\frac{1}{4}\left(  s^{2}\Gamma_{t}%
^{p}\right)  \lambda^{2}=X_{t}^{pa}\lambda_{a}-\frac{1}{2}Y_{t}^{p}\lambda
^{2}\ ,\ \ \ \lambda_{a}\not \equiv \int dt\ s_{a}\Lambda\ .
\]
Let us examine the corresponding even matrix $M=\|M_{q|t^{\prime}%
,t^{\prime\prime}}^{p}\|$ and the related quantity $\Im$%
\[
M_{q|t^{\prime},t^{\prime\prime}}^{p}=\frac{\delta\left(  \Delta
\Gamma_{t^{\prime}}^{p}\right)  }{\delta\Gamma_{t^{\prime\prime}}^{q}%
}\ ,\ \ \ \Im=\mathrm{Str}\ln\left(  \mathbb{I}+M\right)  =-\sum_{n=1}%
^{\infty}\frac{\left(  -1\right)  ^{n}}{n}\mathrm{Str}\left(  M^{n}\right)
\ .
\]
Explicitly, the matrix $M_{q|t^{\prime},t^{\prime\prime}}^{p}$ is given by the
sum of three even matrices:%
\begin{align}
&  M_{q|t^{\prime},t^{\prime\prime}}^{p}=U_{q|t^{\prime},t^{\prime\prime}}%
^{p}+V_{q|t^{\prime},t^{\prime\prime}}^{p}+W_{q|t^{\prime},t^{\prime\prime}%
}^{p}\ ,\,\,\,\mathrm{where}\,\,\,V_{q|t^{\prime},t^{\prime\prime}}%
^{p}=\left(  V_{1}\right)  _{q|t^{\prime},t^{\prime\prime}}^{p}+\left(
V_{2}\right)  _{q|t^{\prime},t^{\prime\prime}}^{p}\ ,\label{Mpqext}\\
&  U_{q|t^{\prime},t^{\prime\prime}}^{p}=X_{t^{\prime}}^{pa}\frac
{\delta\lambda_{a}}{\delta\Gamma_{t^{\prime\prime}}^{q}}\ ,\ \left(
V_{1}\right)  _{q|t^{\prime},t^{\prime\prime}}^{p}=\lambda_{a}\frac{\delta
X_{t^{\prime}}^{pa}}{\delta\Gamma_{t^{\prime\prime}}^{q}}\left(  -1\right)
^{\varepsilon_{p}+1}\ ,\ \left(  V_{2}\right)  _{q|t^{\prime},t^{\prime\prime
}}^{p}=\lambda_{a}Y_{t^{\prime}}^{p}\frac{\delta\lambda^{a}}{\delta
\Gamma_{t^{\prime\prime}}^{q}}\left(  -1\right)  ^{\varepsilon_{p}%
+1}\ ,\ W_{q|t^{\prime},t^{\prime\prime}}^{p}=-\frac{1}{2}\lambda^{2}%
\frac{\delta Y_{t^{\prime}}^{p}}{\delta\Gamma_{t^{\prime\prime}}^{q}}\ .
\label{UVWext}%
\end{align}
Here, the matrix $\left(  V\right)  _{q|t^{\prime},t^{\prime\prime}}^{p}$ of
Subsection \ref{constrained} has been naturally extended by its summation with
the matrix $\left(  V_{2}\right)  _{q|t^{\prime},t^{\prime\prime}}^{p}$, which
has already emerged in the relation (\ref{VU=(1+f)V2}) of the mentioned
subsection. The additional matrix $W_{q|t^{\prime},t^{\prime\prime}}^{p}$ has
also emerged (\ref{W}) in Subsection \ref{constrained}. The matrices $U$, $V$,
$W$ correspond to the matrices $P$, $Q$, $R$\ of Subsection \ref{Yang-Mills}.
This correspondence is given explicitly by Table \ref{table in2}%
.{\ \begin{table}[ptbh]
{\ }
\par
\begin{center}%
\begin{tabular}
[c]{||l|l||}\hline\hline
First-class constraint {systems} & Yang--Mills theories\\\hline\hline
$\Gamma_{t}^{p}\ ,\ \Delta\Gamma_{t}^{p}=\left(  s^{a}\Gamma_{t}^{p}\right)
\lambda_{a}+\frac{1}{4}\left(  s^{2}\Gamma_{t}^{p}\right)  \lambda^{2}$ &
$\phi^{A}\ ,\ \Delta\phi^{A}=\left(  s^{a}\phi^{A}\right)  \lambda_{a}%
+\frac{1}{4}\left(  s^{2}\phi^{A}\right)  \lambda^{2}$\\
$s^{a}\Gamma_{t}^{p}=X_{t}^{pa},\ s^{b}s^{a}\Gamma_{t}^{p}=\varepsilon
^{ab}Y_{t}^{p}\ ,\ s^{c}s^{b}s^{a}\Gamma_{t}^{p}=0$ & $s^{a}\phi^{A}%
=X^{Aa},\ s^{b}s^{a}\phi^{A}=\varepsilon^{ab}Y^{A}\ ,\ s^{c}s^{b}s^{a}\phi
^{A}=0$\\
$\int dt^{\prime}\ \frac{\delta X_{t}^{pa}}{\delta\Gamma_{t^{\prime}}^{q}%
}X_{t^{\prime}}^{qb}=\varepsilon^{ab}Y_{t}^{p},\ Y_{t}^{p}=-\frac{1}%
{2}\varepsilon_{ab}\int dt^{\prime}\frac{\delta X_{t}^{pa}}{\delta
\Gamma_{t^{\prime}}^{q}}X_{t^{\prime}}^{Bb}$ & $\frac{\delta X^{Aa}}%
{\delta\phi^{B}}X^{Bb}=\varepsilon^{ab}Y^{A},\ Y^{A}=-\frac{1}{2}%
\varepsilon_{ab}\frac{\delta X^{Aa}}{\delta\phi^{B}}X^{Bb}$\\
$\int dt\ \frac{\delta X_{t}^{pa}}{\delta\Gamma_{t}^{p}}=\int dt^{\prime
}\ \frac{\delta Y_{t}^{p}}{\delta\Gamma_{t^{\prime}}^{q}}X_{t^{\prime}}%
^{qa}=0$ & $\frac{\delta X^{Aa}}{\delta\phi^{A}}=\frac{\delta Y^{A}}%
{\delta\phi^{B}}X^{Bb}=0$\\
$\frac{\delta\left(  \Delta\Gamma_{t^{\prime}}^{p}\right)  }{\delta
\Gamma_{t^{\prime\prime}}^{q}}=M_{q|t^{\prime},t^{\prime\prime}}%
^{p}=U_{q|t^{\prime},t^{\prime\prime}}^{p}+V_{q|t^{\prime},t^{\prime\prime}%
}^{p}+W_{q|t^{\prime},t^{\prime\prime}}^{p}$ & $\frac{\delta\left(  \Delta
\phi^{A}\right)  }{\delta\phi^{B}}=M_{B}^{A}=P_{B}^{A}+Q_{B}^{A}+R_{B}^{A}$\\
$V_{q|t^{\prime},t^{\prime\prime}}^{p}=\left(  V_{1}\right)  _{q|t^{\prime
},t^{\prime\prime}}^{p}+\left(  V_{2}\right)  _{q|t^{\prime},t^{\prime\prime}%
}^{p}$ & $Q_{B}^{A}=\left(  Q_{1}\right)  _{B}^{A}+\left(  Q_{2}\right)
_{B}^{A}$\\
$\left(  V_{1}\right)  _{q|t^{\prime},t^{\prime\prime}}^{p}=\lambda_{a}%
\frac{\delta X_{t^{\prime}}^{pa}}{\delta\Gamma_{t^{\prime\prime}}^{q}}\left(
-1\right)  ^{\varepsilon_{p}+1}$ & $\left(  Q_{1}\right)  _{B}^{A}=\lambda
_{a}\frac{\delta X^{Aa}}{\delta\phi^{B}}\left(  -1\right)  ^{\varepsilon
_{A}+1}$\\
$\left(  V_{2}\right)  _{q|t^{\prime},t^{\prime\prime}}^{p}=\lambda
_{a}Y_{t^{\prime}}^{p}\frac{\delta\lambda^{a}}{\delta\Gamma_{t^{\prime\prime}%
}^{q}}\left(  -1\right)  ^{\varepsilon_{p}+1}$ & $\left(  Q_{2}\right)
_{B}^{A}=\lambda_{a}Y^{A}\frac{\delta\lambda^{a}}{\delta\phi^{B}}\left(
-1\right)  ^{\varepsilon_{A}+1}$\\
$U_{q|t^{\prime},t^{\prime\prime}}^{p}=X_{t^{\prime}}^{pa}\frac{\delta
\lambda_{a}}{\delta\Gamma_{t^{\prime\prime}}^{q}}\ ,\ W_{q|t^{\prime
},t^{\prime\prime}}^{p}=-\frac{1}{2}\lambda^{2}\frac{\delta Y_{t^{\prime}}%
^{p}}{\delta\Gamma_{t^{\prime\prime}}^{q}}$ & $P_{B}^{A}=X^{Aa}\frac
{\delta\lambda_{a}}{\delta\phi^{B}}\ ,\ R_{B}^{A}=-\frac{1}{2}\lambda^{2}%
\frac{\delta Y^{A}}{\delta\phi^{B}}$\\\hline\hline
\end{tabular}
\end{center}
\par
\vspace{-2ex}\caption{Correspondence of the matrix elements in arbitrary
first-class constraint systems and Yang--Mills theories. Finite
field-dependent BRST-antiBRST transformations with arbitrary parameters.}%
\label{table in2}%
\end{table}}

\noindent In this connection, due to the property $\mathrm{Str}\left(
AB\right)  =\mathrm{Str}\left(  BA\right)  $, Lemmas \ref{lemma M^n}%
--\ref{lemma QP^n} of Subsection\textbf{\ }\ref{Yang-Mills} remain formally
the same (see Footnote \ref{footnote}) in terms of $U_{q|t^{\prime}%
,t^{\prime\prime}}^{p}$, $V_{q|t^{\prime},t^{\prime\prime}}^{p}$,
$W_{q|t^{\prime},t^{\prime\prime}}^{p}$ substituted instead of the respective
matrices $P_{B}^{A}$, $Q_{B}^{A}$, $R_{B}^{A}$, which establishes the following

\begin{proposition}
\label{prop UVW} The matrices $U$, $V$, $W$ possess the properties
\begin{align}
&  \mathrm{Str}\left(  M^{n}\right)  =\mathrm{Str}\left(  U+V\right)
^{n}+n\mathrm{Str}\left(  U^{n-1}W\right)  =\left\{
\begin{array}
[c]{ll}%
\mathrm{Str}\left(  U+V\right)  +\mathrm{Str}\left(  W\right)  \ , & n=1\ ,\\
\mathrm{Str}\left(  U+V\right)  ^{n}\ , & n>1\ .
\end{array}
\right. \label{(U+V)Ham}\\
&  \mathrm{Str}\left(  U+V\right)  ^{n}=\sum_{k=0}^{n}C_{n}^{k}\mathrm{Str}%
\left(  U^{n-k}V^{k}\right)  =\mathrm{Str}\left(  U^{n}+C_{n}^{1}%
U^{n-1}V+C_{n}^{2}U^{n-2}V^{2}\right)  \ ,\ \ \ n=2,3\ ,\label{2,3Ham}\\
&  \mathrm{Str}\left(  U+V\right)  ^{2k}=\sum_{l=0}^{1}C_{2k}^{l}%
\mathrm{Str}\left(  U^{2k-l}V^{l}\right)  +C_{2k}^{1}\sum_{l=0}^{k-2}%
\mathrm{Str}\left[  P^{2(k-l-1)}\left(  U^{l}V\right)  ^{2}\right]  +C_{k}%
^{1}\mathrm{Str}\left[  \left(  U^{k-1}V\right)  ^{2}\right]  ,\ \ \ k\geq
2\ ,\label{U,VHam,2k}\\
&  \mathrm{Str}\left(  U+V\right)  ^{2k+1}=\sum_{l=0}^{1}C_{2k+1}%
^{l}\mathrm{Str}\left(  U^{2k+1-l}V^{l}\right)  +C_{2k+1}^{1}\sum_{l=0}%
^{k-1}\mathrm{Str}\left[  U^{2(k-l)-1}\left(  U^{l}V\right)  ^{2}\right]
\ ,\ \ \ k\geq2\ ,\label{UVHam2k+1}\\
&  \mathrm{Str}\left(  V_{1}\right)  =0\ ,\ \ \ \mathrm{Str}\left(  W\right)
-\frac{1}{2}\mathrm{Str}\left(  V_{1}^{2}\right)  =0\ ,\label{V1W}\\
&  \mathrm{Str}\left(  U^{n}\right)  =-\mathrm{tr}\left[  \left(
m^{n}\right)  _{b}^{a}\right]  \equiv-\mathrm{tr}\left(  m^{n}\right)
=-\left(  m^{n}\right)  _{a}^{a}\ ,\ \ VU^{n}=\mathrm{tr}\left[
m^{n-1}\left(  e+m\right)  Y\right]  \ ,\ \ \ n\geq1\ ,\ \ \mathrm{where}%
\ \ m_{b}^{a}\equiv s^{a}\lambda_{b}\ , \label{Un,UVnHam}%
\end{align}
where $e=\left(  e\right)  _{b}^{a}\equiv\delta_{b}^{a}$, according to the
notation of Subsection\textbf{\ }\ref{Yang-Mills}, and the matrix
$Y=(Y_{b}^{a})_{q|t^{\prime},t^{\prime\prime}}^{p}$ is given by%
\begin{equation}
(Y_{b}^{a})_{q|t^{\prime},t^{\prime\prime}}^{p}\equiv\left(  -1\right)
^{\varepsilon_{p}}\lambda^{a}Y_{t^{\prime},t^{\prime\prime}}^{p}\frac
{\delta\lambda_{b}}{\delta\Gamma_{t^{\prime\prime}}^{q}}\Longrightarrow
(Y_{a}^{a})_{q|t^{\prime},t^{\prime\prime}}^{p}=(V_{2})_{q|t^{\prime
},t^{\prime\prime}}^{p}\ . \label{YHam}%
\end{equation}

\end{proposition}

\noindent From (\ref{(U+V)Ham})--(\ref{YHam}), with allowance for
$\mathrm{Str}\left(  AB\right)  =\mathrm{Str}\left(  BA\right)  $, it follows
that $\Im$ acquires the form, cf. (\ref{J-arb}), (\ref{J-exp}),%
\begin{equation}
\Im=-\mathrm{tr\ln}\left(  e+m\right)  \ ,\ \ \ \mathrm{where}%
\ \ \ \mathrm{\ln}\left[  \left(  e+m\right)  \right]  _{b}^{a}=-\sum
_{n=1}^{\infty}\frac{\left(  -1\right)  ^{n}}{n}\left(  m^{n}\right)  _{b}%
^{a}\ . \label{JHam_arb}%
\end{equation}
The considerations of Subsection \ref{Yang-Mills} following the relation
(\ref{J-lambda}) can now be repeated for the result (\ref{JHam_arb}), with
account taken of the obvious replacement $\lambda_{a}\left(  \phi\right)
=s^{a}\Lambda\left(  \phi\right)  \rightarrow\lambda_{a}\left(  \Gamma\right)
=\int dt\ s_{a}\Lambda\left(  \Gamma\right)  $.

\subsection{General Gauge Theories}

The consideration of general gauge theories in Lagrangian formalism proves
similar to the case of Yang--Mills theories and is based on the lemmas of
Subsection \ref{Yang-Mills}, with minor modifications, necessary to take into
account the facts that in general gauge theories the global BRST-antiBRST
transformations $\mathsf{\Gamma}^{\mathsf{p}}\rightarrow\mathsf{\Gamma
}^{\prime\mathsf{p}}=\mathsf{\Gamma}^{\mathsf{p}}+\delta\mathsf{\Gamma
}^{\mathsf{p}}$, $\delta\mathsf{\Gamma}^{\mathsf{p}}=\left(  \mathsf{s}%
^{a}\mathsf{\Gamma}^{\mathsf{p}}\right)  \lambda_{a}$, $\lambda_{a}%
=\mathrm{const}$, do not respect the invariance of functional integration
measure, $\mathsf{\Gamma}^{\prime\mathsf{p}}\not =\mathsf{\Gamma}^{\mathsf{p}%
}$, and do not possess the anticommutativity of the generators, $\mathsf{s}%
^{a}\mathsf{s}^{b}+\mathsf{s}^{b}\mathsf{s}^{a}\not \equiv 0$. Namely, in the
general case the finite BRST-antiBRST transformations with arbitrary
parameters $\lambda_{a}\left(  \mathsf{\Gamma}\right)  $ have form%
\[
\mathsf{\Gamma}^{\mathsf{p}}\rightarrow\mathsf{\Gamma}^{\prime\mathsf{p}%
}=\mathsf{\Gamma}^{\mathsf{p}}+\Delta\mathsf{\Gamma}^{\mathsf{p}%
}\ ,\ \ \ \Delta\mathsf{\Gamma}^{\mathsf{p}}=\left(  \mathsf{s}^{a}%
\mathsf{\Gamma}^{\mathsf{p}}\right)  \lambda_{a}+\frac{1}{4}\left(
\mathsf{s}^{2}\mathsf{\Gamma}^{\mathsf{p}}\right)  \lambda^{2}=\mathcal{X}%
^{\mathsf{p}a}\lambda_{a}-\frac{1}{2}\mathcal{Y}^{\mathsf{p}}\lambda
^{2}\ ,\ \ \ \lambda_{a}\not \equiv \mathsf{s}_{a}\Lambda\ .
\]
Let us examine the corresponding even matrix $\mathcal{M}=\|\mathcal{M}%
_{\mathsf{q}}^{\mathsf{p}}\|$ and the related quantity $\Im$, namely,%
\[
\mathcal{M}_{\mathsf{q}}^{\mathsf{p}}=\frac{\delta\left(  \Delta
\mathsf{\Gamma}^{p}\right)  }{\delta\mathsf{\Gamma}^{\mathsf{q}}}%
\ ,\ \ \ \Im=\mathrm{Str}\ln\left(  \mathbb{I}+\mathcal{M}\right)
=-\sum_{n=1}^{\infty}\frac{\left(  -1\right)  ^{n}}{n}\mathrm{Str}\left(
\mathcal{M}^{n}\right)  \ ,\ \ \ \mathbb{I}_{\mathsf{q}}^{\mathsf{p}}%
=\delta_{\mathsf{q}}^{\mathsf{p}}\ .
\]
Explicitly, the matrix $\mathcal{M}_{\mathsf{q}}^{\mathsf{p}}$ is given by the
sum of three even matrices:%
\begin{align}
&  \mathcal{M}_{\mathsf{q}}^{\mathsf{p}}=\frac{\delta\left(  \Delta
\mathsf{\Gamma}^{p}\right)  }{\delta\mathsf{\Gamma}^{\mathsf{q}}}%
=\mathcal{U}_{\mathsf{q}}^{\mathsf{p}}+\mathcal{V}_{\mathsf{q}}^{\mathsf{p}%
}+\mathcal{W}_{\mathsf{q}}^{\mathsf{p}}\,,\,\,\,\mathrm{where}%
\,\,\,\mathcal{V}_{\mathsf{q}}^{\mathsf{p}}=\left(  \mathcal{V}_{1}\right)
_{\mathsf{q}}^{\mathsf{p}}+\left(  \mathcal{V}_{2}\right)  _{\mathsf{q}%
}^{\mathsf{p}}\ ,\label{Mpq}\\
&  \mathcal{U}_{\mathsf{q}}^{\mathsf{p}}=\mathcal{X}^{\mathsf{p}a}%
\lambda_{a,q}\ ,\ \ \ \left(  \mathcal{V}_{1}\right)  _{\mathsf{q}%
}^{\mathsf{p}}=\lambda_{a}\mathcal{X}_{,\mathsf{q}}^{\mathsf{p}a}\left(
-1\right)  ^{\varepsilon_{\mathsf{p}}+1}\ ,\ \ \ \left(  \mathcal{V}%
_{2}\right)  _{\mathsf{q}}^{\mathsf{p}}=\lambda_{a}\mathcal{Y}^{\mathsf{p}%
}\lambda_{,\mathsf{q}}^{a}\left(  -1\right)  ^{\varepsilon_{\mathsf{p}}%
+1}\ ,\ \ \ \mathcal{W}_{\mathsf{q}}^{\mathsf{p}}=-\frac{1}{2}\lambda
^{2}\mathcal{Y}_{,\mathsf{q}}^{\mathsf{p}}\ , \label{UVW}%
\end{align}
The matrices $\mathcal{U}_{\mathsf{q}}^{\mathsf{p}}$, $\mathcal{V}%
_{\mathsf{q}}^{\mathsf{p}}$, $\mathcal{W}_{\mathsf{q}}^{\mathsf{p}}$
correspond to the matrices $P_{B}^{A}$, $Q_{B}^{A}$, $R_{B}^{A}$\ of
Subsection \ref{Yang-Mills}. This correspondence is given explicitly by Table
\ref{table in3}.{\ \begin{table}[ptbh]
{\ }
\par
\begin{center}%
\begin{tabular}
[c]{||l|l||}\hline\hline
{General gauge theories} & {Yang--Mills theories}\\\hline\hline
$\mathsf{\Gamma}^{\mathsf{p}},\Delta\mathsf{\Gamma}^{\mathsf{p}}%
=\mathcal{X}^{\mathsf{p}a}\lambda_{a}-\left(  1/2\right)  \mathcal{Y}%
^{\mathsf{p}}\lambda^{2}$ & $\phi^{A},\Delta\phi^{A}=X^{Aa}\lambda_{a}-\left(
1/2\right)  Y^{A}\lambda^{2}$\\
$\mathcal{Y}^{\mathsf{p}}=\left(  1/2\right)  \mathcal{X}_{,\mathsf{q}%
}^{\mathsf{p}a}\mathcal{X}^{\mathsf{q}b}\varepsilon_{ba}$ & $Y^{A}=\left(
1/2\right)  X_{,B}^{Aa}X^{Bb}\varepsilon_{ba}$\\
$\frac{\delta\left(  \Delta\Gamma^{p}\right)  }{\delta\mathsf{\Gamma
}^{\mathsf{q}}}=\mathcal{M}_{q}^{p}$ & $\frac{\delta(\Delta\phi^{A})}%
{\delta\phi^{B}}=M_{B}^{A}$\\
$\mathcal{M}_{\mathsf{q}}^{\mathsf{p}}=\mathcal{U}_{\mathsf{q}}^{\mathsf{p}%
}+\mathcal{V}_{\mathsf{q}}^{\mathsf{p}}+\mathcal{W}_{\mathsf{q}}^{\mathsf{p}}$
& $M_{B}^{A}=P_{B}^{A}+Q_{B}^{A}+R_{B}^{A}$\\
$\mathcal{V}_{\mathsf{q}}^{\mathsf{p}}=\left(  \mathcal{V}_{1}\right)
_{\mathsf{q}}^{\mathsf{p}}+\left(  \mathcal{V}_{2}\right)  _{\mathsf{q}%
}^{\mathsf{p}}$ & $Q_{B}^{A}=\left(  Q_{1}\right)  _{B}^{A}+\left(
Q_{2}\right)  _{B}^{A}$\\
$\left(  \mathcal{V}_{1}\right)  _{\mathsf{q}}^{\mathsf{p}}=\lambda
_{a}\mathcal{X}_{,\mathsf{q}}^{\mathsf{p}a}\left(  -1\right)  ^{\varepsilon
_{\mathsf{p}}+1}$ & $\left(  Q_{1}\right)  _{B}^{A}=\lambda_{a}X_{,B}%
^{Aa}\left(  -1\right)  ^{\varepsilon_{A}+1}$\\
$\left(  \mathcal{V}_{2}\right)  _{\mathsf{q}}^{\mathsf{p}}=\lambda
_{a}\mathcal{Y}^{\mathsf{p}}\lambda_{,\mathsf{q}}^{a}\left(  -1\right)
^{\varepsilon_{\mathsf{p}}+1}$ & $\left(  Q_{2}\right)  _{B}^{A}=\lambda
_{a}Y^{A}\lambda_{,B}^{a}\left(  -1\right)  ^{\varepsilon_{A}+1}$\\
$\mathcal{U}_{\mathsf{q}}^{\mathsf{p}}=\mathcal{X}^{\mathsf{p}a}\lambda
_{a,q},\mathcal{W}_{\mathsf{q}}^{\mathsf{p}}=-\frac{1}{2}\lambda
^{2}\mathcal{Y}_{,\mathsf{q}}^{\mathsf{p}}$ & $P_{B}^{A}=X^{Aa}\lambda
_{a,B},R_{B}^{A}=-\frac{1}{2}\lambda^{2}Y_{,B}^{A}$\\\hline\hline
\end{tabular}
\end{center}
\par
\vspace{-2ex}\caption{Correspondence of the matrix elements in Yang--Mills and
general gauge theories. Finite field-dependent BRST-antiBRST transformations
with arbitrary parameters.}%
\label{table in3}%
\end{table}} In this connection, Lemmas \ref{lemma (P+Q)^n}, \ref{lemma P^n},
\ref{lemma QP^n} and the relations (\ref{R}) of Subsection \ref{Yang-Mills}
remain formally the same in terms of $\mathcal{U}_{\mathsf{q}}^{\mathsf{p}}$,
$\mathcal{V}_{\mathsf{q}}^{\mathsf{p}}$, $\mathcal{W}_{\mathsf{q}}%
^{\mathsf{p}}$, substituted instead of the respective matrices $P_{B}^{A}$,
$Q_{B}^{A}$, $R_{B}^{A}$ , since the relevant considerations do not use any
properties of these objects, except their Grassmann parity and the character
of their dependence on the parameters $\lambda_{a}$, which establishes the following

\begin{proposition}
\label{prop2 UVW} The matrices $\mathcal{U}$, $\mathcal{V}$, $\mathcal{W}$
possess the properties%
\begin{align}
&  \mathrm{Str}\left(  \mathcal{M}^{n}\right)  =\mathrm{Str}\left(
\mathcal{U}+\mathcal{V}\right)  ^{n}+n\mathrm{Str}\left(  \mathcal{U}%
^{n-1}\mathcal{W}\right)  \ ,\ \ \ n\geq1\ ,\label{U+V}\\
&  \mathrm{Str}\left(  \mathcal{U}+\mathcal{V}\right)  ^{n}=\sum_{k=0}%
^{n}C_{n}^{k}\mathrm{Str}\left(  \mathcal{U}^{n-k}\mathcal{V}^{k}\right)
=\mathrm{Str}\left(  \mathcal{U}^{n}+C_{n}^{1}\mathcal{U}^{n-1}\mathcal{V}%
+C_{n}^{2}\mathcal{U}^{n-2}\mathcal{V}^{2}\right)
\ ,\ \ \ n=2,3\ ,\label{(2,3)'}\\
&  \mathrm{Str}\left(  \mathcal{U}+\mathcal{V}\right)  ^{2k}=\sum_{l=0}%
^{1}C_{2k}^{l}\mathrm{Str}\left(  \mathcal{U}^{2k-l}\mathcal{V}^{l}\right)
+C_{2k}^{1}\sum_{l=0}^{k-2}\mathrm{Str}\left[  \mathcal{U}^{2(k-l-1)}\left(
\mathcal{U}^{l}\mathcal{V}\right)  ^{2}\right]  +C_{k}^{1}\mathrm{Str}\left[
\left(  \mathcal{U}^{k-1}\mathcal{V}\right)  ^{2}\right]  ,\ \ \ k\geq
2\ ,\label{(2k)'}\\
&  \mathrm{Str}\left(  \mathcal{U}+\mathcal{V}\right)  ^{2k+1}=\sum_{l=0}%
^{1}C_{2k+1}^{l}\mathrm{Str}\left(  \mathcal{U}^{2k+1-l}\mathcal{V}%
^{l}\right)  +C_{2k+1}^{1}\sum_{l=0}^{k-1}\mathrm{Str}\left[  \mathcal{U}%
^{2(k-l)-1}\left(  \mathcal{U}^{l}\mathcal{V}\right)  ^{2}\right]
\ ,\ \ \ k\geq2\ ,\label{(2k+1)'}\\
&  \mathrm{Str}\left(  \mathcal{U}^{n}\right)  =-\mathrm{tr}\left[  \left(
\mathsf{m}^{n}\right)  _{b}^{a}\right]  \equiv-\mathrm{tr}\left(
\mathsf{m}^{n}\right)  =-\left(  \mathsf{m}^{n}\right)  _{a}^{a}%
\ ,\ \ \mathrm{where}\ \ \mathsf{m}_{b}^{a}\equiv\mathsf{s}^{a}\lambda
_{b}\ ,\label{U^n}\\
&  \mathcal{VU}^{n}=\mathrm{tr}\left[  \mathsf{m}^{n-1}\left(  e+\mathsf{m}%
\right)  \mathcal{Y}\right]  \ ,\ \ \ n\geq1\ , \label{VUgen}%
\end{align}
where $e=\left(  e\right)  _{b}^{a}\equiv\delta_{b}^{a}$, according to the
notation of Subsection\textbf{\ }\ref{Yang-Mills}, and the matrix
$\mathcal{Y}=(\mathcal{Y}_{a}^{b})_{\mathsf{q}}^{\mathsf{p}}$ is given by%
\begin{equation}
(\mathcal{Y}_{b}^{a})_{\mathsf{q}}^{\mathsf{p}}\equiv\left(  -1\right)
^{\varepsilon_{\mathsf{p}}}\lambda^{a}\mathcal{Y}^{\mathsf{p}}\frac
{\delta\lambda_{b}}{\delta\Gamma^{\mathsf{q}}}\Longrightarrow(\mathcal{Y}%
_{a}^{a})_{\mathsf{q}}^{\mathsf{p}}=(\mathcal{V}_{2})_{\mathsf{q}}%
^{\mathsf{p}}\ . \label{matrY'}%
\end{equation}

\end{proposition}

\noindent On the other hand, Lemmas \ref{lemma M^n}, \ref{lemma R-1/2Q^2} use
the explicit structure of functions entering the matrices $\left(
Q_{1}\right)  _{B}^{A}$, $R_{B}^{A}$ and they consequently undergo, in terms
of $\left(  \mathcal{V}_{1}\right)  _{\mathsf{q}}^{\mathsf{p}}$,
$\mathcal{W}_{\mathsf{q}}^{\mathsf{p}}$, the following modifications,
established in respective Appendices \ref{proof_modified lemma 1},
\ref{lemma-zero}:

\setcounter{theorem}{8}

\begin{lemma}
\label{modified lemma 1} There hold the properties%
\begin{equation}
\mathrm{Str}\left(  \mathcal{U}^{n-1}\mathcal{W}\right)  =-\frac{1}{4}%
\frac{\delta\lambda_{a}}{\delta\mathsf{\Gamma}^{\mathsf{p}}}\left(
\mathsf{m}^{n-2}\right)  _{b}^{a}\left(  \mathrm{s}^{b}\mathrm{s}%
^{2}\mathsf{\Gamma}^{\mathsf{p}}\right)  \lambda^{2}\ ,\ \ \ n>1\ .
\label{non-zero}%
\end{equation}

\end{lemma}

\begin{lemma}
\label{modified lemma} The matrices $\mathcal{V}_{1}$ and $\mathcal{W}$ are
related by the equality%
\begin{equation}
\mathrm{Str}\left(  \mathcal{V}_{1}\right)  +\mathrm{Str}\left(
\mathcal{W}\right)  -\frac{1}{2}\mathrm{Str}\left(  \mathcal{V}_{1}%
^{2}\right)  =-\left(  \Delta^{a}S\right)  \lambda_{a}-\frac{1}{4}\left(
\mathsf{s}_{a}\Delta^{a}S\right)  \lambda^{2}\ . \label{zero}%
\end{equation}

\end{lemma}

\noindent Note:\ the properties in (\ref{non-zero}) generalize the equalities
$\mathrm{Str}\left(  P^{n-1}R\right)  =0$, $n>2$, implied by (\ref{P+Q}), due
to the failure of the generators $\mathrm{s}^{a}$ to be nilpotent in the
entire space $\mathsf{\Gamma}^{\mathsf{p}}$, which means that the matrix
$\mathcal{W}$ does not drop out of $\mathrm{Str}\left(  \mathcal{M}%
^{n}\right)  $, $n>1$; the relation (\ref{zero}) extends the properties
(\ref{StrQ=0}) to the case of non-anticommuting generators $\mathrm{s}^{a}$
and a BRST-antiBRST non-invariant integration measure $d\mathsf{\Gamma}$, and
has been established in our paper \cite{MRnew3}; for the sake of completeness
of the present subsection, the corresponding proof is given in Appendix
\ref{lemma-zero}.

In view of the properties (\ref{U+V})--(\ref{matrY'}) and the correspondence
provided by Table \ref{table in3}, the calculation of the quantity $\Im$ here
repeats the considerations of Appendix \ref{Jacobian_arb}, with the
modifications provided by (\ref{non-zero}), (\ref{zero}), in comparison with
(\ref{P+Q}), (\ref{StrQ=0}), which implies the appearance in $\Im$ of an extra
contribution:%
\begin{equation}
\mathrm{Str}\left(  \mathcal{V}_{1}\right)  +\mathrm{Str}\left(
\mathcal{W}\right)  -\frac{1}{2}\mathrm{Str}\left(  \mathcal{V}_{1}%
^{2}\right)  -\sum_{n=2}^{\infty}\frac{\left(  -1\right)  ^{n}}{n}%
n\mathrm{Str}\left(  \mathcal{U}^{n-1}\mathcal{W}\right)  \ . \label{add_cont}%
\end{equation}
Thus, the resulting expression for $\Im$ is given by, cf. (\ref{J-arb}),%
\begin{equation}
\Im=-\left(  \Delta^{a}S\right)  \lambda_{a}-\frac{1}{4}\left(  \mathsf{s}%
_{a}\Delta^{a}S\right)  \lambda^{2}-\mathrm{tr\ln}\left(  e+\mathsf{m}\right)
+\mathbb{\Re}\ , \label{J-arb-gen}%
\end{equation}
where%
\begin{equation}
\mathbb{\Re=}\frac{1}{4}\lambda_{a,\mathsf{p}}[\left(  e+\mathsf{m}\right)
^{-1}]_{b}^{a}\left(  \mathrm{s}^{b}\mathrm{s}^{2}\mathsf{\Gamma}^{\mathsf{p}%
}\right)  \lambda^{2}\ , \label{J-exp-gen}%
\end{equation}
or, explicitly,%
\begin{equation}
\left[  \mathrm{\ln}\left(  e+\mathsf{m}\right)  \right]  _{b}^{a}=-\sum
_{n=1}^{\infty}\frac{\left(  -1\right)  ^{n}}{n}\left(  \mathsf{m}^{n}\right)
_{b}^{a}\ ,\ \ \ [\left(  e+\mathsf{m}\right)  ^{-1}]_{b}^{a}=\sum
_{n=0}^{\infty}\left(  -1\right)  ^{n}\left(  \mathsf{m}^{n}\right)  _{b}%
^{a}\ . \label{explicit}%
\end{equation}
In contrast to our paper \cite{MRnew3}, the result expressed by
(\ref{J-arb-gen}), (\ref{J-exp-gen}) makes no assumption that the functional
parameters $\lambda_{a}$ do not depend on some of their variables from the set
$\mathsf{\Gamma}^{\mathsf{p}}=\left(  \phi^{A},\pi^{Aa},\lambda^{A},\phi
_{Aa}^{\ast},\bar{\phi}_{A}\right)  $, namely, that $\lambda_{a}$ are
restricted to $(\phi^{A},\pi^{Aa},\lambda^{A})$, making it thereby possible to
utilize the anticommutativity of the BRST-antiBRST generators $\mathsf{s}^{a}$
in this subspace. This restriction has now been removed due to the fact that
the considerations of Appendix \ref{Jacobian_arb} do not require, as has been
noticed in Subsection \ref{Yang-Mills}, the BRST-antiBRST generators to be
anticommuting, except for the treatment of the terms $\mathrm{Str}\left(
R\right)  $, $\mathrm{Str}\left(  Q_{1}^{2}\right)  $, $\mathrm{Str}\left(
P^{n-1}R\right)  $, which now correspond to part of the contribution
(\ref{add_cont}). At the same time, by virtue of (\ref{non-zero}),
(\ref{zero}), this contribution has now been calculated for non-anticommuting
generators $\mathsf{s}^{a}$, thereby extending the considerations of Appendix
\ref{Jacobian_arb} to general gauge theories. As a consequence, the result
expressed by (\ref{J-arb-gen}), (\ref{J-exp-gen}) is now presented in terms of
arbitrary anticommuting parameters $\lambda_{a}\left(  \mathsf{\Gamma}\right)
$. In this connection, let us examine the change of the integrand
corresponding to the result (\ref{J-arb-gen}), (\ref{J-exp-gen}):%
\[
\left.  d\mathsf{\Gamma}\exp\left[  \left(  i/\hbar\right)  \mathcal{S}%
_{F}\left(  \mathsf{\Gamma}\right)  \right]  \right\vert _{_{\mathsf{\Gamma
}\rightarrow\mathsf{\Gamma}^{\prime}}}=d\mathsf{\Gamma}\exp\left\{  \left(
i/\hbar\right)  \left[  \mathcal{S}_{F}\left(  \mathsf{\Gamma}+\Delta
\mathsf{\Gamma}\right)  -i\hbar\Im\left(  \mathsf{\Gamma}\right)  \right]
\right\}  \ ,
\]
where, taking into account the relation (\ref{eqinv}) of Appendix
\ref{lemma-zero}, we have%
\begin{align}
&  \ \mathcal{S}_{F}\left(  \mathsf{\Gamma}+\Delta\mathsf{\Gamma}\right)
=\mathcal{S}_{F}\left(  \mathsf{\Gamma}\right)  +\Delta\mathcal{S}_{F}\left(
\mathsf{\Gamma}\right)  \ ,\nonumber\\
&  \ \Delta\mathcal{S}_{F}=\left(  \mathsf{s}^{a}\mathcal{S}_{F}\right)
\lambda_{a}+\frac{1}{4}\left(  \mathsf{s}_{a}\mathsf{s}^{a}\mathcal{S}%
_{F}\right)  \lambda^{2}=-i\hbar\Delta^{a}S\lambda_{a}-\frac{i\hbar}%
{4}\mathsf{s}_{a}\left(  \Delta^{a}S\right)  \lambda^{2}\ ,\nonumber\\
&  \ \Im=-\left(  \Delta^{a}S\right)  \lambda_{a}-\frac{1}{4}\left(
\mathsf{s}_{a}\Delta^{a}S\right)  \lambda^{2}-\mathrm{tr\ln}\left(
e+\mathsf{m}\right)  +\frac{1}{4}\lambda_{a,\mathsf{p}}[\left(  e+\mathsf{m}%
\right)  ^{-1}]_{b}^{a}\left(  \mathrm{s}^{b}\mathrm{s}^{2}\mathsf{\Gamma
}^{\mathsf{p}}\right)  \lambda^{2}\ , \label{wehave}%
\end{align}
whence%
\begin{equation}
\left.  d\mathsf{\Gamma}\exp\left(  \frac{i}{\hbar}\mathcal{S}_{F}\right)
\right\vert _{_{\mathsf{\Gamma}\rightarrow\mathsf{\Gamma}^{\prime}}%
}=d\mathsf{\Gamma}\exp\left\{  \frac{i}{\hbar}\left[  \mathcal{S}_{F}%
+i\hbar\ \mathrm{tr\ln}\left(  e+\mathsf{m}\right)  -\frac{i\hbar}{4}%
\lambda_{a,\mathsf{p}}[\left(  e+\mathsf{m}\right)  ^{-1}]_{b}^{a}\left(
\mathrm{s}^{b}\mathrm{s}^{2}\mathsf{\Gamma}^{\mathsf{p}}\right)  \lambda
^{2}\right]  \right\}  \equiv d\mathsf{\Gamma}\exp\left(  \frac{i}{\hbar
}\mathcal{S}^{\prime}\right)  \ , \label{whence}%
\end{equation}
which implies that, due to the presence in $\mathcal{S}^{\prime}$ of a
non-vanishing contribution with $\lambda^{2}$, the corresponding modified
quantum action $\mathcal{S}^{\prime}$ generally does not describe a change of
gauge-fixing:%
\begin{equation}
\mathcal{S}^{\prime}\not \equiv S+\phi_{Aa}^{\ast}\pi^{Aa}+\bar{\phi}%
_{A}\lambda^{A}-\frac{1}{2}U^{2}F^{\prime}\ ,\ \ \ U^{a}=\left.
\mathrm{s}^{a}\right\vert _{\phi,\pi,\lambda}\ . \label{notdescribe}%
\end{equation}
If we now require that $\mathcal{S}^{\prime}=S+\phi_{Aa}^{\ast}\pi^{Aa}%
+\bar{\phi}_{A}\lambda^{A}-(1/2)U^{2}F^{\prime}$ be indeed the case, then
there arise the conditions%
\begin{align}
&  \lambda_{a,\mathsf{p}}[\left(  e+\mathsf{m}\right)  ^{-1}]_{b}^{a}\left(
\mathrm{s}^{b}\mathrm{s}^{2}\mathsf{\Gamma}^{\mathsf{p}}\right)
=0\ ,\label{cond1}\\
&  i\hbar\ \mathrm{tr\ln}\left(  e+\mathsf{m}\right)  =-(1/2)U^{2}\Delta
F\ ,\ \ \ \Delta F=\left(  F^{\prime}-F\right)  \ . \label{cond2}%
\end{align}
If we furthermore assume that $\Delta F=\Delta F\left(  \phi,\pi
,\lambda\right)  $, which in the case $\Delta F=\Delta F\left(  \phi\right)  $
represents a change of the gauge in the $\mathrm{Sp}(2)$-covariant scheme
\cite{BLT1,BLT2}, then the r.h.s. and l.h.s. of (\ref{cond2}) are independent
of the antifields $\phi_{Aa}^{\ast}$, $\bar{\phi}_{A}$,%
\[
\frac{\delta\lambda_{a}}{\delta\phi_{Aa}^{\ast}}=\frac{\delta\lambda_{a}%
}{\delta\bar{\phi}_{A}}=0\ ,
\]
implying that the condition (\ref{cond1}) is thereby fulfilled:%
\begin{align}
&  \lambda_{a,\mathsf{p}}[\left(  e+\mathsf{m}\right)  ^{-1}]_{b}^{a}\left(
\mathrm{s}^{b}\mathrm{s}^{2}\mathsf{\Gamma}^{\mathsf{p}}\right)
=\lambda_{a,\underline{\mathsf{p}}}[\left(  e+\mathsf{m}\right)  ^{-1}%
]_{b}^{a}\left(  \mathrm{s}^{b}\mathrm{s}^{2}\mathsf{\Gamma}^{\underline
{\mathsf{p}}}\right)  =0\ ,\nonumber\\
&  \mathsf{\Gamma}^{\mathsf{p}}=\left(  \mathsf{\Gamma}^{\underline
{\mathsf{p}}},\mathsf{\Gamma}^{\overline{\mathsf{p}}}\right)
\ ,\ \ \ \mathsf{\Gamma}^{\underline{\mathsf{p}}}=\left(  \phi^{A},\pi
^{Aa},\lambda^{A}\right)  \ ,\ \ \ \mathsf{\Gamma}^{\overline{\mathsf{p}}%
}=\left(  \phi_{Aa}^{\ast},\bar{\phi}_{A}\right)  \ , \label{reduce}%
\end{align}
because the generators $\mathrm{s}^{a}$ are nilpotent in the subspace
$\mathsf{\Gamma}^{\underline{\mathsf{p}}}=\left(  \phi^{A},\pi^{Aa}%
,\lambda^{A}\right)  $, namely, $\mathrm{s}^{a}\mathrm{s}^{2}\mathsf{\Gamma
}^{\underline{\mathsf{p}}}=U^{a}U^{2}\mathsf{\Gamma}^{\underline{\mathsf{p}}%
}\equiv0$. The remaining condition (\ref{cond2}) therefore acquires the form%
\begin{equation}
\Im=\frac{1}{2i\hbar}U^{2}\Delta F\ ,\ \ \ \mathrm{where}\ \ \ \Im
=-\mathrm{tr\ln}\left(  e+\mathsf{\tilde{m}}\right)  \ ,\ \ \ \left(
\mathsf{\tilde{m}}\right)  _{b}^{a}\equiv U^{a}\lambda_{b}\ . \label{acquires}%
\end{equation}
Due to the anticommutativity of $U^{a}$, we can now make use of Lemma
\ref{lemma psi} of Subsection \ref{Yang-Mills} with a subsequent criterion
which can now be represented as follows: the quantity $\mathrm{tr\ln}\left(
e+\mathsf{m}\right)  $ is BRST-antiBRST-exact ($U^{a}$-exact) to all orders of
its expansion in powers of $\lambda_{a}$ if and only if there exists such an
even-valued functional $\Lambda$ that%
\begin{equation}
U^{a}\lambda_{a}=-U^{2}\Lambda\Longrightarrow\lambda_{a}=U_{a}\Lambda+\psi
_{a}\ ,\ \ \ U^{a}\psi_{a}=0\ . \label{exists}%
\end{equation}
Thus, supposing that (\ref{exists}) is indeed the case and taking account of
(\ref{reads}), (\ref{comp_mod}), in terms of $U^{a}$ replacing $s^{a}$, we
find that the condition (\ref{acquires}) is satisfied and reads equivalently%
\begin{equation}
\ln\left(  1-\frac{1}{2}U^{2}\Lambda\right)  ^{-2}-\frac{1}{4}U^{2}\left(
\psi^{2}\right)  \left[  \left(  1-\frac{1}{2}U^{2}\Lambda\right)
^{-2}\right]  =\frac{1}{2i\hbar}U^{2}\Delta F\ ,\ \ \ \psi^{2}\equiv\psi
_{a}\psi^{a}\ , \label{comp_eq_U}%
\end{equation}
which is a compensation equation that expresses $\Lambda$ in terms of a gauge
variation $\Delta F$ and a certain solution $\psi_{a}$ to the equation
$U^{a}\psi_{a}=0$. The relation (\ref{comp_eq_U}) can be accompanied by
comments similar to those which follow (\ref{comp_mod}). In the particular
case $U^{2}\left(  \psi^{2}\right)  =0$, the relation (\ref{comp_eq_U})
reduces to the usual compensation equation.

\section{Relating Gauges in Standard Model and Gribov Ambiguity}

\label{AppA} \renewcommand{\theequation}{\arabic{section}.\arabic{equation}} \setcounter{equation}{0}

Let us consider an application of finite field-dependent BRST-antiBRST
transformations to a fundamental physical model describing almost the entire
variety of the known elementary particles. Namely, we examine the Lagrangian
description of the Standard Model \cite{SM1, SM2, SM3, book2, book3, book4,
Carringt,Kapusta,Okun}, which is an example of a Yang--Mills theory
interacting with spinor and scalar fields.

The classical non-renormalized action of the Standard Model in Minkowski
space-time is given by the sum of several contributions:%
\begin{equation}
S_{\text{SM}}=\int d^{4}x\ \mathcal{L}_{\text{SM}}\ ,\ \ \ \mathcal{L}%
_{\text{SM}}=\mathcal{L}_{\text{gauge fields}}+\mathcal{L}_{\text{leptons}%
}+\mathcal{L}_{\text{quarks}}+\mathcal{L}_{\text{Yukawa}}+\mathcal{L}%
_{\text{Higgs}}\ , \label{SSM}%
\end{equation}
where the Lagrangian density for the even-valued gauge fields $\mathcal{A}%
_{\mu}^{m}(x)=(A_{\mu},A_{\mu}^{\hat{a}},A_{\mu}^{\underline{\alpha}})(x)$ has
the form\footnote{The field $A_{\mu}$ is not to be confused with the
electromagnetic potential.}%
\begin{align}
&  \mathcal{L}_{\text{gauge fields}}=-\frac{1}{4}f_{\mu\nu}f^{\mu\nu}-\frac
{1}{4}F_{\mu\nu}^{\hat{a}}F^{\mu\nu{\hat{a}}}-\frac{1}{4}F_{\mu\nu
}^{\underline{\alpha}}F^{\mu\nu\underline{\alpha}}\ ,\nonumber\\
&  f_{\mu\nu}=\partial_{\lbrack\mu}A_{\nu]}\ ,\ \ \ F_{\mu\nu}^{\hat{a}%
}=\partial_{\lbrack\mu}A_{\nu]}^{\hat{a}}+g\varepsilon_{\hat{b}\hat{c}}%
^{\hat{a}}A_{\mu}^{\hat{b}}A_{\nu}^{\hat{c}}\ ,\ \ \ F_{\mu\nu}^{\underline
{\alpha}}=\partial_{\lbrack\mu}A_{\nu]}^{\underline{\alpha}}+g_{\mathrm{s}%
}f_{\underline{\beta\gamma}}^{\underline{\alpha}}A_{\mu}^{\underline{\beta}%
}A_{\nu}^{\underline{\gamma}}\ ,\nonumber\\
&  A_{\mu}\in u(1)\ ,\ \ \ A_{\mu}^{\hat{a}}\tau_{\hat{a}}\in
su(2)\ ,\ \ \ A_{\mu}^{\underline{\alpha}}\lambda_{\underline{\alpha}}\in
su(3)\ , \label{LSM}%
\end{align}
with $\tau_{\hat{a}}$, ${\hat{a}}=1,2,3$, and $\left(  1/2\right)
\lambda_{\underline{\alpha}}$, $\underline{\alpha}=1,\ldots,8$, being the
$su(2)$ Pauli matrices and Hermitian traceless Gell-Mann matrices, satisfying
the $su(2)$ and $su(3)$ commutation relations%
\begin{equation}
\lbrack\tau_{\hat{a}},\tau_{\hat{a}}]=2i\varepsilon_{{\hat{a}}\hat{b}\hat{c}%
}\tau_{\hat{c}}\ ,\ \ \ \left[  \frac{1}{2}\lambda_{\underline{\alpha}}%
,\frac{1}{2}\lambda_{\underline{\beta}}\right]  =if_{\underline{\alpha
\beta\gamma}}\frac{1}{2}\lambda_{\underline{\gamma}}\ . \label{algsu2su3}%
\end{equation}

The Lagrangian for the odd-valued leptons $l_{L}^{k}$, $l_{R}^{k}$, being
Dirac spinors, reads as follows:%
\begin{equation}
\mathcal{L}_{\text{leptons}}=\sum_{k=1}^{3}\left[  \bar{l}_{L}^{k}i\gamma
^{\mu}\left(  {\partial_{\mu}}-i\frac{g}{2}{A}_{\mu}^{\hat{a}}\tau_{\hat{a}%
}+i\frac{g^{\prime}}{2}{A}_{\mu}\right)  l_{L}^{k}+\bar{l}_{R}^{k}i\gamma
^{\mu}\left(  {\partial_{\mu}}+ig^{\prime}{A}_{\mu}\right)  l_{R}^{k}\right]
\ , \label{Llep}%
\end{equation}
where $l_{L}^{k}$ are left-handed $SU(2)$-doublets, $l_{R}^{k}$ are
right-handed $SU(2)$-singlets, $g$, $g_{\mathrm{s}}$, $g^{\prime}$ are the
coupling constants, and $\gamma^{\mu}$ are the $4\times4$ Dirac matrices
subject to the normalization $\{\gamma^{\mu},\,\gamma^{\nu}\}=2\eta^{\mu\nu}$,
with the Minkowski metric tensor $\eta^{\mu\nu}$. The quantities\footnote{The
explicit form of the Gell-Mann matrices $\lambda_{\underline{\alpha}}$, as
well as the $su(3)$ structure constants $f_{\underline{\alpha\beta\gamma}}$,
may be found, e.g., in \cite{IzyksonZuber}.} $\varepsilon_{\hat{b}\hat{c}%
}^{\hat{a}}=\varepsilon_{\hat{a}\hat{b}\hat{c}}=\varepsilon^{\hat{a}\hat
{b}\hat{c}}$ and $f_{\underline{\beta\gamma}}^{\underline{\alpha}%
}=f_{\underline{\alpha\beta\gamma}}=f^{\underline{\alpha\beta\gamma}}$ in
(\ref{LSM}) and (\ref{algsu2su3}) are completely antisymmetric.

The QCD (quark) sector of the strong interactions described by the $3$ quark
generations $(u,d)$, $(c,s)$, $(t,b)$, organized into Dirac spinors
$(u_{k},d_{k})^{T}$, has the form%
\begin{align}
\mathcal{L}_{\text{quarks}}  &  =\sum_{k=1}^{3}\left\{  \left[
\begin{array}
[c]{c}%
\bar{u}_{k}\\
\bar{d}_{k}^{\prime}%
\end{array}
\right]  _{L}i\gamma^{\mu}\left[  {\partial}_{\mu}-i\frac{g_{\mathrm{s}}}%
{2}{A}_{\mu}^{\underline{\alpha}}\lambda_{\underline{\alpha}}-i\frac{g}{2}%
{A}_{\mu}^{\hat{a}}\tau_{\hat{a}}-i\frac{g^{\prime}}{6}{A}_{\mu}\right]
\left[
\begin{array}
[c]{c}%
u_{k}\\
d_{k}^{\prime}%
\end{array}
\right]  _{L}\right. \nonumber\\
&  +\left.  \bar{u}_{R}^{k}i\gamma^{\mu}\left[  {\partial}_{\mu}%
-i\frac{g_{\mathrm{s}}}{2}{A}_{\mu}^{\underline{\alpha}}\lambda_{\underline
{\alpha}}-i\frac{2g^{\prime}}{3}{A}_{\mu}\right]  u_{R}^{k}+\bar{d}%
_{R}^{\prime k}i\gamma^{\mu}\left[  {\partial}_{\mu}-i\frac{g_{\mathrm{s}}}%
{2}{A}_{\mu}^{\underline{\alpha}}\lambda_{\underline{\alpha}}+i\frac
{g^{\prime}}{3}{A}_{\mu}\right]  d_{R}^{\prime k}\right\}  \ ,\label{QCD}\\
d^{\prime k}  &  =U_{\text{CKM}}^{kk^{\prime}}d^{k^{\prime}}\ ,\ \ \ u^{k}%
=(u,c,t)\ ,\ \ \ d^{k}=(d,s,b)\ , \label{Cabibbo}%
\end{align}
where the respective left- and right-handed $SU(3)$-triplets $(u_{k}%
,d_{k})_{L}^{T}$ and $(u_{k},d_{k})_{R}^{T}$, are $SU(2)$-doublets and
$SU(2)$-singlets, respectively, and $U_{\text{CKM}}$ is the
Cabibbo--Kobayashi--Maskawa matrix \cite{CKM}.

The masses of particles in the Standard Model are generated by the Yukawa
interaction term%
\begin{equation}
\mathcal{L}_{\text{Yukawa}}=-\frac{1}{\sqrt{2}}\sum_{k=1}^{3}\left\{
f_{k}^{u}\left[
\begin{array}
[c]{c}%
\bar{u}^{k}\\
\bar{d}^{k}%
\end{array}
\right]  _{L}\varphi u_{R}^{k}+f_{k}^{d}\left[
\begin{array}
[c]{c}%
\bar{u}^{k}\\
\bar{d}^{k}%
\end{array}
\right]  _{L}\varphi d_{R}^{k}+f_{k}^{l}\bar{l}_{L}^{k}\varphi l_{R}%
^{k}+\text{h.c.}\right\}  \ , \label{Yukawa}%
\end{equation}
where $f_{k}^{u}$, $f_{k}^{d}$, $f_{k}^{l}$ are the Yukawa couplings, and the
Brout--Englert--Higgs Lagrangian is given by%
\begin{equation}
\mathcal{L}_{\text{Higgs}}=\frac{1}{2}\left\vert \left(  i\partial_{\mu
}+\left(  g/2\right)  A_{\mu}^{\hat{a}}\tau_{\hat{a}}+\left(  g^{\prime
}/2\right)  A_{\mu}\right)  \right\vert \varphi^{2}-\frac{\mu^{2}}{2}%
|\varphi|^{2}-\frac{\lambda}{4}|\varphi|^{4}\ , \label{Higgs}%
\end{equation}
where $\varphi$ is a Bosonic field, being an $SU(2)$-doublet, $\mu^{2}$ is a
negative constant, and $\lambda$ is the Higgs self-interaction coupling constant.

We consider the minimal Standard Model, which means that the neutrinos
entering the left-handed $SU(2)$ doublets $l_{L}^{k}$ are assumed to be massless.

The action $S_{\text{SM}}$ in (\ref{SSM}) is invariant with respect to the
following gauge transformations acting in the configuration space
$\mathcal{M}_{\text{SM}}$:%
\begin{equation}
\mathcal{M}_{\text{SM}}=A^{i}=\{\mathcal{A}_{\mu}^{m};\Sigma^{I}\}(x)=\left\{
\mathcal{A}_{\mu}^{m};{l}_{L}^{k\hat{A}},\bar{l}_{L}^{k\hat{B}},{l}_{R}%
^{k},\bar{l}_{R}^{k},\left(  (u_{k},d_{k})_{L}^{p}\right)  ^{T},\left(
(\bar{u}_{k},\bar{d}_{k})_{L}^{q}\right)  ^{T},\left(  (u_{k},d_{k}%
)_{R}^{\underline{A}}\right)  ^{T},\left(  (\bar{u}_{k},\bar{d}_{k}%
)_{R}^{\underline{B}}\right)  ^{T},\varphi^{\hat{C}}\right\}  (x)\ ,
\label{confSM}%
\end{equation}
where $[\hat{A};\hat{A}^{\prime};\hat{C}]=[(\hat{a},1);(\hat{a}^{\prime
},1^{\prime});(\hat{c},1)]$, $[\underline{A};\underline{B}]=[(\underline
{\alpha},1);(\underline{\beta},1)]$, and Dirac-conjugated spinors, such as
$\bar{l}_{L}^{k}$, are assumed to be independent:%
\[
\delta\left(  \mathcal{A}^{\mu m},\Sigma^{I}\right)  (x)=\int d^{4}y\ \left(
R_{n}^{\mu m},R_{n}^{I}\right)  (x,y)\varsigma^{n}(y)\ ,\ \ \ \alpha
=(n,y)\ ,\ \ \ \varsigma^{n}=\left(  \varsigma,\varsigma^{\hat{b}}%
,\varsigma^{\underline{\beta}}\right)  \ .
\]
Here, the generators $\left(  R_{n}^{\mu m},R_{n}^{I}\right)  (x,y)=\left(
R_{n}^{\mu m}(\mathcal{A}),R_{n}^{I}(\Sigma)\right)  \delta(x-y)$ form the Lie
algebra of the gauge transformations and read as follows: for the gauge fields
$\mathcal{A}^{\mu m}$,%
\begin{equation}
R_{n}^{\mu m}(\mathcal{A})=\left\{
\begin{array}
[c]{ll}%
\partial^{\mu}\ , & m=1,\ n=1,\\
\partial^{\mu}\delta^{\hat{a}\hat{b}}+g\varepsilon^{\hat{a}\hat{c}\hat{b}%
}A^{\mu\hat{c}}\ , & m={\hat{a}},\,n=\hat{b},\\
\partial^{\mu}\delta^{\underline{\alpha\beta}}+g_{s}f^{\underline{\alpha
\gamma\beta}}A^{\mu\underline{\gamma}}\ , & m=\underline{\alpha}%
,\,n=\underline{\beta},
\end{array}
\right.  \label{gtranSMegf}%
\end{equation}
and for the matter fields $\Sigma^{{I}}$,
\begin{equation}
R_{n}^{I}(\Sigma)=\left\{
\begin{array}
[c]{ll}%
\left(  0,\ -g\varepsilon^{\hat{a}\hat{c}\hat{b}}({l}_{L}^{k})^{\hat{c}%
},\,\frac{1}{2}g^{\prime}{l}_{L}^{k}\right)  (x), & I=(\hat{a}%
,1),\,n=(\underline{\beta},\hat{b},1),\\
\left(  0,\ -g\varepsilon^{\hat{a}^{\prime}\hat{c}\hat{b}}(\bar{l}_{L}%
^{k})^{\hat{c}},\,\frac{1}{2}g^{\prime}\bar{l}_{L}^{k}\right)  (x), &
I=(\hat{a}^{\prime},1^{\mathrm{I}}),\\
(0,\ 0,\ {g^{\prime}}{l}_{R}^{k})(x), & I=1^{\mathrm{II}},\\
(0,\ 0,\ {g^{\prime}}\bar{l}_{R}^{k})(x), & I=1^{\mathrm{III}},\\
\left(  -g_{s}f^{\underline{\alpha\gamma\beta}}\left(  (u_{k},d_{k})_{L}%
^{T}\right)  ^{\underline{\gamma}},\,-g\varepsilon^{\hat{d}\hat{e}\hat{b}%
}\left(  (u_{k},d_{k})_{L}^{T}\right)  ^{\hat{e}},\,-\frac{1}{6}g^{\prime
}(u_{k},d_{k})_{L}^{T}\right)  (x), & I=(\underline{\alpha},\hat
{d},1^{\mathrm{IV}}),\\
\left(  -g_{s}f^{\underline{\alpha^{\prime}\gamma\beta}}\left(  (\bar{u}%
_{k},\bar{d}_{k})_{L}^{T}\right)  ^{\underline{\gamma}},\,-g\varepsilon
^{\hat{d}^{\prime}\hat{e}\hat{b}}\left(  (\bar{u}_{k},\bar{d}_{k})_{L}%
^{T}\right)  ^{\hat{e}},\,-\frac{1}{6}g^{\prime}(\bar{u}_{k},\bar{d}_{k}%
)_{L}^{T}\right)  (x), & I=(\underline{\alpha}^{\prime},\hat{d}^{\prime
},1^{\mathrm{V}}),\,\\
\left(  -g_{s}f^{\underline{\delta\sigma\beta}}({u_{k}}_{R})^{\underline
{\sigma}},\ 0,\,-\frac{2}{3}g^{\prime}{u_{k}}_{R}\right)  (x), &
I=(\underline{\delta},1^{\mathrm{VI}}),\,\\
\left(  -g_{s}f^{\underline{\delta^{\prime}\sigma\beta}}({d_{k}}%
_{R})^{\underline{\sigma}},\ 0,\,\frac{1}{3}g^{\prime}{d_{k}}_{R}\right)
(x), & I=(\underline{\delta}^{\prime},1^{\mathrm{VII}}),\,\\
\left(  0,\ g\varepsilon^{\hat{c}\hat{e}\hat{b}}\varphi^{\hat{e}},\,\frac
{1}{2}g^{\prime}\varphi\right)  (x), & I=(\hat{c},1^{\mathrm{VIII}}),\,
\end{array}
\right.  \label{gtranSMe}%
\end{equation}
with $k=1,2,3$ and the structure constants $F^{lmn}$ in the sector of the
gauge fields $\mathcal{A}_{\mu}^{m}$ given by%
\begin{equation}
F_{\alpha\beta}^{\gamma}{}=F^{lmn}\delta(x-z)\delta(y-z)\ ,\ \ \ F^{lmn}%
=\left(  0,\ g\varepsilon^{\hat{a}\hat{b}\hat{c}},\ g_{s}f^{\underline
{\alpha\beta\gamma}}\right)  \ , \label{fabg}%
\end{equation}
for the $U(1)$, $SU(2)$, $SU(3)$ gauge subgroups, respectively. The form of
the structure constants for the given model is obviously consistent with
(\ref{R(A)}), taking account of the convention%
\begin{equation}
D_{\mu}^{mn}=\delta^{mn}\partial_{\mu}+F^{mln}\mathcal{A}_{\mu}^{l}\ .
\label{convent}%
\end{equation}
In (\ref{gtranSMe}), we do not expose the explicit structure of the Dirac
spinor indices, implying that it enters the index $I$, except for the gauge
and Higgs fields. Besides,\textbf{ }the scalar indices $1$ and $1^{\mathrm{I}%
},\ldots,1^{\mathrm{VIII}}$ correspond to the $U(1)$ group of the weak hypercharge.

Under the assumption that the vacuum expectation values of all the fields are
zero, we present the Higgs field $\varphi$ as follows:%
\begin{equation}
\varphi=\left[
\begin{array}
[c]{c}%
0\\
\eta+\chi
\end{array}
\right]  +i\zeta^{a}\tau_{a}\left[
\begin{array}
[c]{c}%
0\\
1
\end{array}
\right]  =\left[
\begin{array}
[c]{c}%
\zeta_{2}+i\zeta_{1}\\
\eta+\chi-i\zeta_{3}%
\end{array}
\right]  \ , \label{Higgs-phi}%
\end{equation}
where $\zeta_{a}$ are the Goldstone Bosons, $\eta$ is the vacuum expectation
value of the Higgs field, and $\chi$ are fluctuations of the Higgs field.

Let us choose a gauge Boson $F_{\boldsymbol{\xi}}$ corresponding to an
$R_{\boldsymbol{\xi}}$-like family of gauges, parameterized by a set of
numbers $\boldsymbol{\xi}=\left(  \xi_{1},\xi_{2},\xi_{3}\right)  $ and
related to the Landau and Feynman (covariant) gauges for $\boldsymbol{\xi
}=\left(  0,0,0\right)  $ and $\boldsymbol{\xi}=\left(  1,1,1\right)  $,
respectively:%
\begin{equation}
F_{\boldsymbol{\xi}}\left(  \mathcal{A},C\right)  =-\frac{1}{2}\int
d^{4}x\ \left(  A_{\mu}A^{\mu}+A_{\mu}^{\hat{a}}A^{\hat{a}\mu}+A_{\mu
}^{\underline{\alpha}}A^{\underline{\alpha}\mu}\right)  +\frac{1}%
{4}\varepsilon_{ab}\int d^{4}x\ \left(  \xi_{1}C^{a}C^{b}+\xi_{2}C^{\hat{a}%
a}C^{\hat{a}b}+\xi_{3}C^{\underline{\alpha}a}C^{\underline{\alpha}b}\right)
\ . \label{F(A,C)}%
\end{equation}
Using (\ref{action}), (\ref{solxy})--(\ref{R(A)}), we can now present the
corresponding quantum action $S_{F}\left(  A\right)  $ in the path integral
(\ref{z(j)}). In doing so, we extend the results of \cite{MRnew}, considering
the part that deals with the Yang--Mills theory, in the sense that the
relevant formulae\footnote{Specifically, we use Eqs. (4.11)--(4.16) of
\cite{MRnew}, where $f^{lmn}$ are identified with $F^{lmn}$ in (\ref{fabg}).}
are now written down in the specific cases of the $U\left(  1\right)  $,
$SU\left(  2\right)  $, $SU\left(  3\right)  $ groups and feature
contributions related to the presence of all the three cases, complete with
the corresponding classical fields $(A_{\mu},A_{\mu}^{\hat{a}},A_{\mu
}^{\underline{\alpha}})$, as well as the ghost-antighost $\left(
C^{a},C^{\hat{a}a},C^{\underline{\alpha}a}\right)  $ and Nakanishi--Lautrup
$\left(  B,B^{\hat{a}},B^{\underline{\alpha}}\right)  $ fields. The quantum
action $S_{F}(A,B,C)$ corresponding to the gauge-fixing functional
(\ref{F(A,C)}) reads%
\begin{equation}
S_{F_{\boldsymbol{\xi}}}(A,B,C)=S_{\text{SM}}\left(  A\right)  +\left(
1/2\right)  s^{a}s_{a}\left[  F_{\boldsymbol{\xi}}\left(  \mathcal{A}%
,C\right)  \right]  =S_{\text{SM}}\left(  A\right)  +S_{\mathrm{gf}}\left(
\mathcal{A},B;{\boldsymbol{\xi}}\right)  +S_{\mathrm{gh}}\left(
\mathcal{A},C;{\boldsymbol{\xi}}\right)  +S_{\mathrm{add}}\left(
C;{\boldsymbol{\xi}}\right)  \ , \label{S(A,B,C)}%
\end{equation}
where the gauge-fixing term $S_{\mathrm{gf}}$, the ghost term $S_{\mathrm{gh}%
}$, and the interaction term $S_{\mathrm{add}}$, quartic in the
ghost-antighost fields, are given by%
\begin{align}
S_{\mathrm{gf}}  &  =\int d^{4}x\ \left\{  \left[  \left(  \partial^{\mu
}A_{\mu}\right)  +\frac{\xi_{1}}{2}B\right]  B+\left[  \left(  \partial^{\mu
}A_{\mu}^{\hat{a}}\right)  +\frac{\xi_{2}}{2}B^{\hat{a}}\right]  B^{\hat{a}%
}+\left[  \left(  \partial^{\mu}A_{\mu}^{\underline{\alpha}}\right)
+\frac{\xi_{3}}{2}B^{\underline{\alpha}}\right]  B^{\underline{\alpha}%
}\right\}  \,,\label{Sgf}\\
S_{\mathrm{gh}}  &  =\frac{1}{2}\varepsilon_{ab}\int d^{4}x\ \left[  \left(
\partial^{\mu}C^{a}\right)  \partial_{\mu}C^{b}+\left(  \partial^{\mu}%
C^{\hat{a}a}\right)  D_{\mu}^{\hat{a}\hat{b}}C^{\hat{b}b}+\left(
\partial^{\mu}C^{\underline{\alpha}a}\right)  D_{\mu}^{\underline{\alpha\beta
}}C^{\underline{\beta}b}\right]  \ ,\label{Sgh}\\
S_{\mathrm{add}}  &  =-\frac{1}{48}\varepsilon_{ab}\varepsilon_{cd}\int
d^{4}x\ \left(  \xi_{2}g^{2}\varepsilon^{\hat{a}\hat{b}\hat{e}}\varepsilon
^{\hat{e}\hat{c}\hat{d}}C^{\hat{d}a}C^{\hat{c}c}C^{\hat{b}b}C^{\hat{a}d}%
+\xi_{3}g_{s}^{2}f^{\underline{\alpha\beta\sigma}}f^{\underline{\sigma
\gamma\delta}}C^{\underline{\delta}a}C^{\underline{\gamma}c}C^{\underline
{\beta}b}C^{\underline{\alpha}d}\right)  \,. \label{Sadd}%
\end{align}
In (\ref{F(A,C)}), the gauge-fixing functional $F_{\mathbf{0}}\left(
\mathcal{A},C\right)  $, with $\boldsymbol{\xi}=\left(  0,0,0\right)  $,%
\begin{equation}
F_{\mathbf{0}}\left(  \mathcal{A},C\right)  =-\frac{1}{2}\int d^{4}x\ \left(
A_{\mu}A^{\mu}+A_{\mu}^{\hat{a}}A^{\hat{a}\mu}+A_{\mu}^{\underline{\alpha}%
}A^{\underline{\alpha}\mu}\right)  \ , \label{0}%
\end{equation}
induces the contribution $S_{\mathrm{gf}}\left(  \mathcal{A},B\right)  $ to
the quantum action that arises in the case of the Landau gauge, $\chi
^{m}\left(  \mathcal{A}\right)  =\partial^{\mu}\mathcal{A}_{\mu}^{m}$, whereas
the functional $F_{\mathbf{1}}\left(  \mathcal{A},C\right)  $, with
$\boldsymbol{\xi}=\left(  1,1,1\right)  $,%
\begin{equation}
F_{\mathbf{1}}\left(  \mathcal{A},C\right)  =-\frac{1}{2}\int d^{4}x\ \left(
A_{\mu}A^{\mu}+A_{\mu}^{\hat{a}}A^{\hat{a}\mu}+A_{\mu}^{\underline{\alpha}%
}A^{\underline{\alpha}\mu}\right)  +\frac{1}{4}\varepsilon_{ab}\int
d^{4}x\ \left(  C^{a}C^{b}+C^{\hat{a}a}C^{\hat{a}b}+C^{\underline{\alpha}%
a}C^{\underline{\alpha}b}\right)  \ , \label{1}%
\end{equation}
corresponds to the Feynman gauge, $\chi^{m}\left(  \mathcal{A},B\right)
=\partial^{\mu}\mathcal{A}_{\mu}^{m}+\left(  1/2\right)  B^{m}$.

In order to find the parameters $\lambda_{a}=s_{a}\Lambda$ of a finite
field-dependent BRST-antiBRST transformation that connects an
$R_{\boldsymbol{\xi}}$ gauge with an $R_{\boldsymbol{\xi+}\Delta
\boldsymbol{\xi}}$ gauge, according to (\ref{lambda-Fsol}), we need the
quantities $\Delta F_{\boldsymbol{\xi}}$, $s^{a}\left(  \Delta
F_{\boldsymbol{\xi}}\right)  $, $s^{a}s_{a}\left(  \Delta F_{\boldsymbol{\xi}%
}\right)  $, which are evaluated using%
\begin{align}
&  \Delta F_{\boldsymbol{\xi}}=F_{\boldsymbol{\xi}+\Delta\boldsymbol{\xi}%
}-F_{\boldsymbol{\xi}}=\frac{1}{4}\varepsilon_{ab}\int d^{4}x\ \left(
\Delta\xi_{1}C^{a}C^{b}+\Delta\xi_{2}C^{\hat{a}a}C^{\hat{a}b}+\Delta\xi
_{3}C^{\underline{\alpha}a}C^{\underline{\alpha}b}\right)  \ ,\nonumber\\
&  s^{a}\left(  \Delta F_{\boldsymbol{\xi}}\right)  =\frac{1}{2}\int
d^{4}x\ \left(  \Delta\xi_{1}BC^{a}+\Delta\xi_{2}B^{\hat{a}}C^{\hat{a}%
a}+\Delta\xi_{3}B^{\underline{\alpha}}C^{\underline{\alpha}a}\right)
\ ,\label{using}\\
&  \frac{1}{2}s^{a}s_{a}\left(  \Delta F_{\boldsymbol{\xi}}\right)  =\left.
\left(  S_{\mathrm{gf}}+S_{\mathrm{gh}}+S_{\mathrm{add}}\right)  \right\vert
_{\boldsymbol{\xi}+\Delta\boldsymbol{\xi}}-\left.  \left(  S_{\mathrm{gf}%
}+S_{\mathrm{gh}}+S_{\mathrm{add}}\right)  \right\vert _{\boldsymbol{\xi}%
}\ \ ,\nonumber
\end{align}
with allowance made for (\ref{xy}), (\ref{fabg}) and $B^{m}=\left(
B,B^{\hat{a}},B^{\underline{\alpha}}\right)  $,$\ C^{na}=\left(  C^{a}%
,C^{\hat{a}a},C^{\underline{\alpha}a}\right)  $,%
\begin{align}
&  s^{a}A^{i}(x)=X_{1}^{ia}(x)=\left(  R_{n}^{\mu m}(\mathcal{A}),R_{n}%
^{I}(\Sigma)\right)  C^{na}(x)\ ,\nonumber\\
&  -\frac{1}{2}s^{2}A^{i}(x)=Y_{1}^{i}(x)=\left(  R_{n}^{\mu m}(\mathcal{A}%
),R_{n}^{I}(\Sigma)\right)  B^{n}(x)-\frac{1}{2}\varepsilon_{ab}\left(
F^{mrn}C^{ra}R_{s}^{\mu n}C^{sb},\frac{\partial R_{m}^{I}(\Sigma)}%
{\partial\Sigma^{J}}R_{n}^{J}(\Sigma)C^{na}C^{mb}\right)  (x)\ ,\nonumber\\
&  s^{a}B^{m}(x)=X_{2}^{ma}(x)=-\frac{1}{2}F^{mln}\left(  B^{n}C^{la}+\frac
{1}{6}F^{nrs}C^{sb}C^{ra}C^{lc}\varepsilon_{cb}\right)  (x)\ ,\nonumber\\
&  -\frac{1}{2}s^{2}B^{m}(x)=Y_{2}^{m}(x)=0\,,\nonumber\\
&  s^{a}C^{mb}(x)=X_{3}^{mab}(x)=-\left(  \varepsilon^{ab}B^{m}+\frac{1}%
{2}F^{mln}C^{nb}C^{la}\right)  (x)\ ,\nonumber\\
&  -\frac{1}{2}s^{2}C^{ma}(x)=Y_{3}^{ma}(x)=-2X_{2}^{ma}(x)\ , \label{s^a-phi}%
\end{align}
which determines the finite BRST-antiBRST transformations $\phi^{A}%
\rightarrow\phi^{\prime A}=\phi^{A}\exp(\overleftarrow{s}^{a}\lambda_{a})$ in
the Standard Model.

As a result, the functional parameters $\lambda_{a}=s_{a}\Lambda$ that connect
an $R_{\boldsymbol{\xi}}$-like gauge to an $R_{\boldsymbol{\xi}+\Delta
\boldsymbol{\xi}}$-like gauge are given by an extension of the result
\cite{MRnew}, featuring the contributions related to all the three groups
$U\left(  1\right)  $, $SU\left(  2\right)  $, $SU\left(  3\right)  $:%
\begin{align}
\lambda_{a}  &  =\frac{1}{4i\hbar}\varepsilon_{ab}\int d^{4}x\ \left(
\Delta\xi_{1}BC^{b}+\Delta\xi_{2}B^{\hat{a}}C^{\hat{a}b}+\Delta\xi
_{3}B^{\underline{\alpha}}C^{\underline{\alpha}b}\right) \nonumber\\
&  \times\sum_{n=0}^{\infty}\frac{1}{\left(  n+1\right)  !}\left\{  \frac
{1}{4i\hbar}\int d^{4}y\ \left[  \Delta\xi_{1}B^{2}+\Delta\xi_{2}\left(
B^{\hat{a}}B^{\hat{a}}-\frac{g^{2}}{24}\varepsilon^{\hat{a}\hat{b}\hat{e}%
}\varepsilon^{\hat{e}\hat{c}\hat{d}}C^{\hat{d}c}C^{\hat{c}e}C^{\hat{b}%
d}C^{\hat{a}g}\varepsilon_{cd}\varepsilon_{eg}\right)  \right.  \right.
\nonumber\\
&  +\left.  \left.  \Delta\xi_{3}\left(  B^{\underline{\alpha}}B^{\underline
{\alpha}}-\frac{g_{s}^{2}}{24}f^{\underline{\alpha\beta\sigma}}f^{\underline
{\sigma\gamma\delta}}C^{\underline{\delta}c}C^{\underline{\gamma}%
e}C^{\underline{\beta}d}C^{\underline{\alpha}g}\varepsilon_{cd}\varepsilon
_{eg}\right)  \right]  \right\}  ^{n}\ , \label{lambdaSM}%
\end{align}
where the corresponding potential $\Lambda$ is given by%
\begin{align}
\Lambda &  =\frac{1}{8i\hbar}\varepsilon_{ab}\int d^{4}x\ \left(  \Delta
\xi_{1}C^{a}C^{b}+\Delta\xi_{2}C^{\hat{a}a}C^{\hat{a}b}+\Delta\xi
_{3}C^{\underline{\alpha}a}C^{\underline{\alpha}b}\right)  \times\nonumber\\
&  \times\sum_{n=0}^{\infty}\frac{1}{\left(  n+1\right)  !}\left\{  \frac
{1}{4i\hbar}\int d^{4}y\ \left[  \Delta\xi_{1}B^{2}+\Delta\xi_{2}\left(
B^{\hat{a}}B^{\hat{a}}-\frac{g^{2}}{24}\varepsilon^{\hat{a}\hat{b}\hat{e}%
}\varepsilon^{\hat{e}\hat{c}\hat{d}}C^{\hat{d}c}C^{\hat{c}e}C^{\hat{b}%
d}C^{\hat{a}g}\varepsilon_{cd}\varepsilon_{eg}\right)  \right.  \right.
\nonumber\\
&  +\left.  \left.  \Delta\xi_{3}\left(  B^{\underline{\alpha}}B^{\underline
{\alpha}}-\frac{g_{s}^{2}}{24}f^{\underline{\alpha\beta\sigma}}f^{\underline
{\sigma\gamma\delta}}C^{\underline{\delta}c}C^{\underline{\gamma}%
e}C^{\underline{\beta}d}C^{\underline{\alpha}g}\varepsilon_{cd}\varepsilon
_{eg}\right)  \right]  \right\}  ^{n}\ . \label{corresp_Lambda}%
\end{align}
This solves the problem of reaching any gauge in the family of
$R_{\boldsymbol{\xi}}$-like gauges, starting from a certain gauge encoded in
the path integral by a functional $F_{\boldsymbol{\xi}}$, within the
BRST-antiBRST quantization of the Standard Model, by means of finite
BRST-antiBRST transformations with field-dependent parameters $\lambda_{a}$.

According to (\ref{z(j)}), the generating functionals of Green's functions for
the Standard Model in $R_{\xi}$-like gauges read as follows:%
\begin{align}
&  Z_{\text{SM},\boldsymbol{\xi}}(J,\eta)=\int d\widetilde{\phi}\ \exp\left\{
\frac{i}{\hbar}\left[  S_{\text{SM}}(\widetilde{A})-\frac{1}{2}%
F_{\boldsymbol{\xi}}\overleftarrow{s}{}^{2}+J_{A}\widetilde{\phi}^{A}\right]
\right\}  =\exp\left[  \frac{i}{\hbar}W_{\text{SM},\boldsymbol{\xi}}%
(J,\eta)\right]  \,,\label{z(j)SM}\\
&  \exp\left[  \frac{i}{\hbar}\Gamma_{\text{SM},\xi}(\phi,\eta)\right]  =\int
d\widetilde{\phi}\ \exp\left\{  \frac{i}{\hbar}\left[  S_{\text{SM}%
}(\widetilde{A})-\frac{1}{2}F_{\boldsymbol{\xi}(\widetilde{\mathcal{A}%
},\widetilde{C})}\overleftarrow{s}{}^{2}+\frac{\delta\Gamma_{\text{SM},\xi
}(\phi)}{\delta\phi^{A}}\left(  \phi^{A}-\widetilde{\phi}^{A}\right)  \right]
\right\}  \ , \label{EASM}%
\end{align}
where the effective action\footnote{The minimal Standard Model on a nontrivial
gravitational background with $g_{\mu\nu}(x)=\eta_{\mu\nu}+\ldots$ has been
examined, e.g., in \cite{Kazinski}, where the effective action, depending on
$g_{\mu\nu}$ and $\eta$, was determined on the mass shell.} $\Gamma
_{\text{SM},\boldsymbol{\xi}}(\phi,\eta)$ is the Legendre transform of
$W_{\text{SM}}$ with respect to $J_{A}$, namely,
\begin{equation}
\Gamma_{\text{SM},\boldsymbol{\xi}}(\phi,\eta)=W_{\text{SM},\boldsymbol{\xi}%
}(J,\eta)-J_{A}\phi^{A}\ ,\ \ \ \text{where}\ \ \ \phi^{A}=\frac{\delta
W_{\text{SM}}}{\delta J_{A}}\ ,\ \ \ \frac{\delta\Gamma_{\text{SM}}}%
{\delta\phi^{A}}=-J_{A}\ . \label{EABV}%
\end{equation}
The modified Ward identity (\ref{mWIclalg}) for $Z_{\text{SM},\boldsymbol{\xi
}}(J,\eta)$ depends on field-dependent parameters, $\lambda_{a}\left(
\Delta\boldsymbol{\xi}\right)  =\Lambda\left(  \Delta\boldsymbol{\xi}\right)
\overleftarrow{s}_{a}$ in (\ref{lambdaSM}), and has the form%
\begin{equation}
\left\langle \left\{  1+\frac{i}{\hbar}J_{A}\left[  X^{Aa}\lambda_{a}%
(\Delta\boldsymbol{\xi})-\frac{1}{2}Y^{A}\lambda^{2}(\Delta\boldsymbol{\xi
})\right]  -\frac{1}{4}\left(  \frac{i}{\hbar}\right)  {}^{2}\varepsilon
_{ab}J_{A}X^{Aa}J_{B}X^{Bb}\lambda^{2}(\Delta\boldsymbol{\xi})\right\}
\left(  1-\frac{1}{2}\Lambda\left(  \Delta\boldsymbol{\xi}\right)
\overleftarrow{s}^{2}\right)  {}^{-2}\right\rangle _{F_{\boldsymbol{\xi}}%
,J}=1\ . \label{modWISM}%
\end{equation}

The non-Abelian nature of the gauge group, because of the differential gauges
\cite{Singer} implied\ by the gauge Boson (\ref{F(A,C)}), leads to the Gribov
ambiguity \cite{Gribov}, described initially in the Coulomb gauge, and
controlled in the Gribov--Zwanziger theory \cite{Zwanziger1, Zwanziger2} by
using the horizon functionals $h_{\mathbf{0}}$ and $h_{\mathbf{1}}$ in the
Landau and Feynman gauges, respectively. For contemporary considerations,
justified by lattice calculations of Gribov copies, see, e.g., \cite{lattice,
lattice1, lattice2}. For applications of the Gribov--Zwanziger theory in the
Coulomb, Landau and maximal Abelian gauges, as well as in covariant
$R_{\boldsymbol{\xi}}$-gauges in the pure Yang--Mills theory, see
\cite{Sorellas, Sorellas1, Sorellas2, Sorellas3, Sorellas4, Sorellas5, SS,
MAG, MAG1, HFZwanziger}. Notice that there exist other approaches intended to
eliminate (or bypass) the Gribov ambiguity problem: first, the procedure of
imposing an algebraic (instead of differential) gauge on auxiliary scalar
fields in a theory which is non-perturbatively equivalent to the Yang--Mills
theory \cite{Slavnoveq, Slavnoveq1, Slavnoveq2}, second, the procedure of
averaging over the Gribov copies with a non-uniform weight in the path
integral and the replica trick \cite{Serreau,Serreau1}, third, the
incorporation of the Gribov factor (restricting the functional measure in the
path integral to the first Gribov region) into the Faddeev--Popov matrix,
thereby modifying the gauge algebra of gauge transformations \cite{MRnew5}.

As we turn to the Gribov ambiguity problem and Gribov--Zwanziger theory, it
should be noted, first of all, that the Landau gauge implies, due to the
preservation of the gauge condition when extracting the unique representative
from the gauge orbit of field configurations in terms of the equation%
\begin{equation}
\partial_{\mu}\left(  \mathcal{A}^{\mu m}(x)+\delta\mathcal{A}^{\mu
m}(x)\right)  =\partial_{\mu}\mathcal{A}^{\mu m}(x)\Longrightarrow\int
d^{4}y\ \partial_{\mu}R_{n}^{\mu m}(x,y)\varsigma^{n}(y)=\partial_{\mu}\left(
\partial^{\mu}\varsigma,D_{\mu}^{\hat{a}\hat{b}}\varsigma^{\hat{b}},D_{\mu
}^{\underline{\alpha\beta}}\varsigma^{\underline{\beta}}\right)
(x)=(0,0,0)\ ,\label{grcond}%
\end{equation}
that, in addition to a vanishing solution $\varsigma_{0}^{n}(x)=(0,0,0)$,
there also exist many smooth solutions $\varsigma_{(k_{1},k_{2})}%
^{n}(x)=\left(  0,\varsigma_{k_{1}}^{\hat{b}},\varsigma_{k_{2}}^{\underline
{\beta}}\right)  (x)$ for configurations of the non-Abelian gauge fields
$\mathcal{A}^{\mu\hat{a}},\mathcal{A}^{\mu\underline{\alpha}}$ vanishing at
the spatial infinity in Minkowski space-time. Second, the Gribov--Zwanziger
theory implies the sum of the horizon functionals corresponding to the $SU(2)$
and $SU(3)$ gauge groups\footnote{Further on, the consideration of the
Gribov--Zwanziger theory is based on the assumption that we deal with the
$\mathbb{R}^{4}$ Euclidean space-time.}:
\begin{align}
&  h_{\mathbf{0}}(\mathcal{A})=h_{\boldsymbol{0}}^{SU(2)}+h_{\boldsymbol{0}%
}^{SU(3)}\ ,\label{hfSM}\\
&  h_{\mathbf{0}}^{SU(2)}=\gamma_{1}^{2}g^{2}\int d^{4}x\ d^{4}y\ \varepsilon
^{\hat{a}\hat{b}\hat{c}}A_{\mu}^{\hat{b}}\left(  x\right)  \left(
K_{SU(2)}^{-1}\right)  ^{\hat{a}\hat{d}}\left(  x;y\right)  \varepsilon
^{\hat{d}\hat{e}\hat{c}}A^{\mu\hat{e}}\left(  y\right)  +4\cdot3\gamma_{1}%
^{2}g^{2}\ ,\label{hfSM2}\\
&  h_{\mathbf{0}}^{SU(3)}=\gamma_{2}^{2}g_{s}^{2}\int d^{4}x\ d^{4}%
y\ f^{\underline{\alpha\beta\gamma}}A_{\mu}^{\underline{\beta}}\left(
x\right)  \left(  K_{SU(3)}^{-1}\right)  ^{\underline{\alpha\delta}}\left(
x;y\right)  f^{\underline{\delta\sigma\gamma}}A^{\mu\underline{\sigma}}\left(
y\right)  +4\cdot8\gamma_{2}^{2}g_{s}^{2}\ ,\label{hfSM3}%
\end{align}
where $h_{\boldsymbol{0}}$ does not depend on the matter fields $\Sigma^{I}$,
and $K_{SU(N)}^{-1}$, $N=2,3$, is the inverse,%
\begin{equation}
\int d^{4}z\ \left(  K_{SU(N)}^{-1}\right)  ^{ml}\left(  x;z\right)  \left(
K_{SU(N)}\right)  ^{ln}\left(  z;y\right)  =\int d^{4}z\ \left(
K_{SU(N)}^{-1}\right)  ^{nl}\left(  x;z\right)  \left(  K_{SU(N)}\right)
^{lm}\left(  z;y\right)  =\delta^{mn}\delta\left(  x-y\right)
\ ,\label{K^{-1}}%
\end{equation}
of the (Hermitian) Faddeev--Popov operator, $K_{SU(N)}^{mn}=\partial^{\mu
}D_{\mu}^{mn}$, induced by the gauge-fixing functional $F_{\mathbf{0}}$. Here,
the thermodynamic (\textquotedblleft Gribov mass\textquotedblright) parameters
$\gamma_{1}$ and $\gamma_{2}$ of \cite{Zwanziger1,Zwanziger2} are introduced
in a self-consistent way using the gap equations for the functional
$S_{F_{\mathbf{0}},h}$, being the Gribov--Zwanziger action in the
BRST-antiBRST approach to the Standard Model,%
\begin{equation}
\frac{\partial}{\partial\gamma_{i}}\left\{  \frac{\hbar}{i}\ln\left[  \int
d\phi\ \exp\left(  \frac{i}{\hbar}S_{F_{\mathbf{0}},h}\right)  \right]
\right\}  =\frac{\partial\mathcal{E}_{\mathrm{vac}}}{\partial\gamma_{i}%
}=0\ ,\ \ \ \mathrm{where}\ \ \ i=1,2\ .\label{gapeq}%
\end{equation}
Here, we have used the definition of the vacuum energy $\mathcal{E}%
_{\mathrm{vac}}$ and introduced a modified quantum action for the
Gribov--Zwanziger model as an extension of the Yang--Mills quantum action
$S_{F_{\boldsymbol{0}}}$ in (\ref{S(A,B,C)}), using the Landau gauge:%
\begin{equation}
S_{F_{\mathbf{0}},h}\left(  \phi\right)  =S_{\text{SM}}(A)-\frac{1}{2}\left(
F_{\mathbf{0}}\overleftarrow{s}{}^{2}\right)  \left(  \phi\right)
+h_{\mathbf{0}}(\mathcal{A})\,,\label{GZbab}%
\end{equation}
The action $S_{F_{\mathbf{0}},h}\left(  \phi\right)  $ is non-invariant under
the finite BRST-antiBRST transformations:%
\begin{equation}
S_{F_{\mathbf{0}},h}(\phi\exp[\overleftarrow{s}^{a}\lambda_{a}%
])=S_{F_{\mathbf{0}},h}(\phi)+\Delta h(\phi)=S_{F_{\mathbf{0}},h}%
(\phi)+\left(  h_{\mathbf{0}}\overleftarrow{s}^{a}\lambda_{a}\right)
(\phi)+\frac{1}{4}\left(  h_{\mathbf{0}}\overleftarrow{s}^{2}\lambda
^{2}\right)  (\phi)\neq S_{F_{\mathbf{0}},h}(\phi)\ .\label{nbabh}%
\end{equation}

The covariant gauge implies two options: one of them preserves the gauge
independence of the conventional $S$-matrix, according to the BRST-antiBRST
extension \cite{MRnew} of the Gribov--Zwanziger theory, and the other one
determines the horizon functional in terms of transverse-like non-Abelian
gauge fields \cite{LRquarks2012} (see, as well \cite{caprirxi}). Let us examine the first option, which
implies a finite BRST-antiBRST-transformed functional $h_{\mathbf{0}%
}(\mathcal{A})$:%
\begin{align}
h_{\boldsymbol{\xi}}(\mathcal{A},B,C)  &  =h_{\boldsymbol{\xi}}^{SU(2)}%
(\mathcal{A},B,C)+h_{\boldsymbol{\xi}}^{SU(3)}(\mathcal{A},B,C)\ ,\nonumber\\
h_{\boldsymbol{\xi}}(\mathcal{A},B,C)  &  =h_{\mathbf{0}}+\frac{1}{2i\hbar
}\left(  s^{a}h_{\mathbf{0}}\right)  \left(  s_{a}\Delta F_{\boldsymbol{\xi}%
}\right)  \sum_{n=0}^{\infty}\frac{1}{\left(  n+1\right)  !}\left(  \frac
{1}{4i\hbar}s^{b}s_{b}\Delta F_{\boldsymbol{\xi}}\right)  ^{n}\nonumber\\
&  -\frac{1}{16\hbar^{2}}\left(  s^{2}h_{\mathbf{0}}\right)  \left(  s\Delta
F_{\boldsymbol{\xi}}\right)  ^{2}\left[  \sum_{n=0}^{\infty}\frac{1}{\left(
n+1\right)  !}\left(  \frac{1}{4i\hbar}s^{b}s_{b}\Delta F_{\boldsymbol{\xi}%
}\right)  ^{n}\right]  ^{2}\ . \label{gribovHxi}%
\end{align}
Here, $s^{a}h_{\mathbf{0}}=s^{a}\left(  h_{\mathbf{0}}^{SU(2)}+h_{\mathbf{0}%
}^{SU(3)}\right)  $ are given by the following expressions, with account taken
of the definition (\ref{fabg}) for the structure constants $F^{lmn}$:%
\begin{align}
s^{a}h_{\mathbf{0}}^{SU(2)}  &  =\gamma_{1}^{2}g^{2}\varepsilon^{\hat{a}%
\hat{b}\hat{c}}\varepsilon^{\hat{c}\hat{d}\hat{e}}\int d^{4}x\ d^{4}y\ \left[
2D_{\mu}^{\hat{b}\hat{p}}C^{\hat{p}a}\left(  x\right)  \left(  K_{SU(2)}%
^{-1}\right)  ^{\hat{a}\hat{d}}\left(  x;y\right)  \right. \nonumber\\
&  -g\varepsilon^{\hat{q}\hat{r}\hat{s}}\int d^{4}x^{\prime}\ d^{4}y^{\prime
}\left.  A_{\mu}^{\hat{b}}{\left(  x\right)  }\left(  K_{SU(2)}^{-1}\right)
^{\hat{a}\hat{q}}\left(  x;x^{\prime}\right)  K_{SU(2)}^{\hat{r}\hat{u}%
}\left(  x^{\prime};y^{\prime}\right)  C^{\hat{u}a}\left(  y^{\prime}\right)
\left(  K_{SU(2)}^{-1}\right)  ^{\hat{s}\hat{d}}\left(  y^{\prime};y\right)
\right]  A^{\hat{e}\mu}\left(  y\right)  \ ,\label{sah01}\\
s^{a}h_{\mathbf{0}}^{SU(3)}  &  =\gamma_{2}^{2}g_{s}^{2}f^{\underline
{\alpha\beta\gamma}}f^{\underline{\gamma\delta\epsilon}}\int d^{4}%
x\ d^{4}y\ \left[  2D_{\mu}^{\underline{\beta\rho}}C^{\underline{\rho}%
a}\left(  x\right)  \left(  K_{SU(3)}^{-1}\right)  ^{\underline{\alpha\delta}%
}\left(  x;y\right)  \right. \nonumber\\
&  -g_{s}f^{\underline{\sigma\varsigma\tau}}\int d^{4}x^{\prime}%
\ d^{4}y^{\prime}\left.  A_{\mu}^{\underline{\beta}}{\left(  x\right)
}\left(  K_{SU(3)}^{-1}\right)  ^{\underline{\alpha\sigma}}\left(
x;x^{\prime}\right)  K_{SU(3)}^{\underline{\varsigma\upsilon}}\left(
x^{\prime};y^{\prime}\right)  C^{\underline{\tau}a}\left(  y^{\prime}\right)
\left(  K_{SU(3)}^{-1}\right)  ^{\underline{\upsilon\delta}}\left(  y^{\prime
};y\right)  \right]  A^{\underline{\epsilon}\mu}\left(  y\right)  \ ,
\label{sah02}%
\end{align}
whereas $s^{2}h_{\mathbf{0}}=s^{2}\left(  h_{\mathbf{0}}^{SU(2)}%
+h_{\mathbf{0}}^{SU(3)}\right)  $ is given by%
\begin{align}
s^{2}h_{\mathbf{0}}  &  =s^{2}h_{\mathbf{0}}^{SU(3)}+\gamma_{1}^{2}%
g^{2}\varepsilon^{\hat{a}\hat{b}\hat{c}}\varepsilon^{\hat{c}\hat{d}\hat{e}%
}\int d^{4}x\ d^{4}y\ \left\{  4\left(  -D_{\mu}^{\hat{c}\hat{p}}B^{\hat{p}%
}+\frac{g}{2}\varepsilon^{\hat{c}\hat{p}\hat{q}}C^{\hat{q}a}D_{\mu}^{\hat
{p}\hat{r}}C^{\hat{r}b}\varepsilon_{ab}\right)  \left(  x\right)  \left(
K_{SU(2)}^{-1}\right)  ^{\hat{a}\hat{d}}\left(  x;y\right)  A^{\hat{e}\mu
}\left(  y\right)  \right. \nonumber\\
&  +2\varepsilon_{ab}D_{\mu}^{\hat{c}\hat{p}}C^{\hat{p}a}\left(  x\right)
\left(  K_{SU(2)}^{-1}\right)  ^{\hat{a}\hat{d}}\left(  x;y\right)  D^{\hat
{e}\hat{q}\mu}C^{\hat{q}b}\left(  y\right) \nonumber\\
&  -4\varepsilon_{ab}g\varepsilon^{\hat{p}\hat{q}\hat{r}}\int d^{4}x^{\prime
}\ d^{4}y^{\prime}\ D_{\mu}^{\hat{c}\hat{s}}C^{\hat{s}a}\left(  x\right)
\left(  K_{SU(2)}^{-1}\right)  ^{\hat{a}\hat{p}}\left(  x;x^{\prime}\right)
K_{SU(2)}^{\hat{q}\hat{u}}\left(  x^{\prime};y^{\prime}\right)  C^{\hat{u}%
b}\left(  y^{\prime}\right)  \left(  K_{SU(2)}^{-1}\right)  ^{\hat{r}\hat{d}%
}\left(  y^{\prime};y\right)  A^{\hat{e}\mu}\left(  y\right) \nonumber\\
&  +g\varepsilon^{\hat{p}\hat{q}\hat{r}}\int d^{4}x^{\prime}\ d^{4}y^{\prime
}\ A_{\mu}^{\hat{c}}\left(  x\right)  \left[  -g\varepsilon_{ab}%
\varepsilon^{\hat{s}\hat{u}\hat{v}}\int d^{4}x^{\prime\prime}\ d^{4}%
y^{\prime\prime}\ \left(  K_{SU(2)}^{-1}\right)  ^{\hat{a}\hat{s}}\left(
x;x^{\prime\prime}\right)  K_{SU(2)}^{\hat{u}\hat{w}}\left(  x^{\prime\prime
};y^{\prime\prime}\right)  C^{\hat{w}a}\left(  y^{\prime\prime}\right)
\right. \nonumber\\
&  \times\left(  K_{SU(2)}^{-1}\right)  ^{\hat{v}\hat{p}}\left(
y^{\prime\prime};x^{\prime}\right)  K_{SU(2)}^{\hat{q}\hat{g}}\left(
x^{\prime};y^{\prime}\right)  C^{\hat{g}b}\left(  y^{\prime}\right)  \left(
K_{SU(2)}^{-1}\right)  ^{\hat{r}\hat{d}}\left(  y^{\prime};y\right)
-g\varepsilon_{ab}\varepsilon^{\hat{q}\hat{u}\hat{v}}\left(  K_{SU(2)}%
^{-1}\right)  ^{\hat{a}\hat{p}}\left(  x;x^{\prime}\right)  K_{SU(2)}^{\hat
{v}\hat{w}}\left(  x^{\prime};y^{\prime}\right) \nonumber\\
&  \times C^{\hat{w}a}(y^{\prime})C^{\hat{u}b}\left(  x^{\prime}\right)
\left(  K_{SU(2)}^{-1}\right)  ^{\hat{r}\hat{d}}\left(  y^{\prime};y\right)
+2\left(  K_{SU(2)}^{-1}\right)  ^{\hat{a}\hat{p}}\left(  x;x^{\prime}\right)
K_{SU(2)}^{\hat{q}\hat{g}}\left(  x^{\prime};y^{\prime}\right)  B^{\hat{g}%
}\left(  y^{\prime}\right)  \left(  K_{SU(2)}^{-1}\right)  ^{\hat{r}\hat{d}%
}\left(  y^{\prime};y\right) \nonumber\\
&  +g\varepsilon_{ab}\varepsilon^{\hat{s}\hat{u}\hat{v}}\left(  K_{SU(2)}%
^{-1}\right)  ^{\hat{a}\hat{p}}\left(  x;x^{\prime}\right)  K_{SU(2)}^{\hat
{q}\hat{g}}\left(  x^{\prime};y^{\prime}\right)  C^{\hat{g}a}\left(
y^{\prime}\right) \nonumber\\
&  \times\left.  \left.  \int d^{4}x^{\prime\prime}\ d^{4}y^{\prime\prime
}\ \left(  K_{SU(2)}^{-1}\right)  ^{\hat{r}\hat{s}}\left(  y^{\prime
};x^{\prime\prime}\right)  K_{SU(2)}^{\hat{u}\hat{w}}\left(  x^{\prime\prime
};y^{\prime\prime}\right)  C^{\hat{w}b}\left(  y^{\prime\prime}\right)
\left(  K_{SU(2)}^{-1}\right)  ^{\hat{v}\hat{d}}\left(  y^{\prime\prime
};y\right)  \right]  A^{\hat{e}\mu}\left(  y\right)  \right\}  \ .
\label{s2h0}%
\end{align}
Here, $s^{2}h_{\mathbf{0}}^{SU(3)}$ has the same form as $s^{2}h_{\mathbf{0}%
}^{SU(2)}$, in which one makes a replacement of the expressions $g\varepsilon
^{\hat{a}\hat{b}\hat{c}}$, $(K,K^{-1})_{SU(2)}$ and the $SU(2)$-indices by the
expressions $g_{s}f^{\underline{\alpha\beta\gamma}}$, $(K,K^{-1})_{SU(3)}$ and
the $SU(3)$-indices, respectively. The quantities $s_{a}\Delta
F_{\boldsymbol{\xi}}$, $s^{a}s_{a}\Delta F_{\boldsymbol{\xi}}$ and
$\lambda_{\boldsymbol{\xi}}^{a}(\phi)$ in (\ref{gribovHxi}) are given by
(\ref{using}) and (\ref{lambdaSM}) for $\Delta\boldsymbol{\xi}=\boldsymbol{\xi
}$, which relates the Landau gauge to an arbitrary $R_{\xi}$-like gauge,%
\begin{equation}
Z_{\text{SM},h_{\boldsymbol{\xi}}}=\int d{\phi}\ \exp\left\{  \frac{i}{\hbar
}\left[  S_{\text{SM}}-\frac{1}{2}F_{\boldsymbol{\xi}}\overleftarrow{s}%
^{2}+h_{\mathbf{0}}\exp\left(  \overleftarrow{s}^{a}\lambda_{\boldsymbol{\xi
|}a}\right)  \right]  \right\}  \ , \label{GZZSM}%
\end{equation}
in a manner respecting the gauge-independence of the corresponding $S$-matrix.
In turn, the modified Ward identities (\ref{mWIclalg}) for the generating
functional $Z_{\text{SM},h_{\boldsymbol{0}}}(J,\eta)$ are obtained in the same
way as for the generating functional $Z_{\text{SM},\boldsymbol{\xi}}(J,\eta)$
(\ref{z(j)SM}) without the horizon functional (\ref{modWISMgz}),\textbf{ }%
\begin{align}
&  \left\langle \left\{  1+\frac{i}{\hbar}\left[  J_{A}\phi^{A}+h_{\mathbf{0}%
}\right]  \left[  \overleftarrow{s}^{a}\lambda_{a}(\boldsymbol{\xi})+\frac
{1}{4}\overleftarrow{s}^{2}\lambda^{2}(\boldsymbol{\xi})\right]  -\frac{1}%
{4}\left(  \frac{i}{\hbar}\right)  {}^{2}\left[  J_{A}\phi^{A}+h_{\mathbf{0}%
}\right]  \overleftarrow{s}^{a}\left[  J_{B}\phi^{B}+h_{\mathbf{0}}\right]
\overleftarrow{s}_{a}\lambda^{2}(\boldsymbol{\xi})\right\}  \right.
\nonumber\\
&  \quad\left.  \left(  1-\frac{1}{2}\Lambda\left(  \boldsymbol{\xi}\right)
\overleftarrow{s}^{2}\right)  {}^{-2}\right\rangle _{h_{\mathbf{0}%
},F_{\boldsymbol{0}},J}=1\ . \label{modWISMgz}%
\end{align}
which are reduced, at constant parameters $\lambda_{a}$, to an $\mathrm{Sp}%
(2)$-doublet of the usual Ward identities (at the first order in $\lambda_{a}%
$), as well as to a derivative identity (at the second order in $\lambda_{a}%
$),
\begin{equation}
\left\langle \left(  J_{A}+h_{\mathbf{0},A}\right)  X^{Aa}\right\rangle
_{h_{\mathbf{0}},F_{\boldsymbol{0}},J}=0\ ,\ \ \ \left\langle \left(
J_{A}+h_{\mathbf{0},A}\right)  \left[  2Y^{A}+\left(  i/\hbar\right)
\varepsilon_{ab}X^{Aa}\left(  J_{B}+h_{\mathbf{0},B}\right)  X^{Bb}\right]
\right\rangle _{h_{\mathbf{0}},F_{\boldsymbol{0}},J}=0\ , \label{WIlagSMZ}%
\end{equation}
for the non-renormalized Standard Model in the Gribov--Zwanziger approach.
Here, the symbol \textquotedblleft$\langle\mathcal{O}\rangle
_{h_{\boldsymbol{0}},F_{\boldsymbol{0}},J}$\textquotedblright\ for a quantity
$\mathcal{O}=\mathcal{O}(\phi)$ denotes a source-dependent average expectation
value with respect to $Z_{\text{SM},h_{\boldsymbol{0}}}(J,\eta)$ corresponding
to a gauge-fixing $F_{\boldsymbol{0}}$:\textbf{ }%
\begin{equation}
\left\langle \mathcal{O}\right\rangle _{h_{\boldsymbol{0}},F_{\boldsymbol{0}%
},J}=Z_{\text{SM},h_{\boldsymbol{0}}}^{-1}(J,\eta)\int d\phi\ \mathcal{O}%
\left(  \phi\right)  \exp\left\{  \frac{i}{\hbar}\left[  S_{\text{SM}}%
-\frac{1}{2}F_{\boldsymbol{0}}\overleftarrow{s}^{2}+h_{\boldsymbol{0}}%
+J_{A}\phi^{A}\right]  \right\}  \ ,\ \ \mathrm{with\ \ }\left\langle
1\right\rangle _{h_{\mathbf{0}},F_{\boldsymbol{0}},J}=1\ . \label{aexv1}%
\end{equation}
The modified and standard Ward identities for Green's functions are readily
obtained from (\ref{modWISMgz}) and (\ref{WIlagSMZ}), respectively, using
differentiation over the sources. These identities are fulfilled in the tree
approximation and provide a basis for the study of pa renormalization procedure
using an appropriate gauge-invariant regularization. We intend to study this
problem in separate research.

\section{Discussion\label{Concl}}

\renewcommand{\theequation}{\arabic{section}.\arabic{equation}} \setcounter{equation}{0}

We have extended the results and ideas of our previous study
\cite{MRnew,MRnew1,MRnew2,MRnew3} and have also applied them to the Lagrangian
description of the Standard Model. The main results of the present study are
given by Sections \ref{linear}, \ref{arbitrary}, devoted to the calculation of
functional Jacobians, which requires only the definition of such a Jacobian
and does not have recourse to functional integration in itself. We have
proposed and applied an explicit recipe of exact calculation of the Jacobian
for a change of variables in the vacuum functional corresponding to finite
field-dependent BRST-antiBRST transformations with a linear dependence on
functionally-dependent parameters in Yang--Mills theories and first-class
constraint dynamical systems, given, respectively, in Sections~\ref{linearYM}
and \ref{constrained}, by the relations (\ref{J_lin_final}), (\ref{R_final1})
and (\ref{J_lin_final_Ham})--(\ref{R_final1_Ham2}). This implies that thus
linearized finite BRST-antiBRST transformations can be interpreted neither as
global symmetry transformations of the integrand, nor as field-dependent
transformations inducing an exact change of the gauge-fixing functional,
despite the hope of the authors of \cite{Upadhyay3}; see Eqs. (3.1)--(3.7)
therein. At the same time, we have evaluated the Jacobian for a change of
variables in the vacuum functional corresponding to finite field-dependent
BRST-antiBRST transformations with \emph{arbitrary} functional parameters
$\lambda_{a}(\phi)\not \equiv s_{a}\Lambda(\phi)$, in Yang--Mills theories,
first-class constraint dynamical systems, and general gauge theories,
(\ref{J-arb}), (\ref{J-exp}), (\ref{JHam_arb}), (\ref{J-arb-gen}),
(\ref{J-exp-gen}), which is the main result of the present work. It is
demonstrated that the Jacobians are reduced to the previously known Jacobians
in the case of functionally-dependent odd-valued parameters $\lambda_{a}%
=s_{a}\Lambda$, whereas in general gauge theories the Jacobian
(\ref{J-arb-gen}) has been obtained for the first time. We have demonstrated
that in the general case $\lambda_{a}\not \equiv s_{a}\Lambda$ (more exactly,
$s^{a}\lambda_{a}\not =s^{a}s_{a}\Lambda$)\ the Jacobian fails to be
BRST-antiBRST-invariant, which implies the inconsistency of the compensation
equation with such odd-valued parameters, and thereby entails the appearance,
under such a change of variables, of terms which cannot be absorbed into a
change of the gauge Boson, used in \cite{BLThf, BLTlf} to provide the
consistency of the compensation equations by using a suitable choice of the
parameters in a functionally-dependent form. We have found that the set of
functionally-dependent parameters $\lambda_{a}=s_{a}\Lambda$ generated by an
$s_{a}$-gradient of an $\mathrm{Sp}(2)$-scalar $\Lambda$ can be extended by an
$s_{a}$-divergence of a symmetric $\mathrm{Sp}(2)$-tensor $\Psi_{\{ab\}}$,
namely, $\lambda_{a}=s_{a}\Lambda+s^{b}\Psi_{\{ab\}}$, which, as shown by
(\ref{reads}), (\ref{Jacob-sym}) in Yang--Mills theories and by
(\ref{comp_eq_U}) in general gauge theories, produces a non-trivial
contribution to the Jacobians, thereby modifying the compensation equations,
(\ref{comp_mod}), (\ref{comp_eq_U}), and affecting the change of the
respective gauge Boson. In Yang--Mills theories, we have found the solutions
(\ref{delta_Fnot0})--(\ref{2delta_F=0}) of the modified compensation equation
(\ref{comp_mod}), in particular, a non-trivial solution (\ref{1delta_F=0})
which induces a zero change of gauge-fixing, $\Delta F=0$, resulting in a
Jacobian equal to unity, $\exp\left(  \Im\right)  =1$. We have also presented
(\ref{addact1}) the BRST-antiBRST-non-exact contribution $\Re$ to the Jacobian
induced by finite BRST-antiBRST transformations with arbitrary functional
parameters. This contribution is to be regarded as\ an extra part of the
transformed quantum action in the integrand (\ref{addact}). The same holds
true for general gauge theories, in view of ({\ref{whence})},
(\ref{notdescribe}).

Having applied our results \cite{MRnew} to the evaluation of Jacobians in the
case of Yang--Mills theories, we have explicitly constructed the
functionally-dependent parameters $\lambda_{a}$ in (\ref{lambdaSM}) induced by
a finite change $\Delta F_{\boldsymbol{\xi}}$ of the gauge Boson (\ref{using})
in the quantum action of the Standard Model (\ref{S(A,B,C)}), which generates
a change of the gauge in the path integral within a class of linear
$3$-parameter $R_{\boldsymbol{\xi}}$-like gauges, realized in terms of the
even-valued gauge functionals $F_{\boldsymbol{\xi}}$ in (\ref{F(A,C)}), with
the values $(\xi_{1},\xi_{2},\xi_{3})=\mathbf{0},\mathbf{1}$ corresponding to
the Landau and Feynman (covariant) gauges, respectively. We have obtained a
\emph{modified} Ward identity (\ref{modWISM}) for a generating functional of
Green's functions $Z_{\text{SM},\boldsymbol{\xi}}(J,\eta)$ depending on
field-dependent parameters, $\lambda_{a}=\Lambda\overleftarrow{s}_{a}$, which
reduces to the usual Ward identity for a constant doublet $\lambda_{a}$.

In order to eliminate residual gauge invariance, i.e., Gribov copies, and to
determine a consistent path integral for the Standard Model in the entire set
of field configurations, we have explicitly constructed the Gribov--Zwanziger
theory in the BRST-antiBRST Lagrangian description of the Standard Model. The
construction extends the quantum action in the Landau gauge by a BRST-antiBRST
non-invariant horizon functional $h_{\mathbf{0}}$ in (\ref{hfSM}%
)--({\ref{hfSM3}}). We have found the horizon functional $h_{\boldsymbol{\xi}%
}$ given by (\ref{gribovHxi})--(\ref{s2h0}) in arbitrary $R_{\boldsymbol{\xi}%
}$-like gauges by means of field-dependent BRST-antiBRST transformations with
the parameters $\lambda_{a}$ given by (\ref{lambdaSM}) and providing the
gauge-independence of the conventional $S$-matrix related to the
Gribov--Zwanziger path integral $Z_{\text{SM},h_{\boldsymbol{\xi}}}(\eta)$ in
(\ref{GZZSM}). We have obtained the modified (\ref{modWISMgz}) and usual
(\ref{WIlagSMZ}) Ward identities for the generating functional of Green's
functions $Z_{\text{SM},h_{\boldsymbol{0}}}(J,\eta)$ providing a basis for
renormalization. These are the main results of Section \ref{AppA}.

As has been noticed in Introduction and Section \ref{AppA}, there remains
another option to determine the horizon functional in covariant
$R_{\boldsymbol{\xi}}$-like gauges, which lies in transverse-like non-Abelian
gauge fields \cite{LRquarks2012}, recently examined also in \cite{caprirxi}.
Namely, the Faddeev--Popov operators ${K}_{SU(N)}(\boldsymbol{\xi})$, $N=2,3$,
retain the same formal structure at any values of the gauge parameters
$\boldsymbol{\xi}$. With this in mind, let us consider some extensions
$\bar{K}_{i|SU(2)}^{\hat{a}\hat{b}}$, $\bar{K}_{i|SU(3)}^{\underline
{\alpha\beta}}$, for $i=1,2$, of the Faddeev--Popov operators in
$R_{\boldsymbol{\xi}}$-like gauges (\ref{F(A,C)}),%
\begin{align}
\bar{K}_{i|SU(2)}^{\hat{a}\hat{b}}(A,B;\xi_{2})  &  ={K}_{SU(2)}^{\hat{a}%
\hat{b}}+\frac{g\xi_{2}}{2}\varepsilon^{\hat{a}\hat{c}\hat{b}}\left[
\delta_{i2}\partial_{\mu}\left(  \frac{\partial^{\mu}B^{\hat{c}}}{\partial
^{2}}\right)  +\frac{\delta_{i1}}{2}B^{\hat{c}}\right]  \ ,\ \mathrm{where}%
\ \left(  \bar{K}_{i}^{\hat{a}\hat{b}}(\xi_{2})\right)  ^{\dagger}=\bar{K}%
_{i}^{\hat{a}\hat{b}}(\xi_{2})\footnotemark\ ,\nonumber\\
\bar{K}_{i|SU(3)}^{\underline{\alpha\beta}}(A,B;\xi_{3})  &  ={K}%
_{SU(3)}^{\underline{\alpha\beta}}+\frac{g_{s}\xi_{3}}{2}f^{\underline
{\alpha\gamma\beta}}\left[  \delta_{i2}\partial_{\mu}\left(  \frac
{\partial^{\mu}B^{\underline{\gamma}}}{\partial^{2}}\right)  +\frac
{\delta_{i1}}{4}B^{\underline{\gamma}}\right]  \ ,\ \mathrm{where}\ \left(
\bar{K}_{i}^{\underline{\alpha\beta}}(\xi_{3})\right)  ^{\dagger}=\bar{K}%
_{i}^{\underline{\alpha\beta}}(\xi_{3})\ , \label{FPaug3}%
\end{align}\footnotetext{The Hermitian extended  operator $\bar{K}_{1|SU(N)}^{mn}(A,B;\xi)$ suggested in \cite{LRquarks2012} was written with mistake.}
(${K}_{SU(2)}^{\hat{a}\hat{b}}=\partial^{\mu}D_{\mu}^{\hat{a}\hat{b}}$,
${K}_{SU(3)}^{\underline{\alpha\beta}}=\partial^{\mu}D_{\mu}^{\underline
{\alpha\beta}}$), which are Hermitian with reference to the scalar products in
the spaces of square-integrable functions $L_{2}(\mathbb{R}^{1,3})$ taking
their values in the respective Lie algebras $su(N)$, $N=2,3$,%
\begin{equation}
\left(  f,\bar{K}_{i|SU(N)}g\right)  _{(N)}=\left(  g,\bar{K}_{i|SU(N)}^{\dag
}f\right)  _{(N)}^{\ast},\ \mathrm{where}\ \left(  f,\bar{K}_{i|SU(N)}%
g\right)  _{(N)}=\int d^{4}x\;d^{4}yf^{m}(x)\bar{K}_{i|SU(N)}^{mn}%
(x;y)g^{n}(y)\ , \label{scalprod}%
\end{equation}
with arbitrary test functions $f^{m},g^{n}\in L_{2}(\mathbb{R}^{1,3})$. The
eigenvalues $\lambda_{i|k}^{n}$, $n=(\hat{b},\underline{\beta})$,
$k=0,1,2,\ldots$, in the equation $\bar{K}_{i}^{mn}(\boldsymbol{\xi})u_{k}%
^{n}=\lambda_{i|k}^{n}u_{k}^{n}$ are real-valued and are to determine the Gribov
region $\Omega(\xi_{2},\xi_{3})$ as follows:%
\[
\Omega(\xi_{2},\xi_{3})\equiv\left\{  A_{\mu}^{\hat{a}},A_{\mu}^{\underline
{\alpha}}:\partial^{\mu}\left(  A_{\mu}^{\hat{a}},A_{\mu}^{\underline{\alpha}%
}\right)  =-\frac{1}{2}\left(  \xi_{2}B^{\hat{a}},\xi_{3}B^{\underline{\alpha
}}\right)  ,\left(  {K}_{SU(2)}^{\hat{a}\hat{b}},{K}_{SU(3)}^{\underline
{\alpha\beta}}\right)  >0\right\}  \ .
\]
The Hermitian operators $\bar{K}_{i|SU(N)}^{mn}(x)$ cannot be used
equivalently, i.e., for any $i=1,2$, to determine the eigenvalues of
non-Hermitian operator $K_{SU(N)}^{mn}(\xi_{N})$. Indeed, a definition of the
Gribov region requires that $K_{SU(N)}^{mn}(\xi_{N})$ be positive definite.
The case of $i=2$ does satisfy this condition, and so we propose a form of the
Gribov--Zwanziger functional in $R_{\boldsymbol{\xi}}$\textbf{ }-like gauges,%
\begin{align}
&  h^{T}(A,B;\xi_{2},\xi_{3})=h_{SU(2)}^{T}(A,B;\xi_{2})+h_{SU(3)}^{T}%
(A,B;\xi_{3})\ ,\label{hermGZf}\\
&  h_{SU(2)}^{T}(A,B;\xi_{2})=\gamma_{1}^{2}(\boldsymbol{\xi})g^{2}\int
d^{4}x\ d^{4}y\ \varepsilon^{\hat{a}\hat{b}\hat{c}}A_{\mu}^{\hat{b}T}\left(
x\right)  \left(  \bar{K}_{2|SU(2)}^{-1}\right)  ^{\hat{a}\hat{d}}\left(
x;y\right)  \varepsilon^{\hat{d}\hat{e}\hat{c}}A^{\mu\hat{e}T}\left(
y\right)  +4\cdot3g^{2}\gamma_{1}^{2}(\boldsymbol{\xi})\ ,\label{thfSM2}\\
&  h_{SU(3)}^{T}(A,B;\xi_{3})=\gamma_{2}^{2}(\boldsymbol{\xi})g_{s}^{2}\int
d^{4}x\ d^{4}y\ f^{\underline{\alpha\beta\gamma}}A_{\mu}^{\underline{\beta}%
T}\left(  x\right)  \left(  \bar{K}_{2|SU(3)}^{-1}\right)  ^{\underline
{\alpha\delta}}\left(  x;y\right)  f^{\underline{\delta\sigma\gamma}}%
A^{\mu\underline{\sigma}T}\left(  y\right)  +4\cdot8g_{s}^{2}\gamma_{2}%
^{2}(\boldsymbol{\xi})\ . \label{thfSM3}%
\end{align}
which determines the Gribov region ${\Omega}(\boldsymbol{\xi})$. The
thermodynamic Gribov parameters $\gamma_{i}^{2}(\boldsymbol{\xi})$ must depend
on the gauge parameters $\boldsymbol{\xi}$ so as to be determined in a
self-consistent way from the relations (\ref{gapeq}), (\ref{GZbab}), involving
the functional $S_{F_{\mathbf{0}},h}\left(  \phi\right)  $ and the vacuum
energy $\mathcal{E}_{\mathrm{vac}}(\boldsymbol{\xi})$. In fact, the parameters
$\gamma_{i}^{2}(\boldsymbol{\xi})$, $i=1,2$, must depend on $\xi_{2},\xi_{3}$,
albeit with the horizon functional $h^{T}(A,B;\xi_{2},\xi_{3})$ in
(\ref{hermGZf}), instead of $h_{\boldsymbol{0}}$ given by the Landau gauge.
The suggested introduction of the Gribov--Zwanziger horizon functional is
based on a representation of the Yang--Mills connection by using the
transverse, $A_{\mu}^{\hat{a}\mathrm{T}},A_{\mu}^{\underline{\alpha}%
\mathrm{T}}$, and longitudinal, $A_{\mu}^{\hat{a}\mathrm{L}},A_{\mu
}^{\underline{\alpha}\mathrm{L}}$, components:%
\begin{align}
\left(  A_{\mu}^{\hat{a}\mathrm{T}},A_{\mu}^{\underline{\alpha}\mathrm{T}%
}\right)   &  =\left(  \delta_{\mu}^{\nu}-\frac{\partial_{\mu}\partial^{\nu}%
}{\partial^{2}}\right)  \left(  A_{\nu}^{\hat{a}},A_{\nu}^{\underline{\alpha}%
}\right)  =\left(  A_{\mu}^{\hat{a}},A_{\mu}^{\underline{\alpha}}\right)
+\frac{\partial_{\mu}}{2\partial^{2}}(\xi_{2}B^{\hat{a}},\xi_{3}%
B^{\underline{\alpha}})\ ,\nonumber\\
\left(  A_{\mu}^{\hat{a}\mathrm{L}},A_{\mu}^{\underline{\alpha}\mathrm{L}%
}\right)   &  =\frac{\partial_{\mu}\partial^{\nu}}{\partial^{2}}\left(
A_{\nu}^{\hat{a}},A_{\nu}^{\underline{\alpha}}\right)  =-\frac{\partial_{\mu}%
}{2\partial^{2}}(\xi_{2}B^{\hat{a}},\xi_{3}B^{\underline{\alpha}})\ ,
\label{divide}%
\end{align}
so that the $R_{\boldsymbol{\xi}}$-like gauge induced by the gauge Boson
$F_{\boldsymbol{\xi}}$ in (\ref{F(A,C)}) is equivalent to the conditions%
\[
\partial^{\mu}\left(  A_{\mu}^{\hat{a}\mathrm{T}},A_{\mu}^{\underline{\alpha
}\mathrm{T}}\right)  =0\ ,\ \ \ \partial^{\mu}\left(  A_{\mu}^{\hat
{a}\mathrm{L}},A_{\mu}^{\underline{\alpha}\mathrm{L}}\right)  =-\frac{1}%
{2}(\xi_{2}B^{\hat{a}},\xi_{3}B^{\underline{\alpha}})\ .
\]
As a consequence, the operators $\bar{K}_{2|SU(2)}^{\hat{a}\hat{b}}$ and
$\bar{K}_{2|SU(3)}^{\underline{\alpha\beta}}$ are nothing else than the
Faddeev--Popov operators for the transverse components of the gauge fields
$\left(  A_{\mu}^{\hat{a}\mathrm{T}},A_{\mu}^{\underline{\alpha}\mathrm{T}%
}\right)  $, which determine the physical degrees of freedom,
\begin{equation}
\left(  \bar{K}_{2|SU(2)}^{\hat{a}\hat{b}},\bar{K}_{2|SU(3)}^{\underline
{\alpha\beta}}\right)  =\partial^{\mu}\left[  \left(  \partial_{\mu}%
\delta^{\hat{a}\hat{b}},\partial_{\mu}\delta^{\underline{\alpha\beta}}\right)
+\left(  g\varepsilon^{\hat{a}\hat{c}\hat{b}}A_{\mu}^{\hat{c}\mathrm{T}}%
,g_{s}f^{\underline{\alpha\gamma\beta}}A_{\mu}^{\underline{\gamma}\mathrm{T}%
}\right)  \right]  =\left(  {K}_{SU(2)}^{\hat{a}\hat{b}},{K}_{SU(3)}%
^{\underline{\alpha\beta}}\right)  (A^{\mathrm{T}})\ . \label{FPatrans}%
\end{equation}
To provide a justification of the horizon functional (\ref{hermGZf}),
(\ref{thfSM3}), we examine the following \setcounter{theorem}{6}

\begin{proposition}
\label{gribreg} For the transverse components $\left(  A_{\mu}^{\hat
{a}\mathrm{T}},A_{\mu}^{\underline{\alpha}\mathrm{T}}\right)  \in\Omega
(\xi_{2},\xi_{3})$ of the gauge fields, the equations
\begin{equation}
\left(  {K}_{SU(2)}^{\hat{a}\hat{b}}(A)\varsigma^{\hat{b}},{K}_{SU(3)}%
^{\underline{\alpha\beta}}(A)\varsigma^{\underline{\beta}}\right)  =(0,0)
\label{equnuq}%
\end{equation}
for arbitrary field configurations $\left(  A_{\mu}^{\hat{a}},A_{\mu
}^{\underline{\alpha}}\right)  $ admit only the vanishing solutions
$(\varsigma^{\hat{b}},\varsigma^{\underline{\beta}})=(0,0)$ in the class of
functions regular in $\xi_{2},\xi_{3}$.
\end{proposition}

A proof is based on the hermiticity of ${K}_{SU(2)}^{\hat{a}\hat{b}%
}(A^{\mathrm{T}})$, ${K}_{SU(3)}^{\underline{\alpha\beta}}(A^{\mathrm{T}})$,
due to the relations (\ref{FPaug3}), (\ref{FPatrans}), which implies their
invertibility and positive definitiveness. The regularity imposed on the
respective zero-mode parameters $\varsigma^{\hat{b}}(x,\xi_{2})=\varsigma
^{\hat{b}}(\xi_{2})$, $\varsigma^{\underline{\beta}}(x,\xi_{3})=\varsigma
^{\underline{\beta}}(\xi_{3})$ for the operators ${K}_{SU(2)}^{\hat{a}\hat{b}%
}(A)$, ${K}_{SU(3)}^{\underline{\alpha\beta}}$ implies the possibility of
their representation as power series in the respective gauge parameters
$\xi_{2},\xi_{3}$,%
\begin{equation}
\varsigma^{\hat{b}}(\xi_{2})=\sum_{n\geq0}\varsigma_{n}^{\hat{b}}(\xi_{2}%
)^{n}\ ,\ \ \ \varsigma^{\underline{\beta}}(\xi_{3})=\sum_{n\geq0}%
\varsigma_{n}^{\underline{\beta}}(\xi_{3})^{n}\ , \label{represconv}%
\end{equation}
which converge within certain convergence radiuses $R_{2}$, $R_{3}$,
respectively. From (\ref{equnuq}) it follows that
\begin{align}
&  \left(  \varsigma^{\hat{b}}(\xi_{2}),\varsigma^{\underline{\beta}}(\xi
_{3})\right)  =-\frac{1}{2}\left(  g\xi_{2}\varepsilon^{\hat{a}\hat{c}\hat{d}%
}\left(  \bar{K}_{2|SU(2)}^{-1}\right)  ^{\hat{b}\hat{a}}\partial_{\mu}\left(
\frac{\partial^{\mu}B^{\hat{c}}}{\partial^{2}}\right)  \varsigma^{\hat{d}}%
(\xi_{2})\ ,\ g_{s}\xi_{3}f^{\underline{\alpha\gamma\delta}}\left(  \bar
{K}_{2|SU(3)}^{-1}\right)  ^{\underline{\beta\alpha}}\partial_{\mu}\left(
\frac{\partial^{\mu}B^{\underline{\gamma}}}{\partial^{2}}\right)
\varsigma^{\underline{\delta}}(\xi_{3})\right) \label{dereq}\\
&  \overset{\ref{represconv}}{\Longrightarrow}\left\{
\begin{array}
[c]{c}%
\sum_{n\geq0}\varsigma_{n}^{\hat{b}}(\xi_{2})^{n}=-\frac{g}{2}\varepsilon
^{\hat{a}\hat{c}\hat{d}}\sum_{n\geq0}(\xi_{2})^{n+1}\left(  \bar{K}%
_{2|SU(2)}^{-1}\right)  ^{\hat{b}\hat{a}}\partial_{\mu}\left[  \frac
{\partial^{\mu}B^{\hat{c}}}{\partial^{2}}\right]  \varsigma_{n}^{\hat{d}}%
=\sum_{n\geq0}(\xi_{2})^{n+1}\varphi_{n}^{\hat{b}},\\
\sum_{n\geq0}\varsigma_{n}^{\underline{\beta}}(\xi_{3})^{n}=-\frac{g_{s}}%
{2}f^{\underline{\alpha\gamma\delta}}\sum_{n\geq0}(\xi_{3})^{n+1}\left(
\bar{K}_{2|SU(3)}^{-1}\right)  ^{\underline{\beta\alpha}}\partial_{\mu}\left[
\frac{\partial^{\mu}B^{\underline{\gamma}}}{\partial^{2}}\right]
\varsigma_{n}^{\underline{\delta}}=\sum_{n\geq0}(\xi_{3})^{n+1}\varphi
_{n}^{\underline{\beta}}%
\end{array}
\right.  . \label{dereq1}%
\end{align}
The system of equations (\ref{dereq1}) for unknowns functions $\varsigma
_{n}^{\hat{b}}$, $\varsigma_{n}^{\underline{\beta}}$ at a fixed order in $n$,
starting from $n=0$, yields the solution%
\begin{equation}
\left(  \varsigma_{0}^{\hat{b}},\ \varsigma_{0}^{\underline{\beta}}\right)
=\left(  0,0\right)  \overset{\ref{dereq1}}{\Longrightarrow}\left(
\varphi_{0}^{\hat{b}},\ \varphi_{0}^{\underline{\beta}}\right)  =(0,0).
\label{dereq2}%
\end{equation}
Therefore, $\left(  \varsigma_{n}^{\hat{b}},\ \varsigma_{n}^{\underline{\beta
}}\right)  =\left(  \xi_{2}\varphi_{n-1}^{\hat{b}},\ \xi_{3}\varphi
_{n-1}^{\underline{\beta}}\right)  $ implies subsequently\ $\left(
\varsigma_{n}^{\hat{b}},\ \varsigma_{n}^{\underline{\beta}}\right)  =0$,
$\left(  \varsigma_{n}^{\hat{b}},\ \varsigma_{n}^{\underline{\beta}}\right)
=(0,0)$, for $n=1,2,...$. As a result, the series (\ref{represconv}) in their
respective convergence regions $R_{2}$, $R_{3}$ vanish identically, which
thereby proves the proposition.

Notice that the choice for the zero-modes of the respective Faddeev--Popov
operators to be regular in $\xi_{2}$, $\xi_{3}$\ is based on the assumption
that we obtain the Gribov region for the Landau gauge in the limit $(\xi
_{2},\xi_{3})\rightarrow(0,0)$. At the same time, Proposition~\ref{gribreg}
means that the Gribov region $\Omega(\xi_{2},\xi_{3})$ contains only the
transverse components of the gauge fields:
\[
\Omega(\xi_{2},\xi_{3})\equiv\left\{  A_{\mu}^{\hat{a}\mathrm{T}},A_{\mu
}^{\underline{\alpha}\mathrm{T}}:\partial^{\mu}\left(  A_{\mu}^{\hat
{a}\mathrm{T}},A_{\mu}^{\underline{\alpha}\mathrm{T}}\right)  =(0,0),\ \left(
\bar{K}_{2|SU(2)}^{\hat{a}\hat{b}},{\bar{K}}_{2|SU(3)}^{\underline{\alpha
\beta}}\right)  >0\right\}  \ .
\]
We can thereby construct the Gribov--Zwanziger theory in arbitrary (covariant)
$R_{\boldsymbol{\xi}}$-like gauges, suggested earlier
\cite{LRquarks2012,caprirxi}, as an extension of the BRST-invariant
Faddeev--Popov action for the Yang--Mills theory in the case of a
BRST-antiBRST-invariant quantum action for the Standard Model, in a way
different from the one suggested by the equation (\ref{GZZSM}) for
$S_{F_{\boldsymbol{\xi}},h_{\boldsymbol{\xi}}}$. Our proposal has the form%
\begin{equation}
S_{F_{\boldsymbol{\xi}},h_{\boldsymbol{\xi}}^{T}}=S_{\text{SM}}\left(
{A}\right)  -\left(  1/2\right)  F_{\boldsymbol{\xi}}\overleftarrow{s}{}%
^{2}+h^{\mathrm{T}}(A,B;\xi_{2},\xi_{3})\ . \label{GZSsm}%
\end{equation}
There remains the question of establishing the coincidence of $Z_{\text{SM}%
,h_{\boldsymbol{\xi}}}(\eta)$ in (\ref{GZZSM}) with $Z_{\text{SM}%
,h_{\boldsymbol{\xi}}^{T}}(\eta)$ determined using $h^{\mathrm{T}}(A,B;\xi
_{2},\xi_{3})$, namely,%
\begin{equation}
Z_{\text{SM},h_{\boldsymbol{\xi}}}(\eta)\overset{?}{=}Z_{\text{SM}%
,h_{\boldsymbol{\xi}}^{\mathrm{T}}}(\eta)\ . \label{GZSsmprob}%
\end{equation}
We intend to study this problem in separate research.

In addition, there are various lines of research for extending the results of
the present work. First, the study of finite field-dependent BRST
transformations in the multilevel formalism \cite{bt1, bt2} involving
non-Abelian hypergauges and a non-trivial geometry. Second, the study of the
Gribov ambiguity in generalized Hamiltonian formalism, as well as the study of
a Hamiltonian Gribov--Zwanziger theory -- see \cite{HFZwanziger} -- for
Yang--Mills theories in the Lagrangian description using different gauges by
means of finite field-dependent BRST(-antiBRST) transformations. Third, the
study of an explicit relation between the two approaches using the finite
field-dependent BRST transformations in the Yang--Mills theory \cite{JM,LL1}
and general gauge theories \cite{Reshetnyak}. Fourth, the influence of
renormalizability on the properties of the ingredients of BRST-antiBRST
quantization at finite BRST-antiBRST transformations is also an open problem.
We are, however, convinced that the presence of a gauge-invariant
regularization which respects the Ward identities will replicate the
properties of the non-renormalized theory by the properties of the
renormalized one.

In the Standard Model, due to the presence of chiral Fermions in the lepton
sector, described by the Lagrangian (\ref{Llep}), one can adopt a
gauge-invariant regularization as the higher derivative regularization
\cite{Slavnov:1971aw,Slavnov:1972sq}, which is the Pauli--Villars
regularization extended by higher-derivative terms. The first successful
application of this regularization to the calculation of the one-loop
effective action in the BRST-invariant Yang--Mills theory has been given by
\cite{Martin:1994cg,Asorey:1995tq,Bakeyev:1996is}. This regularization, when
adapted to $N=1$ supersymmetric field theory models
\cite{hderregN1,Westhdern1}, preserves explicit supersymmetry, unlike the
standard dimensional regularization, and has been recently elaborated in the
$N=2$ supersymmetric Yang--Mills theory interacting with matter
\cite{buchstepanyantz}, thereby respecting gauge invariance and $N=2$
supersymmetry. In its turn, the dimensional regularization has been recently
used \cite{anselmi} to study the problem of gauge-dependence in terms of the
Ward identities, including the case of beta-functions, for renormalizable and
non-renormalizable general chiral gauge theories in the BV quantization
method. This regularization can also be implemented, but only in those parts
of the Standard Model which do not include the lepton fields. The dimensional
regularization has been partially applied \cite{Krauss} to the electroweak
sector described by the Lagrangian (\ref{LSM}). This is done using the method
of algebraic renormalization \cite{PiquetSorella} and aiming to describe
electroweak interactions in the Standard Model to all orders of perturbation
theory under BRST symmetry, with the infrared-finiteness of the off-shell
Green functions, however, without the fulfilment of the Gribov
\textquotedblleft no-pole condition\textquotedblright\ \cite{Gribov} for the
ghost Green functions. Therefore, a mathematically rigorous renormalization of
the Standard Model in BRST and BRST-antiBRST quantization remains a topical problem.

Let us finally mention the search for an equivalent local description of the
Gribov horizon functional by using a set of auxiliary fields, as in
\cite{Zwanziger2}, such that it should be consistent with both the
infinitesimal and finite forms of BRST-antiBRST invariance.

\section*{Acknowledgments}

We are grateful to R.R. Metsaev and K.V. Stepanyantz for useful discussions.
The study was carried out in 2015 within the Tomsk State University
Competitiveness Improvement Program and was also supported by the grant of
Leading Scientific Schools of the Russian Federation under Project No. 88.2014.2.

\appendix

\section*{Appendix}

\section{Linearized Transformations}

\label{AppB} \renewcommand{\theequation}{\Alph{section}.\arabic{equation}} \setcounter{equation}{0}

In this appendix, we make an explicit calculation of the Jacobian
corresponding to linearized finite BRST-antiBRST transformations, i.e.,
transformations corresponding to the part of finite BRST-antiBRST
transformations being linear in parameters of a special form, $\lambda
_{a}=s_{a}\Lambda$. To this end, notice that, in virtue of (\ref{transp}%
)--(\ref{Str(P+Q)^n}), the quantity $\Im$ can be subsequently transformed as
follows:%
\begin{align}
\Im &  =-\sum_{n=1}^{\infty}\frac{\left(  -1\right)  ^{n}}{n}\mathrm{Str}%
\left(  M^{n}\right)  =-\sum_{n=1}^{\infty}\frac{\left(  -1\right)  ^{n}}%
{n}\mathrm{Str}\left(  P+Q\right)  ^{n}=-\sum_{n=1}^{3}\frac{\left(
-1\right)  ^{n}}{n}\mathrm{Str}\left(  P+Q\right)  ^{n}-\sum_{n=4}^{\infty
}\frac{\left(  -1\right)  ^{n}}{n}\mathrm{Str}\left(  P+Q\right)
^{n}\nonumber\\
&  =-\sum_{n=1}^{\infty}\frac{\left(  -1\right)  ^{n}}{n}\mathrm{Str}\left(
P^{n}\right)  -\sum_{n=2}^{\infty}\left(  -1\right)  ^{n}\mathrm{Str}\left(
P^{n-1}Q\right)  -\sum_{n=2}^{3}\frac{\left(  -1\right)  ^{n}}{n}C_{n}%
^{2}\mathrm{Str}\left(  P^{n-2}Q^{2}\right)  -\sum_{n=4}^{\infty}\frac{\left(
-1\right)  ^{n}}{n}K_{n}\mathrm{Str}\left(  P^{n-3}QPQ\right) \nonumber\\
&  =-\sum_{n=1}^{\infty}\frac{\left(  -1\right)  ^{n}}{n}\mathrm{Str}\left(
P^{n}\right)  -\frac{1}{2}\mathrm{Str}\left(  Q^{2}\right)  +\mathrm{Str}%
\left(  PQ^{2}\right)  +\sum_{n=1}^{\infty}\left(  -1\right)  ^{n}%
\mathrm{Str}\left(  P^{n}Q\right)  +\frac{1}{2}\sum_{n=1}^{\infty}\left(
-1\right)  ^{n}n\mathrm{Str}\left(  P^{n}QPQ\right)  \ . \label{lin1}%
\end{align}
whence%
\begin{equation}
\Im=-\sum_{n=1}^{\infty}\frac{\left(  -1\right)  ^{n}}{n}f^{n-1}%
\mathrm{Str}\left(  P\right)  -\frac{1}{2}\mathrm{Str}\left(  Q^{2}\right)
+\mathrm{Str}\left(  QPQ\right)  +\sum_{n=1}^{\infty}\left(  -1\right)
^{n}f^{n-1}\mathrm{Str}\left(  QP\right)  +\frac{1}{2}\sum_{n=1}^{\infty
}\left(  -1\right)  ^{n}nf^{n-1}\mathrm{Str}\left(  QPQP\right)  \ .
\label{lin2}%
\end{equation}
Finally,%
\begin{align}
&  \Im=2\sum_{n=1}^{\infty}\frac{\left(  -1\right)  ^{n}}{n}f^{n}%
-\mathrm{Str}\left(  R\right)  +\left(  1+f\right)  \mathrm{Str}\left(
Q_{2}Q\right)  +\mathrm{Str}\left(  Q_{2}\right)  \sum_{n=1}^{\infty}\left(
-1\right)  ^{n}f^{n-1}\left(  1+f\right) \nonumber\\
&  +\frac{1}{2}\mathrm{Str}\left(  Q_{2}^{2}\right)  \sum_{n=1}^{\infty
}\left(  -1\right)  ^{n}nf^{n-1}\left(  1+f\right)  ^{2}\equiv-2\ln\left(
1+f\right)  +\Re\ , \label{finally}%
\end{align}
where%
\begin{align}
&  \ \Re=-\mathrm{Str}\left(  R\right)  +\left(  1+f\right)  \cdot
\mathrm{Str}\left(  Q_{2}Q\right)  +\varphi\left(  f\right)  \cdot
\mathrm{Str}\left(  Q_{2}\right)  +\frac{1}{2}\psi\left(  f\right)
\cdot\mathrm{Str}\left(  Q_{2}^{2}\right)  \ ,\nonumber\\
&  \ \varphi\left(  x\right)  =\left(  1+x\right)  \sum_{n=1}^{\infty}\left(
-1\right)  ^{n}x^{n-1}\ ,\ \ \ \psi\left(  x\right)  =\left(  1+x\right)
^{2}\sum_{n=1}^{\infty}\left(  -1\right)  ^{n}nx^{n-1}\ . \label{where}%
\end{align}
Let us study the formal series $\varphi\left(  x\right)  $ and $\psi\left(
x\right)  $:%
\begin{align}
\varphi\left(  x\right)   &  =\left(  1+x\right)  \sum_{n=1}^{\infty}\left(
-1\right)  ^{n}x^{n-1}=-\left(  1+x\right)  \sum_{m=0}^{\infty}\left(
-1\right)  ^{m}x^{m}=-\left(  1+x\right)  \frac{1}{\left(  1+x\right)
}=-1\ ,\nonumber\\
\psi\left(  x\right)   &  =\left(  1+x\right)  ^{2}\sum_{n=1}^{\infty}\left(
-1\right)  ^{n}nx^{n-1}=\left(  1+x\right)  ^{2}\frac{\partial}{\partial
x}\sum_{n=1}^{\infty}\left(  -1\right)  ^{n}x^{n}=\left(  1+x\right)
^{2}\frac{\partial}{\partial x}\left(  \frac{1}{1+x}-1\right) \nonumber\\
&  =\left(  1+x\right)  ^{2}\frac{\partial}{\partial x}\left(  \frac{1}%
{1+x}\right)  =-\left(  1+x\right)  ^{2}\frac{1}{\left(  1+x\right)  ^{2}%
}=-1\ , \label{formal}%
\end{align}
Therefore,%
\[
\Re=-\mathrm{Str}\left(  R\right)  +\left(  1+f\right)  \mathrm{Str}\left(
Q_{2}Q\right)  -\mathrm{Str}\left(  Q_{2}\right)  -\frac{1}{2}\mathrm{Str}%
\left(  Q_{2}^{2}\right)  =-\mathrm{Str}\left[  R+Q_{2}+\frac{1}{2}Q_{2}%
^{2}-\left(  1+f\right)  \left(  Q_{2}Q\right)  \right]  \ ,
\]
which proves the relation (\ref{J_lin_final}).

\section{Transformations with Arbitrary Parameters}

\label{AppC} \renewcommand{\theequation}{\Alph{section}.\arabic{equation}} \setcounter{equation}{0}

In this appendix, we prove Lemmas \ref{lemma M^n}--\ref{lemma QP^n} and
present explicit calculations related to the Jacobian of finite BRST-antiBRST
transformations with arbitrary field-dependent parameters $\lambda_{a}$.

\subsection{Proof of Lemma \ref{lemma M^n}\label{proof_lemma M^n}}

Considering the relations (\ref{R}), we examine the quantities $\mathrm{Str}%
\left(  P^{n-1}R\right)  $ which obey%
\begin{equation}
\mathrm{Str}\left(  P^{n-1}R\right)  =\left\{
\begin{array}
[c]{ll}%
\mathrm{Str}\left(  R\right)  \ , & n=1\ ,\\
0\ , & n>1\ .
\end{array}
\right.  \label{StrPR}%
\end{equation}
\bigskip Indeed, due to $Y_{,B}^{A}X^{Bb}=0$, we can write down a chain of
relations:%
\begin{align}
&  \left(  RP\right)  _{B}^{A}=R_{D}^{A}P_{B}^{D}=-\frac{1}{2}\lambda
^{2}\left(  \frac{\delta Y^{A}}{\delta\phi^{D}}X^{Db}\right)  \frac
{\delta\lambda_{b}}{\delta\phi^{B}}=0\ ,\nonumber\\
&  \left(  RP^{2}\right)  _{B}^{A}=\left(  RP\right)  _{D}^{A}P_{B}%
^{D}=0\ ,\nonumber\\
&  \ldots\nonumber\\
&  \left(  RP^{n-1}\right)  _{B}^{A}=\left(  RP\right)  _{D}^{A}\left(
P^{n-1}\right)  _{B}^{D}=0\ ,\ \ \ n>1\ . \label{PnR}%
\end{align}
Using the property $\mathrm{Str}\left(  AB\right)  =\mathrm{Str}\left(
BA\right)  $ for even matrices,%
\begin{equation}
\mathrm{Str}\left(  P^{n-1}R\right)  =\mathrm{Str}\left(  RP^{n-1}\right)
=0\ , \label{PR=RP}%
\end{equation}
we arrive at%
\[
\mathrm{Str}\left(  M^{n}\right)  =\mathrm{Str}\left(  P+Q\right)
^{n}+n\mathrm{Str}\left(  P^{n-1}R\right)  =\left\{
\begin{array}
[c]{ll}%
\mathrm{Str}\left(  P+Q\right)  +\mathrm{Str}\left(  R\right)  \ , & n=1\ ,\\
\mathrm{Str}\left(  P+Q\right)  ^{n}\ , & n>1\ ,
\end{array}
\right.
\]
which thereby proves Lemma \ref{lemma M^n}.

\subsection{Proof of Lemma \ref{lemma (P+Q)^n}\label{proof_lemma (P+Q)^n}}

Considering the contribution $\mathrm{Str}\left(  P+Q\right)  ^{n}$ in
(\ref{P+Q}), we notice that an occurrence of $Q\sim\lambda_{a}$ more then
twice yields zero, $\lambda_{a}\lambda_{b}\lambda_{c}\equiv0$. A direct
calculation for $n=2,3$ leads to the binomial rule%
\begin{equation}
\mathrm{Str}\left(  P+Q\right)  ^{n}=\sum_{k=0}^{n}C_{n}^{k}\mathrm{Str}%
\left(  P^{n-k}Q^{k}\right)  =\mathrm{Str}\left(  P^{n}+nP^{n-1}Q+C_{n}%
^{2}P^{n-2}Q^{2}\right)  \ , \label{StrP+Q}%
\end{equation}
whereas the case $n=4$ fails to conform to this rule due to the presence of
the products $PQPQ$ and $QPQP$, which cannot be rearranged to the form
$Q^{2}P^{2}$ under the symbol of supertrace by using the property
$\mathrm{Str}\left(  AB\right)  =\mathrm{Str}\left(  BA\right)  $. On the
other hand, this property allows one to present the case $n=4$ as follows:%
\begin{equation}
\mathrm{Str}\left(  P+Q\right)  ^{4}=\mathrm{Str}\left(  P^{4}+4P^{3}%
Q+4P^{2}Q^{2}+2PQPQ\right)  \ . \label{(P+Q)^4}%
\end{equation}
The consideration of the case $n>4$ is simplified by the fact that one needs
to keep track of the products that contain the matrix $Q$ no more than twice,
i.e., we only need to retain $P^{n}$, $P^{n-1}Q$ and pairs of $Q$'s, while
separating the expressions reduced to $P^{n-2}Q^{2}$ from those containing
pairs of $Q$'s so \textquotedblleft sandwiched\textquotedblright\ between
$P$'s as not to allow their rearrangement into $P^{n-2}Q^{2}$ by using the
property (\ref{transp}). Starting from the case $n=4$, given by (\ref{(P+Q)^4}%
), and considering a monomial $p^{2}q^{2}$ composed by $c$-numbers $p$, $q$,
we find that under the symbol of supertrace the coefficient $C_{4}^{2}$
decomposes into $C_{4}^{1}$ for $P^{2}Q^{2}$ and $C_{2}^{1}$ for $(PQ)^{2}$,
$C_{4}^{2}=C_{4}^{1}+C_{2}^{1}$, so that%
\begin{equation}
\mathrm{Str}\left(  P+Q\right)  ^{4}=\sum_{k=0}^{1}C_{4}^{k}\mathrm{Str}%
\left(  P^{4-k}Q^{k}\right)  +C_{4}^{1}\mathrm{Str}\left(  P^{2}Q^{2}\right)
+C_{2}^{1}\mathrm{Str}\left(  PQPQ\right)  \ . \label{P+Q4}%
\end{equation}
For $n=5$, we consider a $c$-number monomial $p^{3}q^{2}$ and find that the
coefficient $C_{5}^{2}$ decomposes into $C_{5}^{1}$ for $P^{3}Q^{2}$ and
$C_{5}^{1}$ for $P(PQ)^{2}$, $C_{5}^{2}=C_{5}^{1}+C_{5}^{1}$, so that
\begin{equation}
\mathrm{Str}\left(  P+Q\right)  ^{5}=\sum_{k=0}^{1}C_{5}^{k}\mathrm{Str}%
\left(  P^{5-k}Q^{k}\right)  +C_{5}^{1}\mathrm{Str}\left(  P^{3}Q^{2}\right)
+C_{5}^{1}\mathrm{Str}\left[  P\left(  PQ\right)  ^{2}\right]  \ .
\label{P+Q5}%
\end{equation}
For $n=6$, we consider a $c$-number monomial $p^{4}q^{2}$ and find that the
coefficient $C_{6}^{2}$ decomposes into $C_{6}^{1}$ for $P^{4}Q^{2}$,
$C_{6}^{1}$ for $P^{2}(PQ)^{2}$, and $C_{3}^{1}$ for $(P^{2}Q)^{2}$,
$C_{6}^{2}=C_{6}^{1}+C_{6}^{1}+C_{3}^{1}$, so that%
\begin{equation}
\mathrm{Str}\left(  P+Q\right)  ^{6}=\sum_{k=0}^{1}C_{6}^{k}\mathrm{Str}%
\left(  P^{6-k}Q^{k}\right)  +C_{6}^{1}\mathrm{Str}\left(  P^{4}Q^{2}\right)
+C_{6}^{1}\mathrm{Str}\left[  P^{2}\left(  PQ\right)  ^{2}\right]  +C_{3}%
^{1}\mathrm{Str}\left[  \left(  P^{2}Q\right)  ^{2}\right]  \ . \label{P+Q6}%
\end{equation}
For $n=7$, we consider a $c$-number monomial $p^{5}q^{2}$ and find that the
coefficient $C_{7}^{2}$ decomposes into $C_{7}^{1}$ for $P^{5}Q^{2}$,
$C_{7}^{1}$ for $P^{3}(PQ)^{2}$, and $C_{7}^{1}$ for $P(P^{2}Q)^{2}$,
$C_{7}^{2}=C_{7}^{1}+C_{7}^{1}+C_{7}^{1}$, so that
\begin{align*}
\mathrm{Str}\left(  P+Q\right)  ^{7}  &  =\sum_{k=0}^{1}C_{7}^{k}%
\mathrm{Str}\left(  P^{7-k}Q^{k}\right)  +C_{7}^{1}\mathrm{Str}\left(
P^{5}Q^{2}\right) \\
&  +C_{7}^{1}\mathrm{Str}\left[  P^{3}\left(  PQ\right)  ^{2}\right]
+C_{7}^{1}\mathrm{Str}\left[  P\left(  P^{2}Q\right)  ^{2}\right]  \ .
\end{align*}
For $n=8$, we consider a $c$-number monomial $p^{6}q^{2}$ and find that the
coefficient $C_{8}^{2}$ decomposes into $C_{8}^{1}$ for $P^{6}Q^{2}$,
$C_{8}^{1}$ for $P^{4}(PQ)^{2}$, $C_{8}^{1}$ for $P^{2}(P^{2}Q)^{2}$, and
$C_{4}^{1}$ for $(P^{3}Q)^{2}$, $C_{8}^{2}=3C_{8}^{1}+C_{4}^{1}$, so that
\begin{align*}
\mathrm{Str}\left(  P+Q\right)  ^{8}  &  =\sum_{k=0}^{1}C_{8}^{k}%
\mathrm{Str}\left(  P^{8-k}Q^{k}\right)  +C_{8}^{1}\mathrm{Str}\left(
P^{6}Q^{2}\right) \\
&  +C_{8}^{1}\mathrm{Str}\left[  P^{4}\left(  PQ\right)  ^{2}\right]
+C_{8}^{1}\mathrm{Str}\left[  P^{2}\left(  P^{2}Q\right)  ^{2}\right]
+C_{4}^{1}\mathrm{Str}\left[  \left(  P^{3}Q\right)  ^{2}\right]  .
\end{align*}
Proceeding by induction for $n=2k$ and considering $c$-number monomials
$p^{2(k-1)}q^{2}$, we find that the coefficient $C_{2k}^{2}$ decomposes into
$C_{2k}^{1}$ for $P^{2(k-1)}Q^{2}$, $C_{2k}^{1}$ for $P^{2(k-2)}(PQ)^{2}$,
$C_{2k}^{1}$ for $P^{2(k-3)}(P^{2}Q)^{2}$,\ldots,$C_{2k}^{1}$ for
$P^{2}(P^{(k-2)}Q)^{2}$, and $C_{k}^{1}$ for $(P^{(k-1)}Q)^{2}$, $C_{2k}%
^{2}=(k-1)C_{2k}^{1}+C_{k}^{1}$, so that%
\begin{align}
\mathrm{Str}\left(  P+Q\right)  ^{2k}  &  =\sum_{l=0}^{1}C_{2k}^{l}%
\mathrm{Str}\left(  P^{2k-l}Q^{l}\right)  +C_{2k}^{1}\sum_{l=0}^{k-2}%
\mathrm{Str}\left[  P^{2(k-l-1)}\left(  P^{l}Q\right)  ^{2}\right] \nonumber\\
&  +C_{k}^{1}\mathrm{Str}\left[  \left(  P^{k-1}Q\right)  ^{2}\right]
,\ \ \ k\geq2\ . \label{P+Q2k}%
\end{align}
For $n=2k+1$, we consider $c$-number monomials $p^{2k-1}q^{2}$ and find that
the coefficient $C_{2k+1}^{2}$ decomposes into $C_{2k+1}^{1}$ for
$P^{2k-1}Q^{2}$, $C_{2k+1}^{1}$ for $P^{2k-3}(PQ)^{2}$, $C_{2k+1}^{1}$ for
$P^{2k-5}(P^{2}Q)^{2}$,\ldots, $C_{2k+1}^{1}$ for $P^{3}(P^{(k-2)}Q)^{2}$, and
$C_{2k+1}^{1}$ for $P(P^{(k-1)}Q)^{2}$, $C_{2k+1}^{2}=kC_{2k+1}^{1}$, so that%
\begin{equation}
\mathrm{Str}\left(  P+Q\right)  ^{2k+1}=\sum_{l=0}^{1}C_{2k+1}^{l}%
\mathrm{Str}\left(  P^{2k+1-l}Q^{l}\right)  +C_{2k+1}^{1}\sum_{l=0}%
^{k-1}\mathrm{Str}\left[  P^{2(k-l)-1}\left(  P^{l}Q\right)  ^{2}\right]
\ ,\ \ \ k\geq2\ . \label{2k+1}%
\end{equation}
Formulae (\ref{P+Q2k}), (\ref{2k+1}) thereby prove Lemma \ref{lemma (P+Q)^n}.

\subsection{Proof of Lemma \ref{lemma R-1/2Q^2}\label{proof R-1/2Q^2}}

Due to the relations $X_{,A}^{Aa}=0$, we have%
\begin{equation}
\mathrm{Str}\left(  Q_{1}\right)  =\left(  Q_{1}\right)  _{A}^{A}\left(
-1\right)  ^{\varepsilon_{A}}=\frac{\delta X^{Aa}}{\delta\phi^{A}}\lambda
_{a}=0\ . \label{Q1}%
\end{equation}
We also observe the following:%
\begin{equation}
\mathrm{Str}\left(  Q_{1}^{2}\right)  =\left(  Q_{1}^{2}\right)  _{A}%
^{A}\left(  -1\right)  ^{\varepsilon_{A}}=\frac{\delta X^{Aa}}{\delta\phi^{B}%
}\lambda_{a}\frac{\delta X^{Bb}}{\delta\phi^{A}}\lambda_{b}\left(  -1\right)
^{\varepsilon_{B}}=\frac{\delta X^{Aa}}{\delta\phi^{B}}\frac{\delta X^{Bb}%
}{\delta\phi^{A}}\lambda_{b}\lambda_{a}\left(  -1\right)  ^{\varepsilon_{A}%
}\ . \label{Q12rel}%
\end{equation}
Differentiating the relation $X_{,B}^{Aa}X^{Bb}=\varepsilon^{ab}Y^{A}$\ with
respect to $\phi^{A}$, we find%
\begin{equation}
\frac{\delta}{\delta\phi^{B}}\left(  \frac{\delta X^{Aa}}{\delta\phi^{A}%
}\right)  X^{Bb}\left(  -1\right)  ^{\varepsilon_{B}}+\frac{\delta X^{Aa}%
}{\delta\phi^{B}}\frac{\delta X^{Bb}}{\delta\phi^{A}}+\varepsilon^{ba}%
\frac{\delta Y^{A}}{\delta\phi^{A}}=0\ . \label{derivative}%
\end{equation}
Once again, using the relation $X_{,A}^{Aa}=0$, we have%
\begin{equation}
\frac{\delta X^{Aa}}{\delta\phi^{B}}\frac{\delta X^{Bb}}{\delta\phi^{A}%
}=\varepsilon^{ab}\frac{\delta Y^{A}}{\delta\phi^{A}}\ , \label{conseq}%
\end{equation}
whence%
\begin{equation}
\mathrm{Str}\left(  Q_{1}^{2}\right)  =\varepsilon^{ab}\frac{\delta Y^{A}%
}{\delta\phi^{A}}\lambda_{b}\lambda_{a}\left(  -1\right)  ^{\varepsilon_{A}%
}=-\frac{\delta Y^{A}}{\delta\phi^{A}}\lambda^{2}\left(  -1\right)
^{\varepsilon_{A}}=2\mathrm{Str}\left(  R\right)  \ . \label{2R}%
\end{equation}
Relations (\ref{Q1}) and (\ref{2R}) thereby prove Lemma \ref{lemma R-1/2Q^2}.

\subsection{Proof of Lemma \ref{lemma P^n}\label{proof_lemma P^n}}

Let us write down a chain of relations:%
\begin{align}
&  P_{B}^{A}=X^{Aa}\frac{\delta\lambda_{a}}{\delta\phi^{A}}\ ,\nonumber\\
&  \left(  P^{2}\right)  _{B}^{A}=P_{D}^{A}P_{B}^{D}=X^{Aa}\left(
\frac{\delta\lambda_{a}}{\delta\phi^{B}}X^{Bb}\right)  \frac{\delta\lambda
_{b}}{\delta\phi^{A}}=X^{Aa}m_{a}^{b}\frac{\delta\lambda_{b}}{\delta\phi^{B}%
}\ ,\nonumber\\
&  \left(  P^{3}\right)  _{B}^{A}=\left(  P^{2}\right)  _{D}^{A}P_{B}%
^{D}=X^{Aa}m_{a}^{b}\left(  \frac{\delta\lambda_{b}}{\delta\phi^{D}}%
X^{Dd}\right)  \frac{\delta\lambda_{d}}{\delta\phi^{B}}=X^{Aa}\left(
m^{2}\right)  _{a}^{b}\frac{\delta\lambda_{b}}{\delta\phi^{B}}\ ,\nonumber\\
&  \ldots\nonumber\\
&  \left(  P^{n}\right)  _{B}^{A}=\left(  P^{n-1}\right)  _{D}^{A}P_{B}%
^{D}=X^{Aa}\left(  m^{n-2}\right)  _{a}^{b}\left(  \frac{\delta\lambda_{b}%
}{\delta\phi^{D}}X^{Dd}\right)  \frac{\delta\lambda_{d}}{\delta\phi^{B}%
}=X^{Aa}\left(  m^{n-1}\right)  _{a}^{b}\frac{\delta\lambda_{b}}{\delta
\phi^{B}}\ , \label{chain}%
\end{align}
whence%
\begin{equation}
\mathrm{Str}\left(  P^{n}\right)  =\left(  P^{n}\right)  _{A}^{A}\left(
-1\right)  ^{\varepsilon_{A}}=-\left(  m^{n-1}\right)  _{a}^{b}\left(
\frac{\delta\lambda_{b}}{\delta\phi^{A}}X^{Aa}\right)  =-\left(
m^{n-1}\right)  _{a}^{b}m_{b}^{a}=-\left(  m^{n}\right)  _{a}^{a}\ ,
\label{thereby}%
\end{equation}
which thereby proves Lemma \ref{lemma P^n}.

\subsection{Proof of Lemma \ref{lemma QP^n}\label{proof_lemma QP^n}}

Let us consider the matrix%
\begin{align}
\left(  QP\right)  _{B}^{A}  &  \equiv Q_{D}^{A}P_{B}^{D}=\left(  -1\right)
^{\varepsilon_{A}+1}\lambda_{a}\left(  \frac{\delta X^{Aa}}{\delta\phi^{D}%
}+Y^{A}\frac{\delta\lambda^{a}}{\delta\phi^{D}}\right)  X^{Dd}\frac
{\delta\lambda_{d}}{\delta\phi^{B}}\nonumber\\
&  =\left(  -1\right)  ^{\varepsilon_{A}+1}\lambda_{a}\left[  \varepsilon
^{ab}Y^{A}+Y^{A}\left(  s^{b}\lambda^{a}\right)  \right]  \frac{\delta
\lambda_{b}}{\delta\phi^{B}}\equiv\left(  Q_{2}\right)  _{B}^{A}+\left(
-1\right)  ^{\varepsilon_{A}+1}m^{ba}\lambda_{a}Y^{A}\frac{\delta\lambda_{b}%
}{\delta\phi^{B}}\ . \label{fQ2}%
\end{align}
Since in the case of arbitrary $\lambda_{a}$ there is no information on the
symmetry properties of $m^{ab}=s^{a}\lambda^{b}$, we thus arrive at a new
matrix:%
\begin{equation}
(Q_{2}^{(1)})_{B}^{A}\equiv\left(  -1\right)  ^{\varepsilon_{A}+1}%
m^{ba}\lambda_{a}Y^{A}\frac{\delta\lambda_{b}}{\delta\phi^{B}}\ , \label{Q12}%
\end{equation}
which is not contained among the matrices $P$, $Q$. If we now consider the
matrix $Q_{2}+Q_{2}^{(1)}$ acting on $P$,%
\begin{align}
(Q_{2}+Q_{2}^{(1)})_{D}^{A}P_{B}^{D}  &  =\left(  -1\right)  ^{\varepsilon
_{A}+1}\lambda_{a}\left(  Y^{A}\frac{\delta\lambda^{a}}{\delta\phi^{D}}%
+m^{ba}Y^{A}\frac{\delta\lambda_{b}}{\delta\phi^{D}}\right)  X^{Dd}%
\frac{\delta\lambda_{d}}{\delta\phi^{B}}\nonumber\\
&  =\left(  -1\right)  ^{\varepsilon_{A}+1}\lambda_{a}\left[  Y^{A}\left(
s^{d}\lambda^{a}\right)  +m^{ba}Y^{A}\left(  s^{d}\lambda_{b}\right)  \right]
\frac{\delta\lambda_{d}}{\delta\phi^{B}}\nonumber\\
&  =\left(  -1\right)  ^{\varepsilon_{A}+1}m^{da}\lambda_{a}Y^{A}\frac
{\delta\lambda_{d}}{\delta\phi^{B}}+\left(  -1\right)  ^{\varepsilon_{A}%
+1}m_{b}^{d}m^{ba}\lambda_{a}Y^{A}\frac{\delta\lambda_{d}}{\delta\phi^{B}}\ ,
\label{Q2+Q2(1)}%
\end{align}
it follows that,%
\begin{equation}
(Q_{2}+Q_{2}^{(1)})_{D}^{A}P_{B}^{D}=(Q_{2}^{(1)})_{B}^{A}+\left(  -1\right)
^{\varepsilon_{A}+1}m_{b}^{d}m^{ba}\lambda_{a}Y^{A}\frac{\delta\lambda_{d}%
}{\delta\phi^{B}}\ , \label{Q2Q2(1)result}%
\end{equation}
so we have another new matrix:
\begin{equation}
(Q_{2}^{(2)})_{B}^{A}\equiv\left(  -1\right)  ^{\varepsilon_{A}+1}m_{b}%
^{d}m^{ba}\lambda_{a}Y^{A}\frac{\delta\lambda_{d}}{\delta\phi^{B}}\ .
\label{newmatrix}%
\end{equation}
Let us, once again, consider a similar construction:%
\begin{align}
(Q_{2}^{(1)}+Q_{2}^{(2)})_{D}^{A}P_{B}^{D}  &  =\left[  \left(  -1\right)
^{\varepsilon_{A}+1}m^{ba}\lambda_{a}Y^{A}\frac{\delta\lambda_{b}}{\delta
\phi^{D}}+\left(  -1\right)  ^{\varepsilon_{A}+1}m_{b}^{c}m^{ba}\lambda
_{a}Y^{A}\frac{\delta\lambda_{c}}{\delta\phi^{D}}\right]  X^{Dd}\frac
{\delta\lambda_{d}}{\delta\phi^{B}}\nonumber\\
&  =\left(  -1\right)  ^{\varepsilon_{A}+1}m^{ba}\lambda_{a}Y^{A}\frac
{\delta\lambda_{b}}{\delta\phi^{D}}X^{Dd}\frac{\delta\lambda_{d}}{\delta
\phi^{B}}+\left(  -1\right)  ^{\varepsilon_{A}+1}m_{b}^{c}m^{ba}\lambda
_{a}Y^{A}\frac{\delta\lambda_{c}}{\delta\phi^{D}}X^{Dd}\frac{\delta\lambda
_{d}}{\delta\phi^{B}}\nonumber\\
&  =\left(  -1\right)  ^{\varepsilon_{A}+1}m_{b}^{d}m^{ba}\lambda_{a}%
Y^{A}\frac{\delta\lambda_{d}}{\delta\phi^{B}}+\left(  -1\right)
^{\varepsilon_{A}+1}m_{c}^{d}m_{b}^{c}m^{ba}\lambda_{a}Y^{A}\frac
{\delta\lambda_{d}}{\delta\phi^{B}}\equiv(Q_{2}^{(2)}+Q_{2}^{(3)})_{B}^{A}\ .
\label{similar}%
\end{align}
Generally, the above process leads to
\begin{equation}
(Q_{2}^{(n)}+Q_{2}^{(n+1)})P=Q_{2}^{(n+1)}+Q_{2}^{(n+2)}\ ,\ \ \ n\geq
0\ ,\ \ \ Q_{2}^{(0)}\equiv Q_{2}\ , \label{generally}%
\end{equation}
so that there emerges an infinite sequence of objects $Q_{2}^{(n)}$
constructed by multiplication of the matrix with the elements $m_{b}^{a}%
=s^{a}\lambda_{b}$. Using this observation and the fact that $m^{ba}%
\lambda_{a}=\left(  s^{b}\lambda^{a}\right)  \lambda_{a}=-\left(  s^{b}%
\lambda_{a}\right)  \lambda^{a}=-m_{a}^{b}\lambda^{a}$, let us rewrite the
above relations containing $m^{ab}$ in terms of $m_{b}^{a}$:%
\begin{align}
&  \left(  QP\right)  _{B}^{A}=\left(  Q_{2}\right)  _{B}^{A}+\left(
-1\right)  ^{\varepsilon_{A}}m_{a}^{b}\lambda^{a}Y^{A}\frac{\delta\lambda_{b}%
}{\delta\phi^{B}}=(Q_{2}+Q_{2}^{(1)})_{B}^{A}\ ,\nonumber\\
&  (Q_{2}+Q_{2}^{(1)})_{B}^{A}P_{B}^{D}=\left(  -1\right)  ^{\varepsilon_{A}%
}m_{a}^{b}\lambda^{a}Y^{A}\frac{\delta\lambda_{b}}{\delta\phi^{B}}+\left(
-1\right)  ^{\varepsilon_{A}}m_{b}^{d}m_{a}^{b}\lambda^{a}Y^{A}\frac
{\delta\lambda_{d}}{\delta\phi^{B}}=(Q_{2}^{(1)}+Q_{2}^{(2)})_{B}%
^{A}\ ,\nonumber\\
&  (Q_{2}^{(1)}+Q_{2}^{(2)})_{B}^{A}P_{B}^{D}=\left(  -1\right)
^{\varepsilon_{A}}m_{b}^{d}m_{a}^{b}\lambda^{a}Y^{A}\frac{\delta\lambda_{d}%
}{\delta\phi^{B}}+\left(  -1\right)  ^{\varepsilon_{A}}m_{c}^{d}m_{b}^{c}%
m_{a}^{b}\lambda^{a}Y^{A}\frac{\delta\lambda_{d}}{\delta\phi^{B}}=(Q_{2}%
^{(2)}+Q_{2}^{(3)})_{B}^{A}\ ,\nonumber\\
&  \ldots\nonumber\\
&  (Q_{2}^{(n)}+Q_{2}^{(n+1)})_{D}^{A}P_{B}^{D}=\left(  -1\right)
^{\varepsilon_{A}}\left(  m^{n+1}\right)  _{a}^{b}\lambda^{a}Y^{A}\frac
{\delta\lambda_{b}}{\delta\phi^{B}}+\left(  -1\right)  ^{\varepsilon_{A}%
}\left(  m^{n+2}\right)  _{a}^{b}\lambda^{a}Y^{A}\frac{\delta\lambda_{b}%
}{\delta\phi^{B}}=(Q_{2}^{(n+1)}+Q_{2}^{(n+2)})_{B}^{A}\ ,\ n\geq0\ ,
\label{matrix^a_b}%
\end{align}
which implies%
\begin{equation}
(Q_{2}^{(n)})_{B}^{A}=\left(  -1\right)  ^{\varepsilon_{A}}(m^{n})_{b}%
^{a}\lambda^{b}Y^{A}\frac{\delta\lambda_{a}}{\delta\phi^{B}}\ ,\ \ \ n\geq0\ ,
\label{Q2n}%
\end{equation}
Using the matrix $Y=(Y_{a}^{b})_{B}^{A}$ given by%
\[
(Y_{a}^{b})_{B}^{A}\equiv\left(  -1\right)  ^{\varepsilon_{A}}\lambda^{b}%
Y^{A}\frac{\delta\lambda_{a}}{\delta\phi^{B}}\ ,\ \ (Y_{a}^{a})_{B}^{A}%
=(Q_{2})_{A}^{A}\ ,\ \ \ \mathrm{tr}\left(  Y\right)  =Q_{2}\ ,
\]
we can represent the above sequence as follows:%
\begin{equation}
Q_{2}^{(n)}=\mathrm{tr}\left(  m^{n}Y\right)  \ ,\ \ \ n\geq0\ ,
\label{sequence}%
\end{equation}
Hence, taking account of the property (\ref{generally}), we have%
\begin{align}
&  \left(  Q_{1}+Q_{2}\right)  P=Q_{2}+Q_{2}^{(1)}=\mathrm{tr}\left(
Y+mY\right)  =\mathrm{tr}\left[  \left(  e+m\right)  Y\right]  \ ,\nonumber\\
&  \left(  Q_{1}+Q_{2}\right)  P^{2}=Q_{2}^{(1)}+Q_{2}^{(2)}=\mathrm{tr}%
\left(  mY+m^{2}Y\right)  =\mathrm{tr}\left[  m\left(  e+m\right)  Y\right]
\ ,\nonumber\\
&  \left(  Q_{1}+Q_{2}\right)  P^{3}=Q_{2}^{(2)}+Q_{2}^{(3)}=\mathrm{tr}%
\left(  m^{2}Y+m^{3}Y\right)  =\mathrm{tr}\left[  m^{2}\left(  e+m\right)
Y\right]  \ ,\nonumber\\
&  \ldots\nonumber\\
&  \left(  Q_{1}+Q_{2}\right)  P^{n}=Q_{2}^{(n-1)}+Q_{2}^{(n)}=\mathrm{tr}%
\left(  m^{n-1}Y+m^{n}Y\right)  =\mathrm{tr}\left[  m^{n-1}\left(  e+m\right)
Y\right]  \ ,\ \ \ n\geq1\ . \label{(Q1+Q2)R^n}%
\end{align}
Recalling that $Q=Q_{1}+Q_{2}$, we finally have%
\[
QP^{n}=\mathrm{tr}\left[  m^{n-1}\left(  e+m\right)  Y\right]  \ ,\ \ \ n\geq
1\ ,
\]
which completes the proof of Lemma \ref{lemma QP^n}.

\subsection{Calculation of Jacobian \label{Jacobian_arb}}

Let us consider a calculation of the quantity $\Im$ on the basis of the
relations (\ref{Strgen})--(\ref{matrY}). First of all, we have%
\begin{equation}
\Im=-\sum_{n=1}^{\infty}\frac{\left(  -1\right)  ^{n}}{n}\mathrm{Str}\left(
M^{n}\right)  =\mathrm{Str}\left(  R\right)  -\sum_{n=1}^{3}\frac{\left(
-1\right)  ^{n}}{n}\mathrm{Str}\left(  P+Q\right)  ^{n}-\sum_{n=4}^{\infty
}\frac{\left(  -1\right)  ^{n}}{n}\mathrm{Str}\left(  P+Q\right)  ^{n}\ .
\label{R_start}%
\end{equation}
Decomposing the summation number $n\geq4$ into odd and even components,
$n=\left(  2k+1,2k\right)  $, $k\geq2$, we have, according to (\ref{(2k)}),
(\ref{(2k+1)}),%
\begin{align}
\Im &  =\mathrm{Str}\left(  R\right)  -\sum_{n=1}^{3}\frac{\left(  -1\right)
^{n}}{n}\mathrm{Str}\left(  P+Q\right)  ^{n}\nonumber\\
&  -\sum_{k=2}^{\infty}\frac{\left(  -1\right)  ^{2k+1}}{2k+1}\left\{
\sum_{l=0}^{1}C_{2k+1}^{l}\mathrm{Str}\left(  P^{2k+1-l}Q^{l}\right)
+C_{2k+1}^{1}\sum_{l=0}^{k-1}\mathrm{Str}\left[  P^{2(k-l)-1}\left(
P^{l}Q\right)  ^{2}\right]  \right\} \nonumber\\
&  -\sum_{k=2}^{\infty}\frac{\left(  -1\right)  ^{2k}}{2k}\left\{  \sum
_{l=0}^{1}C_{2k}^{l}\mathrm{Str}\left(  P^{2k-l}Q^{l}\right)  +C_{2k}^{1}%
\sum_{l=0}^{k-2}\mathrm{Str}\left[  P^{2(k-l-1)}\left(  P^{l}Q\right)
^{2}\right]  +C_{k}^{1}\mathrm{Str}\left[  \left(  P^{k-1}Q\right)
^{2}\right]  \right\}  \ , \label{decomp0}%
\end{align}
whence%
\begin{align}
\Im &  =\mathrm{Str}\left(  R\right)  -\sum_{n=1}^{3}\frac{\left(  -1\right)
^{n}}{n}\mathrm{Str}\left(  P+Q\right)  ^{n}-\sum_{n=4}^{\infty}\frac{\left(
-1\right)  ^{n}}{n}\sum_{l=0}^{1}C_{n}^{l}\mathrm{Str}\left(  P^{n-l}%
Q^{l}\right)  -\frac{1}{2}\sum_{k=2}^{\infty}\mathrm{Str}\left[  \left(
P^{k-1}Q\right)  ^{2}\right] \nonumber\\
&  +\sum_{k=2}^{\infty}\sum_{l=0}^{k-1}\mathrm{Str}\left[  P^{2(k-l)-1}\left(
P^{l}Q\right)  ^{2}\right]  -\sum_{k=2}^{\infty}\sum_{l=0}^{k-2}%
\mathrm{Str}\left[  P^{2(k-l-1)}\left(  P^{l}Q\right)  ^{2}\right]  \ .
\label{decomp}%
\end{align}
\bigskip It should be noted that%
\begin{equation}
\sum_{k=2}^{\infty}\sum_{l=0}^{k-1}\mathrm{Str}\left[  P^{2(k-l)-1}\left(
P^{l}Q\right)  ^{2}\right]  =\sum_{k=2}^{\infty}\mathrm{Str}\left[  P\left(
P^{k-1}Q\right)  ^{2}\right]  +\sum_{k=2}^{\infty}\sum_{l=0}^{k-2}%
\mathrm{Str}\left[  P^{2(k-l)-1}\left(  P^{l}Q\right)  ^{2}\right]  \ ,
\label{formula0}%
\end{equation}
whereas%
\begin{align}
&  \ \sum_{k=2}^{\infty}\sum_{l=0}^{k-2}\mathrm{Str}\left[  P^{2(k-l)-1}%
\left(  P^{l}Q\right)  ^{2}\right]  =\mathrm{Str}\left[  P^{2(2-0)-1}\left(
P^{0}Q\right)  ^{2}\right]  +\sum_{k=3}^{\infty}\sum_{l=0}^{k-2}%
\mathrm{Str}\left[  P^{2(k-l)-1}\left(  P^{l}Q\right)  ^{2}\right] \nonumber\\
&  \ =\mathrm{Str}\left(  P^{3}Q^{2}\right)  +\sum_{k=3}^{\infty}\left\{
\mathrm{Str}\left[  P^{2(k-0)-1}\left(  P^{0}Q\right)  ^{2}\right]
+\sum_{l=1}^{k-2}\mathrm{Str}\left[  P^{2(k-l)-1}\left(  P^{l}Q\right)
^{2}\right]  \right\} \nonumber\\
&  \ =\sum_{k=2}^{\infty}\mathrm{Str}\left(  P^{2k-1}Q^{2}\right)  +\sum
_{k=3}^{\infty}\sum_{l=1}^{k-2}\mathrm{Str}\left[  P^{2(k-l)-1}\left(
P^{l}Q\right)  ^{2}\right]  \ . \label{formula1}%
\end{align}
It should also be noted that%
\begin{align}
&  \ -\sum_{k=2}^{\infty}\sum_{l=0}^{k-2}\mathrm{Str}\left[  P^{2(k-l-1)}%
\left(  P^{l}Q\right)  ^{2}\right]  =-\mathrm{Str}\left[  P^{2(2-0-1)}\left(
P^{0}Q\right)  ^{2}\right]  -\sum_{k=3}^{\infty}\sum_{l=0}^{k-2}%
\mathrm{Str}\left[  P^{2(k-l-1)}\left(  P^{l}Q\right)  ^{2}\right] \nonumber\\
&  \ =-\mathrm{Str}\left(  P^{2}Q^{2}\right)  -\sum_{k=3}^{\infty}\left\{
\mathrm{Str}\left[  P^{2(k-0-1)}\left(  P^{0}Q\right)  ^{2}\right]
+\sum_{l=1}^{k-2}\mathrm{Str}\left[  P^{2(k-l-1)}\left(  P^{l}Q\right)
^{2}\right]  \right\} \nonumber\\
\  &  =-\sum_{k=2}^{\infty}\mathrm{Str}\left(  P^{2(k-1)}Q^{2}\right)
-\sum_{k=3}^{\infty}\sum_{l=1}^{k-2}\mathrm{Str}\left[  P^{2(k-l-1)}\left(
P^{l}Q\right)  ^{2}\right]  \ . \label{formula2}%
\end{align}
From (\ref{decomp}), (\ref{formula0}), (\ref{formula1}), (\ref{formula2}), it
follows that%
\begin{align}
\Im &  =\mathrm{Str}\left(  R\right)  -\sum_{n=1}^{3}\frac{\left(  -1\right)
^{n}}{n}\mathrm{Str}\left(  P+Q\right)  ^{n}-\sum_{n=4}^{\infty}\frac{\left(
-1\right)  ^{n}}{n}\sum_{l=0}^{1}C_{n}^{l}\mathrm{Str}\left(  P^{n-l}%
Q^{l}\right) \nonumber\\
&  +\sum_{k=2}^{\infty}\mathrm{Str}\left[  P\left(  P^{k-1}Q\right)
^{2}-\frac{1}{2}\left(  P^{k-1}Q\right)  ^{2}\right]  +\sum_{k=2}^{\infty
}\mathrm{Str}\left[  \left(  P^{2k-1}-P^{2(k-1)}\right)  Q^{2}\right]
\nonumber\\
&  +\sum_{k=3}^{\infty}\sum_{l=1}^{k-2}\mathrm{Str}\left[  \left(
P^{2(k-l)-1}-P^{2(k-l-1)}\right)  \left(  P^{l}Q\right)  ^{2}\right]  \ .
\label{formula4}%
\end{align}
By virtue of (\ref{(2,3)}),%
\begin{align}
&  \ -\sum_{n=1}^{3}\frac{\left(  -1\right)  ^{n}}{n}\mathrm{Str}\left(
P+Q\right)  ^{n}=\mathrm{Str}\left(  P+Q\right)  -\sum_{n=2}^{3}\frac{\left(
-1\right)  ^{n}}{n}\mathrm{Str}\left(  P+Q\right)  ^{n}\ ,\nonumber\\
&  \ -\sum_{n=2}^{3}\frac{\left(  -1\right)  ^{n}}{n}\mathrm{Str}\left(
P+Q\right)  ^{n}=-\sum_{n=2}^{3}\frac{\left(  -1\right)  ^{n}}{n}%
\mathrm{Str}\left(  P^{n}+nP^{n-1}Q+C_{n}^{2}P^{n-2}Q^{2}\right)
\ ,\ \ \ n=2,3\ , \label{intermed1}%
\end{align}
we have%
\begin{equation}
-\sum_{n=1}^{3}\frac{\left(  -1\right)  ^{n}}{n}\mathrm{Str}\left(
P+Q\right)  ^{n}=\mathrm{Str}\left(  P+Q\right)  -\sum_{n=2}^{3}\frac{\left(
-1\right)  ^{n}}{n}\mathrm{Str}\left(  P^{n}+C_{n}^{1}P^{n-1}Q\right)
-\sum_{n=2}^{3}\frac{\left(  -1\right)  ^{n}}{n}\mathrm{Str}\left(  C_{n}%
^{2}P^{n-2}Q^{2}\right)  \ , \label{interrmed2}%
\end{equation}
which implies%
\begin{align}
&  \ -\sum_{n=1}^{3}\frac{\left(  -1\right)  ^{n}}{n}\mathrm{Str}\left(
P+Q\right)  ^{n}-\sum_{n=4}^{\infty}\frac{\left(  -1\right)  ^{n}}{n}%
\sum_{l=0}^{1}C_{n}^{l}\mathrm{Str}\left(  P^{n-l}Q^{l}\right) \nonumber\\
&  \ =-\sum_{n=1}^{\infty}\frac{\left(  -1\right)  ^{n}}{n}\mathrm{Str}\left(
P^{n}\right)  -\sum_{n=1}^{\infty}\frac{\left(  -1\right)  ^{n}}%
{n}\mathrm{Str}\left(  C_{n}^{1}P^{n-1}Q\right)  -\sum_{n=2}^{3}\frac{\left(
-1\right)  ^{n}}{n}\mathrm{Str}\left(  C_{n}^{2}P^{n-2}Q^{2}\right)  \ .
\label{intermed3}%
\end{align}
Consequently, using (\ref{formula4}), we arrive at the representation%
\begin{equation}
\Im=-\sum_{n=1}^{\infty}\frac{\left(  -1\right)  ^{n}}{n}\mathrm{Str}\left(
P\right)  ^{n}+\Re\ , \label{represent}%
\end{equation}
where, in virtue of the obvious relations%
\begin{align}
&  \sum_{n=2}^{3}\frac{\left(  -1\right)  ^{n}}{n}\mathrm{Str}\left(
C_{n}^{2}P^{n-2}Q^{2}\right)  =\frac{1}{2}\mathrm{Str}\left(  Q^{2}\right)
-\mathrm{Str}\left(  QPQ\right)  \ ,\nonumber\\
&  \sum_{n=1}^{\infty}\left(  -1\right)  ^{n}\mathrm{Str}\left(
P^{n-1}Q\right)  =-\mathrm{Str}\left(  Q\right)  +\sum_{n=2}^{\infty}\left(
-1\right)  ^{n}\mathrm{Str}\left(  QP^{n-1}\right)  \ , \label{obvious}%
\end{align}
and, due to the property $\mathrm{Str}\left(  AB\right)  =\mathrm{Str}\left(
BA\right)  $, we have%
\begin{align}
\Re &  =\mathrm{Str}\left(  R\right)  -\frac{1}{2}\mathrm{Str}\left(
Q^{2}\right)  +\mathrm{Str}\left(  QPQ\right)  +\mathrm{Str}\left(  Q\right)
-\sum_{n=2}^{\infty}\left(  -1\right)  ^{n}\mathrm{Str}\left(  QP^{n-1}\right)
\nonumber\\
&  +\sum_{k=2}^{\infty}\mathrm{Str}\left[  Q\left(  P^{k}-\frac{1}{2}%
P^{k-1}\right)  QP^{k-1}\right]  +\sum_{k=2}^{\infty}\mathrm{Str}\left[
Q\left(  P^{2k-1}-P^{2k-2}\right)  Q\right] \nonumber\\
&  +\sum_{k=3}^{\infty}\sum_{l=1}^{k-2}\mathrm{Str}\left[  Q\left(
P^{2k-l-1}-P^{2k-l-2}\right)  QP^{l}\right]  \ . \label{formula5}%
\end{align}
Let us show that the quantity $\Re$ is zero. To this end, let us recall the
properties (\ref{MABext}), (\ref{StrQ=0}), (\ref{QPgen}), (\ref{matrY}),%
\begin{align*}
&  \mathrm{Str}\left(  Q_{1}\right)  \equiv0\ ,\ \ \ \mathrm{Str}\left(
R\right)  -\frac{1}{2}\mathrm{Str}\left(  Q_{1}^{2}\right)  \equiv0\ ,\\
&  Q=Q_{1}+Q_{2\ ,\ \ \ }Q_{2}=\mathrm{tr}\left(  Y\right)  \ ,\ \ \ QP^{n}%
=\mathrm{tr}\left[  m^{n-1}\left(  e+m\right)  Y\right]  \ ,\ \ \ n\geq1\ ,
\end{align*}
which imply the relations%
\begin{align}
&  \mathrm{Str}\left(  Q\right)  =\mathrm{Str}\left[  \mathrm{tr}\left(
Y\right)  \right]  \ ,\nonumber\\
&  \mathrm{Str}\left(  R\right)  -\frac{1}{2}\mathrm{Str}\left(  Q^{2}\right)
=-\mathrm{Str}\left[  Q_{1}\mathrm{tr}\left(  Y\right)  \right]  -\frac{1}%
{2}\mathrm{Str}\left[  \mathrm{tr}\left(  Y\right)  \mathrm{tr}\left(
Y\right)  \right]  \label{imlpy}%
\end{align}
and%
\begin{align}
&  QPQ=\mathrm{tr}\left[  \left(  e+m\right)  Y\right]  \left[  Q_{1}%
+\mathrm{tr}\left(  Y\right)  \right]  \ ,\nonumber\\
&  QP=\mathrm{tr}\left[  \left(  e+m\right)  Y\right]  \ ,\ \ \ QP^{n-1}%
=\mathrm{tr}\left[  m^{n-2}\left(  e+m\right)  Y\right]  \ ,\nonumber\\
&  Q\left[  P^{k}-\left(  1/2\right)  P^{k-1}\right]  =\mathrm{tr}\left[
m^{k-2}\left(  m-e/2\right)  \left(  e+m\right)  Y\right]  \ ,\nonumber\\
&  QP^{k-1}=\mathrm{tr}\left[  m^{k-2}\left(  e+m\right)  Y\right]
\ ,\nonumber\\
&  Q\left(  P^{2k-1}-P^{2k-2}\right)  =\mathrm{tr}\left[  m^{2k-3}\left(
m^{2}-e\right)  Y\right]  \ ,\nonumber\\
&  Q\left(  P^{2k-l-1}-P^{2k-l-2}\right)  =\mathrm{tr}\left[  \left(
m^{2k-l-2}-m^{2k-l-3}\right)  \left(  e+m\right)  Y\right]  \ ,\nonumber\\
&  QP^{l}=\mathrm{tr}\left[  m^{l-1}\left(  e+m\right)  Y\right]  \ ,
\label{auxil}%
\end{align}
As a consequence, we arrive at the following representation of (\ref{formula5}%
):%
\begin{equation}
\Re=\Re_{1}+\Re_{2}+\Re_{3}\ , \label{R=sum}%
\end{equation}
where the contributions $\Re_{1}$, $\Re_{2}$,$\Re_{3}$ are given by%
\begin{align}
\Re_{1}  &  =\mathrm{Str}\left[  \mathrm{tr}\left(  Y\right)  \right]
-\sum_{n=2}^{\infty}\left(  -1\right)  ^{n}\mathrm{Str}\left\{  \mathrm{tr}%
\left[  m^{n-2}\left(  e+m\right)  Y\right]  \right\}  \ ,\label{R1_contrib}\\
\Re_{2}  &  =-\mathrm{Str}\left[  Q_{1}\mathrm{tr}\left(  Y\right)  \right]
+\mathrm{Str}\left\{  Q_{1}\mathrm{tr}\left[  \left(  e+m\right)  Y\right]
\right\}  +\sum_{k=2}^{\infty}\mathrm{Str}\left\{  Q_{1}\mathrm{tr}\left[
m^{2k-3}\left(  m^{2}-e\right)  Y\right]  \right\}  \ ,\label{R2_contrib}\\
\Re_{3}  &  =-\frac{1}{2}\mathrm{Str}\left[  \mathrm{tr}\left(  Y\right)
\mathrm{tr}\left(  Y\right)  \right]  +\mathrm{Str}\left\{  \mathrm{tr}\left[
\left(  e+m\right)  Y\right]  \mathrm{tr}\left(  Y\right)  \right\}
\nonumber\\
&  +\sum_{k=2}^{\infty}\mathrm{Str}\left\{  \mathrm{tr}\left[  m^{2k-3}\left(
m^{2}-e\right)  Y\right]  \mathrm{tr}\left(  Y\right)  \right\} \nonumber\\
&  +\sum_{k=2}^{\infty}\mathrm{Str}\left\{  \mathrm{tr}\left[  m^{k-2}\left(
m-e/2\right)  \left(  e+m\right)  Y\right]  \mathrm{tr}\left[  m^{k-2}\left(
e+m\right)  Y\right]  \right\} \nonumber\\
&  +\sum_{k=3}^{\infty}\sum_{l=1}^{k-2}\mathrm{Str}\left\{  \mathrm{tr}\left[
\left(  m^{2k-l-2}-m^{2k-l-3}\right)  \left(  e+m\right)  Y\right]
\mathrm{tr}\left[  m^{l-1}\left(  e+m\right)  Y\right]  \right\}  \ .
\label{R3_contrib}%
\end{align}
Notice that the operation $\mathrm{tr}$ enters $\Re_{1}$, $\Re_{2}$ linearly,
whereas $\Re_{3}$ contains the operation $\mathrm{tr}$ quadratically. Let us
show that $\Re_{1}$, $\Re_{2}$, $\Re_{3}$ are equal to zero.

The contribution $\Re_{1}$, linear in the elements of the matrix $Y$, reads
equivalently%
\begin{equation}
\Re_{1}=\mathrm{Str}\left[  \mathrm{tr}\left(  Y\right)  \right]
+\mathrm{Str}\sum_{k=1}^{\infty}\left(  -1\right)  ^{k}\mathrm{tr}\left[
m^{k-1}\left(  e+m\right)  Y\right]  \equiv\mathrm{Str}\left[  \mathrm{tr}%
\left(  \mathcal{A}Y\right)  \right]  \ , \label{R1_equiv}%
\end{equation}
where%
\[
\mathcal{A}=e+\sum_{k=1}^{\infty}\left(  -1\right)  ^{k}m^{k-1}\left(
e+m\right)  =e-\sum_{k=0}^{\infty}\left(  -1\right)  ^{k}m^{k}+\sum
_{k=1}^{\infty}\left(  -1\right)  ^{k}m^{k}=e-e\equiv0\ .
\]
Therefore, the contribution $\Re_{1}$ vanishes identically, $\Re_{1}\equiv0$.

The contribution $\Re_{2}$, bilinear in the elements of the matrices $Y$ and
$Q_{1}$, reads equivalently%
\begin{align}
&  \Re_{2}=-\mathrm{Str}\left[  Q_{1}\mathrm{tr}\left(  Y\right)  \right]
+\mathrm{Str}\left(  Q_{1}\mathrm{tr}\left[  \left(  e+m\right)  Y\right]
\right) \nonumber\\
&  +\sum_{k=2}^{\infty}\mathrm{Str}\left\{  Q_{1}\mathrm{tr}\left[
m^{2k-1}Y\right]  \right\}  -\sum_{k=2}^{\infty}\mathrm{Str}\left\{
Q_{1}\mathrm{tr}\left[  m^{2k-3}Y\right]  \right\}  \equiv\mathrm{Str}\left[
Q_{1}\mathrm{tr}\left(  \mathcal{B}Y\right)  \right]  \ , \label{R2_equiv}%
\end{align}
where%
\[
\mathcal{B}=-e+\left(  e+m\right)  +\sum_{k=2}^{\infty}m^{2k-1}-\sum
_{k=2}^{\infty}m^{2k-3}\equiv0\ ,
\]
since%
\begin{equation}
\sum_{k=2}^{\infty}\left(  m^{2k-1}-m^{2k-3}\right)  =\sum_{k=2}^{\infty
}m^{2k-1}-\sum_{k=1}^{\infty}m^{2k-1}=-m\ . \label{-m}%
\end{equation}
Therefore, the contribution $\Re_{2}$ vanishes identically, $\Re_{2}\equiv0$.

The contribution $\Re_{3}$, quadratic in the elements of the matrix $Y$, reads
equivalently%
\begin{align}
\Re_{3}  &  =\left(  1/2\right)  \mathrm{Str}\left[  \mathrm{tr}\left(
Y\right)  \mathrm{tr}\left(  Y\right)  \right]  +\mathrm{Str}\left[
\mathrm{tr}\left(  mY\right)  \mathrm{tr}\left(  Y\right)  \right] \nonumber\\
&  +\sum_{k=2}^{\infty}\mathrm{Str}\left\{  \mathrm{tr}\left[  \left(
m^{2k-1}-m^{2k-3}\right)  Y\right]  \mathrm{tr}\left(  Y\right)  \right\}
\nonumber\\
&  +\sum_{k=2}^{\infty}\mathrm{Str}\left\{  \mathrm{tr}\left[  \left(
m^{k}+m^{k-1}/2-m^{k-2}/2\right)  Y\right]  \mathrm{tr}\left[  \left(
m^{k-1}+m^{k-2}\right)  Y\right]  \right\} \nonumber\\
&  +\sum_{k=3}^{\infty}\sum_{l=1}^{k-2}\mathrm{Str}\left\{  \mathrm{tr}\left[
\left(  m^{2k-l-1}-m^{2k-l-3}\right)  Y\right]  \mathrm{tr}\left[  \left(
m^{l-1}+m^{l}\right)  Y\right]  \right\}  \ , \label{quadr_contr}%
\end{align}
and therefore the expression for $\Re_{3}$ has the structure%
\begin{equation}
\Re_{3}=\sum_{n=0}^{\infty}\Re_{3}^{\left(  n\right)  }\ ,\ \ \ \mathrm{where}%
\ \ \ \Re_{3}^{\left(  n\right)  }=\mathrm{Str}\sum_{k,l}a_{kl}\mathrm{tr}%
\left(  m^{k}Y\right)  \mathrm{tr}\left(  m^{l}Y\right)  \ ,\ \ \ n=k+l\ .
\label{stucture}%
\end{equation}
Let us examine the following contribution, taking into account the property
$\mathrm{Str}\left(  AB\right)  =\mathrm{Str}\left(  BA\right)  $:%
\begin{align}
&  \ \mathrm{Str}\sum_{k=2}^{\infty}\mathrm{tr}\left[  \left(  m^{k-1}%
/2-m^{k-2}/2\right)  Y\right]  \mathrm{tr}\left[  \left(  m^{k-1}%
+m^{k-2}\right)  Y\right] \nonumber\\
&  =\frac{1}{2}\mathrm{Str}\sum_{k=2}^{\infty}\mathrm{tr}\left(
m^{k-1}Y\right)  \mathrm{tr}\left(  m^{k-1}Y\right)  -\frac{1}{2}%
\mathrm{Str}\sum_{k=2}^{\infty}\mathrm{tr}\left(  m^{k-2}Y\right)
\mathrm{tr}\left(  m^{k-2}Y\right) \nonumber\\
&  \ =\frac{1}{2}\mathrm{Str}\sum_{k=2}^{\infty}\mathrm{tr}\left(
m^{k-1}Y\right)  \mathrm{tr}\left(  m^{k-1}Y\right)  -\frac{1}{2}%
\mathrm{Str}\sum_{k=1}^{\infty}\mathrm{tr}\left(  m^{k-1}Y\right)
\mathrm{tr}\left(  m^{k-1}Y\right) \nonumber\\
&  =-\frac{1}{2}\mathrm{Str}\left[  \mathrm{tr}\left(  m^{0}Y\right)
\mathrm{tr}\left(  m^{0}Y\right)  \right]  =-\frac{1}{2}\mathrm{Str}\left[
\mathrm{tr}\left(  Y\right)  \mathrm{tr}\left(  Y\right)  \right]  \ .
\label{contrib1}%
\end{align}
Let us also examine the contribution%
\begin{equation}
\mathrm{Str}\sum_{k=2}^{\infty}\mathrm{tr}\left[  \left(  m^{2k-1}%
-m^{2k-3}\right)  Y\right]  \mathrm{tr}\left(  Y\right)  =-\mathrm{Str}%
\left\{  \left[  \mathrm{tr}\left(  mY\right)  \right]  \mathrm{tr}\left(
Y\right)  \right\}  \ , \label{contrib2}%
\end{equation}
where account has been taken of (\ref{-m}). As a consequence of
(\ref{contrib1}), (\ref{contrib2}), the terms of $\Re_{3}$ containing
$\mathrm{Str}\left[  \left(  \mathrm{tr}\left(  Y\right)  \right)  {}^{2}
\right]  $ and $\mathrm{Str}\left[  \mathrm{tr}\left(  mY\right)
\mathrm{tr}\left(  Y\right)  \right]  $ are cancelled out, and therefore the
expression becomes simplified:%
\begin{align}
\Re_{3}  &  =\mathrm{Str}\sum_{k=2}^{\infty}\mathrm{tr}\left[  \left(
m^{k-1}+m^{k-2}\right)  Y\right]  \mathrm{tr}\left(  m^{k}Y\right) \nonumber\\
&  +\mathrm{Str}\sum_{k=3}^{\infty}\sum_{l=1}^{k-2}\mathrm{tr}\left[  \left(
m^{2k-l-1}-m^{2k-l-3}\right)  Y\right]  \mathrm{tr}\left[  \left(
m^{l-1}+m^{l}\right)  Y\right]  \ , \label{simplif}%
\end{align}
which also means that $\Re_{3}$ formally starts with the second order in the
elements of the matrix $m$. In more detail, let us examine the constituents of
$\Re_{3}$ in their relation to the order $n$ of expansion in powers of the
matrix elements $m_{b}^{a}$:%
\begin{align}
&  k\geq2:\nonumber\\
&
\begin{tabular}
[c]{|l|l|}\hline
$\mathrm{tr}\left(  m^{k-2}Y\right)  \mathrm{tr}\left(  m^{k}Y\right)  $ &
$\mathrm{tr}\left(  m^{k-1}Y\right)  \mathrm{tr}\left(  m^{k}Y\right)
$\\\hline
$n=2k-2$ & $n=2k-1$\\\hline
\end{tabular}
\ ,\nonumber\\
&  k\geq3:\label{nthorder}\\
&
\begin{tabular}
[c]{|l|l|l|l|}\hline
$\mathrm{tr}\left(  m^{2k-l-1}Y\right)  \mathrm{tr}\left(  m^{l-1}Y\right)  $
& $\mathrm{tr}\left(  m^{2k-l-3}Y\right)  \mathrm{tr}\left(  m^{l-1}Y\right)
$ & $\mathrm{tr}\left(  m^{2k-l-1}Y\right)  \mathrm{tr}\left(  m^{l}Y\right)
$ & $\mathrm{tr}\left(  m^{2k-l-3}Y\right)  \mathrm{tr}\left(  m^{l}Y\right)
$\\\hline
$n=2k-2$ & $n=2k-4$ & $n=2k-1$ & $n=2k-3$\\\hline
\end{tabular}
\ .\nonumber
\end{align}
For even degrees $n=2r$:%
\begin{equation}%
\begin{tabular}
[c]{lllll}%
$+\mathrm{tr}\left(  m^{k-2}Y\right)  \mathrm{tr}\left(  m^{k}Y\right)  \ ,$ &
$2k-2=2r\ ,$ & $k\geq2\ ,$ & $k=r+1\ ,$ & $r\geq1\ ,$\\
$+\mathrm{tr}\left(  m^{2k-l-1}Y\right)  \mathrm{tr}\left(  m^{l-1}Y\right)
\ ,$ & $2k-2=2r\ ,$ & $k\geq3\ ,$ & $k=r+1\ ,$ & $r\geq2\ ,\ \ \ l=1,\ldots
,r-1\ ,$\\
$-\mathrm{tr}\left(  m^{2k-l-3}Y\right)  \mathrm{tr}\left(  m^{l-1}Y\right)
\ ,$ & $2k-4=2r\ ,$ & $k\geq3\ ,$ & $k=r+2\ ,$ & $r\geq1\ ,\ \ \ l=1,\ldots
,r\ .$%
\end{tabular}
\label{even}%
\end{equation}
For odd degrees $n=2r+1$:%
\begin{equation}%
\begin{tabular}
[c]{lllll}%
$+\mathrm{tr}\left(  m^{k-1}Y\right)  \mathrm{tr}\left(  m^{k}Y\right)  \ ,$ &
$2k-1=2r+1\ ,$ & $k\geq2\ ,$ & $k=r+1\ ,$ & $r\geq1\ ,$\\
$+\mathrm{tr}\left(  m^{2k-l-1}Y\right)  \mathrm{tr}\left(  m^{l}Y\right)
\ ,$ & $2k-1=2r+1\ ,$ & $k\geq3\ ,$ & $k=r+1\ ,$ & $r\geq2\ ,\ \ \ l=1,\ldots
,r-1\ ,$\\
$-\mathrm{tr}\left(  m^{2k-l-3}Y\right)  \mathrm{tr}\left(  m^{l}Y\right)
\ ,$ & $2k-3=2r+1\ ,$ & $k\geq3\ ,$ & $k=r+2\ ,$ & $r\geq1\ ,\ \ \ l=1,\ldots
,r\ .$%
\end{tabular}
\label{odd}%
\end{equation}
In the case $n=2$ ($r=1$) we have%
\begin{equation}%
\begin{tabular}
[c]{lll}%
$+\mathrm{tr}\left(  m^{k-2}Y\right)  \mathrm{tr}\left(  m^{k}Y\right)  \ ,$ &
$k=r+1=2\ ,$ & \\
$+\mathrm{tr}\left(  m^{2k-l-1}Y\right)  \mathrm{tr}\left(  m^{l-1}Y\right)
\ ,$ & $k=r+1=2\ ,$ & $k\not \geq 3\ ,$\\
$-\mathrm{tr}\left(  m^{2k-l-3}Y\right)  \mathrm{tr}\left(  m^{l-1}Y\right)
\ ,$ & $k=r+2=3\ ,$ & $l=1\ ,$%
\end{tabular}
\ \ \ \ \ \ \ \ \ \label{n=2}%
\end{equation}
which implies%
\begin{equation}
\Re_{3}^{\left(  2\right)  }=\mathrm{Str}\left[  \mathcal{R}_{3}^{\left(
2\right)  }\right]  \equiv0\ ,\ \ \ \mathcal{R}_{3}^{\left(  2\right)  }%
\equiv\mathrm{tr}\left(  Y\right)  \mathrm{tr}\left(  m^{2}Y\right)
-\mathrm{tr}\left(  m^{2}Y\right)  \mathrm{tr}\left(  Y\right)  \ , \label{R2}%
\end{equation}
whereas in the case $n=2r\geq4$ we have%
\begin{align}
\Re_{3}^{\left(  2r\right)  }  &  =\mathrm{Str}\left[  \mathcal{R}%
_{3}^{\left(  2k\right)  }\right]  \equiv0\ ,\nonumber\\
\mathcal{R}_{3}^{\left(  2k\right)  }  &  \equiv\mathrm{tr}\left(  m^{\left(
r+1\right)  -2}Y\right)  \mathrm{tr}\left(  m^{\left(  r+1\right)  }Y\right)
+\sum_{l=1}^{r-1}\mathrm{tr}\left(  m^{2\left(  r+1\right)  -l-1}Y\right)
\mathrm{tr}\left(  m^{l-1}Y\right)  -\sum_{l=1}^{r}\mathrm{tr}\left(
m^{2\left(  r+2\right)  -l-3}Y\right)  \mathrm{tr}\left(  m^{l-1}Y\right)
\nonumber\\
&  +\mathrm{tr}\left(  m^{r-2}Y\right)  \mathrm{tr}\left(  m^{r+1}Y\right)
+\sum_{l=1}^{r-1}\mathrm{tr}\left(  m^{2r+1-l}Y\right)  \mathrm{tr}\left(
m^{l-1}Y\right)  -\sum_{l=1}^{r}\mathrm{tr}\left(  m^{2r+1-l}Y\right)
\mathrm{tr}\left(  m^{l-1}Y\right) \nonumber\\
&  =\mathrm{tr}\left(  m^{r-1}Y\right)  \mathrm{tr}\left(  m^{r+1}Y\right)
-\mathrm{tr}\left(  m^{r+1}Y\right)  \mathrm{tr}\left(  m^{r-1}Y\right)
\ ,\ \ \ \mathrm{for\ \ \ }r\geq2\ . \label{>4}%
\end{align}
In the case $n=3$ ($r=1$) we have%
\begin{equation}%
\begin{tabular}
[c]{lll}%
$+\mathrm{tr}\left(  m^{k-1}Y\right)  \mathrm{tr}\left(  m^{k}Y\right)  \ ,$ &
$k=r+1=2\ ,$ & \\
$+\mathrm{tr}\left(  m^{2k-l-1}Y\right)  \mathrm{tr}\left(  m^{l}Y\right)
\ ,$ & $k=r+1=2\ ,$ & $k\not \geq 3\ ,$\\
$-\mathrm{tr}\left(  m^{2k-l-3}Y\right)  \mathrm{tr}\left(  m^{l}Y\right)
\ ,$ & $k=r+2=3\ ,$ & $l=1\ ,$%
\end{tabular}
\label{n=3}%
\end{equation}
which implies%
\begin{equation}
\Re_{3}^{\left(  3\right)  }=\mathrm{Str}\left[  \mathcal{R}_{3}^{\left(
3\right)  }\right]  \equiv0\ ,\ \ \ \mathcal{R}_{3}^{\left(  3\right)  }%
\equiv\mathrm{tr}\left(  mY\right)  \mathrm{tr}\left(  m^{2}Y\right)
-\mathrm{tr}\left(  m^{2}Y\right)  \mathrm{tr}\left(  mY\right)  \ ,
\label{R3}%
\end{equation}
whereas in the case $n=2r+1\geq5$ we have%
\begin{align}
\Re_{3}^{\left(  2r+1\right)  }  &  =\mathrm{Str}\left[  \mathcal{R}%
_{3}^{\left(  2r+1\right)  }\right]  \equiv0\ ,\nonumber\\
\mathcal{R}_{3}^{\left(  2k\right)  }  &  \equiv\mathrm{tr}\left(  m^{\left(
r+1\right)  -1}Y\right)  \mathrm{tr}\left(  m^{r+1}Y\right)  +\sum_{l=1}%
^{r-1}\mathrm{tr}\left(  m^{2\left(  r+1\right)  -l-1}Y\right)  \mathrm{tr}%
\left(  m^{l}Y\right)  -\sum_{l=1}^{r}\mathrm{tr}\left(  m^{2\left(
r+2\right)  -l-3}Y\right)  \mathrm{tr}\left(  m^{l}Y\right) \nonumber\\
&  +\mathrm{tr}\left(  m^{r}Y\right)  \mathrm{tr}\left(  m^{r+1}Y\right)
+\sum_{l=1}^{r-1}\mathrm{tr}\left(  m^{2r+1-l}Y\right)  \mathrm{tr}\left(
m^{l}Y\right)  -\sum_{l=1}^{r}\mathrm{tr}\left(  m^{2r+1-l}Y\right)
\mathrm{tr}\left(  m^{l}Y\right) \nonumber\\
&  =\mathrm{tr}\left(  m^{r}Y\right)  \mathrm{tr}\left(  m^{r+1}Y\right)
-\mathrm{tr}\left(  m^{r+1}Y\right)  \mathrm{tr}\left(  m^{r}Y\right)
\ ,\ \ \ \mathrm{for\ \ \ }r\geq2\ . \label{>5}%
\end{align}
Collecting the above results, we can state that%
\begin{equation}
\Re_{3}^{\left(  2r\right)  }=\Re_{3}^{\left(  2r+1\right)  }\equiv
0\ ,\ \ \ r=0,1,2\ldots\ , \label{R3cases}%
\end{equation}
which implies that the contribution $\Re_{3}$ is an identical zero:%
\begin{equation}
\Re_{3}=\sum_{n=0}^{\infty}\Re_{3}^{\left(  n\right)  }\equiv0\ .
\label{R3final}%
\end{equation}

\subsection{Proof of Lemma \ref{lemma psi}\label{proof_lemma psi}}

Let us suppose $s^{a}\lambda_{a}=-s^{2}\Lambda$ with anticommuting $s^{a}$ and
a certain even-valued $\Lambda$. Using the consequent nilpotency $s^{a_{1}%
}\cdots s^{a_{n}}\equiv0$, $n\geq3$, the obvious property $s^{2}\lambda
_{a}\equiv0$, and the general relation%
\[
s^{a}\left(  AB\right)  =\left(  s^{a}A\right)  B\left(  -1\right)
^{\varepsilon_{B}}+A\left(  s^{a}B\right)  \ ,
\]
we can write down identically%
\begin{align}
s^{a_{1}}s^{a_{2}}\left(  \lambda_{a_{1}}\lambda_{a_{2}}\right)   &  =\left(
s^{a_{1}}\lambda_{a_{1}}\right)  \left(  s^{a_{2}}\lambda_{a_{2}}\right)
-\left(  s^{a_{2}}\lambda_{a_{1}}\right)  \left(  s^{a_{1}}\lambda_{a_{2}%
}\right)  =\left(  s^{a}\lambda_{a}\right)  ^{2}-\mathrm{tr}\left(
m^{2}\right)  \ ,\label{s2}\\
s^{a_{1}}s^{a_{2}}s^{a_{3}}\left(  \lambda_{a_{1}}\lambda_{a_{2}}%
\lambda_{a_{3}}\right)   &  =\left(  s^{a_{1}}\lambda_{a_{1}}\right)  \left(
s^{a_{2}}\lambda_{a_{2}}\right)  \left(  s^{a_{3}}\lambda_{a_{3}}\right)
+\left(  s^{a_{3}}\lambda_{a_{1}}\right)  \left(  s^{a_{1}}\lambda_{a_{2}%
}\right)  \left(  s^{a_{2}}\lambda_{a_{3}}\right) \nonumber\\
&  +\left(  s^{a_{3}}\lambda_{a_{2}}\right)  \left(  s^{a_{2}}\lambda_{a_{1}%
}\right)  \left(  s^{a_{1}}\lambda_{a_{3}}\right)  -\left(  s^{a_{2}}%
\lambda_{a_{2}}\right)  \left(  s^{a_{3}}\lambda_{a_{1}}\right)  \left(
s^{a_{1}}\lambda_{a_{3}}\right) \nonumber\\
&  -\left(  s^{a_{1}}\lambda_{a_{1}}\right)  \left(  s^{a_{3}}\lambda_{a_{2}%
}\right)  \left(  s^{a_{2}}\lambda_{a_{3}}\right)  -\left(  s^{a_{3}}%
\lambda_{a_{3}}\right)  \left(  s^{a_{2}}\lambda_{a_{1}}\right)  \left(
s^{a_{1}}\lambda_{a_{2}}\right) \nonumber\\
&  =\left(  s^{a}\lambda_{a}\right)  ^{3}+2\mathrm{tr}\left(  m^{3}\right)
-3\left(  s^{a}\lambda_{a}\right)  \mathrm{tr}\left(  m^{2}\right)
\equiv0\ ,\label{s3}\\
&  \ldots\nonumber\\
s^{a_{1}}\cdots s^{a_{n}}\left(  \lambda_{a_{1}}\cdots\lambda_{a_{n}}\right)
&  =\left(  s^{a}\lambda_{a}\right)  ^{n}+K_{n|0}\mathrm{tr}\left(
m^{n}\right)  +\sum_{k=1}^{n-2}K_{n|k}\left(  s^{a}\lambda_{a}\right)
^{k}\mathrm{tr}\left(  m^{n-k}\right)  \equiv0\ ,\ \ \ n\geq3\ , \label{sn}%
\end{align}
where $K_{n|k}$, $k=1,\ldots,n-1$, are certain combinatorial coefficients,
whose specific form is not essential here, and $K_{n|0}$ is given by%
\begin{equation}
K_{n|0}=\left(  n-1\right)  !\left(  -1\right)  ^{n-1}\ . \label{Kn0}%
\end{equation}
Indeed, the term $\left(  s^{a}\lambda_{a}\right)  ^{n}$ is generated as
follows:%
\begin{align}
s^{a_{1}}\cdots s^{a_{n}}\left(  \lambda_{a_{1}}\cdots\lambda_{a_{n}}\right)
&  =s^{a_{2}}\cdots s^{a_{n}}s^{a_{1}}\left(  \lambda_{a_{2}}\cdots
\lambda_{a_{n}}\lambda_{a_{1}}\right)  =\left(  s^{a_{2}}\cdots s^{a_{n}%
}\lambda_{a_{2}}\cdots\lambda_{a_{n}}\right)  \left(  s^{a_{1}}\lambda_{a_{1}%
}\right)  +\cdots\nonumber\\
&  =\left(  s^{a_{1}}\lambda_{a_{1}}\right)  s^{a_{3}}\cdots s^{a_{n}}%
s^{a_{2}}\left(  \lambda_{a_{3}}\cdots\lambda_{a_{n}}\lambda_{a_{2}}\right)
+\cdots\nonumber\\
&  =\left(  s^{a_{1}}\lambda_{a_{1}}\right)  \left(  s^{a_{2}}\lambda_{a_{2}%
}\right)  s^{a_{3}}\cdots s^{a_{n}}\left(  \lambda_{a_{3}}\cdots\lambda
_{a_{n}}\right)  +\cdots\nonumber\\
&  \ldots\nonumber\\
&  =\left(  s^{a_{1}}\lambda_{a_{1}}\right)  \left(  s^{a_{2}}\lambda_{a_{2}%
}\right)  \cdots\left(  s^{a_{n}}\lambda_{a_{n}}\right)  +\cdots=\left(
s^{a}\lambda_{a}\right)  ^{n}+\cdots\ , \label{(spsi)n}%
\end{align}
whereas the term $K_{n|0}\mathrm{tr}\left(  m^{n}\right)  $ can be traced back
to%
\begin{align}
s^{a_{1}}\cdots s^{a_{n}}\left(  \lambda_{a_{1}}\cdots\lambda_{a_{n}}\right)
&  =s^{a_{2}}\cdots s^{a_{n}}s^{a_{1}}\left(  \lambda_{a_{1}}\cdots
\lambda_{a_{n}}\right)  \left(  -1\right)  ^{n-1}=\left(  s^{a_{1}}%
\lambda_{a_{n}}\right)  s^{a_{2}}\cdots s^{a_{n}}\left(  \lambda_{a_{1}}%
\cdots\lambda_{a_{n-1}}\right)  \left(  -1\right)  ^{n-1}+\cdots\nonumber\\
&  =\left(  s^{a_{1}}\lambda_{a_{n}}\right)  \left(  s^{a_{n}}\lambda
_{a_{n-1}}\right)  s^{a_{2}}\cdots s^{a_{n-1}}\left(  \lambda_{a_{1}}%
\cdots\lambda_{a_{n-2}}\right)  \left(  -1\right)  ^{n-1}+\cdots\nonumber\\
&  =\left(  s^{a_{1}}\lambda_{a_{n}}\right)  \left(  s^{a_{n}}\lambda
_{a_{n-1}}\right)  \left(  s^{a_{n-1}}\lambda_{a_{n-2}}\right)  \cdots\left(
s^{a_{2}}\lambda_{a_{1}}\right)  \left(  -1\right)  ^{n-1}+\cdots\ ,
\label{trmn-0}%
\end{align}
and therefore, collecting equal contributions with the leading terms as above,
we arrive at%
\begin{align}
&  K_{n|0}\mathrm{tr}\left(  m^{n}\right)  =\sum_{P}P\left(  A_{n\cdots
2}\right)  =\left(  n-1\right)  !A_{n\cdots2}\ ,\nonumber\\
&  A_{n\cdots k\cdots2}=\left(  s^{a_{1}}\lambda_{a_{n}}\right)  \left(
s^{a_{n}}\lambda_{a_{n-1}}\right)  \left(  s^{a_{n-1}}\lambda_{a_{n-2}%
}\right)  \cdots\left(  s^{a_{k+1}}\lambda_{a_{k}}\right)  \cdots\left(
s^{a_{3}}\lambda_{a_{2}}\right)  \left(  s^{a_{2}}\lambda_{a_{1}}\right)  \ ,
\label{trmn}%
\end{align}
where $P\left(  A_{n\cdots2}\right)  $ is an arbitrary permutation of the
indices in $A_{n\cdots2}$ corresponding to $a_{n},\ldots,a_{2}$ in (\ref{trmn}).

From (\ref{s2}), (\ref{s3}), it follows that the contributions $\mathrm{tr}%
(m^{2})$, $\mathrm{tr}(m^{3})$ are BRST-antiBRST-exact:%
\begin{align}
\mathrm{tr}\left(  m^{2}\right)   &  =\left(  s^{2}\Lambda\right)  ^{2}%
-\frac{1}{2}s^{2}\left(  \lambda^{2}\right)  \equiv-s^{2}\Lambda
_{2}\ ,\label{m2}\\
\mathrm{tr}\left(  m^{3}\right)   &  =\frac{1}{2}\left(  s^{2}\Lambda\right)
^{3}-\frac{3}{2}\left(  s^{2}\Lambda\right)  \mathrm{tr}\left(  m^{2}\right)
=-\left(  s^{2}\Lambda\right)  ^{3}+\frac{3}{4}\left(  s^{2}\Lambda\right)
s^{2}\left(  \lambda^{2}\right)  \equiv-s^{2}\Lambda_{3}\ . \label{m3}%
\end{align}
Proceeding by induction in the general case $n\geq2$ and assuming
$\mathrm{tr}(m^{k})$, $k=1,\ldots,n$, to be BRST-antiBRST-exact,
$\mathrm{tr}(m^{k})=-s^{2}\Lambda_{k}$, we can now prove, by using the
relation (\ref{sn}) and the identity $s^{a_{1}}\cdots s^{a_{n+1}}\left(
\lambda_{a_{1}}\cdots\lambda_{a_{n+1}}\right)  \equiv0$, the fact that%
\begin{equation}
\mathrm{tr}\left(  m^{n+1}\right)  =\left(  -1\right)  ^{n}K_{n+1|0}%
^{-1}\left(  s^{2}\Lambda\right)  ^{n+1}+\sum_{k=1}^{n-1}\left(  -1\right)
^{k}K_{n+1|0}^{-1}K_{n+1|k}\left(  s^{2}\Lambda\right)  ^{k}\left(
s^{2}\Lambda_{n+1-k}\right)  \equiv-s^{2}\Lambda_{n+1}\ , \label{m(n+1)}%
\end{equation}
whence the contribution $\mathrm{tr}(m^{n+1})$ is also BRST-antiBRST-exact,
which proves Lemma \ref{lemma psi}.

\subsection{Proof of Lemma \ref{lemma psi 2} \label{proof_lemma psi2}}

Let us consider an odd-valued doublet $\psi_{a}$ subject to the condition
$s^{a}\psi_{a}=0$. Making in (\ref{m2})--(\ref{m(n+1)}) the substitution
$\lambda_{a}=\psi_{a}$,$\ \Lambda=0$, $m\equiv m_{\psi}$, we obtain%
\begin{align}
&  \mathrm{tr}\left(  m_{\psi}^{2}\right)  =-\frac{1}{2}s^{2}\left(  \psi
^{2}\right)  \ ,\nonumber\\
&  \mathrm{tr}\left(  m_{\psi}^{3}\right)  =0\ ,\nonumber\\
&  \ldots\nonumber\\
&  \mathrm{tr}\left(  m_{\psi}^{n}\right)  =0\ ,\ \ \ n\geq3\ , \label{m_psi}%
\end{align}
which proves the relations (\ref{implies}) of Lemma \ref{lemma psi 2}. This
allows one to make an explicit calculation of the corresponding quantity $\Im
$, parameterized by the functional parameters $\left(  \Lambda,\psi
_{a}\right)  $. Indeed, due to the relations%
\begin{equation}
\mathrm{tr}\left(  m_{\Lambda}+m_{\psi}\right)  ^{n}=\mathrm{tr}\sum_{k=0}%
^{n}C_{n}^{k}f^{n-k}m_{\psi}^{k}\ ,\ \ \ \left(  m_{\Lambda}\right)  _{b}%
^{a}=s^{a}s_{b}\Lambda=\delta_{b}^{a}f\ ,\ \ \ \left(  m_{\psi}\right)
_{b}^{a}=s^{a}\psi_{b}\ , \label{trmm}%
\end{equation}
the corresponding quantity $\Im=\Im\left(  \Lambda,\psi\right)  $ reads%
\begin{equation}
\Im\left(  \Lambda,\psi\right)  =\ln\left(  1-\frac{1}{2}s^{2}\Lambda\right)
^{-2}+M\left(  \Lambda,\psi\right)  \ ,\ \ \ M\left(  \Lambda,\psi\right)
=\sum_{n=1}^{\infty}\frac{\left(  -1\right)  ^{n}}{n}\sum_{k=1}^{n}C_{n}%
^{k}f^{n-k}\mathrm{tr}\left[  \left(  m_{\psi}\right)  ^{k}\right]
_{\ f=-\frac{1}{2}s^{2}\Lambda}\ . \label{J-lambdapsi}%
\end{equation}
The only nontrivial quantity $\mathrm{tr}(m_{\psi}^{2})\not \equiv 0$ amongst
$\mathrm{tr}(m_{\psi}^{n})$ leads to%
\begin{align}
M\left(  \Lambda,\psi\right)   &  =-\mathrm{tr}\left[  \left(  m_{\psi
}\right)  \right]  +\mathrm{tr}\left(  m_{\psi}^{2}\right)  \sum_{n=2}%
^{\infty}\frac{\left(  -1\right)  ^{n}}{n}\left.  C_{n}^{2}f^{n-2}\right\vert
_{f=-\frac{1}{2}s^{2}\Lambda}\nonumber\\
&  =\frac{1}{2}\mathrm{tr}\left(  m_{\psi}^{2}\right)  +\mathrm{tr}\left(
m_{\psi}^{2}\right)  \sum_{n=3}^{\infty}\frac{\left(  -1\right)  ^{n}}%
{n}\left.  C_{n}^{2}f^{n-2}\right\vert _{f=-\frac{1}{2}s^{2}\Lambda}=\frac
{1}{2}\mathrm{tr}\left(  m_{\psi}^{2}\right)  +\frac{1}{2}\mathrm{tr}\left(
m_{\psi}^{2}\right)  \sum_{k=1}^{\infty}\left(  -1\right)  ^{k}\left(
k+1\right)  \left.  f^{k}\right\vert _{f=-\frac{1}{2}s^{2}\Lambda}\ ,
\label{M(Lambda,psi)}%
\end{align}
where%
\begin{equation}
\sum_{k=1}^{\infty}\left(  -1\right)  ^{k}\left(  k+1\right)  x^{k}%
=-\sum_{k=2}^{\infty}\left(  -1\right)  ^{k}kx^{k-1}=-\frac{\partial}{\partial
x}\sum_{k=2}^{\infty}\left(  -1\right)  ^{k}x^{k}=-\frac{\partial}{\partial
x}\left[  \left(  1+x\right)  ^{-1}-1+x\right]  =\left(  1+x\right)
^{-2}-1\ . \label{series}%
\end{equation}
Therefore,%
\begin{align}
M\left(  \Lambda,\psi\right)   &  =\frac{1}{2}\mathrm{tr}\left(  m_{\psi}%
^{2}\right)  +\mathrm{tr}\left(  m_{\psi}^{2}\right)  \sum_{n=3}^{\infty}%
\frac{\left(  -1\right)  ^{n}}{n}\left.  C_{n}^{2}f^{n-2}\right\vert
_{f=-\frac{1}{2}s^{2}\Lambda}=\frac{1}{2}\mathrm{tr}\left(  m_{\psi}%
^{2}\right)  +\frac{1}{2}\mathrm{tr}\left(  m_{\psi}^{2}\right)  \sum
_{k=1}^{\infty}\left(  -1\right)  ^{k}\left(  k+1\right)  \left.
f^{k}\right\vert _{f=-\frac{1}{2}s^{2}\Lambda}\nonumber\\
&  =\frac{1}{2}\mathrm{tr}\left(  m_{\psi}^{2}\right)  \left[  1+\sum
_{k=1}^{\infty}\left(  -1\right)  ^{k}\left(  k+1\right)  f^{k}\right]
_{f=-\frac{1}{2}s^{2}\Lambda}=\frac{1}{2}\mathrm{tr}\left(  m_{\psi}%
^{2}\right)  \left[  \left(  1-\frac{1}{2}s^{2}\Lambda\right)  ^{-2}\right]
\ . \label{M(Lambda,psi)2}%
\end{align}
This result describes the contribution to $\Im\left(  \Lambda,\psi\right)  $
caused by the arbitrariness in the solutions of $s^{a}\left(  \lambda
_{a}-s_{a}\Lambda\right)  =0$, with a given $\Lambda$. This contribution is
BRST-antiBRST-exact due to the fact that $\mathrm{tr}(m_{\psi}^{2})=-\left(
1/2\right)  s^{2}\left(  \psi^{2}\right)  $:%
\begin{equation}
M\left(  \Lambda,\psi\right)  =s^{2}N\left(  \Lambda,\psi\right)  \ .
\label{exactness}%
\end{equation}
The relations (\ref{J-lambdapsi}), (\ref{M(Lambda,psi)2}) prove (\ref{reads}),
which finishes the proof of Lemma \ref{lemma psi 2}.

\subsection{Proof of Lemma \ref{lemma psi 2 solution}%
\label{proof_psi 2 solution}}

Let us examine the equation (\ref{comp_mod}),%
\[
\ln\left(  1-\frac{1}{2}s^{2}\Lambda\right)  ^{-2}-\frac{1}{4}s^{2}\left(
\psi^{2}\right)  \left(  1-\frac{1}{2}s^{2}\Lambda\right)  ^{-2}=\frac
{1}{2i\hbar}s^{2}\Delta F\ ,
\]
for an unknown functional $\Lambda=\Lambda\left(  \Delta F,\psi\right)  $ and
introduce the following notation:%
\begin{align}
&  \Lambda_{0}:\ln\left(  1-\frac{1}{2}s^{2}\Lambda_{0}\right)  ^{-2}=\frac
{1}{2i\hbar}s^{2}\Delta F\ ,\nonumber\\
&  \frac{1}{4}s^{2}\left(  \psi^{2}\right)  \equiv\gamma\ ,\ \ \ X\equiv
\left(  1-\frac{1}{2}s^{2}\Lambda\right)  ^{-2}=X_{0}+\Delta X\ ,\nonumber\\
&  X\equiv\left(  1-\frac{1}{2}s^{2}\Lambda\right)  ^{-2}\Longrightarrow\ln
X_{0}=\frac{1}{2i\hbar}s^{2}\Delta F\ , \label{foll_not}%
\end{align}
whence follows a chain of relations:%
\begin{align}
&  \ln X-\gamma X=\frac{1}{2i\hbar}s^{2}\Delta F\ ,\nonumber\\
&  \ln\left(  X_{0}+\Delta X\right)  -\gamma\left(  X_{0}+\Delta X\right)
=\frac{1}{2i\hbar}s^{2}\Delta F\ ,\nonumber\\
&  \ln X_{0}+\ln\left(  1+\frac{\Delta X}{X_{0}}\right)  -\gamma X_{0}\left(
1+\frac{\Delta X}{X_{0}}\right)  =\frac{1}{2i\hbar}s^{2}\Delta F\ ,\nonumber\\
&  \ln\left(  1+\frac{\Delta X}{X_{0}}\right)  =\gamma X_{0}\left(
1+\frac{\Delta X}{X_{0}}\right)  \ . \label{ch_rel}%
\end{align}
Let us introduce a new function,%
\begin{equation}
\theta\left(  x\right)  =\frac{\ln\left(  1+x\right)  }{\left(  1+x\right)
}\ ,\ \ \ \theta\left(  0\right)  =0\ , \label{theta}%
\end{equation}
and the inverse function:%
\begin{equation}
\vartheta\left(  y\right)  :\vartheta\left(  \theta\left(  x\right)  \right)
=x\ ,\ \ \ \vartheta\left(  0\right)  =0\ . \label{vartheta}%
\end{equation}
Hence,%
\begin{equation}
\theta\left(  \Delta X/X_{0}\right)  =\gamma X_{0}\Longrightarrow\Delta
X/X_{0}=\vartheta\left(  \gamma X_{0}\right)  \Longrightarrow\Delta
X=X_{0}\cdot\vartheta\left(  \gamma X_{0}\right)  \ , \label{Hence}%
\end{equation}
which implies%
\begin{equation}
X=X_{0}+\Delta X=X_{0}+X_{0}\cdot\vartheta\left(  \gamma X_{0}\right)
=X_{0}\left[  1+\vartheta\left(  \gamma X_{0}\right)  \right]
\label{whichimp}%
\end{equation}
and ensures%
\begin{equation}
X\left(  X_{0},\gamma\right)  =X_{0}\left[  1+\vartheta\left(  \gamma
X_{0}\right)  \right]  \Longrightarrow X\left(  X_{0},0\right)  =X_{0}\ .
\label{andensur}%
\end{equation}
This implies%
\begin{equation}
\ln X=\ln X_{0}+\ln\left[  1+\vartheta\left(  \gamma X_{0}\right)  \right]
=\frac{1}{2i\hbar}s^{2}\Delta F+\ln\left[  1+\vartheta\left(  \gamma
X_{0}\right)  \right]  \label{thisimply}%
\end{equation}
Recalling (\ref{foll_not}),%
\[
X=\left(  1-\frac{1}{2}s^{2}\Lambda\right)  ^{-2}\ ,
\]
we arrive at a chain of relations:%
\begin{align}
&  \ln\left(  1-\frac{1}{2}s^{2}\Lambda\right)  =-\frac{1}{4i\hbar}s^{2}\Delta
F-\frac{1}{2}\ln\left[  1+\vartheta\left(  \gamma X_{0}\right)  \right]
\ ,\nonumber\\
&  1-\frac{1}{2}s^{2}\Lambda=\exp\left(  -\frac{1}{4i\hbar}s^{2}\Delta
F\right)  \exp\left\{  \ln\left[  1+\vartheta\left(  \gamma X_{0}\right)
\right]  ^{-\frac{1}{2}}\right\}  \ ,\nonumber\\
&  1-\frac{1}{2}s^{2}\Lambda=\exp\left(  -\frac{1}{4i\hbar}s^{2}\Delta
F\right)  \left[  1+\vartheta\left(  \gamma X_{0}\right)  \right]  ^{-\frac
{1}{2}}\ , \label{ch_rel2}%
\end{align}
whence%
\begin{equation}
s^{2}\Lambda\left(  \Delta F,\gamma\right)  =2\left\{  1-\left[
1+\vartheta\left(  \gamma X_{0}\right)  \right]  ^{-\frac{1}{2}}\exp\left[
\left(  i/4\hbar\right)  s^{2}\Delta F\right]  \right\}  \ . \label{whence2}%
\end{equation}
In the case $s^{2}\Delta F\not =0$, a solution $\Lambda$ to this equation can
be found as%
\begin{equation}
\Lambda\left(  \Delta F,\psi\right)  =\frac{2\Delta F}{s^{2}\Delta F}\left\{
1-\left[  1+\vartheta\left(  \gamma X_{0}\right)  \right]  ^{-\frac{1}{2}}%
\exp\left[  \left(  i/4\hbar\right)  s^{2}\Delta F\right]  \right\}  \ ,
\label{foundas}%
\end{equation}
where it must be recalled that (\ref{foll_not})%
\[
X_{0}=\exp\left(  \frac{1}{2i\hbar}s^{2}\Delta F\right)  \ ,\ \ \ \gamma
=\frac{1}{4}s^{2}\left(  \psi^{2}\right)  \ .
\]
Let us now examine the case $s^{2}\Delta F=0$:%
\begin{align}
s^{2}\Lambda &  =2\left.  \left\{  1-\left[  1+\vartheta\left(  \gamma
X_{0}\right)  \right]  ^{-\frac{1}{2}}\exp\left[  \left(  i/4\hbar\right)
s^{2}\Delta F\right]  \right\}  \right\vert _{s^{2}\Delta F=0}\nonumber\\
&  =2\left\{  1-\left[  1+\vartheta\left(  \gamma\right)  \right]  ^{-\frac
{1}{2}}\right\}  \ , \label{case_zero}%
\end{align}
whence there are two possibilities:%
\begin{align}
\gamma &  =0:\ \ \ s^{2}\Lambda=0\Longrightarrow\Lambda=s^{a}\tilde{\lambda
}_{a}+s^{2}\tilde{\Lambda}\ ,\label{two-poss0}\\
\gamma &  \not =0:\ \ \ s^{2}\Lambda=2\left\{  1-\left[  1+\vartheta\left(
\gamma\right)  \right]  ^{-\frac{1}{2}}\right\}  \ , \label{two-poss}%
\end{align}
which, in the latter case, implies%
\begin{equation}
\Lambda\left(  \psi\right)  =\frac{2\psi^{2}}{s^{2}\left(  \psi^{2}\right)
}\left\{  1-\left[  1+\vartheta\left(  \gamma\right)  \right]  ^{-\frac{1}{2}%
}\right\}  _{\gamma=\frac{1}{4}s^{2}(\psi^{2})}\ . \label{latter-case}%
\end{equation}
Summarizing the relations (\ref{foundas}), (\ref{two-poss0}), (\ref{two-poss}%
), (\ref{latter-case}) and the respective cases $\Lambda=0$, $s^{2}\left(
\psi^{2}\right)  =0$ of (\ref{comp_mod}), (\ref{foundas}), we have%
\begin{equation}%
\begin{tabular}
[c]{ll}%
$\mathrm{a)}\ s^{2}\Delta F\not =0:$ & $\Lambda\left(  \Delta F,\psi\right)
=\frac{2\Delta F}{s^{2}\Delta F}\left.  \left\{  1-\left[  1+\vartheta\left(
\gamma X_{0}\right)  \right]  ^{-\frac{1}{2}}X_{0}^{-\frac{1}{2}}\right\}
\right\vert _{X_{0}=\exp\left(  \frac{1}{2i\hbar}s^{2}\Delta F\right)
,\ \gamma=\frac{1}{4}s^{2}(\psi^{2})}\ $\\
$\mathrm{b)}\ s^{2}\Delta F=0\ ,\ s^{2}\left(  \psi^{2}\right)  =0:$ &
$\Lambda=s^{a}\tilde{\lambda}_{a}+s^{2}\tilde{\Lambda}\ ,$\\
$\mathrm{c)}\ s^{2}\Delta F=0,\ s^{2}\left(  \psi^{2}\right)  \not =0:$ &
$\Lambda\left(  \psi\right)  =\frac{2\psi^{2}}{s^{2}\left(  \psi^{2}\right)
}\left.  \left\{  1-\left[  1+\vartheta\left(  \gamma\right)  \right]
^{-\frac{1}{2}}\right\}  \right\vert _{\gamma=\frac{1}{4}s^{2}(\psi^{2})}%
\ ,$\\
$\mathrm{d)}\ \Lambda\left(  \Delta F,\psi\right)  =0:$ & $-\frac{1}{4}%
s^{2}\left(  \psi^{2}\right)  =\frac{1}{2i\hbar}s^{2}\Delta F\ ,$\\
$\mathrm{e)}\ s^{2}\left(  \psi^{2}\right)  =0,\ s^{2}\Delta F\not =0:$ &
$\Lambda\left(  \Delta F,0\right)  =\frac{2\Delta F}{s^{2}\Delta F}\left[
1-\exp\left(  \frac{i}{4\hbar}s^{2}\Delta F\right)  \right]  \ ,$%
\end{tabular}
\label{summarize}%
\end{equation}
where the relations a), b), c) in (\ref{summarize}) thereby prove Lemma
\ref{lemma psi 2 solution}.

\subsection{Proof of Lemma \ref{modified lemma 1}
\label{proof_modified lemma 1}}

Using the property $\mathrm{Str}\left(  AB\right)  =\mathrm{Str}\left(
BA\right)  $ for even matrices, we examine the quantities $\mathrm{Str}\left(
\mathcal{U}^{n-1}\mathcal{W}\right)  =\mathrm{Str}\left(  \mathcal{WU}%
^{n-1}\right)  $, $n>1$, where%
\[
\mathcal{U}_{\mathsf{q}}^{\mathsf{p}}=\mathcal{X}^{\mathsf{p}a}\lambda
_{a,q}=\left(  \mathsf{s}^{a}\mathsf{\Gamma}^{\mathsf{p}}\right)
\lambda_{a,q}\ ,\ \ \ \mathcal{W}_{\mathsf{q}}^{\mathsf{p}}=-\frac{1}%
{2}\lambda^{2}\mathcal{Y}_{,\mathsf{q}}^{\mathsf{p}}=\frac{1}{4}\lambda
^{2}\left(  \mathsf{s}^{2}\mathsf{\Gamma}^{\mathsf{p}}\right)  _{,\mathsf{q}%
}\ ,
\]
and write down a chain of relations, taking account of $\lambda_{b,p}%
\mathcal{X}^{\mathsf{p}a}=\mathsf{s}^{a}\lambda_{b}=\mathsf{m}_{b}^{a}$:%
\begin{align}
n  &  =2\text{: }\left(  \mathcal{WU}\right)  _{q}^{p}=\mathcal{W}_{r}%
^{p}\mathcal{U}_{q}^{r}=\frac{1}{4}\lambda^{2}\left(  \mathsf{s}%
^{2}\mathsf{\Gamma}^{\mathsf{p}}\right)  _{,\mathsf{r}}\mathcal{X}%
^{\mathsf{r}a}\lambda_{a,q}=\frac{1}{4}\lambda^{2}\left(  \mathsf{s}%
^{a}\mathsf{s}^{2}\mathsf{\Gamma}^{\mathsf{p}}\right)  \lambda_{a,q}%
\nonumber\\
n  &  =3\text{:\ }\left(  \mathcal{WU}^{2}\right)  _{q}^{p}=\left(
\mathcal{WU}\right)  _{r}^{p}\mathcal{U}_{q}^{r}=\frac{1}{4}\lambda^{2}\left(
\mathsf{s}^{a}\mathsf{s}^{2}\mathsf{\Gamma}^{\mathsf{p}}\right)  \left(
\lambda_{a,r}\mathcal{X}^{\mathsf{r}b}\right)  \lambda_{b,q}=\frac{1}%
{4}\lambda^{2}\left(  \mathsf{s}^{a}\mathsf{s}^{2}\mathsf{\Gamma}^{\mathsf{p}%
}\right)  \mathsf{m}_{a}^{b}\lambda_{b,q}\ ,\nonumber\\
n  &  =4\text{:\ }\left(  \mathcal{WU}^{3}\right)  _{q}^{p}=\left(
\mathcal{WU}^{2}\right)  _{r}^{p}\mathcal{U}_{q}^{r}=\frac{1}{4}\lambda
^{2}\left(  \mathsf{s}^{a}\mathsf{s}^{2}\mathsf{\Gamma}^{\mathsf{p}}\right)
\mathsf{m}_{a}^{b}\left(  \lambda_{b,r}\mathcal{X}^{\mathsf{r}c}\right)
\lambda_{c,q}=\frac{1}{4}\lambda^{2}\left(  \mathsf{s}^{a}\mathsf{s}%
^{2}\mathsf{\Gamma}^{\mathsf{p}}\right)  \left(  \mathsf{m}^{2}\right)
_{a}^{b}\lambda_{b,q}\nonumber\\
&  \ldots\nonumber\\
n  &  \geq2\text{:\ }\left(  \mathcal{WU}^{n-1}\right)  _{q}^{p}=\left(
\mathcal{WU}^{2}\right)  _{r}^{p}\mathcal{U}_{q}^{r}=\frac{1}{4}\lambda
^{2}\left(  \mathsf{s}^{a}\mathsf{s}^{2}\mathsf{\Gamma}^{\mathsf{p}}\right)
\left(  \mathsf{m}^{n-2}\right)  _{a}^{b}\lambda_{b,q}\ , \label{chain_new}%
\end{align}
whence%
\[
\mathrm{Str}\left(  \mathcal{U}^{n-1}\mathcal{W}\right)  =\left(
\mathcal{WU}^{n-1}\right)  _{p}^{p}\left(  -1\right)  ^{\varepsilon_{p}}%
=\frac{1}{4}\lambda^{2}\left(  \mathsf{s}^{a}\mathsf{s}^{2}\mathsf{\Gamma
}^{\mathsf{p}}\right)  \left(  \mathsf{m}^{n-2}\right)  _{a}^{b}\lambda
_{b,p}\left(  -1\right)  ^{\varepsilon_{p}}=-\frac{1}{4}\lambda_{a,p}\left(
\mathsf{m}^{n-2}\right)  _{b}^{a}\left(  \mathsf{s}^{b}\mathsf{s}%
^{2}\mathsf{\Gamma}^{\mathsf{p}}\right)  \lambda^{2}\ ,\ \ \ n>1\ ,
\]
which thereby proves Lemma \ref{modified lemma 1}.

\subsection{Proof of Lemma \ref{modified lemma}\label{lemma-zero}}

Let us establish the relation (\ref{zero}) between the matrices $\mathcal{V}%
_{1}$ and $\mathcal{W}$ in (\ref{MABext}). To do so, we use the generating
equations (\ref{3.3}) and represent the condition of invariance of the
integrand $\mathcal{I}_{\Gamma}^{\left(  _{F}\right)  }$ in (\ref{z(0)}) under
the BRST-antiBRST transformations $\delta\Gamma^{p}=\left(  \mathsf{s}%
^{a}\Gamma^{p}\right)  \mu_{a}=\mathcal{X}^{pa}\mu_{a}$ in the form, being a
reformulation of (\ref{identsg}),
\begin{equation}
\mathcal{S}_{F,p}\mathcal{X}^{\mathsf{p}a}=i\hbar\mathcal{X}_{,\mathsf{p}%
}^{\mathsf{p}a}\ ,\ \ \ \mathrm{where}\ \ \ \mathcal{X}_{,\mathsf{p}%
}^{\mathsf{p}a}=-\Delta^{a}S\ . \label{eqinv}%
\end{equation}
Let us write down identically:%
\begin{align}
\mathrm{Str}\left(  \mathcal{V}_{1}\right)  +\mathrm{Str}\left(
\mathcal{W}\right)  -\frac{1}{2}\mathrm{Str}\left(  \mathcal{V}_{1}%
^{2}\right)   &  =\left[  \left(  \mathcal{V}_{1}\right)  _{\mathsf{p}%
}^{\mathsf{p}}+\mathcal{W}_{\mathsf{p}}^{\mathsf{p}}-\frac{1}{2}\left(
\mathcal{V}_{1}\right)  _{\mathsf{q}}^{\mathsf{p}}\left(  \mathcal{V}%
_{1}\right)  _{\mathsf{p}}^{\mathsf{q}}\right]  \left(  -1\right)
^{\varepsilon_{\mathsf{p}}}\nonumber\\
&  =\mathcal{X}_{,\mathsf{p}}^{\mathsf{p}a}\lambda_{a}-\frac{1}{2}\left(
-1\right)  ^{\varepsilon_{\mathsf{p}}}\left(  \mathcal{Y}_{,\mathsf{p}%
}^{\mathsf{p}}-\frac{1}{2}\mathcal{X}_{,\mathsf{q}}^{\mathsf{p}a}%
\mathcal{X}_{,\mathsf{p}}^{\mathsf{q}b}\varepsilon_{ba}\right)  \lambda^{2}\ .
\label{identic}%
\end{align}
Considering%
\begin{align}
\mathcal{Y}_{,\mathsf{p}}^{\mathsf{p}}-\frac{1}{2}\mathcal{X}_{,\mathsf{q}%
}^{\mathsf{p}a}\mathcal{X}_{,\mathsf{p}}^{\mathsf{q}b}\varepsilon_{ba}  &
=\frac{1}{2}\varepsilon_{ba}\left(  \mathcal{X}_{,\mathsf{qp}}^{\mathsf{p}%
a}\mathcal{X}^{\mathsf{q}b}\left(  -1\right)  ^{\varepsilon_{\mathsf{p}%
}\left(  \varepsilon_{\mathsf{q}}+1\right)  }+\mathcal{X}_{,\mathsf{q}%
}^{\mathsf{p}a}\mathcal{X}_{,\mathsf{p}}^{\mathsf{q}b}\right)  -\frac{1}%
{2}\varepsilon_{ba}\mathcal{X}_{,\mathsf{q}}^{\mathsf{p}a}\mathcal{X}%
_{,\mathsf{p}}^{\mathsf{q}b}\nonumber\\
&  =\frac{1}{2}\varepsilon_{ba}\left(  \mathcal{X}_{,\mathsf{qp}}%
^{\mathsf{p}a}\mathcal{X}^{\mathsf{q}b}\left(  -1\right)  ^{\varepsilon
_{\mathsf{p}}\left(  \varepsilon_{\mathsf{q}}+1\right)  }+\mathcal{X}%
_{,\mathsf{q}}^{\mathsf{p}a}\mathcal{X}_{,\mathsf{p}}^{\mathsf{q}%
b}-\mathcal{X}_{,\mathsf{q}}^{\mathsf{p}a}\mathcal{X}_{,\mathsf{p}%
}^{\mathsf{q}b}\right)  =\frac{1}{2}\varepsilon_{ba}\mathcal{X}_{,\mathsf{pq}%
}^{\mathsf{p}a}\mathcal{X}^{\mathsf{q}b}\left(  -1\right)  ^{\varepsilon
_{\mathsf{p}}}\ , \label{consstrmpq}%
\end{align}
we arrive at%
\begin{equation}
\mathrm{Str}\left(  \mathcal{V}_{1}\right)  +\mathrm{Str}\left(
\mathcal{W}\right)  -\frac{1}{2}\mathrm{Str}\left(  \mathcal{V}_{1}%
^{2}\right)  =\mathcal{X}_{,\mathsf{p}}^{\mathsf{p}a}\lambda_{a}+\frac{1}%
{4}\varepsilon_{ab}\mathcal{X}_{,\mathsf{pq}}^{\mathsf{p}a}\mathcal{X}%
^{\mathsf{q}b}\lambda^{2}\ , \label{lnjacob}%
\end{equation}
where (\ref{eqinv}) implies%
\begin{equation}
\mathcal{X}_{,\mathsf{p}}^{\mathsf{p}a}=-\Delta^{a}S\ ,\ \ \ \mathcal{X}%
_{,\mathsf{pq}}^{\mathsf{p}a}\mathcal{X}^{\mathsf{q}b}=-\left(  \Delta
^{a}S\right)  _{,\mathsf{p}}\mathcal{X}^{\mathsf{p}b}=-\mathsf{s}^{b}\left(
\Delta^{a}S\right)  \ ,\ \ \mathrm{where}\ \ G_{,\mathsf{p}}\mathcal{X}%
^{\mathsf{p}a}=G_{,\mathsf{p}}\left(  \mathsf{s}^{a}\mathsf{\Gamma}%
^{p}\right)  =\mathsf{s}^{a}G\ . \label{diffcons}%
\end{equation}
Hence,%
\[
\mathrm{Str}\left(  \mathcal{V}_{1}\right)  +\mathrm{Str}\left(
\mathcal{W}\right)  -\frac{1}{2}\mathrm{Str}\left(  \mathcal{V}_{1}%
^{2}\right)  =-\left(  \Delta^{a}S\right)  \lambda_{a}-\frac{1}{4}\left(
\mathsf{s}_{a}\Delta^{a}S\right)  \lambda^{2}\ ,
\]
which thereby proves Lemma \ref{modified lemma}.


\begin{thebibliography}{999}                                                                                              %


\bibitem {MRnew}Moshin, P.Yu., Reshetnyak, A.A.: Field-dependent BRST-antiBRST
transformations in Yang--Mills and Gribov--Zwanziger theories. Nucl. Phys.
B888, 92 (2014), arXiv:1405.0790 [hep-th]

\bibitem {MRnew1}Moshin, P.Yu., Reshetnyak, A.A.: Field-dependent
BRST-antiBRST transformations in generalized Hamiltonian formalism. Int. J.
Mod. Phys. A29, 1450159 (2014), arXiv:1405.7549 [hep-th]

\bibitem {MRnew2}Moshin, P.Yu., Reshetnyak, A.A.: Finite BRST-antiBRST
transformations in Lagrangian formalism. Phys. Lett. B739, 110 (2014),
arXiv:1406.0179 [hep-th]

\bibitem {MRnew3}Moshin, P.Yu., Reshetnyak, A.A.: Field-dependent
BRST-antiBRST Lagrangian transformations. Int. J. Mod. Phys. A30, 1550021
(2015), arXiv:1406.5086 [hep-th]

\bibitem {aBRST1}Curci, G., Ferrari, R.: Slavnov transformation and
supersymmetry. Phys. Lett. {B63, }91 (1976)\newline Ojima, I.: Another BRS
transformation. Prog. Theor. Phys. Suppl. {64, }625 (1980)

\bibitem {aBRST2}Alvarez-Gaume, L., Baulieu, L.: The two quantum symmetries
associated with a classical symmetry. Nucl. Phys. B212, 255 (1983)

\bibitem {aBRST3}Hwang, S.: Properties of the anti-BRS symmetry in a general
framework. Nucl. Phys. B231, 386 (1984)

\bibitem {aBRST4}Spiridonov, V.P.: Sp(2)-covariant ghost fields in gauge
theories. Nucl. Phys. B308, 527 (1988)

\bibitem {BLT1h}Batalin, I.A., Lavrov, P.M., Tyutin, I.V.: Extended BRST
quantization of gauge theories in generalized canonical formalism. J. Math.
Phys. 31, 6 (1990)

\bibitem {BLT2h}Batalin, I.A., Lavrov, P.M., Tyutin, I.V.: An Sp(2)-covariant
version of generalized canonical quantization of dynamical systems with
linearly dependent constraints. J. Math. Phys. 31, 2708 (1990)

\bibitem {GH1}Gregoire, P., Henneaux, M.: Hamiltonian BRST--anti-BRST theory.
Comm. Math. Phys. 157, 279 (1993)

\bibitem {BLT1}Batalin, I.A., Lavrov, P.M., Tyutin, I.V.: Covariant
quantization of gauge theories in the framework of extended BRST symmetry. J.
Math. Phys. 31, 1487 (1990)

\bibitem {BLT2}Batalin, I.A., Lavrov, P.M., Tyutin, I.V.: An Sp(2)-covariant
quantization of gauge theories with linearly dependent generators, J. Math.
Phys. 32, 532 (1991)

\bibitem {Hull}Hull, C.M.: The BRST-anti-BRST invariant quantization of
general gauge theories. Mod. Phys. Lett. A5, 1871 (1990)

\bibitem {BLThf}Batalin, I.A., Lavrov, P.M., Tyutin, I.V.: A systematic study
of finite BRST--BFV Transformations in Sp(2)-extended generalized Hamiltonian
formalism. Int. J. Mod. Phys. A29, 1450128 (2014), arXiv:1405.7218 [hep-th]

\bibitem {BLTlf}Batalin, I.A., Lavrov, P.M., Tyutin, I.V.: A systematic study
of finite field dependent BRST--BV transformations in Sp(2) extended
field-antifield formalism. Int. J. Mod. Phys. A29, 1450167 (2014),
arXiv:1406.4695 [hep-th]

\bibitem {BRST1}Becchi, C., Rouet, A., Stora, R.: The Abelian Higgs--Kibble,
unitarity of the S-operator. Phys. Lett. B52, 344 (1974); Renormalization of
gauge theories, Ann. Phys. (N.Y.) 98, 287 (1976)

\bibitem {BRST2}Tyutin, I.V.: Gauge invariance in field theory and ststistical
mechanics. Lebedev Inst. preprint No. 39 (1975), arXiv:0812.0580 [hep-th]

\bibitem {BRST3}Fradkin, E.S., Vilkovisky, G.A.: Quantization of relativistic
systems with constraints. Phys. Lett. {B55, }224 (1975)\newline Batalin, I.A.,
Vilkovisky, G.A.: Relativistic S-matrix of dynamical systems with boson and
fermion constraints. Phys. Lett. B69, 309 (1977)

\bibitem {deWittH}de Witt, B., van Holten, J.W.: Covariant quantization of
gauge theories with open gauge algebra. Phys. Lett. B79, 389 (1979)

\bibitem {BV}Batalin, I.A., Vilkovisky, G.A.: Gauge algebra and quantization,
Phys. Lett. B102, 27 (1981); Quantization of gauge theories with linearly
dependent generators. Phys. Rev. D28, 2567 (1983)

\bibitem {BFV}Batalin, I.A., Vilkovisky, G.A.: A generalized canonical
formalism and quantization of reducible gauge theories. Phys. Lett. B122, 157 (1983)

\bibitem {Henneaux1}Henneaux, M.: Hamiltonian form of the path integral for
theories with a gauge freedom. Phys. Pep. 126, 1 (1985)

\bibitem {JM}Joglekar, S.D., Mandal, B.P.: Finite field dependent BRS
transformations. Phys. Rev. D51, 1919 (1995)

\bibitem {RM}Rai, S.K., Mandal, B.P.: Finite nilpotent BRST transformations in
Hamiltonian formalism. Int. J. Theor. Phys. 52, 3512 (2013), arXiv:1204.5365 [hep-th]

\bibitem {Upadhyay1}Upadhyay, S., Rai, S.K., Mandal, B.P.: Off-shell nilpotent
finite BRST/anti-BRST transformations. J. Math. Phys. 52, 022301 (2011),
arXiv:1002.1373 [hep-th]

\bibitem {Upadhyay3}Upadhyay, S., Rai, S.K., Mandal, B.P.: Field dependent
nilpotent symmetry for gauge theories. Eur. Phys. J. C72, 2065 (2012),
arXiv:1201.0084 [hep-th]

\bibitem {FP}Faddeev, L.D., Popov, V.N.: Feynman diagrams for the Yang--Mills
field. Phys. Lett. B25 (1967) 29

\bibitem {LL1}Lavrov, P., Lechtenfeld, O.: Field-dependent BRST
transformations in Yang--Mills theory. Phys. Lett. B725, 382 (2013),
arXiv:1305.0712 [hep-th]

\bibitem {BLThfbrst}Batalin, I.A., Lavrov, P.M., Tyutin, I.V.: A systematic
study of finite BRST--BFV Transformations in generalized Hamiltonian
formalism. Int. J. Mod. Phys. A29, 1450127 (2014), arXiv:1404.4154 [hep-th]

\bibitem {Reshetnyak}Reshetnyak, A.: On gauge independence for gauge models
with soft breaking of BRST symmetry. Int. J. Mod. Phys. A29, 1450184 (2014),
arXiv:1312.2092 [hep-th]

\bibitem {BLTfin}Batalin, I.A., Lavrov, P.M., Tyutin, I.V.: A systematic study
of finite BRST--BV transformations in field-antifield formalism. Int. J. Mod.
Phys. A29, 1450166 (2014), arXiv:1405.2621 [hep-th]

\bibitem {MRnew4}Moshin, P.Yu., Reshetnyak, A.A.: Finite BRST-antiBRST
transformations for the theories with gauge group. TSPU Bulletin 12, 192
(2014), arXiv:1412.0226 [hep-th]

\bibitem {MRnew5}Moshin, P.Yu., Reshetnyak, A.A.: On consistent Lagrangian
quantization of Yang--Mills theories without Gribov copies. Proc. of 18th Int.
Seminar Quarks'2014, 2--8 June 2014, Suzdal, Russia, arXiv:1412.8428 [hep-th]

\bibitem {llr1}Lavrov, P., Lechtenfeld, O., Reshetnyak, A.: Is soft breaking
of BRST symmetry consistent? JHEP 1110, 043 (2011), arXiv:1108.4820 [hep-th]

\bibitem {lrr}Lavrov, P., Radchenko, O., Reshetnyak, A.: Soft breaking of BRST
symmetry and gauge dependence. Mod. Phys. Lett. A27, 1250067 (2012),
arXiv:1201.4720 [hep-th]

\bibitem {rl}Radchenko, O., Reshetnyak, A.: Notes on soft breaking of BRST
symmetry in the Batalin-Vilkovisky formalism. Russ. Phys. J. 55 (2013) 1005,
arXiv:1210.6140 [hep-th]

\bibitem {Wett-Reu-1}Reuter, M., Wetterich, C.: Average action for the Higgs
model with abelian gauge symmetry. Nucl. Phys. B391, 147 (1993)

\bibitem {Wett-Reu-2}Reuter, M., Wetterich, C.: Effective average action for
gauge theories and exact evolution equations. Nucl. Phys. B417, 181 (1994)

\bibitem {Wett-1}Wetterich, C.: Average action and the renormalization group
equations. Nucl. Phys. B352, 529 (1991)

\bibitem {Polch}Polchinski, J.: Renormalization and effective Lagrangians.
Nucl. Phys. B231, 269 (1984)

\bibitem {LS}Lavrov, P., Shapiro, I.: On the Functional renormalization group
approach for Yang--Mills fields. JHEP, 1306, 086 (2013), arXiv:1212.2577 [hep-th]

\bibitem {Salmhofer}Metzner, W., Salmhofer, M., Honerkamp, C., Meden, V.,
Schoenhammer, K.: Functional renormalization group approach to correlated
fermion systems. Rev. Mod. Phys. 84, 299 (2012), arXiv:1105.5289 [cond-mat.str-el]

\bibitem {Gribov}Gribov, V.N.: Quantization of nonabelian gauge theories.
Nucl. Phys. B139, 1 (1978)

\bibitem {Zwanziger1}Zwanziger, D.: Action from the Gribov horizon. Nucl.
Phys. B321, 591 (1989)

\bibitem {Zwanziger2}Zwanziger, D.: Local and renormalizable action from the
Gribov horizon. Nucl. Phys. B323, 513 (1989)

\bibitem {LL2}Lavrov, P., Lechtenfeld, O.: Gribov horizon beyond the Landau
gauge. Phys. Lett. B725, 386 (2013), arXiv:1305.2931 [hep-th]

\bibitem {Reshetnyak2}Reshetnyak, A.: On composite fields approach to Gribov
copies elimination in Yang--Mills theories. arXiv:1402.3060 [hep-th]

\bibitem {FRTnsend}Freedman, D.Z., Townsend, P.K.: Antisymmetric tensor gauge
theories and non-linear\emph{ }$\sigma$-models. Nucl. Phys. B177, 282 (1981)

\bibitem {SM1}Weinberg, S.: Conceptual foundations of the unified theory of
weak and electromagnetic interactions. Rev. Mod. Phys. 52, 515 (1980); Science
210, 1212--1218 (1980)

\bibitem {SM2}Salam, A.: Gauge unification of fundamental forces. Rev. Mod.
Phys. 52, 525 (1980); Science 210, 723 (1980)

\bibitem {SM3}Glashow, S.L.: Towards a unified theory: threads in a tapestry.
Rev. Mod. Phys. 52, 539 (1980); Science 210, 1319 (1980).

\bibitem {book1}Weinberg, S.: The Quantum Theory of Fields. Vol. 1:
Foundations. Cambridge, UK: Univercity Press (1995)

\bibitem {book2}Weinberg, S.: The Quantum Theory of Fields. Vol. 2: Modern
applications. Cambridge, UK: Univercity Press (1996)

\bibitem {book3}Nagashima, Y.: Elementary Particle Physics: Foundations of the
Standard Model. Vol. 2. Wiley (2013)

\bibitem {book4}Schwartz, M.D.: Quantum Field Theory and the Standard Model.
Cambridge University Press (2013)

\bibitem {BE}Englert, F., Brout R.: Broken symmetry and the mass of gauge
vector mesons. Phys. Rev. Lett. 13, 321 (1964)

\bibitem {Higgs1}Higgs, P.W.: Broken symmetries, massless particles and gauge
fields. Phys. Lett. 12, 132 (1964)

\bibitem {Higgs2}Higgs, P.W.: Broken symmetries and the masses of gauge
bosons. Phys. Rev. Lett. 13, 508 (1964)

\bibitem {Gural'nik}Guralnik, G.S., Hagen, C.R., Kibble, T.W.B.: Global
conservation laws and massless particles. Phys. Rev. Lett. 13, 585 (1964)

\bibitem {CMS}CMS Collaboration: Observation of a new boson at a mass of 125
GeV with the CMS experiment at the LHC. Phys. Lett. B716, 30 (2012),
arXiv:1207.7235 [hep-ex]

\bibitem {Atlas}ATLAS Collaboration: Observation of a new particle in the
search for the standard model Higgs boson with the ATLAS detector at the LHC.
Phys. Lett. B716, 1 (2012), arXiv:1207.7214 [hep-ex]

\bibitem {CMSATlas}ATLAS, CMS Collaborations: Combined measurement of the
Higgs boson mass in pp collisions at $\sqrt{s}$=7 and 8 TeV with the ATLAS and
CMS experiments. Phys. Rev. Lett. 114, 191803 (2015), arXiv:1503.07589 [hep-ex]

\bibitem {LRquarks2012}Lavrov, P., Reshetnyak, A.: Gauge dependence of vacuum
expectation values of gauge invariant operators from soft breaking of BRST
symmetry. Example of Gribov--Zwanziger action. In:\ Khlebnikov, V., Matveev,
V., Rubakov, V. (eds.) Proc. of the 17th Int. Seminar QUARKS'2012, Yaroslavl,
Russia, June 4--10, 2012, p. 233. Moscow (2013), arXiv:1210.5651 [hep-th]

\bibitem {SlavnovFI}Slavnov, A.A.: Continual Integral in Perturbation Theory,
Theor. Math. Phys. 22, 177 (1975)

\bibitem {GitmanTyutin}Gitman D.M., Tyutin, I.V.: Quantization of fields with
constraints. Berlin, Germany: Springer (1990)

\bibitem {KalloshTyutin}Kallosh, K.E., Tyutin, I.V.: The equivalence theorem
and gauge invariance in renormalizable theories, Sov. J. Nucl. Phys. 17, 98 (1973)

\bibitem {BSemikhatov}Batalin, I.A., Marnelius, R., Semikhatov, A.M.:
Triplectic quantization: a geometrically covariant description of the Sp(2)
symmetric Lagrangian formalism, Nucl. Phys. B446, 249 (1995), arXiv:hep-th/9502031

\bibitem {Carringt}Carrington, M.E.: Effective potential at finite temperature
in the standard model. Phys. Rev. D45, 2933 (1992)

\bibitem {Kapusta}Kapusta, J.I., Gale, Ch.: Finite-temperature Field Theory,
Cambridge University Press, Cambridge (2006)

\bibitem {Okun}Okun, L.B.: Leptons and Quarks. North Holland, New York (1982)

\bibitem {IzyksonZuber}Itzykson, C., Zuber, J.-B.: Quantum Field Theory.
McGraw-Hill Inc., New York (1980)

\bibitem {CKM}Cabibbo, N.: Unitary symmetry and leptonic decays. Phys. Rev.
Lett. 10 (12), 531 (1963)\newline Kobayashi, M., Maskawa, T.: CP-violation in
the renormalizable theory of weak interaction. Prog. Theor. Phys. 49, 652 (1973)

\bibitem {Kazinski}Kazinski, P.O., Gravitational mass-shift effect in the
Standard Model. Phys. Rev. D85, 044008, (2012), arXiv:1107.4714 [gr-qc]

\bibitem {Singer}Singer, I.M.: Some remarks on the Gribov ambiguity, Comm.
Math. Phys. 60 (1978) 7

\bibitem {lattice}Bogolubsky, I.L., Ilgenfritz, E.M., Muller-Preussker, M.,
Sternbeck, A.: Lattice gluodynamics computation of Landau gauge Green's
functions in the deep infrared. Phys. Lett. B676 (2009) 69, arXiv:0901.0736 [hep-lat]

\bibitem {lattice1}Bornyakov, V., Mitrjushkin, V., Muller-Preussker, M.: SU(2)
lattice gluon propagator: continuum limit, finite-volume effects and infrared
mass scale m(IR). Phys. Rev. D81, 054503, (2010), arXiv:0912.4475 [hep-lat]

\bibitem {lattice2}Bornyakov, V.G., Mitrushkin, V.K., Rogalyov, R.N.: Gluon
propagators in 3D SU(2) theory and effects of Gribov copies. arXiv:1112.4975 [hep-lat]

\bibitem {Sorellas}Capri, M.A.L., G\'{o}mes, A.J., Guimaraes, M.S., Lemes,
V.E.R., Sorella, S.P., Tedesco, D.G.: A remark on the BRST symmetry in the
Gribov--Zwanziger theory, Phys. Rev. D82, 105019 (2010), arXiv:1009.4135 [hep-th]

\bibitem {Sorellas1}Baulieu, L., Capri, M.A.L., G\'{o}mes, A.J., Guimaraes,
M.S., Lemes, V.E.R., Sobreiro, R.F., Sorella, S.P.: Renormalizability of a
quark-gluon model with soft BRST breaking in the infrared region. Eur. Phys.
J. C66, 451 (2010), arXiv:0901.3158 [hep-th]

\bibitem {Sorellas2}Dudal, D., Sorella, S.P., Vandersickel, N., Verschelde,
H.: Gribov no-pole condition, Zwanziger horizon function, Kugo--Ojima
confinement criterion, boundary conditions, BRST breaking and all that. Phys.
Rev. D79, 121701 (2009), arXiv:0904.0641 [hep-th]

\bibitem {Sorellas3}Baulieu L., Sorella, S.P.: Soft breaking of BRST
invariance for introducing non-perturbative infrared effects in a local and
renormalizable way. Phys. Lett. B671, 481 (2009), arXiv:0808.1356 [hep-th]

\bibitem {Sorellas4}Capri, M.A.L., G\'{o}mes, A.J., Guimaraes, M.S., Lemes,
V.E.R., Sorella, S.P., Tedesco, D.G.: Renormalizability of the linearly broken
formulation of the BRST symmetry in presence of the Gribov horizon in Landau
gauge Euclidean Yang--Mills theories. arXiv:1102.5695 [hep-th]

\bibitem {Sorellas5}Dudal, D., Sorella, S.P., Vandersickel, N.: The dynamical
origin of the refinement of the Gribov--Zwanziger theory, arXiv:1105.3371 [hep-th].

\bibitem {SS}Sobreiro, R.F., Sorella, S.P.: A study of the Gribov copies in
linear covariant gauges in Euclidean Yang--Mills theories. JHEP 0506, 054
(2005) arXiv:hep-th/0506165

\bibitem {MAG}Dudal, D., Capri, M.A.L., Gracey, J.A., et al.: Gribov
ambiguities in the maximal Abelian gauge. Braz. J. Phys. 37, 320 (2007), arXiv:hep-th/0609160

\bibitem {MAG1}Gongyo, Sh., Iida, H.: Gribov--Zwanziger action in $SU(2)$
maximally Abelian gauge with $U(1)_{3}$ Landau gauge. Phys. Rev. D89, 025022
(2014), arXiv:1310.4877 [hep-th]

\bibitem {HFZwanziger}Zwanziger, D.: Equation of state of gluon plasma from
local action. Phys. Rev. D76, 125014 (2007), [arXiv:hep-ph/0610021]

\bibitem {Slavnoveq}Slavnov, A.A.: The study of ambiguity in non-Abelian gauge
theories. Theor. Math. Phys. 170, 198 (2012)

\bibitem {Slavnoveq1}Quadri, A., Slavnov, A.A.: Renormalization of the
Yang--Mills theory in the ambiguity-free gauge. JHEP 07, 087 (2010),
arXiv:1002.2490 [hep-th]

\bibitem {Slavnoveq2}Slavnov, A.A.: New approach to the quantization of the
Yang--Mills field. Theor. Math. Phys. 183:2, 585 (2015), arXiv:1503.03380 [hep-th]

\bibitem {Serreau}Serreau, J., Tissier, M., Tresmontant, A.: Covariant gauges
without Gribov ambiguities in Yang--Mills theories. arXiv:1307.6019 [hep-th]

\bibitem {Serreau1}Serreau, J., Tissier, M., Tresmontant, A.: On the influence
of Gribov ambiguities in a class of nonlinear covariant gauges.
arXiv:1505.07270 [hep-th]

\bibitem {caprirxi}Capri, M.A.L., Pereira, A.D., Sobreiro, R.F., Sorella,
S.P.: Non-perturbative treatment of the linear covariant gauges by taking into
account the Gribov copies. arXiv:1505.05467 [hep-th]

\bibitem {bt1}Batalin, I.A., Tyutin, I.V.: On possible generalizations of
field-antifield formalism. Int. J. Mod. Phys. A8, 2333 (1993), arXiv:hep-th/9211096

\bibitem {bt2}Batalin, I.A., Tyutin, I.V.: On the multilevel generalization of
the field-antifield formalism, Mod. Phys. Lett. A8, 3673 (1993), arXiv:hep-th/9309011

\bibitem {Slavnov:1971aw}Slavnov, A.A.: Invariant regularization of nonlinear
chiral theories. Nucl.\ Phys.\ B31, 301 (1971)

\bibitem {Slavnov:1972sq}Slavnov, A.A.: Invariant regularization of gauge
theories. Theor. Math. Phys. 13, 1064 (1972)

\bibitem {Martin:1994cg}Martin, C.P., Ruiz, Ruiz F.: Higher covariant
derivative Pauli--Villars regularization does not lead to a consistent QCD.
Nucl. Phys. B436, 545 (1995)

\bibitem {Asorey:1995tq}Asorey, M., Falceto, F.: On the consistency of the
regularization of gauge theories by high covariant derivatives. Phys. Rev.
D54, 5290 (1996), arxiv:hep-th/9502025

\bibitem {Bakeyev:1996is}Bakeyev, T.D.,\ Slavnov, A.A.: Higher covariant
derivative regularization revisited. Mod. Phys. Lett. A11, 1539 (1996), arXiv:hep-th/9601092

\bibitem {hderregN1}Krivoshchekov, V.K.: Invariant regularization for
supersymmetric gauge theories. Theor. Math. Phys. 36, 745 (1978) [Teor. Mat.
Fiz. 36, 291 (1978)]

\bibitem {Westhdern1}West, P.C.: Higher derivative regulation of
supersymmetric theories. Nucl. Phys. B268, 113 (1986)

\bibitem {buchstepanyantz}Buchbinder, I.L., Pletnev, N.G., Stepanyantz, K.V.:
Manifestly N = 2 supersymmetric regularization for N = 2 supersymmetric field
theories. arXiv:1509.08055 [hep-th]

\bibitem {anselmi}Anselmi, D.: Ward identities and gauge independence in
general chiral gauge theories, Phys. Rev. D92, 025027 (2015), arXiv:1501.06692[hep-th]

\bibitem {Krauss}Krauss, E.: Renormalization of the electroweak Standard Model
to all orders. Annals Phys. 262, 155 (1998), arXiv:hep-th/9709154

\bibitem {PiquetSorella}Piguet, O., Sorella, S.: Algebraic Renormalization,
Lecture Notes in Physics 28, Springer Verlag (1995)
\end{thebibliography}
\end{document}